\documentclass{aa}
\usepackage[varg]{txfonts}
\usepackage{graphicx}
\usepackage{array, booktabs, makecell}
\usepackage{siunitx}
\usepackage[version=4]{mhchem}
\usepackage{textcomp, gensymb}
\usepackage{amsmath}
\usepackage{amsfonts}
\usepackage[normalem]{ulem}
\usepackage{comment}
\usepackage{multicol}
\usepackage[utf8]{inputenc}
\usepackage[english]{babel}
\usepackage{natbib}
\bibpunct{(}{)}{;}{a}{}{,}
\usepackage[hidelinks,colorlinks=true,linkcolor=blue,citecolor=blue]{hyperref}
\newcommand{\DS}{\displaystyle}

\begin{document}

\title{Magnetic and thermal acceleration in extragalactic jets}
\subtitle{An application to NGC~315}

\author{L.\ Ricci \inst{1,2}, 
        M.\ Perucho \inst{3,4},
        J.\ López-Miralles \inst{3,5},
        J.\ M.\ Martí \inst{3,4},
        B.\ Boccardi \inst{1}
     } 

\institute{
\inst{1} Max-Planck-Institut fu{\"u}r Radioastronomie, Auf dem H{\"u}gel 69, D-53121 Bonn, Germany \\
\inst{2} Julius-Maximilians-Universität Würzburg, Fakult{\"a}t für Physik und 
Astronomie, Institut f{\"u}r Theoretische Physik und Astrophysik, 
Lehrstuhl f{\"u}r Astronomie, Emil-Fischer-Str. 31, D-97074 Würzburg, 
Germany \\
\inst{3} Departament d’Astronomia i Astrofísica, Universitat de València, C/ Dr. Moliner, 50, 46100, Burjassot, València, Spain \\
\inst{4} Observatori Astronòmic, Universitat de València, C/ Catedràtic José Beltrán 2, 46980, Paterna, València, Spain \\
\inst{5} Aurora Technology for the European Space Agency, ESAC/ESA, Camino Bajo del Castillo s/n, Urb. Villafranca del Castillo, 28691 Villanueva de la Cañada, Madrid, Spain\\
}

\date{Received 11 May 2023 / Accepted 08 December 2023}

  \abstract
   {}
{Relativistic jets launched from active galactic nuclei accelerate up to highly relativistic velocities within a few parsecs to tens of parsecs.
The precise way in which this process takes place is still under study. While magnetic acceleration is known to be able to accelerate relativistic outflows, little attention has been paid to the role of thermal acceleration. The latter has been assumed to act only on compact regions, very close to the central engine, and to become negligible on parsec scales. However, this holds under the assumption of small internal energies as compared to the magnetic ones, and whether this is true or what happens when we drop this assumption is currently uncertain.}
{We use a 2D relativistic magnetohydrodynamical code to explore jet acceleration from sub-parsec to parsec scales.
As initial conditions for our models, we use observational constraints on jet properties derived by means of very long baseline interferometry observations for a Fanaroff Riley I radio galaxy, NGC\,315. 
We investigate the parameter space established for this source and perform a number of simulations of magnetically, thermally or kinetically dominated jets at injection, and compare our results with the observed ones.
Furthermore, we employ different models to characterize our jets, involving different magnetic field configurations, i.e., force-free vs. non-force-free configurations, as well as varying shear layer thicknesses.}
{Our simulated jets show that when thermal energy is comparable to or exceeds magnetic energy, thermal acceleration becomes significant at parsec scales.
This result has important consequences, potentially extending the acceleration region far beyond the collimation scales, as thermal acceleration can effectively operate within a conically expanding jet.
In all the models, we observe acceleration to be driven by expansion, as expected. A number of our models allow us to reproduce the acceleration and opening angles observed in NGC\,315.
Finally, our results indicate that disk-launched winds might play an important role in the jet propagation.
Namely, when the jet has an initial force-free magnetic field configuration, thicker shear layers are needed to shield the internal spine from the action of the external medium, delaying the growth of instabilities.
}
   {}
   
\titlerunning{Magnetic and thermal acceleration in extragalactic jets}
\authorrunning{L. Ricci et al.}
\maketitle

\section{Introduction} \label{sec:introduction}

Active galactic nuclei (AGN) power relativistic outflows that propagate up to hundreds and even thousands of kiloparsecs \citep{Begelman_1984, Blandford_2019}.
A fundamental ingredient for their capacity to extend through nine orders of magnitude in distance is their inertia \citep[e.g.,][and references therein]{Perucho_2019}, achieved by means of the investment of energy into the acceleration of the flow to relativistic velocities. The mechanisms behind the establishment of such outflows have been largely studied in the past years, and substantial progress has been made \citep[e.g.,][]{Vlahakis_2004,Komissarov_2007}. However, such studies have focused on the assumption that acceleration is, in essence, magnetically driven, ignoring thermal acceleration.

The crucial region in which this process can be explored in the jet is the so-called acceleration and collimation region. There, jets collimate evolving from a parabolic geometry (i.e., 
$r \propto \mathrm{z}^{\sim 0.5}$, where $\mathrm{z}$ is the radial distance from the core and $r$ is the jet radius), to a conical one $(r \propto \mathrm{z}^{\sim 1})$, and accelerate up to high relativistic velocities along the same distance \citep[see e.g.,][and references therein]{Boccardi_2017}.
Therefore, in cold jets, the nature of the collimation and acceleration phenomena seems to be strictly correlated. 
The current common view of this process suggests that equipartition between magnetic field and particle energy is reached \citep{Nokhrina_2019, Nokhrina_2020} by the end of the acceleration region, and that external conditions, such as the crossing of the Bondi radius \citep{Kovalev_2020}, may play a relevant role in shaping the jet expansion profile. It is often considered that equipartition refers to an equilibrium between magnetic and kinetic energy densities in the jet but, from synchrotron radiative output, only the equipartition between magnetic and internal energies can be inferred. It is therefore plausible that even though magnetic and internal energy are in or close to equipartition at the end of the acceleration region, it is the kinetic energy that dominates jet dynamics from that point on.

Jets in Fanaroff-Riley I \citep[FR~I;][]{Faranoff_Riley} radio galaxies, which are used as example of this study, are expected to reach maximum Lorentz factors of $\gamma_\mathrm{max} \sim 10$, as seen from statistical studies of BL Lac objects, their blazar counterpart \citep{Hovatta_2009}.
As mentioned above, acceleration can be related to two main possible sources of energy \citep[see, e.g.,][]{Komissarov_2012}: i) thermal energy via the Bernoulli mechanism, or ii) magnetic energy via differential collimation mechanism.
The former allows the jet to reach a maximum Lorentz factor 
$\gamma_\mathrm{max} =\gamma[1 + 4 \varepsilon/(3 c^2)]$, where $\varepsilon$ is the jet specific internal energy at injection and $\gamma$ is the initial Lorentz factor, whereas for the latter $\gamma_\mathrm{max} =\gamma[1 + B^2/(4\pi \rho c^2)]$ where $B$ and $\rho$ are, respectively, the jet magnetic field and rest-mass density.

The current paradigm of magnetic jet acceleration relies on the theoretical work of \citet{Vlahakis_2003_a, Vlahakis_2003_b,Vlahakis_2004}. 
The authors suggest that thermal acceleration would be relevant only at very compact scales, and magnetic acceleration would take over and prolong the accelerating process up to parsec scales. Magnetic acceleration was further studied by \citet{Komissarov_2007}, who showed how the conversion of magnetic energy is a viable way to accelerate the bulk flow by means of numerical simulations. 
A relevant aspect revealed by these works is that magnetic acceleration does not occur when the outflow is following a conical expansion, but it becomes effective when the jet expands following a parabolic law. 
In this process, the toroidal field acts not only as a collimating agent, but also as the main driver of jet bulk acceleration. The basic theory behind this process is summarized in \citet{Komissarov_2012}. 

 However, this view is founded on a more fundamental assumption that the jet internal energy is rather small as compared to the total jet energy flux. \citet{Vlahakis_2004} used classical expressions to derive this conclusion, which implicitly means that the rest-mass energy of the particles is significantly larger than their internal energy. It is nevertheless unclear that jet thermodynamics can be dealt without considering their relativistic nature \citep[see, e.g.,][]{Perucho_2017}. On the one hand, jets can be thermodynamically relativistic (i.e., $\varepsilon \geq c^2$).
 On the other hand, strong jet expansion is prevented by either the ambient pressure or the magnetic tension provided by a toroidal field, which can keep the jet plasma at relativistic internal energies, avoiding its cooling to $\varepsilon \ll c^2$.
 This is, at least, observed in numerical simulations \citep[e.g.,][]{Marti_2016,Moya-T_2021,Angles_2021}. 
As a consequence, internal energy could still be relevant for the overall process of jet acceleration up to parsec scales.
Thermal acceleration (together with magnetic acceleration) in jets is suggested in \citet{Lopez2022} for the case of outflows from high-mass X-ray binaries.
In this paper, we relax the assumption of cold jets and explore the different roles that both accelerating mechanisms can play.


As initial template for the physical conditions, we use the 
radio galaxy NGC~315 as prototype.
NGC~315 is a nearby \citep[$z = 0.0165$,][]{Trager_2000}, giant FR~I \citep[e.g.,][]{Laing_2006} radio galaxy whose acceleration and collimation properties have been extensively studied by employing very long baseline interferometry (VLBI) observations. Such observations showed the jet in NGC\,315 to accelerate up to $v \sim (0.8-0.9) \, c$, and to collimate on similar scales, with the highest Lorentz factor $\gamma \sim 4$ reached at $\sim 2 \, \mathrm{pc}$ \citep{Boccardi_2021, Park_2021, Ricci_2022}.
Other physical properties such as the magnetic field strength, the jet power and the jet width have been estimated \citep{Boccardi_2021, Ricci_2022} and are also used as initial conditions in our models.
The complete characterization of the jet at injection would require initial values for density and pressure. However, they are unknown and we use them as free parameters that allow us to vary the jet configuration and dominant energy channel at the boundary condition of our numerical simulations at the initial state.

The goal of this paper is twofold: while exploring the acceleration in different types of jets, i.e., hot and cold, and the interplay between the separate energy channels along the propagation, we aim at reproducing the final part of the acceleration in NGC\,315 and at testing whether the inferred physical properties on sub-parsec scales allow for the jet to remain stable on parsec scales and accelerate up to $\gamma \sim 4$.
To achieve the goal of our work, we investigate a large number of setups using the 2D relativistic magnetohydrodinamical (RHMD) code presented in \cite{Marti_2015,Marti_2016}. 

The paper is structured as follows. In Sect.\, \ref{sec:numerical_simulations} we describe the numerical setup; in Sect.\, \ref{sec:results} we present our results; in Sect.\, \ref{sec:discussion} we discuss them and in Sect.\, \ref{sec:Conclusions} we draw our conclusions.

\section{Numerical simulations} \label{sec:numerical_simulations}


\subsection{Numerical methods}\label{sec:numerical_methods}

The simulations presented in this paper are performed using the conservative, 2D axisymmetric RMHD code presented in \cite{Marti_2015}. 
While for an in-depth study on the jet propagation, a 3D approach could allow us to study the phenomenon with fewer assumptions \citep[allowing for example to explore the effects of the development of helical modes, see, e.g.,][]{Lopez2022}, we consider the 2D axisymmetric approach well suited for our goals.
Indeed, jet acceleration is connected with jet expansion, which is an axisymmetric process. 
Furthermore, a 2D grid allows to test a large number of models due to the lower computational resources required.

Regarding the effects of our assumption on jet stability, the instability modes that we observe are mostly associated with small-scale structures (see Sect.~\ref{sec:results}). 
In the long term, the oscillation of the jet outer radius could trigger the development of pinching modes, which are, in general, not disruptive for relativistic flows \citep[e.g.,][]{Perucho_2005}.
Moreover, the dynamics of the simulated jets at injection is completely dominated by expansion and, in such a situation, the growth of instability modes is negligible \citep{Hardee_1986}.

In our code, we employ a second-order Godunov-type scheme with the HLL Riemann solver \citep{harten83} and the piecewise linear method (PLM) for cell reconstruction, where the VanLeer slope limiter preserves monotonicity \citep{vanLeer1974, Mignone06}. 
The limiter is degraded to MinMod\citep{Roe86} when strong shocks are detected in order to avoid numerical oscillations. 
Time integration follows the third-order TVD-preserving Runge-Kutta scheme of \cite{shu89} with CFL = 0.2. 
The relativistic correction algorithm CA2 of \cite{marti152} is used to correct the conserved variables after each time iteration. 
The magnetic field divergence-free constraint is preserved with the constrained transport (CT) method 
\citep{balsara99}.

\subsection{Model assumptions and transversal equilibrium} \label{sec:model_assumptions}

We study the acceleration problem by means of the ideal RMHD equations, to which we impose axisymmetry. We describe the system using cylindrical coordinates ($r$,$\phi$,$\mathrm{z}$) where $r$, $\phi$ and $\mathrm{z}$ are the radial, azimuthal and axial coordinates. Axisymmetry allows us to drop the dependence on the azimuthal cylindrical coordinate in the RMHD equations. We assume both the jet and ambient plasma to be perfect gas with a constant adiabatic index of $\Gamma = 4/3$. Taking into account that the ambient medium is used as a passive element in the simulations and that the jet plasma is hot, we consider this approach sufficient for our purposes.




The code evolves the rest-mass, momentum and total energy densities, and the magnetic field in a conservative way. Besides this, and in order to minimize the perturbation of the flow, the jet is injected in conditions of transversal equilibrium. As a consequence, the radial components of the magnetic field and the flow velocity, $B^r(r)$, $v^r(r)$, at injection are set to zero, and the equilibrium solutions are characterized by six functions, namely the density and pressure, $\rho(r)$, $p(r)$, and the two remaining components of the flow velocity, $v^\phi(r)$, $v^\mathrm{z}(r)$, and of the magnetic field, $B^\phi(r)$, $B^\mathrm{z}(r)$.

In what follows, all the quantities will be expressed in code units, in which the initial jet radius, $r_j$, the ambient density at the jet base, $\rho_a$, and the speed of light, $c$, are equal to 1 (and a factor $1/\sqrt{4 \pi}$ is absorbed in the definition of the magnetic field).

The transversal equilibrium condition is given by a single ordinary differential equation \citep[see e.g.,][]{Marti_2015}:
\begin{equation}
    \frac{dp^*}{dr} = \frac{\rho h^* \gamma^2 (v^\phi)^2 - (b^\phi)^2}{r},
\label{eq:pressure_equilibrium}
\end{equation}
in which, $\gamma= 1 / \sqrt{1 - v^iv_i}$ ($i=r,\phi,\mathrm{z}$, and summation over repeated indices is assumed) is the Lorentz factor, $p^*$ and $h^*$ are the total (gas and magnetic) pressure and total specific enthalpy of the jet, namely $p^* = p + b^2/2$ and $h^* = 1 + \varepsilon + p/\rho + b^2/\rho$.
In the previous expressions, $b^2 = b^\mu b_\mu$ ($\mu=t,r,\phi,\mathrm{z}$), where $b^\mu$ is the magnetic field four-vector in the fluid rest frame.
Its components are defined as $b^0 = \gamma B^i v_i$ and $b^i = B^i / \gamma + b^0 v^i$ which leads to
\begin{equation}
    b^2 = \frac{B^2}{\gamma^2} + (B^i v_i)^2 \, .
\end{equation}

Moreover, we assume constant initial density and axial velocity across the jet, and a non-rotating jet model
\begin{equation}
    \rho(r) = 
    \begin{cases}
      \rho_j, \, \, \, \, \, 0 \leq r \leq 1 \\
      1,  \, \, \, \, \, \, \, \, \, \,  r > 1, \\
    \end{cases}
\end{equation}
\begin{equation}
    v^\mathrm{z}(r) = 
    \begin{cases}
      v^\mathrm{z}_j, \, \, \, \, \, 0 \leq r \leq 1 \\
      0, \, \, \, \, \, \, \, \, \, \, r > 1, \\
    \end{cases}
\end{equation}
and $v^\phi(r) = 0$. 
With these definitions and the corresponding profiles for the azimuthal and axial components of the magnetic field, $B^\phi(r)$, $B^\mathrm{z}(r)$, Eq.~\ref{eq:pressure_equilibrium} gives the equilibrium pressure profile of the jet at injection.

\subsection{Physical model} \label{sec:phy_model}

\begin{figure*}[h]
    \centering
\begin{multicols}{2}
    \includegraphics[width=0.85\linewidth]{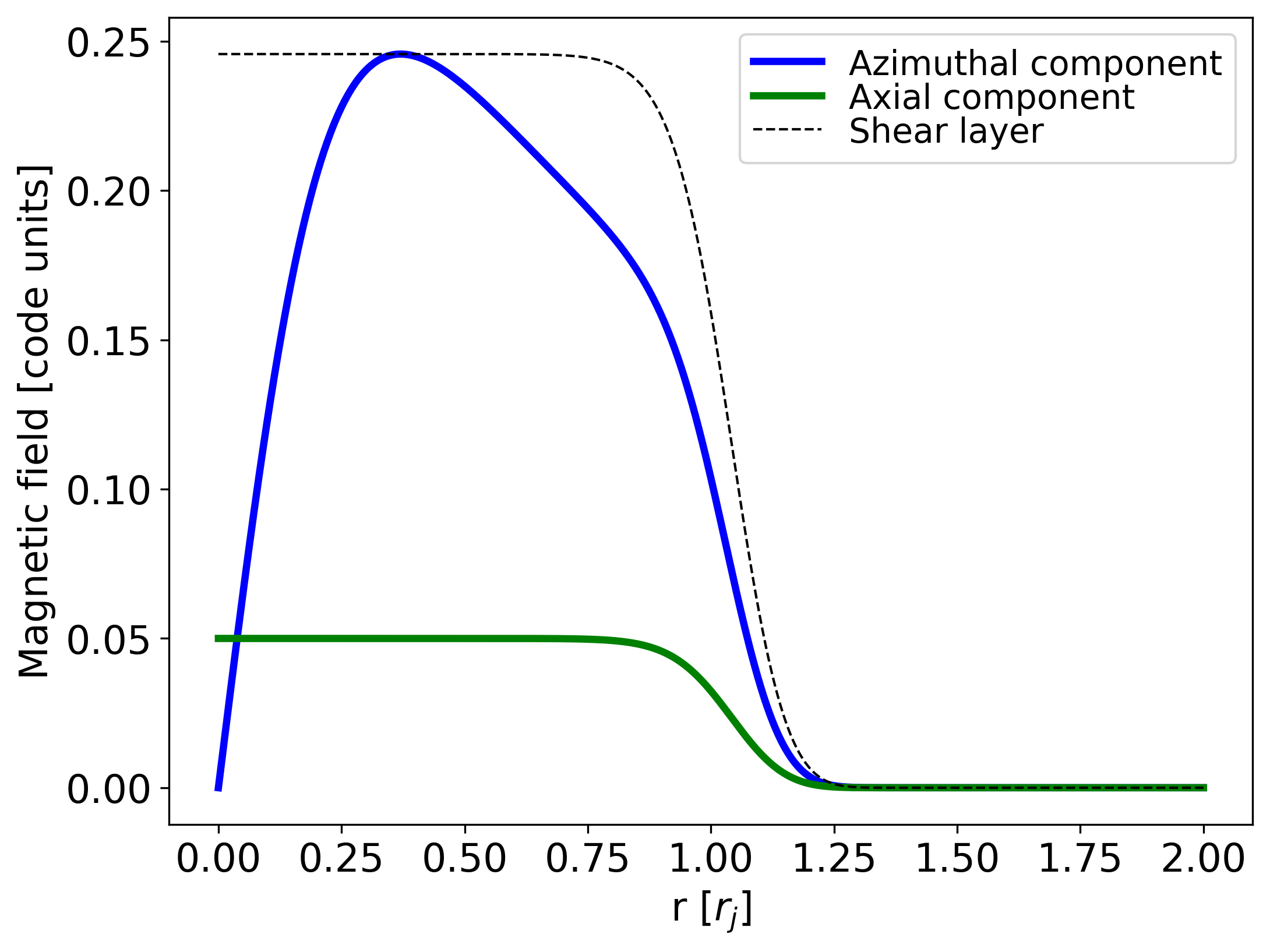}\par
    \includegraphics[width=0.85\linewidth]{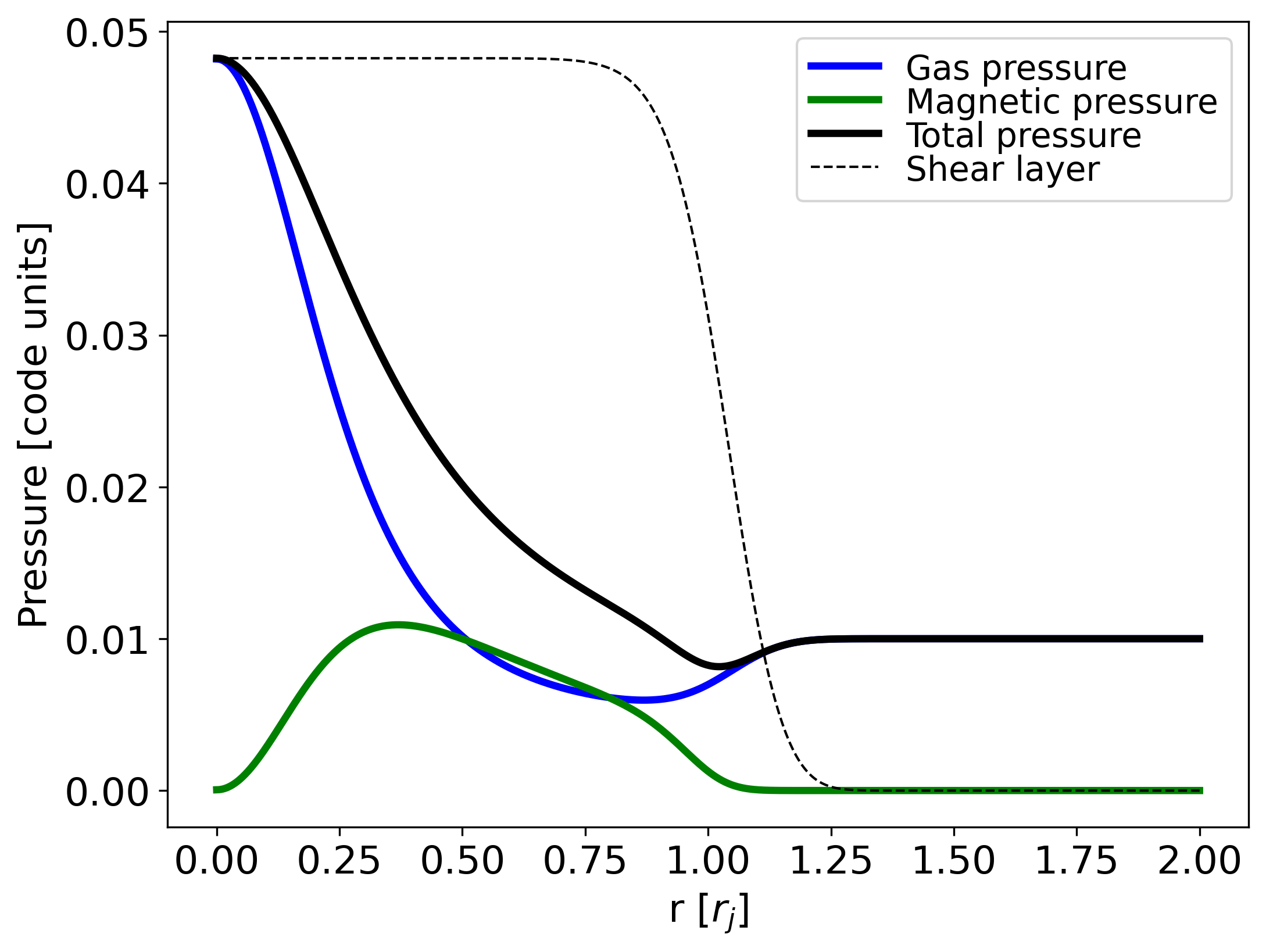}\par
\end{multicols}
\begin{multicols}{2}
    \includegraphics[width=0.85\linewidth]{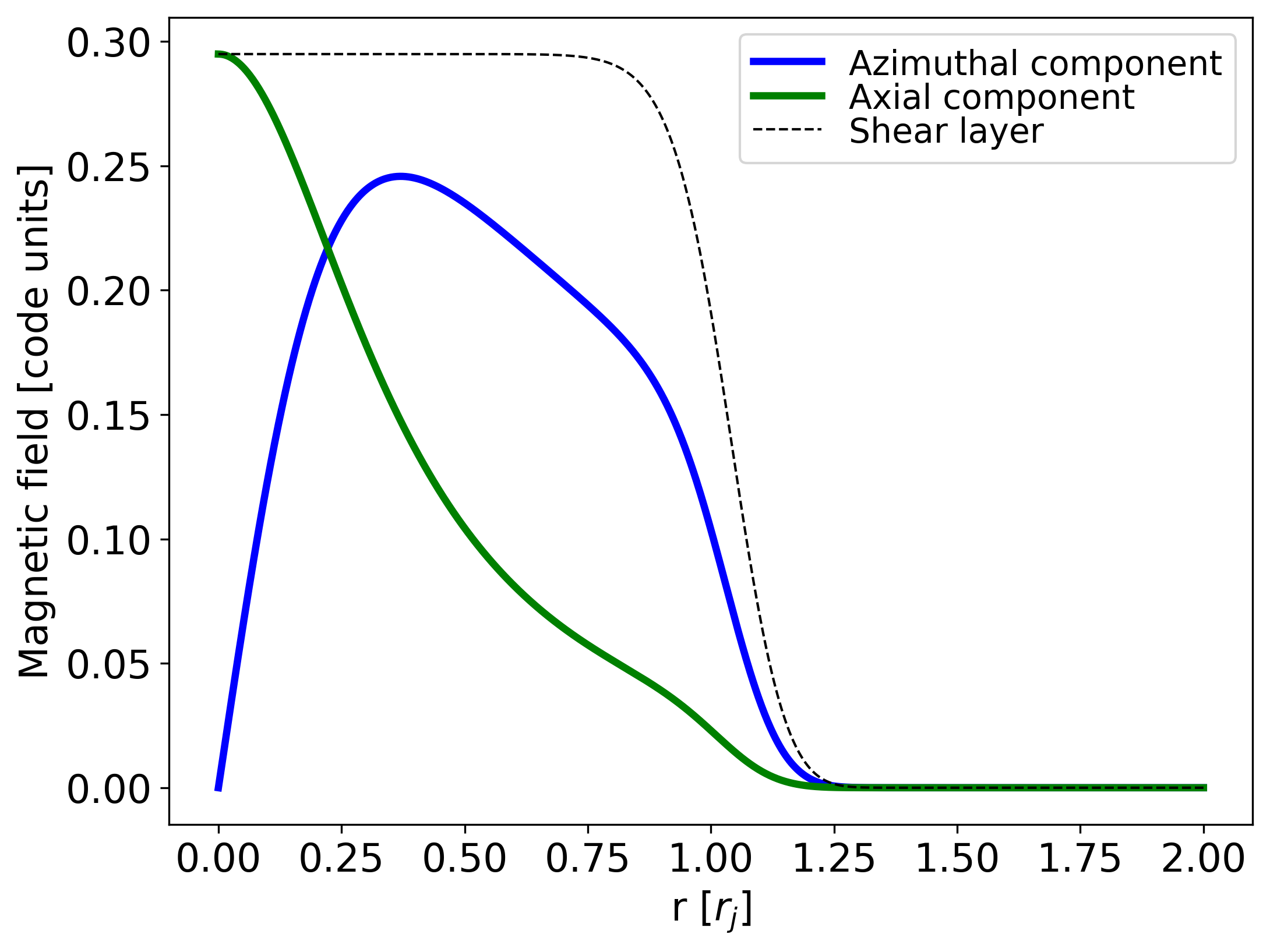}\par
    \includegraphics[width=0.85\linewidth]{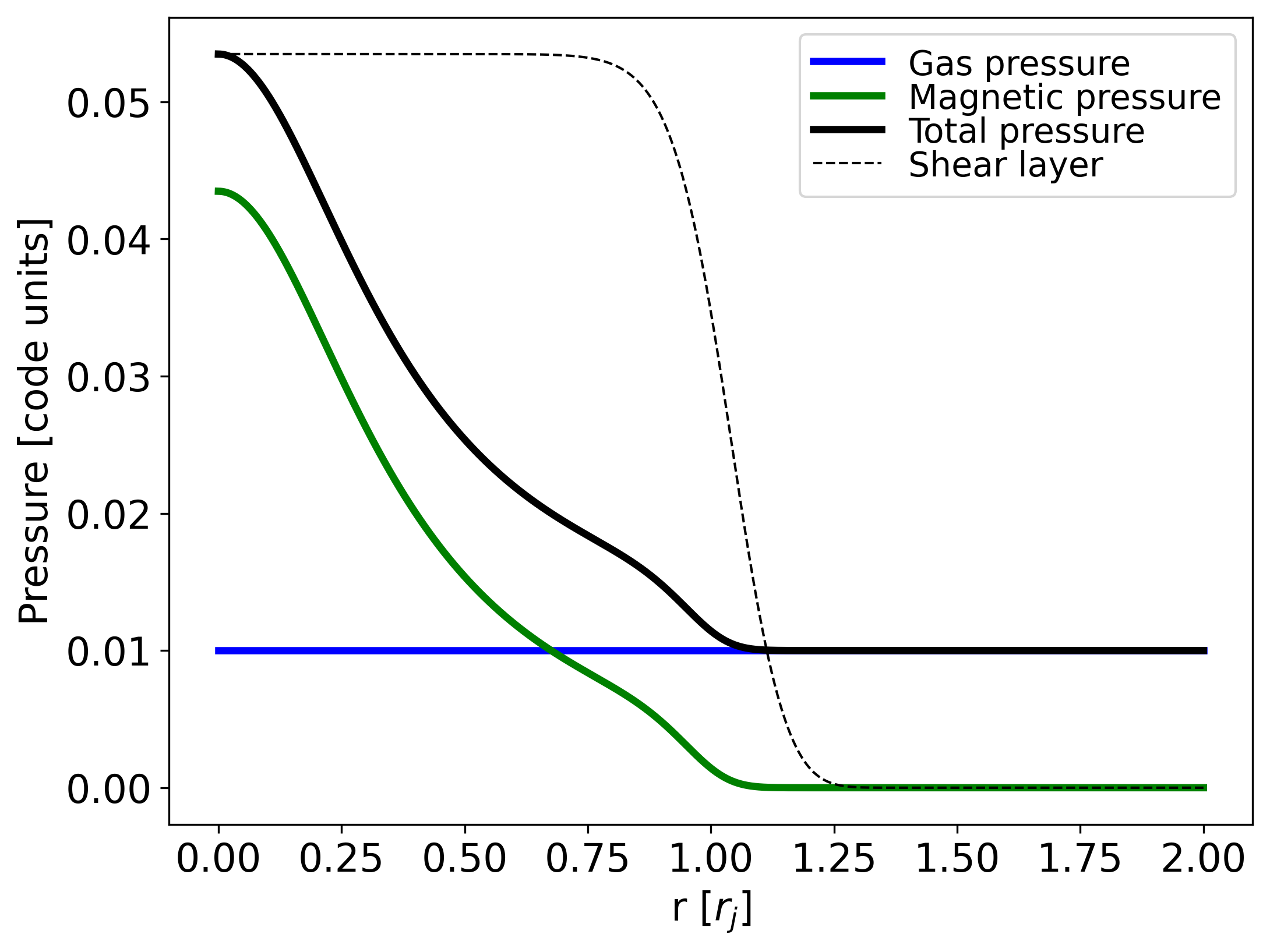}\par
\end{multicols}
    \caption{Initial profiles along the radial distance in the different magnetic field configurations. We use the representative values of $B^\phi = 0.24$, $R_m = 0.37$, $v_j = 0.8$, $p_a = 0.01$, and $B^\mathrm{z} = 0.05$ (for the non-force free configuration) in code units. Upper panels: non-force-free configuration. Lower panel: force-free configurations. Left panels: Magnetic field distribution of the azimuthal (blue line) and axial (green line) components. Right panels: profile distribution of the gas pressure (blue line), magnetic pressure (green line), and total pressure (black line). The dotted black line indicates the shear layer with $m = 8$ normalized at the higher value in the plot for an easier representation.}
    \label{fig:setup_profiles}
\end{figure*}

In code units, the total jet power is given by
\begin{equation}
    L_j = \pi \left[ \rho_j h_j \gamma^2_j + (B^\phi_j)^2 \right] v^\mathrm{z}_j \, .
    \label{eq:totaljet}
\end{equation}
The specific enthalpy is defined as:
\begin{equation}
    h_j = 1 + \varepsilon_j + p_j / \rho_j
\end{equation}
with pressure, density and specific internal energy related through an ideal gas equation of state, $\varepsilon_j = p_j / \big[ \rho_j (\Gamma - 1) \big]$, where $\Gamma$ is the corresponding adiabatic index.

The different energy flux channels that account for the total jet power (Eq.~\ref{eq:totaljet}) are computed as follows:
\begin{itemize}
    \item Magnetic: $F_\mathrm{M} = (B^\phi_j)^2 v^\mathrm{z}_j$ ;
    \item Internal: $F_\mathrm{I} = \rho_j \gamma^2_j v^\mathrm{z}_j ( h_j - 1)$ ;
    \item Kinetic: $F_\mathrm{K}  = \rho_j \gamma^2_j v^\mathrm{z}_j$ .
\end{itemize}
In addition, we compute the total rest mass energy as $F_\mathrm{R}  = \rho_j \gamma_j v^\mathrm{z}_j$.
As far as the magnetic field configuration is concerned, we use two different approaches: the non-force-free configuration, as defined in \citet{Marti_2015}, and the force-free configuration as defined in \citet{Moya-T_2021}.
The two models are described in Sect.\ \ref{sec:non_force_free_conf} and in Sect.\ \ref{sec:force_free_conf}, respectively.

We note that the transition of the initial magnetic field and pressure profiles at the jet/ambient medium interface is smoothed out by convolving the original profiles with a continuous function mimicking a shear layer.
In this way, we avoid discontinuities and abrupt changes in our initial conditions, making them more stable.
As shear layer, we consider the function $f(r) = 1 / \mathrm{cosh}(r^m)$ \citep{Bodo_1994,Perucho_2004}, in which $m = 2,4, \mathrm{and} \, 8$ with $m = 8$ being the value used the most in our models.

Since our simulations are axisymmetric, we assume reflecting boundary conditions (b.c.) on the axis
($r=0$), while we assume zero gradient b.c. at large values of the radial coordinate.
Along the z direction, we use inflow/outflow b.c. at the inner boundary, for the jet and the ambient medium, respectively, and outflow b.c. through the outer one.

In this study, we do not take into account the effects associated with resistivity and, consequently, magnetic reconnection.
Although the length scales at which reconnection occurs are expected to be significantly smaller than those addressed in this work, its impact might still be relevant.
In particular, the conversion of magnetic energy into thermal energy may elevate the pressure within the jet, increasing its internal energy budget. 
This effect could potentially manifest on larger scales when an important fraction of magnetic energy across the jet is involved, potentially altering the physical characteristics of the bulk flow and enhancing the role of thermal acceleration, due to the increased internal energy.
Although investigating the role of resistivity and magnetic reconnection would be of great interest (see e.g. \citealt{Medina2021,Mattia2023}), it is beyond the scope of this work.

\begin{table*}[]
\caption{Initial conditions of the simulated models.}
\centering
\begin{tabular}{c|c|c|c|c|c|c|c}
\hline
\begin{tabular}[c]{@{}c@{}}Model\\ code\end{tabular} & 
\begin{tabular}[c]{@{}c@{}}$p_j$\\ {[}$p_a${]}\end{tabular} & 
\begin{tabular}[c]{@{}c@{}}$\rho_j$\\ {[}$\rho_a${]}\end{tabular} & 
\begin{tabular}[c]{@{}c@{}}$B$\\ {[}G{]}\end{tabular} & 
\begin{tabular}[c]{@{}c@{}}$\phi_\mathrm{B}$\\ {[}deg{]}\end{tabular} &
\begin{tabular}[c]{@{}c@{}}$v_z$\\ {[}c{]}\end{tabular} &
\begin{tabular}[c]{@{}c@{}}$\mathcal{M}_\mathrm{ms}$\\ {}\end{tabular} &
\begin{tabular}[c]{@{}c@{}}$F_j$\\ {[}$\mathrm{10^{43} erg/s}${]} \end{tabular} \\
\hline
FMH1 & 1.0 & 0.01 & 0.42 & 70.7 & 0.87 & 1.12 & 4.8 \\ 
FMH1\_m4 & 1.0 & 0.01 & 0.42 & 70.7 & 0.87 & 1.12 & 4.8 \\ 
FIH2 & 2.0 & 0.01 & 0.46 & 68.9 & 0.84 & 1.05 & 5.6  \\ 
FIH3 & 3.0 & 0.01 & 0.49 & 67.3 & 0.81 & 1.01 & 6.1  \\ 
FIH4 & 4.0 & 0.01 & 0.52 & 66.4 & 0.79 & 1.00 & 6.7  \\ 
FMH1b & 1.0 & 0.01 & 0.59 & 74.4 & 0.92 & 1.00 & 12.7  \\ 
FMH1.3 & 1.3 & 0.05 & 0.55 & 67.8 & 0.82 & 1.00 & 6.3 \\ 
FKS1 & 1.0 & 0.1 & 0.39 & 61.3 & 0.64 & 1.00 & 2.54  \\ 
FML1 & 1.0 & 0.001 & 0.56 & 75.3 & 0.93 & 1.01 & 12.5  \\ 
FMH1c & 1.0 & 0.01 & 0.68 & 75.3 & 0.93 & 1.00 & 17.8 \\ 
FIH3b & 3.0 & 0.01 & 0.13 & 60.6 & 0.61 & 1.04 & 1.6  \\ 
FMH0.3 & 0.3 & 0.05 & 0.66 & 72.7 & 0.90 & 1.03 & 12.4  \\ 

\hline
NIH4 & 4.0 & 0.01 & 0.29 & 89.9 & 0.70 & 1.06 & 3.6  \\ 
NIH1 & 1.0 & 0.01 & 0.22 & 88.2 & 0.75 & 1.09 & 1.6  \\ 

\hline 
\end{tabular}
\label{tab:setups}

\begin{flushleft} 
\textbf{Notes.} Column 1: model names; Column 2: initial jet thermal overpressure factor; Column 3: initial jet density with respect to the ambient one; Column 4: initial magnetic field strength in G; Column 5: initial magnetic pitch angle; Column 6: initial axial velocity; Column 7: average initial Magnetosonic number; Column 8: total jet flux in units of $\mathrm{10^{43} erg/s}$.

The model names obey the following logic. i) The first letter is for the magnetic field configuration: F for force-free and N for non-force-free. ii) The second letter is for the dominant energy channel: M for magnetic, I for internal, and K for kinetic. iii) The third letter is for the jet density: S for $\rho_j \geq 0.1 \, \rho_a$, H for $0.01 \, \rho_a \leq \rho_j < 0.1 \, \rho_a$, and L for $\rho_j < 0.01 \, \rho_a$. iv) The last number is the same as the jet thermal overpressure factor at injection.
In the case of two models with the same nomenclature, we add a letter in progressively alphabetical order at the end of the name.
Models with \_m4 have the shear layer exponent set to $m = 4$ while for all the other models the exponent is $m = 8$.
\end{flushleft}
\end{table*}

\begin{table*}[]
\caption{Modeled, simulated, and final jet flux ratios for the different energy channels. The differences between them are due to the convolution of the initial profiles with the shear layer. The final values are the different ratios in the row of the last cells.}
\centering
\begin{tabular}{c|c|c|c|c||c|c|c|c||c|c|c|c}

\hline
\begin{tabular}[c]{@{}c@{}}Model\\ code\end{tabular} & 
\begin{tabular}[c]{@{}c@{}}$F_K^{i}/F_{j}$\\ {[}\%{]}\end{tabular} &
\begin{tabular}[c]{@{}c@{}}$F_R^{i}/F_K^{i}$\\ {[}\%{]}\end{tabular} &
\begin{tabular}[c]{@{}c@{}}$F_M^{i}/F_{j}$\\ {[}\%{]}\end{tabular} &
\begin{tabular}[c]{@{}c@{}}$F_I^{i}/F_{j}$\\ {[}\%{]}\end{tabular} &
\begin{tabular}[c]{@{}c@{}}$F_K^{s}/F_{j}$\\ {[}\%{]}\end{tabular} &
\begin{tabular}[c]{@{}c@{}}$F_R^{s}/F_K^{s}$\\ {[}\%{]}\end{tabular} &
\begin{tabular}[c]{@{}c@{}}$F_M^{s}/F_{j}$\\ {[}\%{]}\end{tabular} &
\begin{tabular}[c]{@{}c@{}}$F_I^{s}/F_{j}$\\ {[}\%{]}\end{tabular} &
\begin{tabular}[c]{@{}c@{}}$F_K^{f}/F_{j}$\\ {[}\%{]}\end{tabular} &
\begin{tabular}[c]{@{}c@{}}$F_R^{f}/F_K^{f}$\\ {[}\%{]}\end{tabular} &
\begin{tabular}[c]{@{}c@{}}$F_M^{f}/F_{j}$\\ {[}\%{]}\end{tabular} &
\begin{tabular}[c]{@{}c@{}}$F_I^{f}/F_{j}$\\ {[}\%{]}\end{tabular} \\
\hline

FMH1 & 9.03 & 49.31 & 54.86 & 36.11 & 24.33 & 65.47 & 47.92 & 27.75 & 35.19 & 44.95 & 36.92 & 27.89 \\ 
FMH1\_m4 & 9.03 & 49.31 & 54.86 & 36.11 & 33.10 & 67.27 & 43.40 & 23.50 & 47.53 & 52.56 & 34.56 & 17.91 \\ 
FIH2 & 6.11 & 54.26 & 44.99 & 48.90 & 18.97 & 68.99 & 41.01 & 40.02 & 34.98 & 38.41 & 36.71 & 28.31 \\ 
FIH3 & 4.63 & 58.64 & 39.81 & 55.56 & 15.83 & 72.04 & 36.94 & 47.23 & 31.66 & 36.48 & 35.75 & 32.59 \\ 
FIH4 & 3.79 & 61.31 & 35.60 & 60.61 & 13.72 & 73.81 & 33.55 & 52.73 & 30.84 & 33.17 & 29.88 & 39.28 \\ 
FMH1b & 5.73 & 39.19 & 71.33 & 22.94 & 13.53 & 56.65 & 68.31 & 18.16 & 25.92 & 30.17 & 60.96 & 13.12 \\ 
FMH1.3 & 23.99 & 57.24 & 51.06 & 24.95 & 31.50 & 65.62 & 47.49 & 21.01 & 51.58 & 42.62 & 35.29 & 13.13 \\ 
FKS1 & 51.78 & 76.84 & 27.51 & 20.71 & 57.98 & 81.25 & 24.58 & 17.44 & 76.48 & 68.39 & 15.10 & 8.42 \\ 
FML1 & 0.66 & 36.76 & 72.74 & 26.60 & 10.20 & 60.76 & 69.15 & 20.65 & 19.88 & 30.58 & 62.64 & 17.48 \\ 
FMH1c & 4.68 & 36.76 & 76.58 & 18.74 & 10.69 & 54.31 & 74.38 & 14.93 & 22.62 & 24.82 & 66.81 & 10.57 \\ 
FIH3b & 7.34 & 79.24 & 4.61 & 88.05 & 27.92 & 87.02 & 3.82 & 68.26 & 47.86 & 55.33 & 3.96 & 48.18 \\ 
FMH0.3 & 23.16 & 43.59 & 71.28 & 5.56 & 26.82 & 53.22 & 68.38 & 4.80 & 46.36 & 32.43 & 47.73 & 5.91 \\ 

\hline
NIH4 & 4.63 & 71.41 & 21.26 & 74.11 & 17.41 & 81.48 & 18.96 & 63.63 & 29.79 & 46.76 & 14.55 & 55.66 \\ 
NIH1 & 13.11 & 66.14 & 34.44 & 52.45 & 36.57 & 78.81 & 24.74 & 38.69 & 46.75 & 60.08 & 16.86 & 36.39 \\ 

\hline 
\end{tabular}
\label{tab:fluxes}

\begin{flushleft} 
\textbf{Notes.} Column 1: model names; Column 2: modeled kinetic flux ratio; Column 3: modeled rest mass flux over modeled kinetic flux ratio; Column 4: modeled magnetic flux ratio; Column 5: modeled thermal flux ratio; Column 6: simulated kinetic flux ratio; Column 7: simulated rest mass flux over simulated kinetic flux; Column 8: simulated magnetic flux ratio; Column 9: simulated thermal flux ratio. Column 10: final kinetic flux ratio; Column 11: final rest mass flux over final kinetic flux; Column 12: final magnetic flux ratio; Column 13: final thermal flux ratio.
\end{flushleft} 

\end{table*}

\subsubsection{Non-force-free configuration} \label{sec:non_force_free_conf}

In the non-force free configuration, the azimuthal magnetic field component is expressed as \citep[][]{Marti_2015}:

in which $B^\phi_\mathrm{j,m}$ is the maximum toroidal field strength reached at the distance $R_\mathrm{m} = 0.37$.
In this configuration, the azimuthal magnetic field grows linearly in the region $r \ll R_\mathrm{m}$, reaches its maximum at $r = R_\mathrm{m}$, and then decreases as $\propto r^{-2}$ when $r \gg R_\mathrm{m}$ as shown in the upper left panel of Fig.\ \ref{fig:setup_profiles}.
The axial component is assumed to be constant across the jet radius,
\begin{equation}
    B^\mathrm{z}(r) = 
    \begin{cases}
      \DS B^\mathrm{z}_j, \, \, \, \, \, 0 \leq r \leq 1 \\
      0, \, \, \, \, \, \, \, r > 1. \\
    \end{cases}
\end{equation}
Therefore, the mean\footnote{The mean quantities are computed as 
$\bar{q_j} = \int_0^1 \! q(r) r \, \mathrm{d}r \Big{/} \int_0^1 \! r \, \mathrm{d}r \, .$}
azimuthal magnetic field component within the jet is:
\begin{equation}
    \bar{B^\phi_j} = 4 B^\phi_\mathrm{j,m} R_\mathrm{m} \Bigg[ 1 - R_\mathrm{m} \mathrm{atan} \bigg(\frac{1}{R_m}\bigg) \Bigg]
\end{equation}
while the axial one is simply $\bar{B^\mathrm{z}_j} = B^\mathrm{z}_j$.

By solving Eq.\,\ref{eq:pressure_equilibrium} with these magnetic field components, we obtain a gas pressure profile expressed as

\begin{equation}
    p(r) = 
    \begin{cases}
      \DS 2 \Bigg[ \frac{B^\phi_\mathrm{j,m}}{ \gamma_j(1 + (r/R_\mathrm{m})^2)}\Bigg]^2 + C, \, \, \, \, \, 0 \leq r \leq 1 \\
      \\
      p_a, \, \, \, \, \, \, \, \, \, \, \, \, \, \, \, \, \, \, \, \, \, \, \, \, \, \, \, r > 1. \\
    \end{cases}
\end{equation}
in which the integration constant can be computed using the boundary condition at the jet/ambient medium interface $p^* (r = 1) = p_a$, leading to 
\begin{equation}
     C = p_a - \frac{(B^\mathrm{z}_j)^2}{2} - \frac{(B^\phi_1)^2}{2 \gamma_j^2} \big[1 + (R_m)^2 \big]
\end{equation}
(in this expression, $B^\phi_1 = B^\phi (r=1)$).
The mean gas pressure is $\bar{p_j} = p_a - (B^\mathrm{z}_j)^2 / 2$ and the magnetic pressure
$\bar{p_m}=\bar{b}_j^2 / 2$. 
Fiducial magnetic and thermal pressure profiles are shown in the lower left panel of Fig.\ \ref{fig:setup_profiles}.

The magnetic tension \citep[$\tau_\mathrm{m} = - (b^\phi)^2/r$, see, e.g.,][]{Marti_2016,Moya-T_2021} is:
\begin{equation}
    \tau_\mathrm{m}(r) = - \frac{4 (B^\phi_\mathrm{j,m})^2}{\gamma^2 R_\mathrm{m}^2} \frac{r}{( 1 + (r/R_\mathrm{m})^2)^2}.
\label{eq:magnetic_tension}
\end{equation}

\subsubsection{Force-free configuration} \label{sec:force_free_conf}

In the force-free configuration, the azimuthal magnetic field component has the same profile as in the non-force-free configuration, while the axial component becomes:
\begin{equation}
    B^\mathrm{z}(r) = 
    \begin{cases}
      \DS \frac{2 B^\phi_\mathrm{j,m} / \gamma_j}{1 + (r/R_\mathrm{m})^2}, \, \, \, \, \, \, \, \, 0 \leq r \leq 1 \\
      0, \, \, \, \, \, \, \, \, \, \, \, \, \, \, \, \, \, \, \, \, \, \, \, \, \, \, \, r > 1. \\
    \end{cases}
\end{equation}
The axial magnetic field has its maximum at the jet center and decreases towards the outer radius as $\propto r^{-2}$. Its mean value is:
\begin{equation}
    \bar{B^\mathrm{z}_j} = \frac{2 B^\phi_{j,m}R_m^2}{\gamma_j} \left[ \mathrm{ln}(R_m^2 + 1) - \mathrm{ln}(R_m^2) \right].
\end{equation}

In this configuration, the gas pressure is constant in the jet, with $p(r) = \bar{p_j} = p_a$, $0 \leq r \leq 1$, and the pressure equilibrium is achieved by compensating the magnetic tension produced by the toroidal component of the field (see Eq.~\ref{eq:magnetic_tension}) with the magnetic pressure produced by the axial one. 

The advantage of this magnetic configuration is that the gas pressure becomes independent from the jet magnetization, allowing the field to increase without generating non-physical, negative thermal pressures at the jet/ambient boundary \citep[][]{Moya-T_2021}. This configuration is thus necessary to simulate models of highly magnetized jets in pressure equilibrium. 
The resulting initial radial magnetic field and pressure profiles are shown in the bottom panels of Fig.\ \ref{fig:setup_profiles}.

\subsection{Jet and ambient parameters} \label{sec:jet_ambient_params_2D}

With the aim to study jet acceleration from sub-relativistic to relativistic outflow velocities, we inject the jets with the lowest possible axial velocity. 
The injection of low velocity, highly magnetized flows makes them submagnetosonic, which allows waves to propagate upstream and generate inconsistencies between the jet in the grid and its injection boundary condition. In particular, the axial magnetic field is changed and a jump appears between the boundary and the first cell in the grid, which represents an unphysical situation.

To overcome this problem, the initial jet velocity $v^\mathrm{z}$ was adjusted in the different models to result in magnetosonic Mach numbers of at least 1. While this approach prevents us from exploring the acceleration starting from sub-relativistic velocities, it still allows us to achieve our main goal, which is understanding the role played by internal and magnetic energy fluxes in accelerating jets. 
Specifically, FR~I jets can reach velocities up to $\gamma \sim 10$ \citep[e.g.,][]{Hovatta_2009} at parsec scales, and from our injection velocities, which imply $\gamma \sim 1-2$, we can explore the increase from mildly relativistic to relativistic speeds for different types of jets, i.e., magnetically, internally or kinetically dominated, and the velocity structures that they generate.

Using our information on NGC\,315 as starting point, we design our grids to simulate outflows along the z-coordinate at $\mathrm{z}_i = 0.3 \, \mathrm{pc}$, since this corresponds to a distance at which the jet has already velocities in the range $v^\mathrm{z}_j \sim 0.3 \, c$ to $v^\mathrm{z}_j \sim 0.8 \, c$, depending on the frequencies and epochs considered \citep[see Fig.\, 3,][]{Ricci_2022}. 
The grid extends up to $\mathrm{z}_o = 3.3 \, \mathrm{pc}$, covering a distance that is large enough to let us explore the acceleration phenomenon from sub-parsec to parsec scales. 
In the radial direction, the grid extends from $r_i = 0$ up to $r_o = 0.9 \, \mathrm{pc}$, with an initial jet radius of $r_j = 0.03 \, \mathrm{pc}$ \citep[see Fig.\, 3,][]{Boccardi_2021}. 
Taking into account that jets are built with large overpressure factors (see below), this radial size of the grid is large enough to allow for jet expansion.
With our choice of the initial jet radius, our numerical domain consists of a grid of $30\,r_j\times100\,r_j$ with $n_r \times n_\mathrm{z} = 600\times2000$ cells.

\begin{figure}[t]
    \centering
    \includegraphics[width=0.9\linewidth]{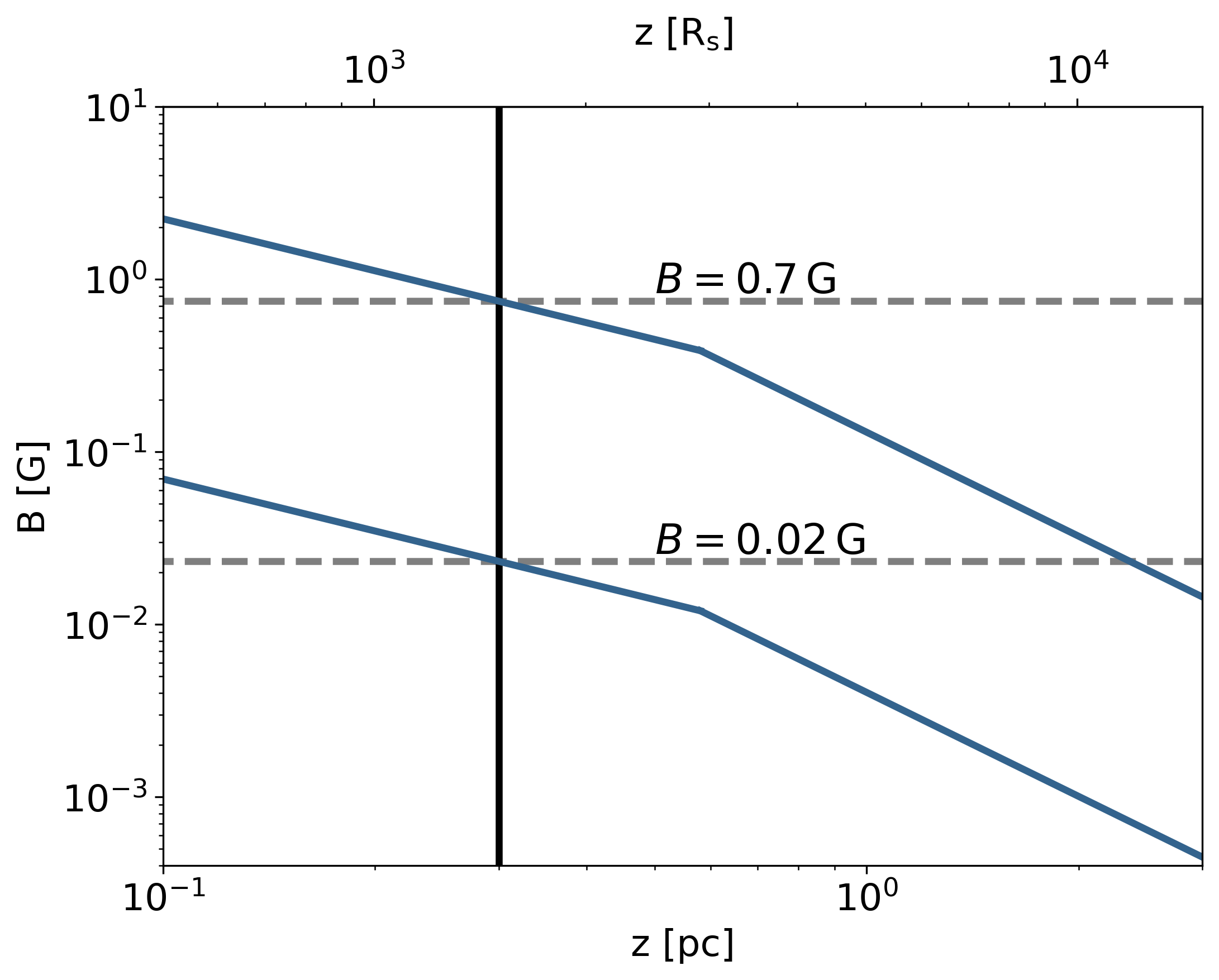}
    \caption{Expected magnetic field distribution between $0.1 \, \mathrm{pc}$ and $3 \, \mathrm{pc}$. The upper and lower magnetic field limits are extrapolated from the core shift measurement under the different assumption \citep[for the details see][]{Ricci_2022}.
    The black vertical line highlights the starting point of $\mathrm{z}_i = 0.3 \, \mathrm{pc}$ of our simulations, while the horizontal dashed lines are the expected range of magnetic field at $\mathrm{z}_i$.}
    \label{fig:Magnetic_field_initial}
\end{figure}

\begin{figure}[t]
    \centering
    \includegraphics[width=0.8\linewidth]{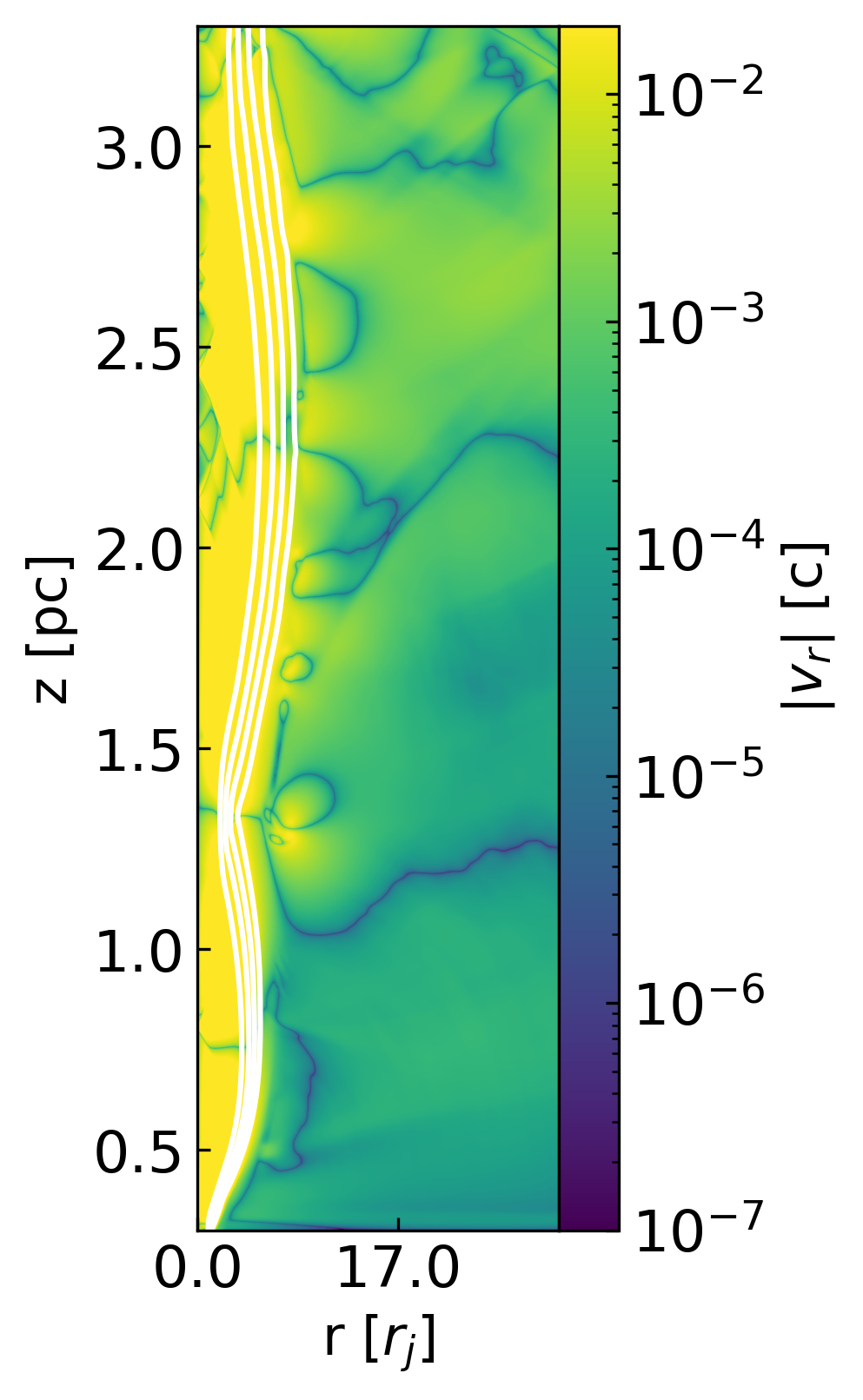}\par
    \caption{Absolute value of radial velocity for the model FMH1\_m4. The white contours highlight the tracer values at levels of 0.2, 0.4, 0.6, and 0.8 (the tracer levels are in decreasing order, i.e., from 0.8 to 0.2, from left to right). Outside of the jet structure delineated by the shear layer (white contours), the velocity is smaller than 0.01 everywhere in the grid, meaning that this can be considered a steady solution.}
    \label{fig:velx_equilibrium}
\end{figure}

To determine the initial magnetic field we refer to Fig.~\ref{fig:Magnetic_field_initial}, which highlights the possible magnetic field values, in our region of interest, extracted from the results published in \citet{Ricci_2022}.
According to the different models and different magnetic field evolution, the initial total strength is allowed to span between $B = 0.02 \, \mathrm{G}$ and $B = 0.7 \, \mathrm{G}$. The magnetic field structure considered in this paper ranges from helical in the force-free configurations (with the mean pitch angle defined as $\bar\phi_{\mathrm{B}_j} = \mathrm{tan}^{-1} (\bar{B}^\phi_j / \bar{B}^\mathrm{z}_j$)), to a dominating toroidal field in the non-force-free configurations.

The numerical grid is initially filled with a jet with the injected properties (see Table~\ref{tab:setups}) in the region $(r, \mathrm{z}) \in [0,1\, r_j] \times [0, 100 \, r_j]$ and an ambient medium that occupies the rest of it. 
The initial jet overpressure forces its expansion into this ambient medium, which we let evolve until an equilibrium, steady solution is reached. The description of the criteria by which we decide when equilibrium has been reached are described in Sect.\ \ref{sec:init_to_eq}.

By establishing a jet across the whole grid, we avoid jet injection and the generation of a bow shock that would arise from the interaction of the newly ejected jet with the external medium, while focusing on jet physics alone.
 
We assume jets with kinetic power in the range $L_j \sim 10^{43} - 10^{44} \, \mathrm{erg} \, \mathrm{s^{-1}}$, in accordance with the estimate given by \citet{Morganti_2009, Ricci_2022} for NGC~315.
To build the required jet powers, we distribute the energy among the different channels (see Section \ref{sec:phy_model}).
Regarding the external medium, we refer to the X-ray Chandra observations for the environment surrounding NGC~315. 
The measurements for the galactic core region, i.e., the inner 0.3 kpc, give pressure and density of $p_0 = 4.5 \times 10^{-10} \, \mathrm{dyn} \, \mathrm{cm^{-2}}$ and $\rho_0 = 0.46 \times 10^{-24} \, \mathrm{g} \, \mathrm{cm^{-3}}$, respectively \citep{Worrall_2007}. 
However, jet configuration for the aforementioned powers requires pressures that can be six to eight orders of magnitude over these values. 

While this raises questions about jet equilibrium and collimation at these scales, it is well possible that the jet is indeed strongly overpressured with respect to its environment, but that this extreme overpressure has been slowly achieved with time. It is relevant to recall that when formed and injected for the first time, jets are surrounded by their own shocked plasma, which is probably overpressured with respect to the jet itself. As it expands, the pressure around the jet falls gradually \citep[see, e.g.][]{Perucho_2014b,Perucho_2019,Perucho_2022} and the jet keeps its collimation due to its own velocity, which limits opening angles to $\sim 1/ \gamma$. In conclusion, in the case of evolved jets (like those in radio galaxies) one could expect free expansion with large opening angles while the jet accelerates, asymptotically reaching $\sim 1/ \gamma$. This is in agreement with the transition from parabolic to conical jet expansion, precisely in the acceleration region. 

Unfortunately, in our case, we cannot let the jet to adapt to such extreme conditions in a gradual way, and we establish the overpressure since the beginning of our runs. 
Furthermore, such large overpressure causes the code to crash as the jet violently expands into the ambient medium. 
However, since we only need the jet to freely expand, it is enough to give the ambient medium a pressure that allows this, even if the simulated ambient pressure is much higher than that estimated by Chandra. Therefore, we assume arbitrarily initial high pressure and density, with values of $p_a = 1.0 \times 10^{-3} \, \mathrm{dyn} \, \mathrm{cm^{-2}}$ and $\rho_a = 1.67 \times 10^{-22} \, \mathrm{g} \, \mathrm{cm^{-3}}$, respectively. This medium is set as isothermal, and the pressure (and density) are given a gradient in the $\mathrm{z}$ direction to fall with distance as $\propto \mathrm{z}^{-1}$.
In this way, at injection the jet overpressure factor spans from 1 (i.e., jet in thermal pressure equilibrium with the environment) to 4 and reaches values up to $10^2$ at the end of the grid.
The initial jet density spans between $\rho_j = 1.67 \times 10^{-25} - 1.67 \times 10^{-23} \, \mathrm{g} \, \mathrm{cm^{-3}}$.

Jet pressure and density, together with the magnetic field strength, are modified from model to model in order to explore a variety of initial jet conditions, from highly magnetized, cold jets (with a magnetic energy flux $F_M$, representing $\sim 75\%$ of the total) to hot jets (with an internal energy flux, $F_I$, representing $\sim 88\%$ of the total). 
The parameters that determine each of the simulated models are given in Table \ref{tab:setups}. 
The model names are given using the following code: the first letter represents the magnetic field configuration, with F for force-free and N for non-force free, the second letter stands for the type of jet, i.e.\ magnetic (M), internal (I) or kinetic (K) energy dominated, the third letter gives the code for the jet density, with S for $\rho_j \geq 0.1 \, \rho_a$, H for $0.01 \, \rho_a \leq \rho_j < 0.1 \, \rho_a$, and L for $\rho_j < 0.01 \, \rho_a$, while the last numbers refer to the jet thermal overpressure factor at injection.
When two different models are expected to have the same nomenclature, we add at the end of the name a letter in progressively alphabetical order (e.g.\ FMH1b, FMH1c).
Models with \_m4 have the shear layer exponent set to $m = 4$, i.e., a wider (thicker) shear layer, while for all the other models the exponent is set to $m = 8$.

Table \ref{tab:fluxes} displays the ratio of the different flux channels over the total one (index $i$) as derived from the jet parameters, together with the simulated ones, which are convolved to the shear layer (index $s$), and final ones computed at the final row of cells at 3.3 pc (index $f$).
In Appendix \ref{app:radial_fluxes}, we discuss the reason and implications in the initial radial profile of the fluxes within the jet, considering their convolution with the shear layer.

\section{Results} \label{sec:results}
\subsection{From initial conditions to equilibrium} \label{sec:init_to_eq}

\begin{figure*}[htpb]
    \centering
\begin{multicols}{3}
    \includegraphics[width=0.8\linewidth]{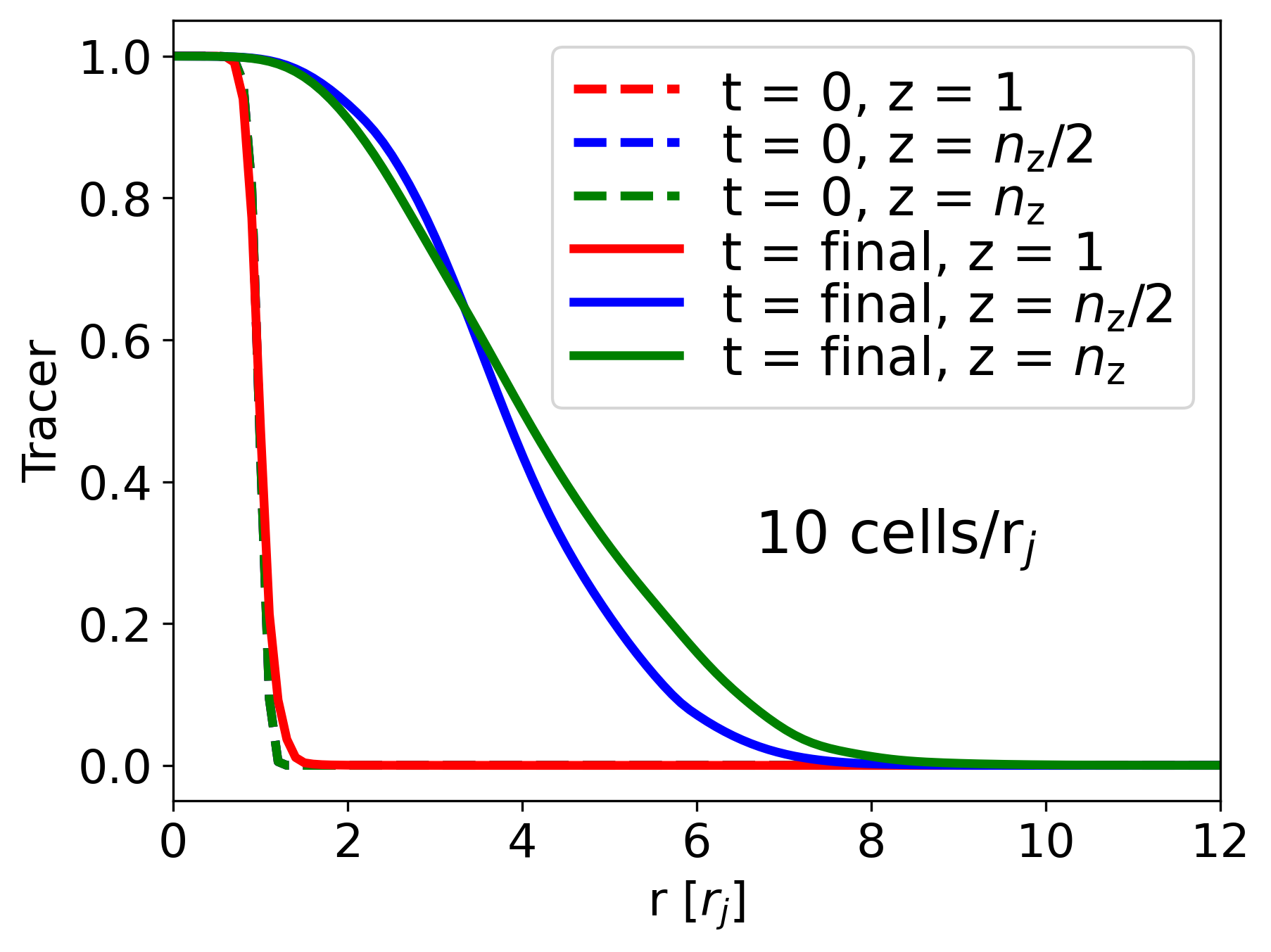}\par
    \includegraphics[width=0.8\linewidth]{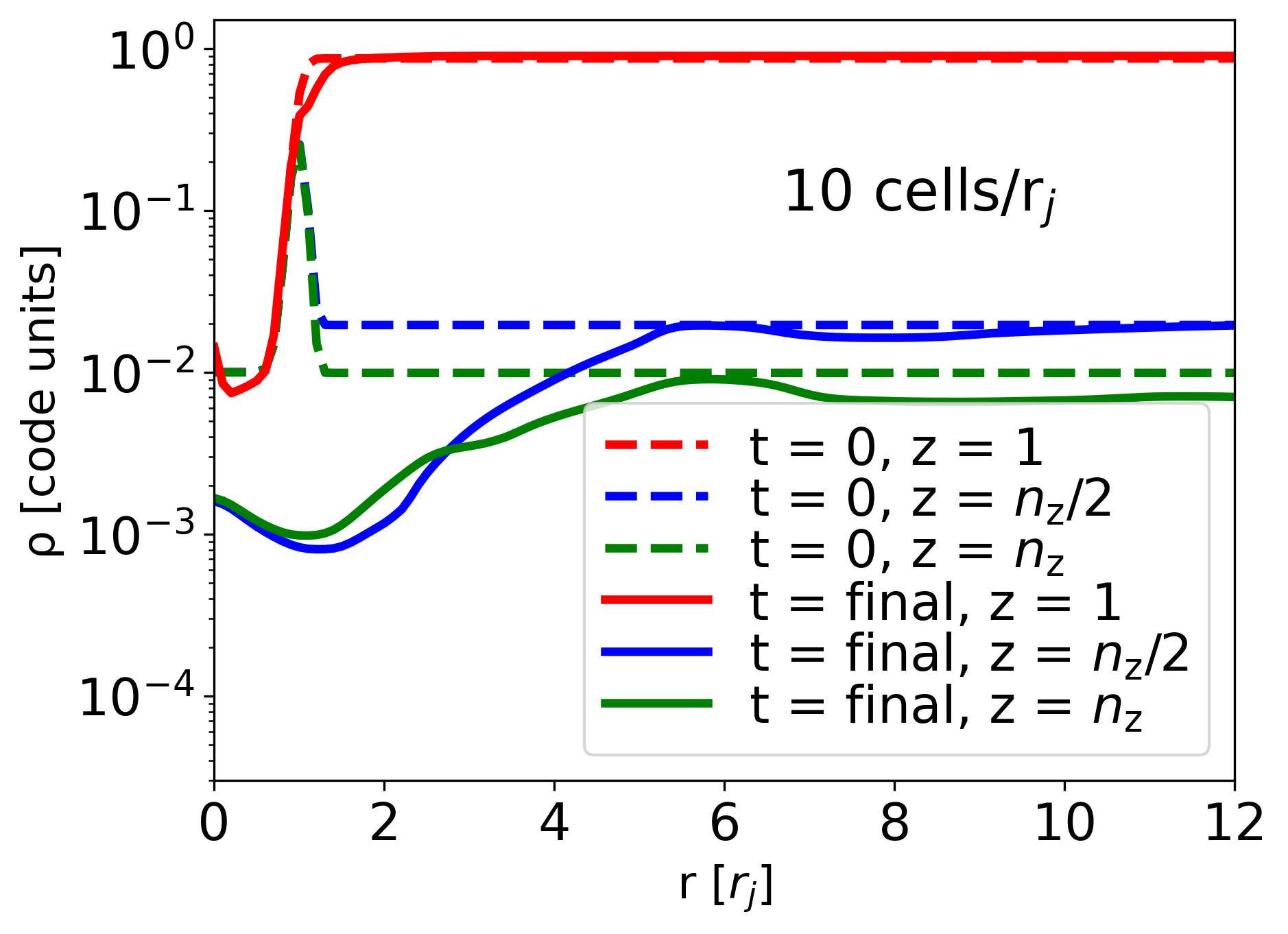}\par
    \includegraphics[width=0.8\linewidth]{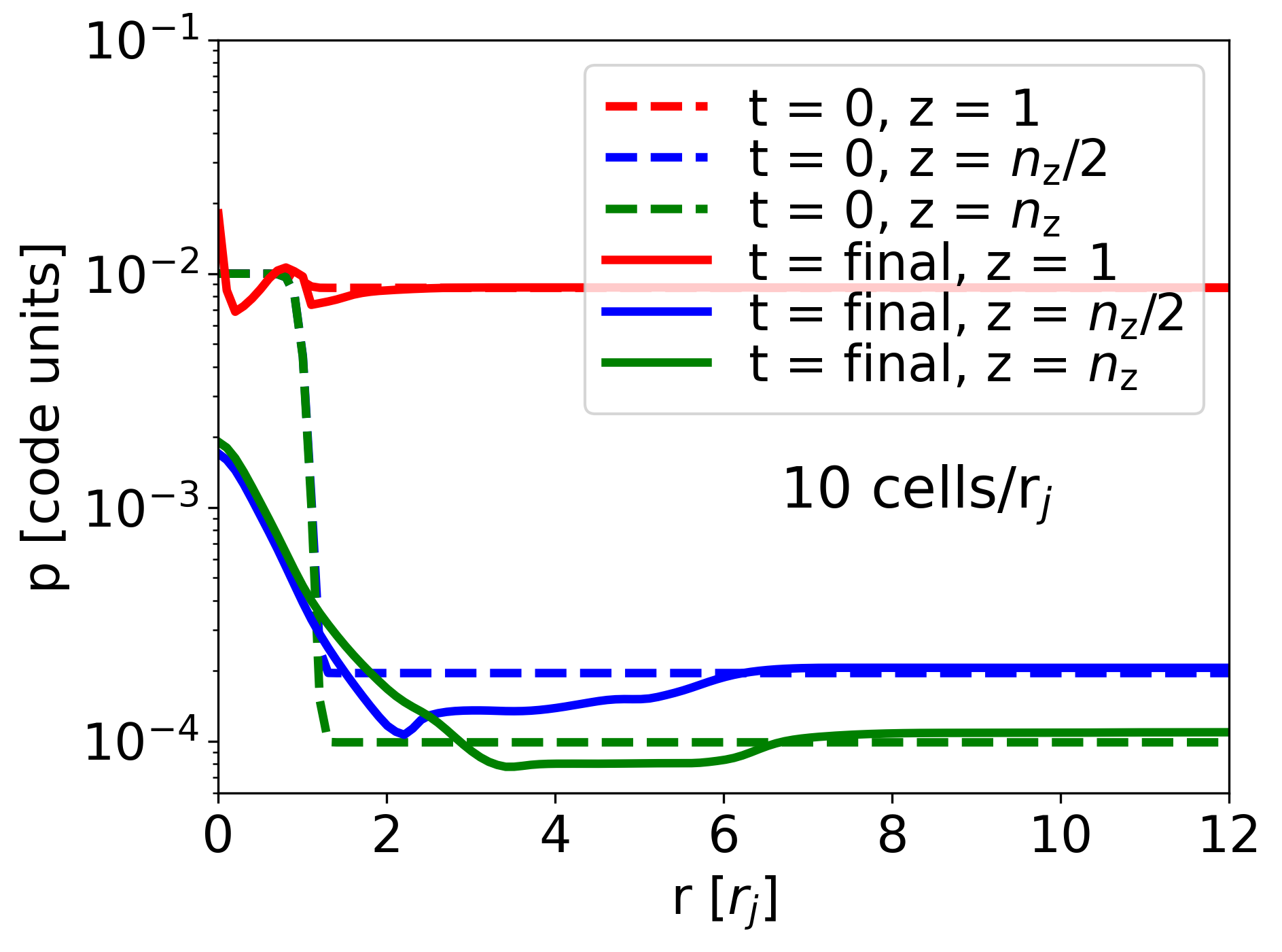}\par
\end{multicols}
\vspace{-0.8cm}
\begin{multicols}{3}
    \includegraphics[width=0.8\linewidth]{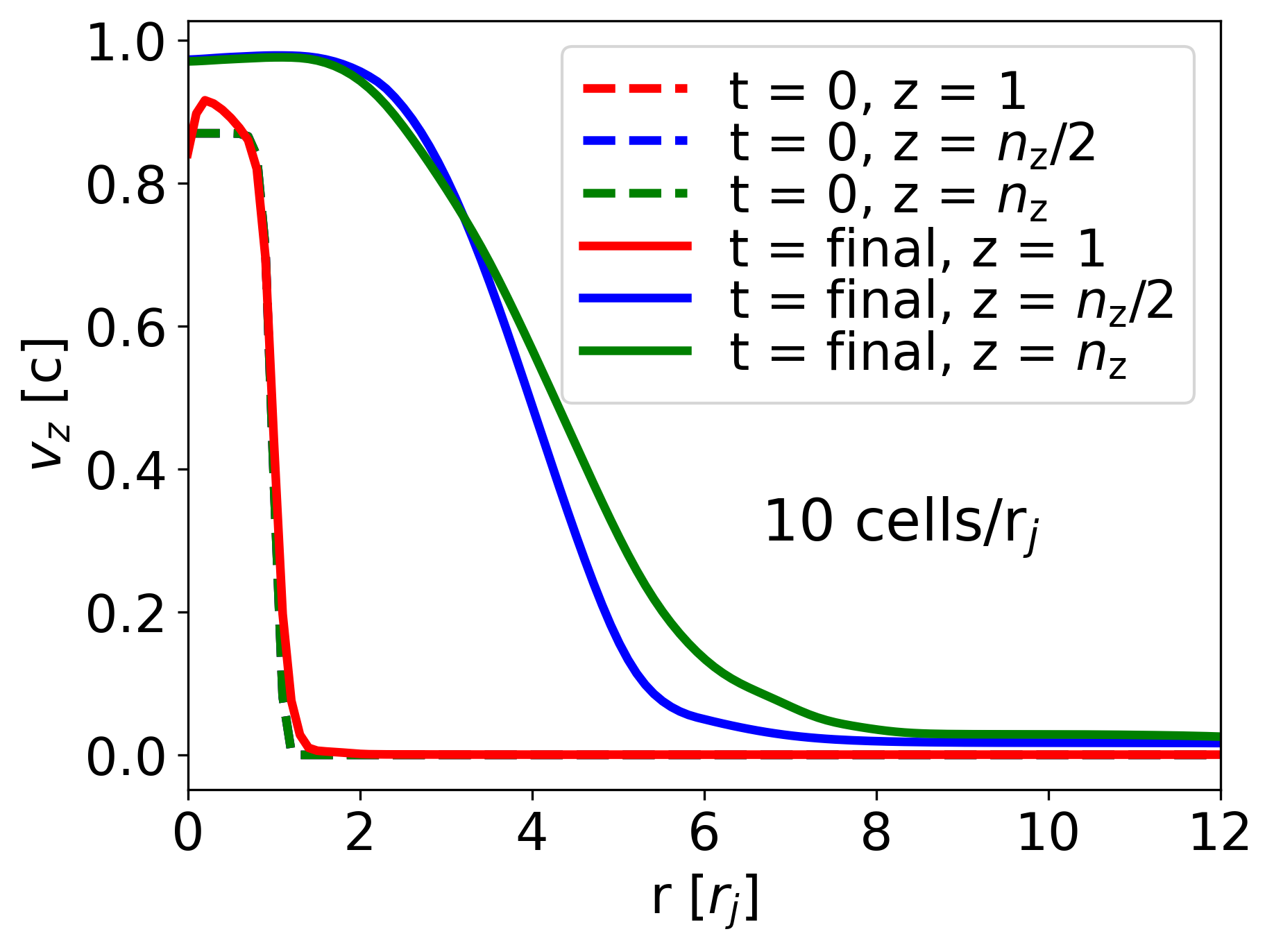}\par
    \includegraphics[width=0.8\linewidth]{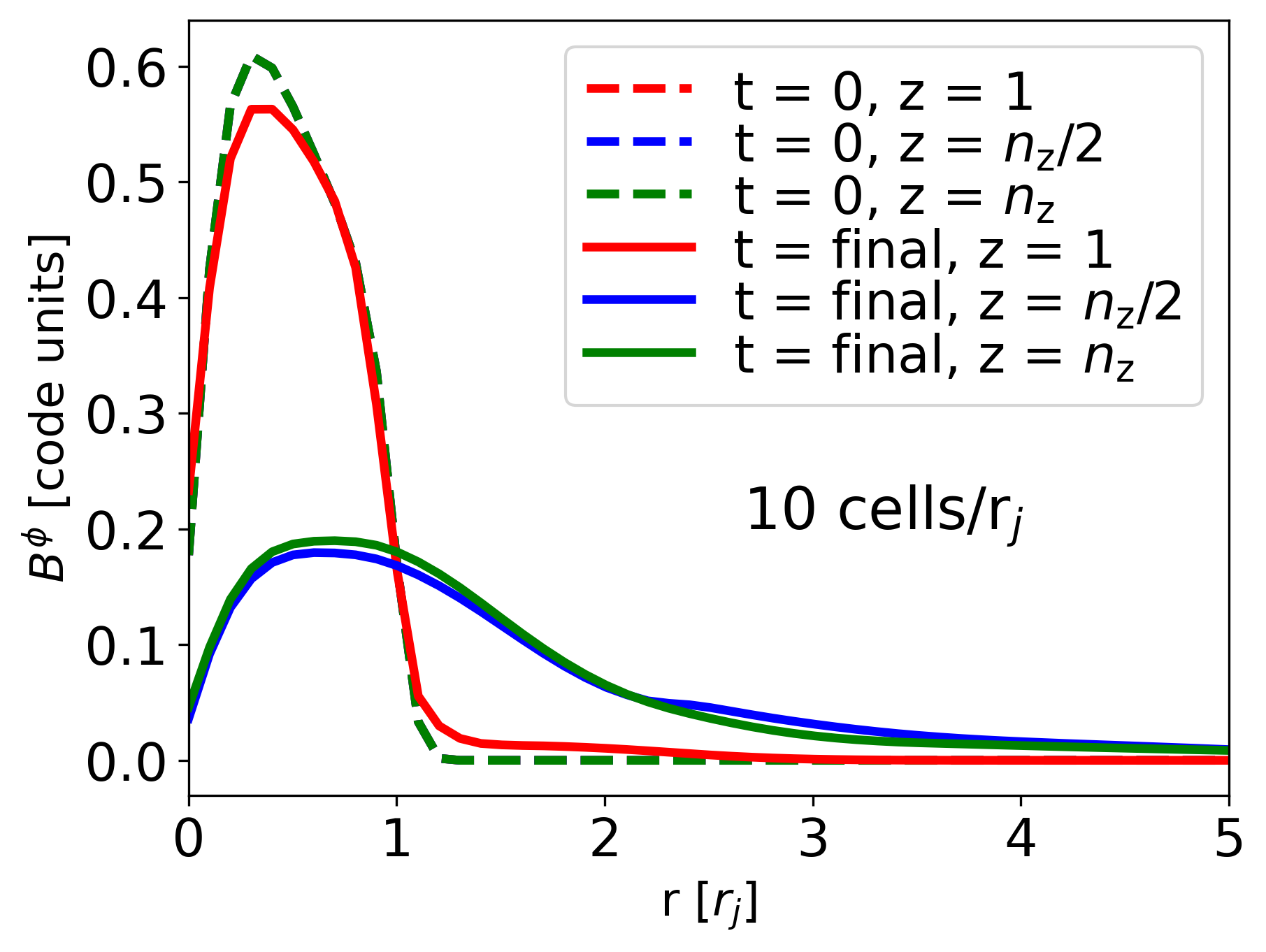}\par
    \includegraphics[width=0.8\linewidth]{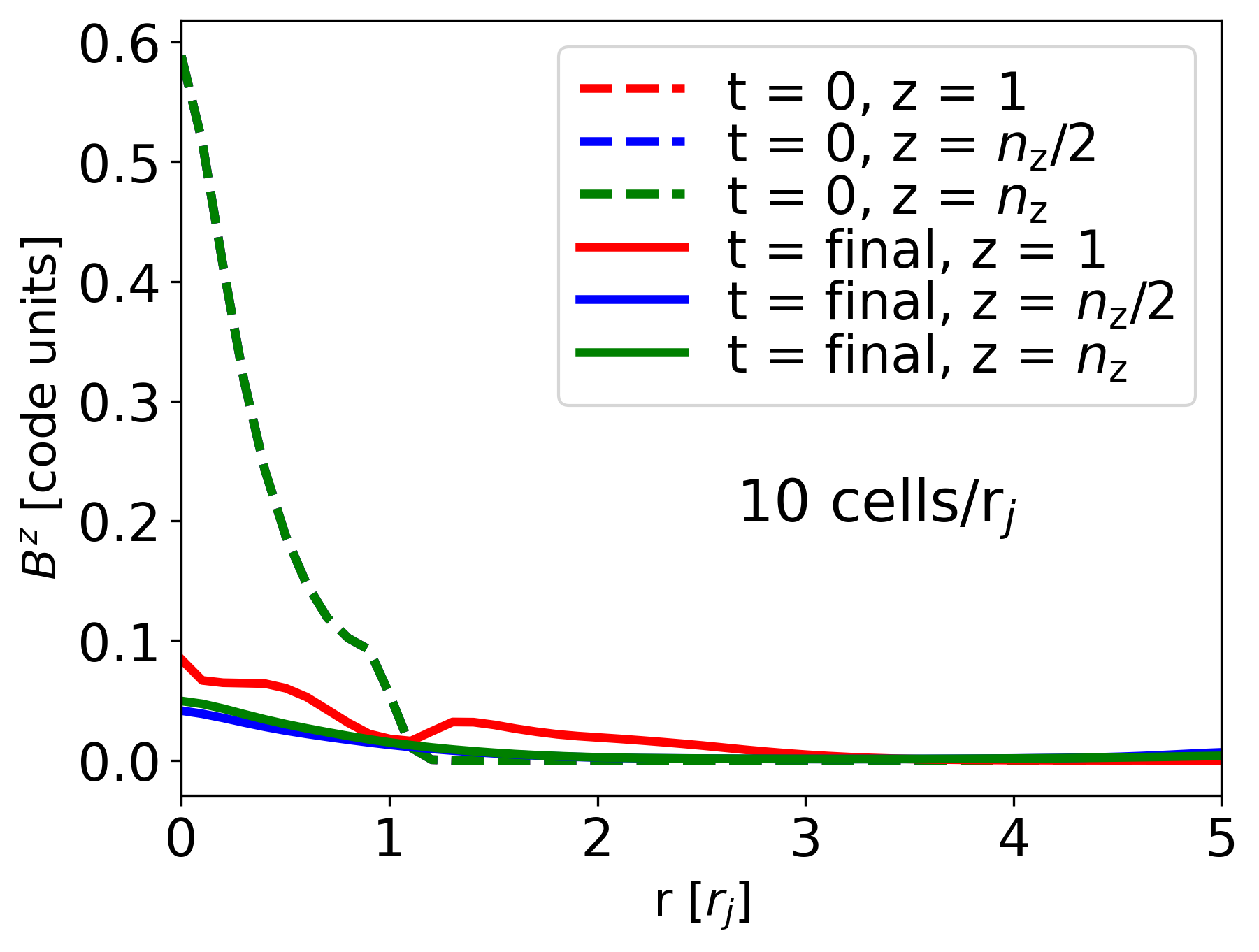}\par
\end{multicols}
\vspace{-0.4cm}
\noindent\hrulefill
\vspace{-0.3cm}
\begin{multicols}{3}
    \includegraphics[width=0.8\linewidth]{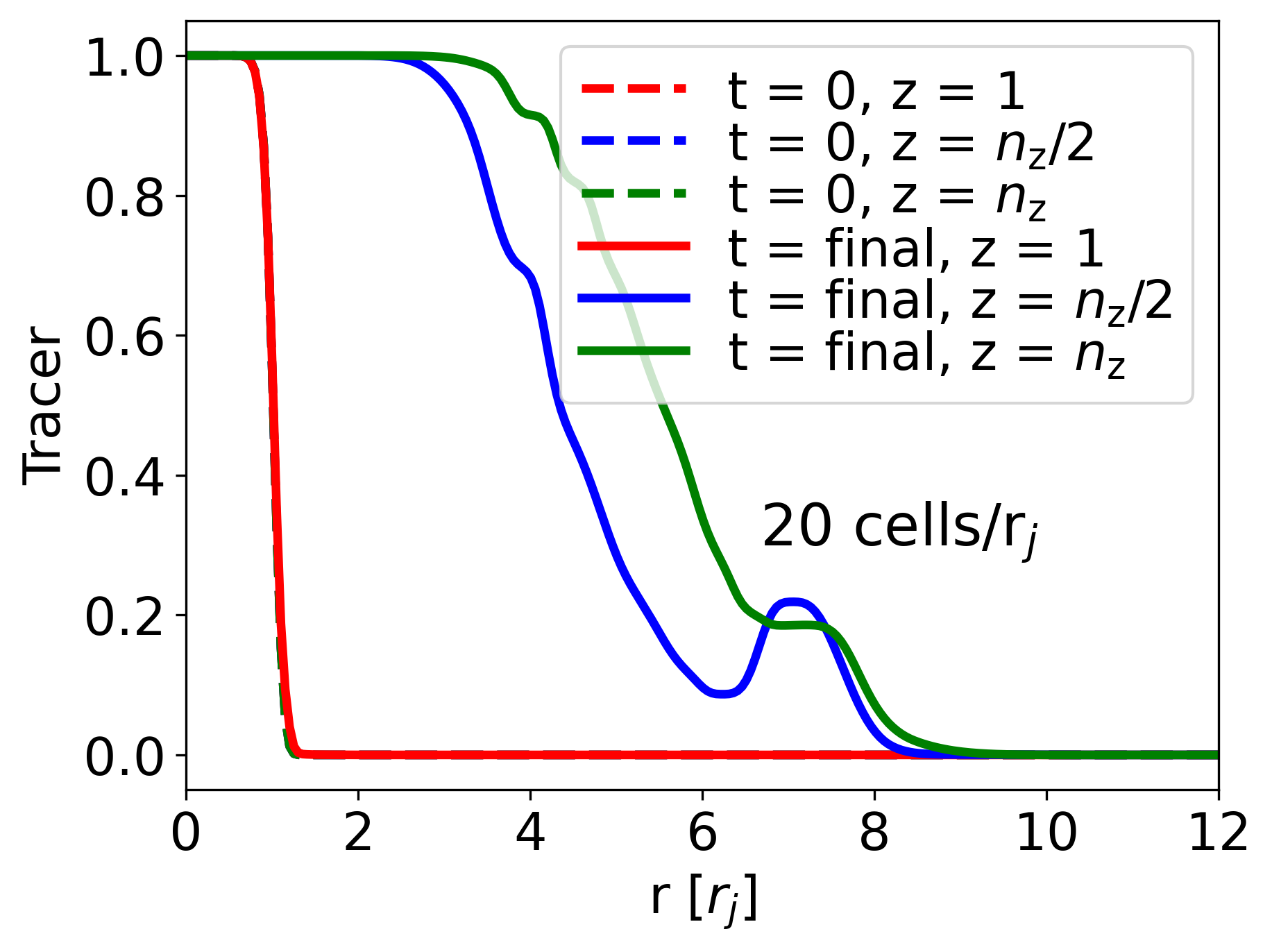}\par
    \includegraphics[width=0.8\linewidth]{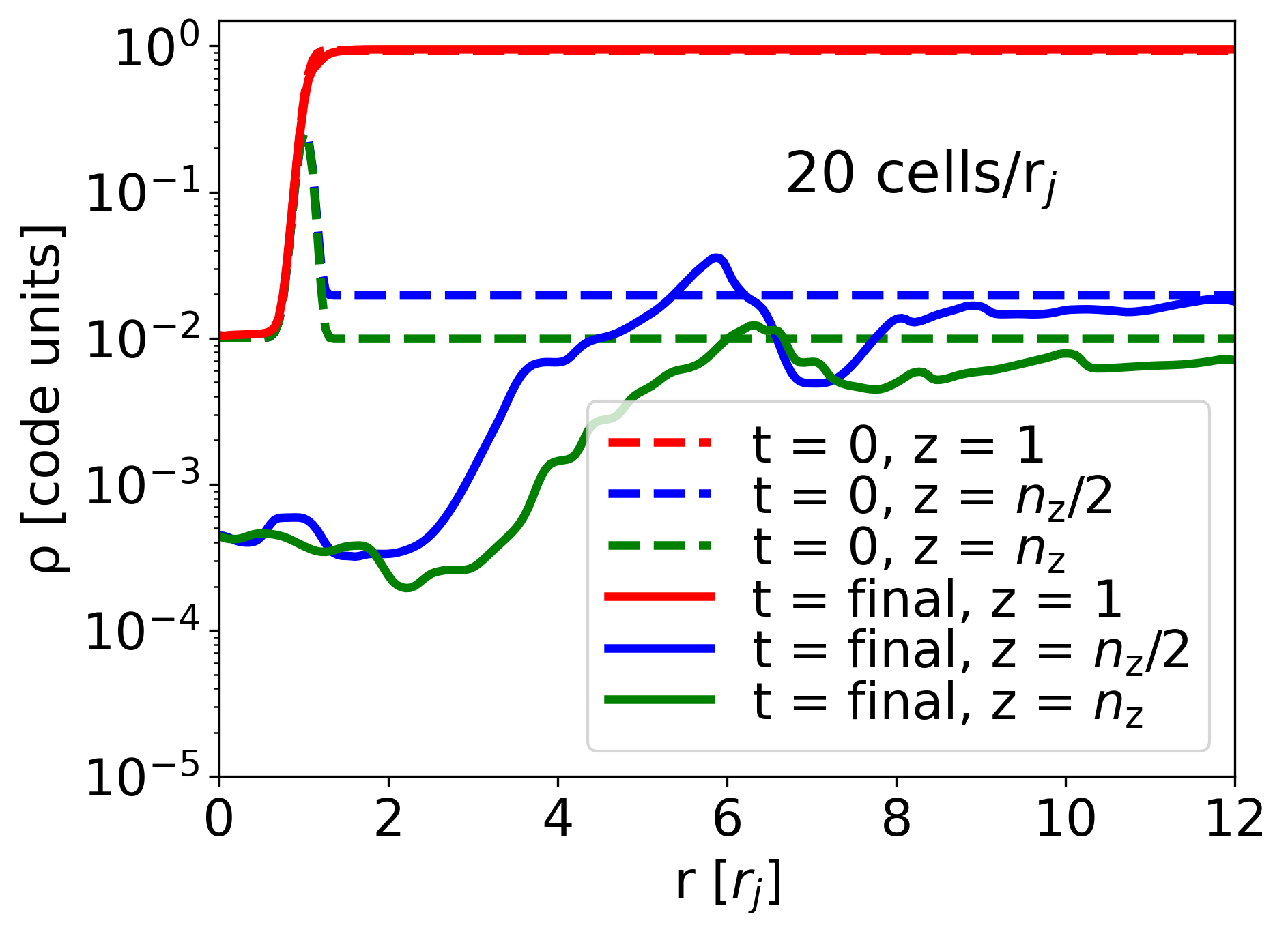}\par
    \includegraphics[width=0.8\linewidth]{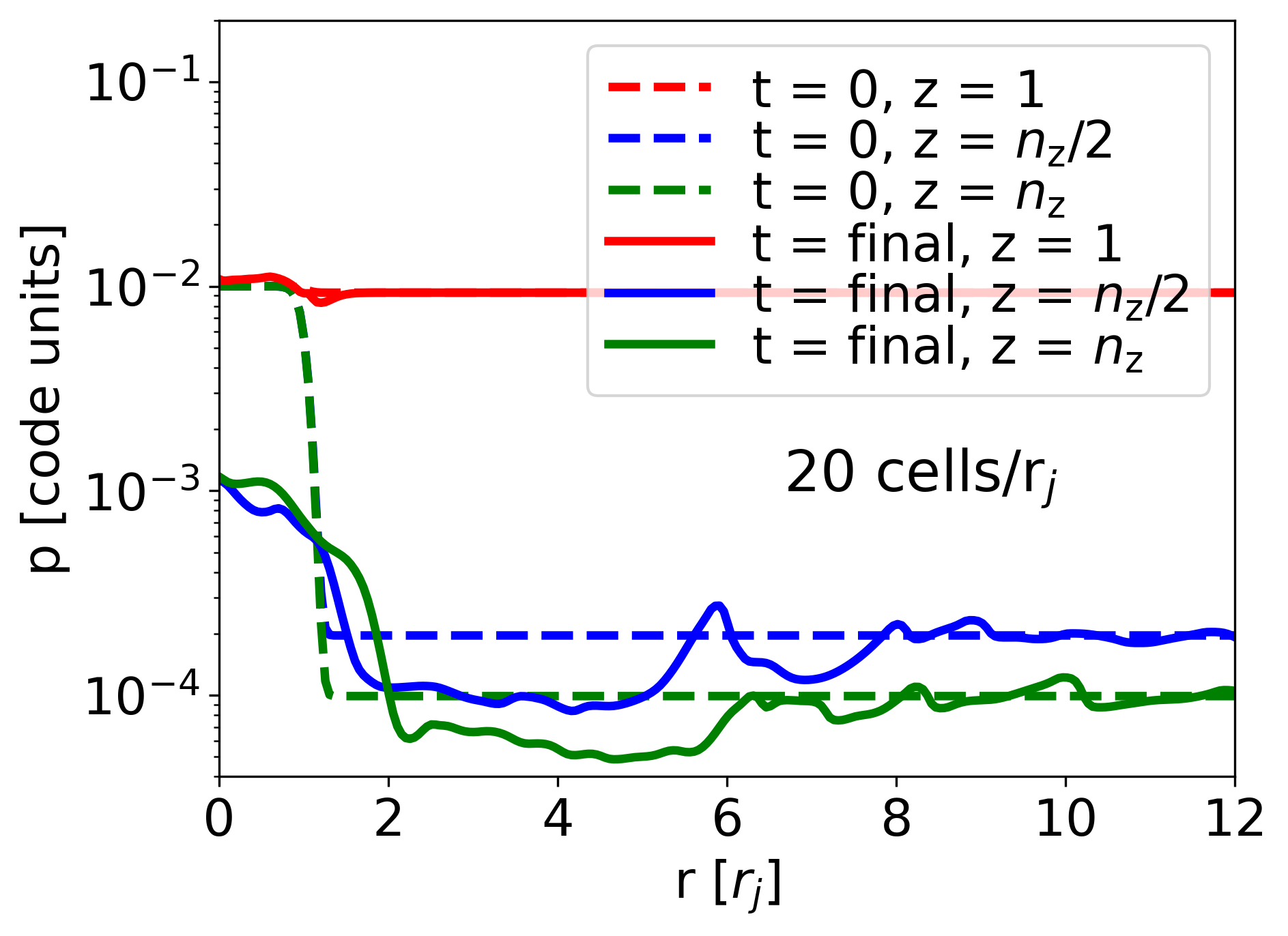}\par
\end{multicols}
\vspace{-0.8cm}
\begin{multicols}{3}
    \includegraphics[width=0.8\linewidth]{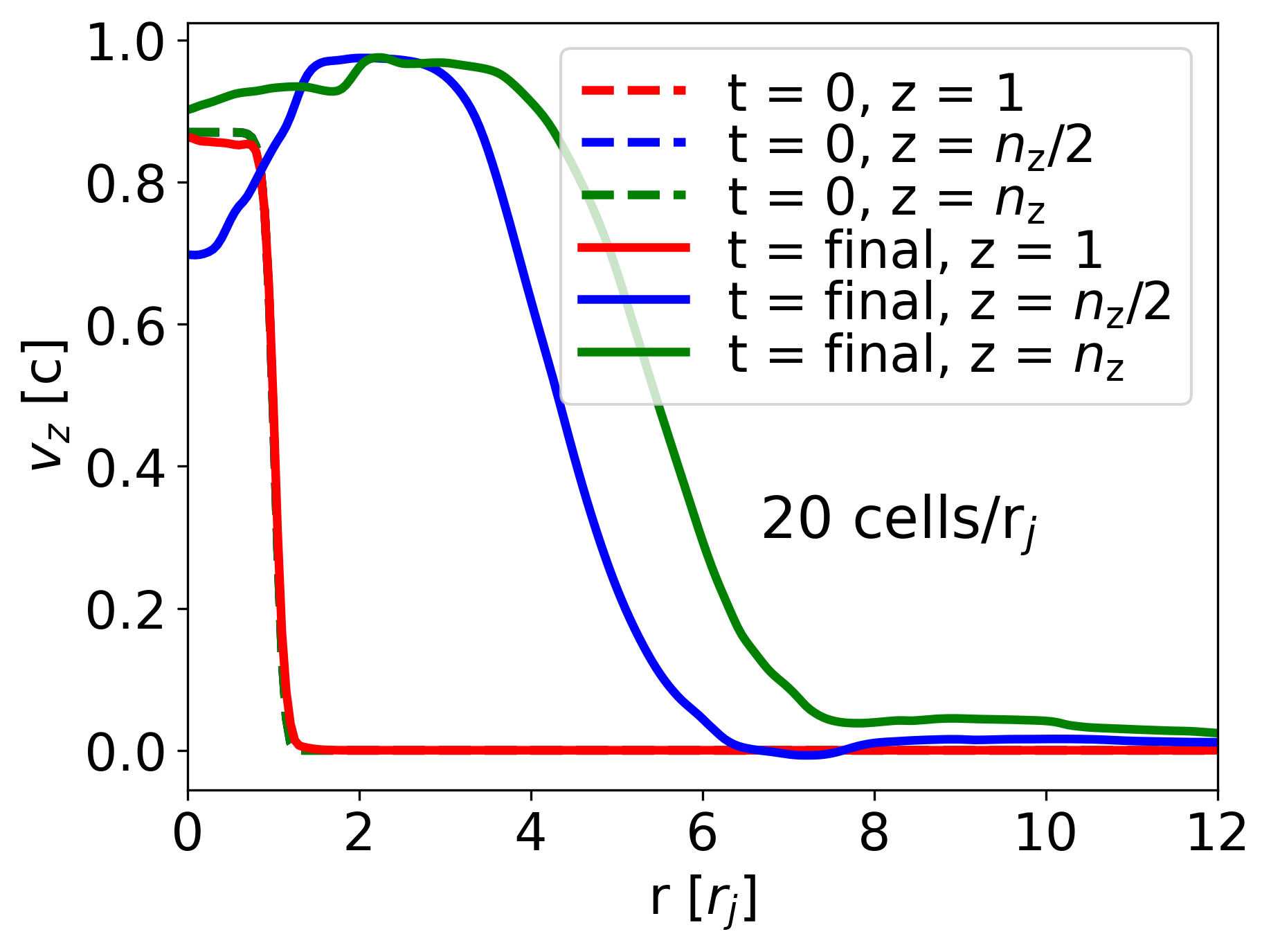}\par
    \includegraphics[width=0.8\linewidth]{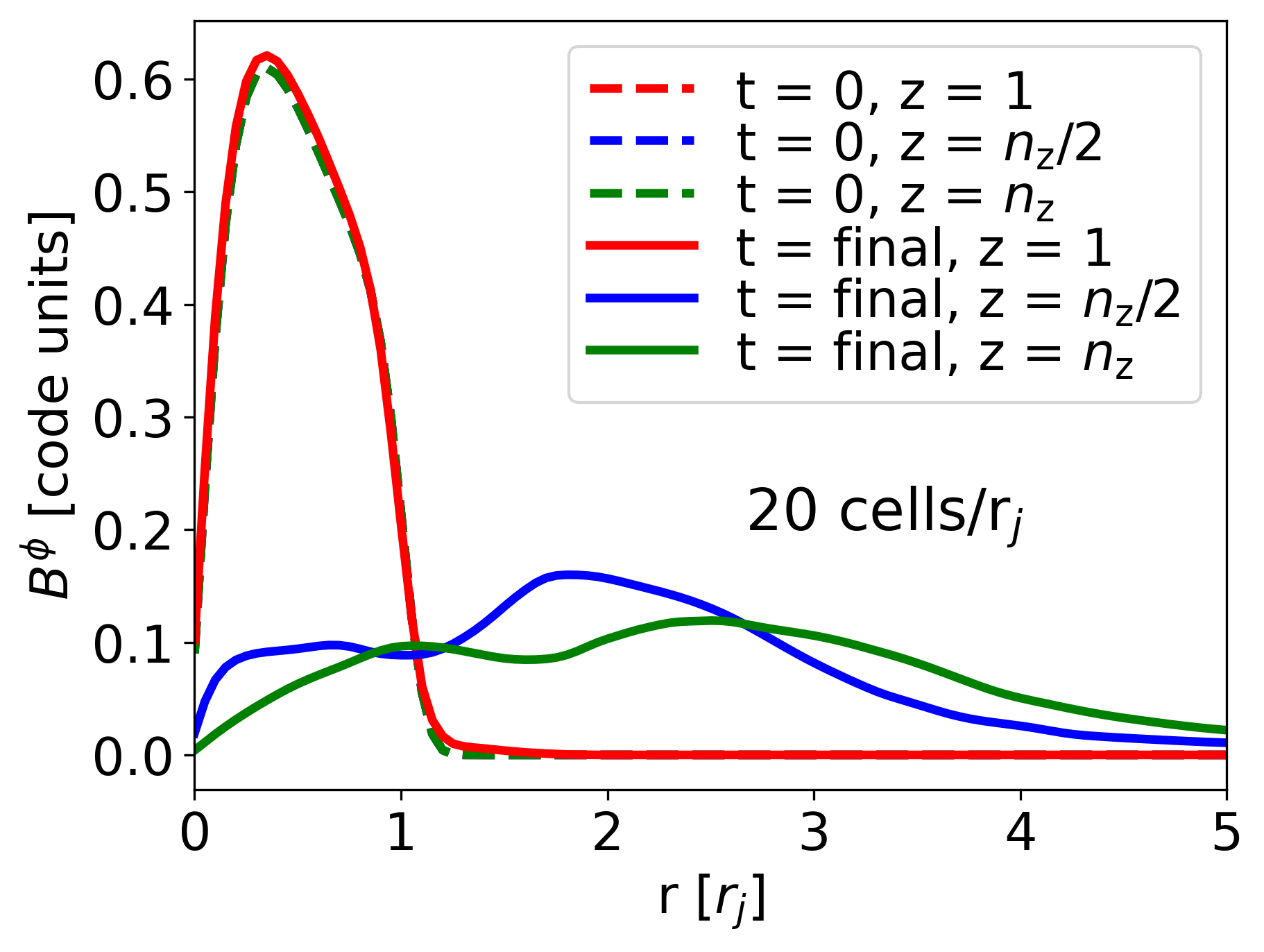}\par
    \includegraphics[width=0.8\linewidth]{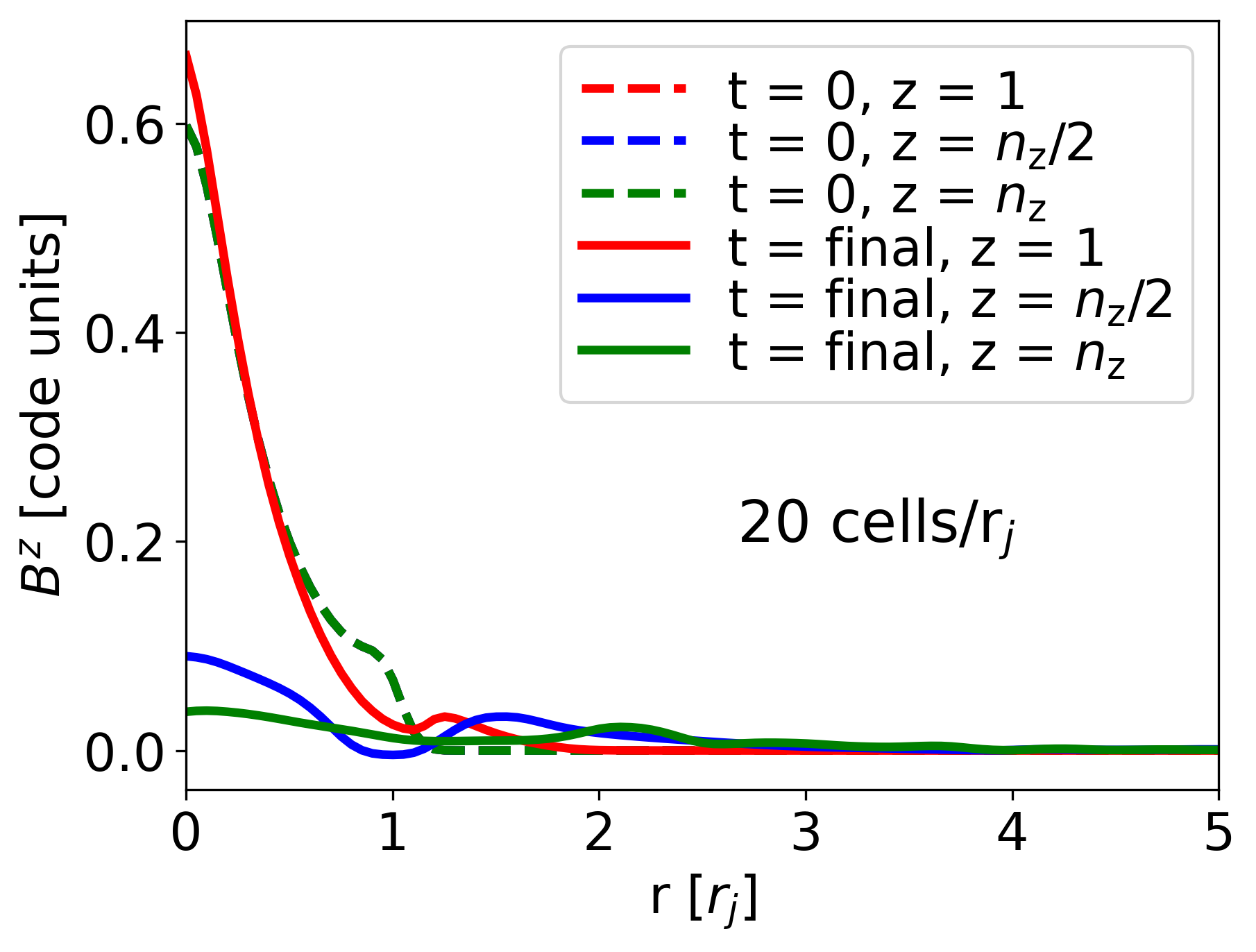}\par
\end{multicols}
\vspace{-0.4cm}
\noindent\hrulefill
\vspace{-0.3cm}
\begin{multicols}{3}
    \includegraphics[width=0.8\linewidth]{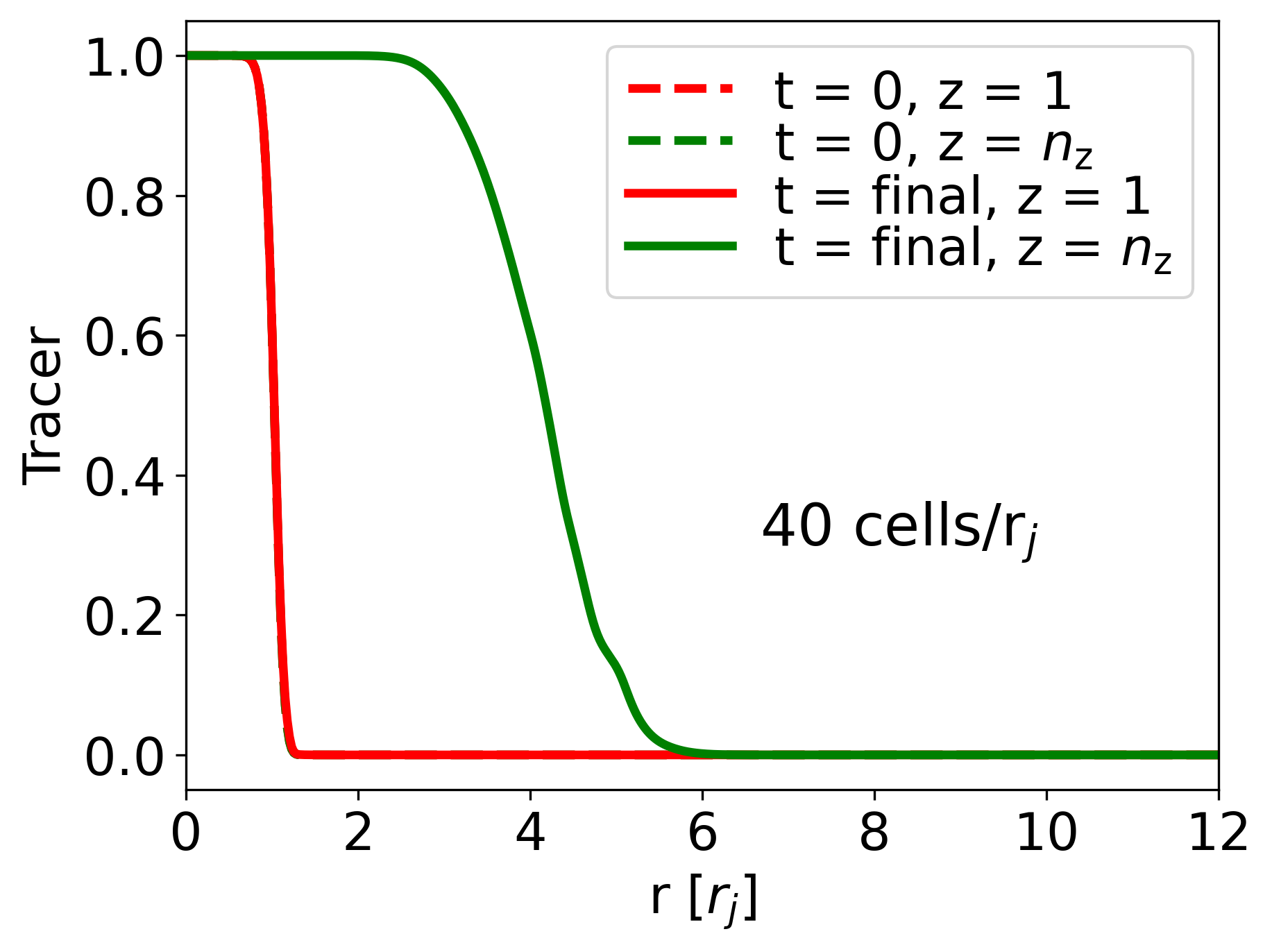}\par
    \includegraphics[width=0.8\linewidth]{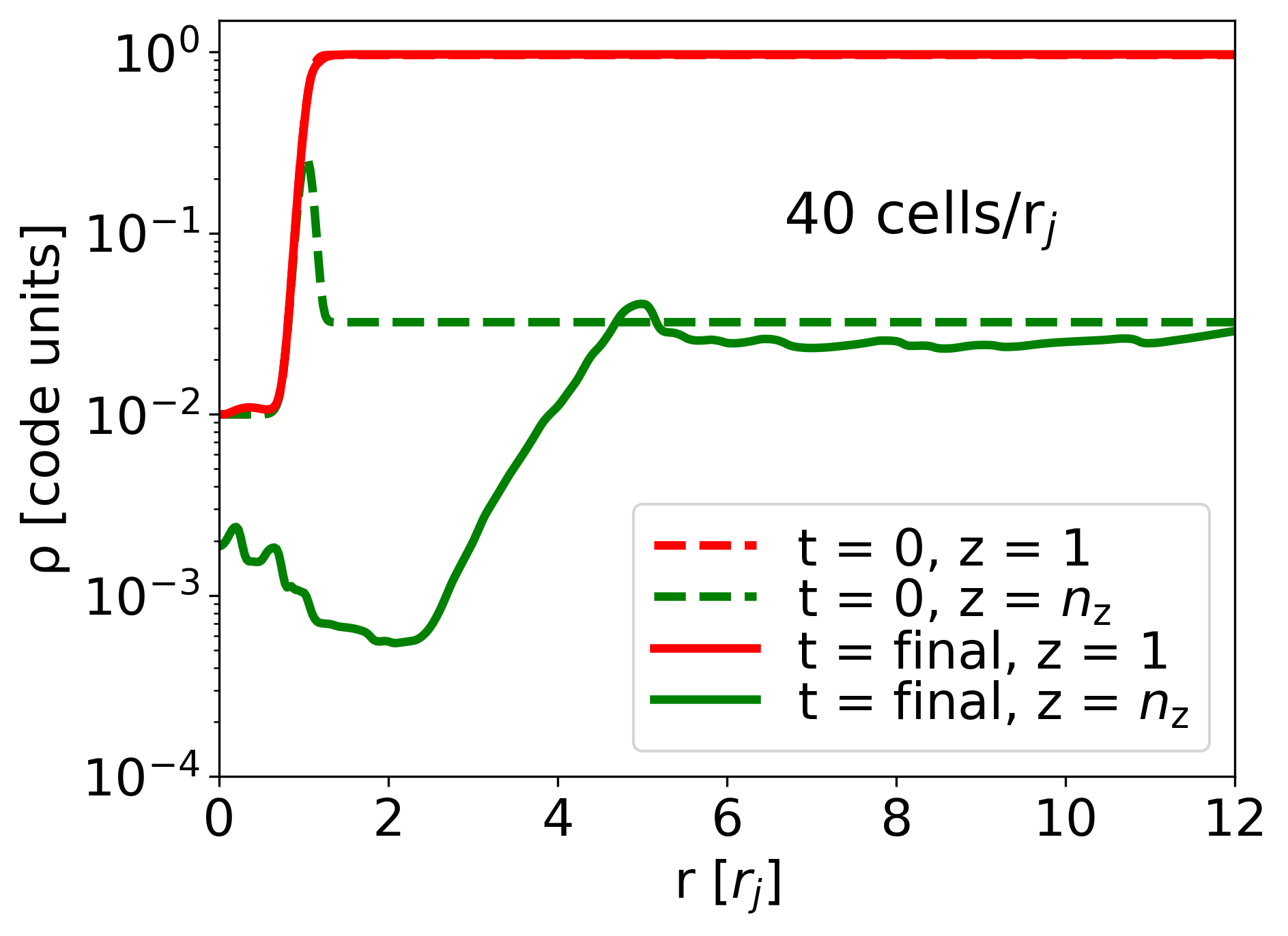}\par
    \includegraphics[width=0.8\linewidth]{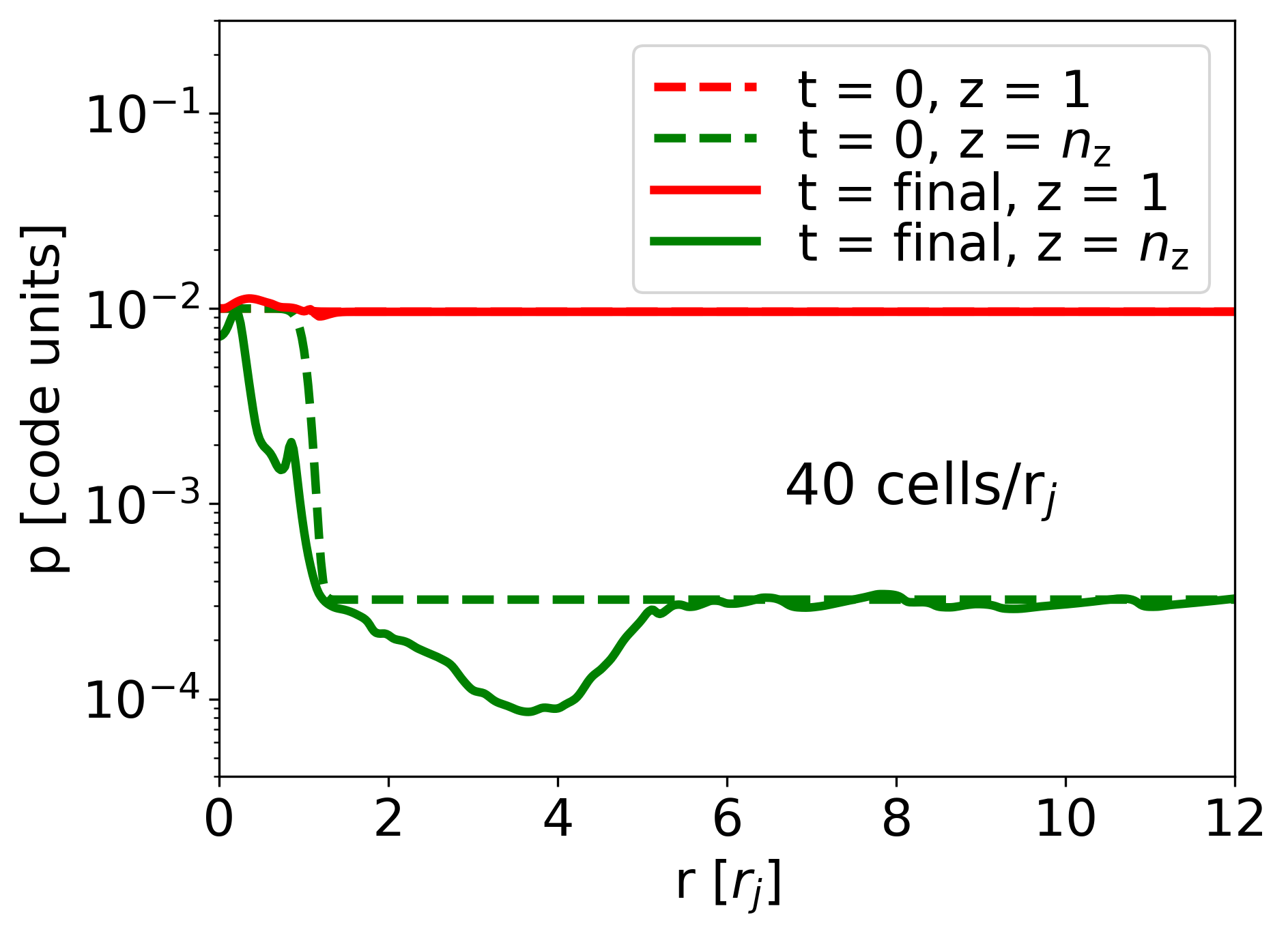}\par
\end{multicols}
\vspace{-0.8cm}
\begin{multicols}{3}
    \includegraphics[width=0.8\linewidth]{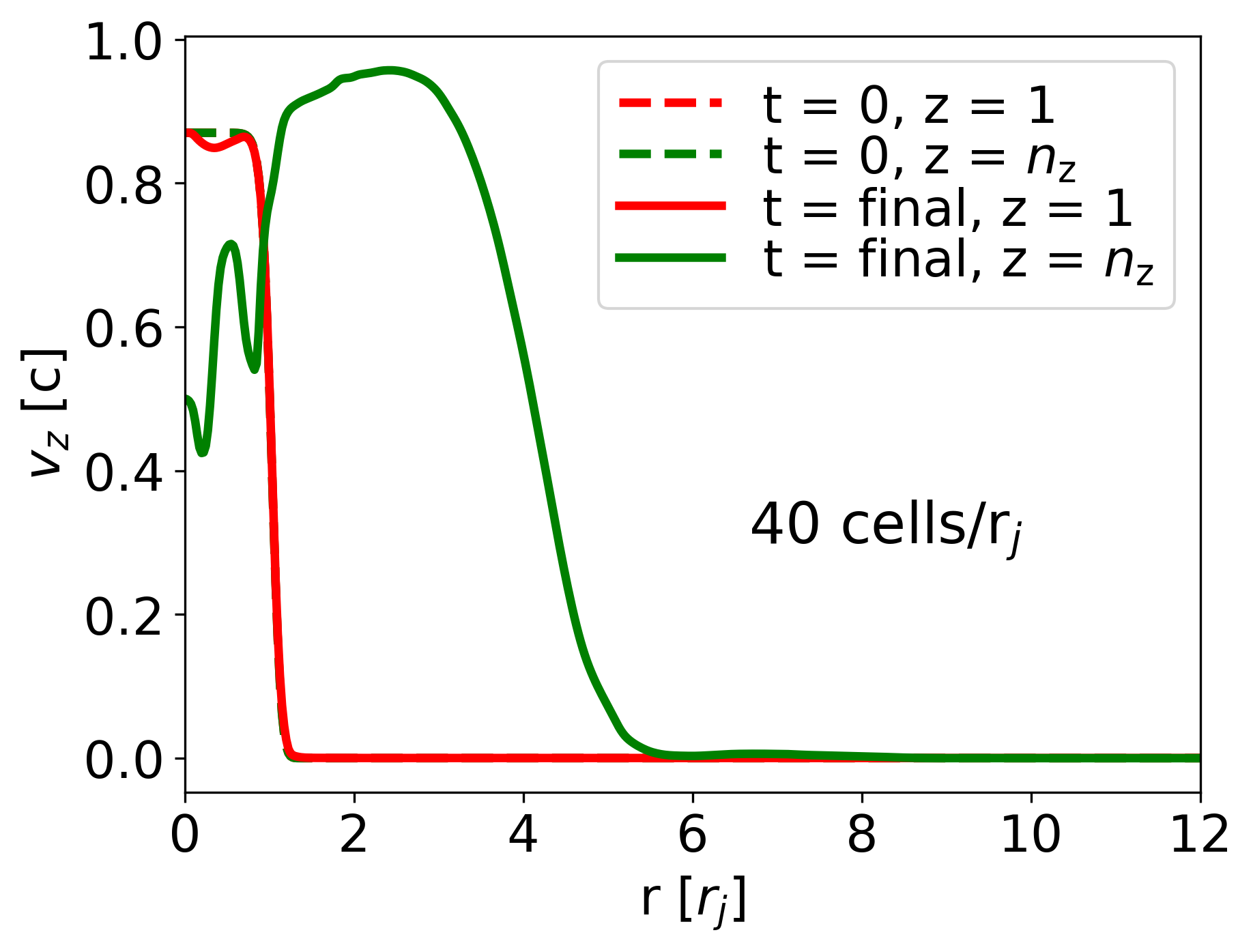}\par
    \includegraphics[width=0.8\linewidth]{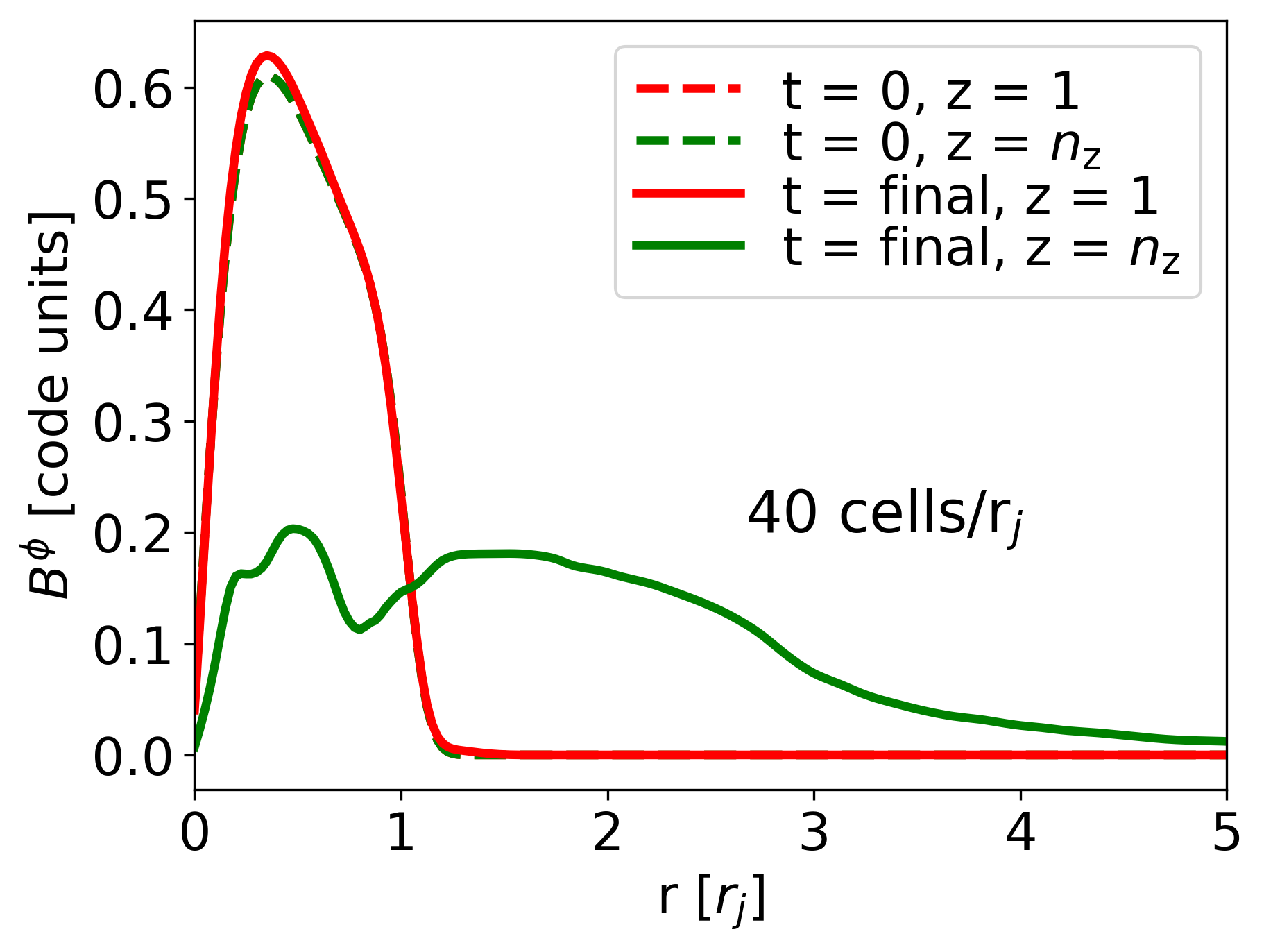}\par
    \includegraphics[width=0.8\linewidth]{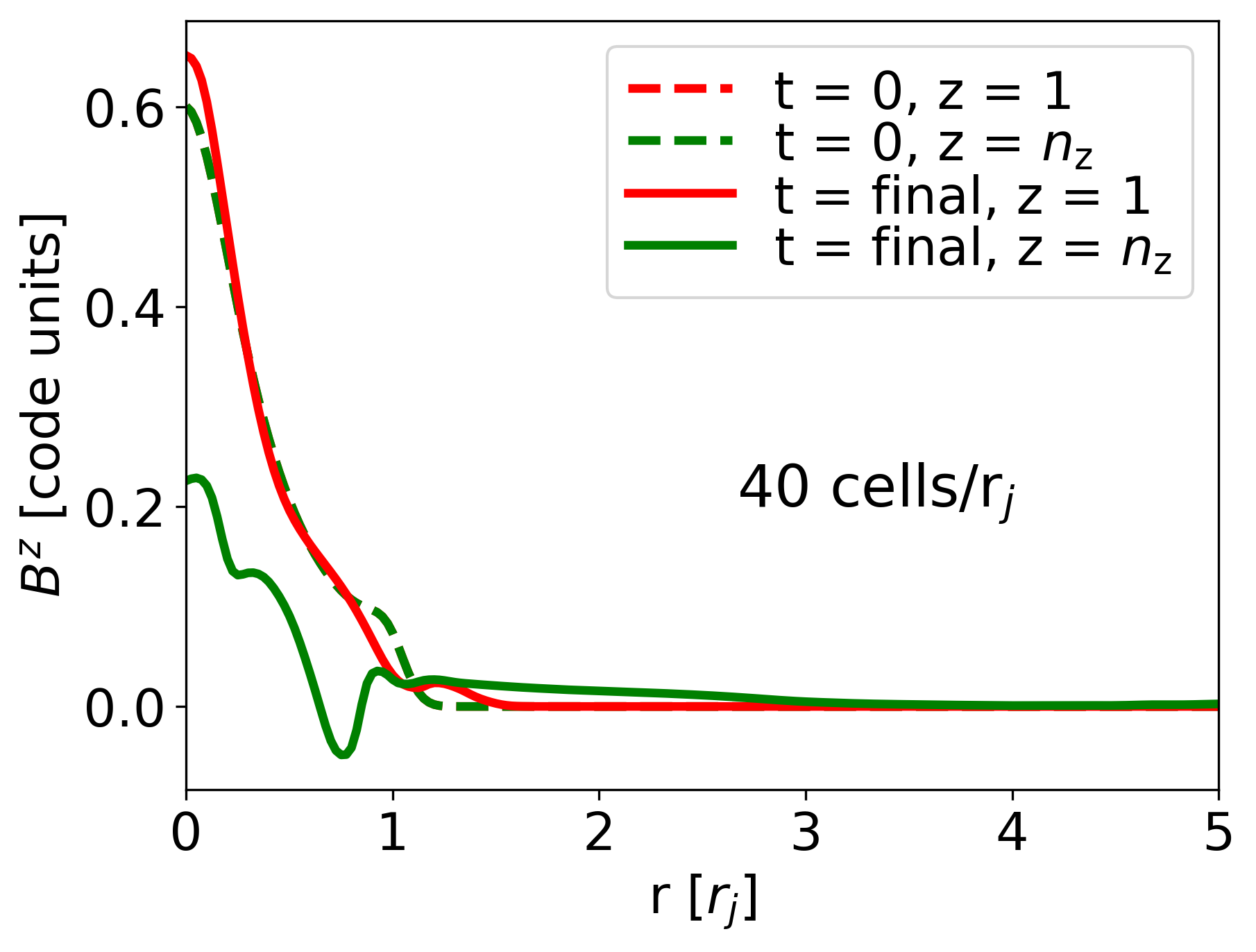}\par
\end{multicols}
    \caption{Radial profiles of the different physical properties for model FMH1 at different distances from the injection (cell 1 in red, half grid in blue, and final cell in green) and at different times (initial in dotted lines and final in continuous lines). In the top panels, we show the radial profiles with a resolution of 10 points, in the middle the ones at a resolution of 20 points, and in the bottom ones the profiles with 40 points (here, the half grid is not shown due to the smaller simulated grid). The convergence in the profiles between 20 and 40 points led us to use resolutions of 20 points for our simulations.}
    \label{fig:radial_profiles_FMH1}
\end{figure*}

The simulation setup consists of a jet that extends the injection conditions along the numerical grid with constant radius $r\,=\,r_j$, with a parameter smoothing function out towards the ambient medium, i.e., a shear layer, as described in Sect.\ \ref{sec:jet_ambient_params_2D}. 
The jet structure is computed considering pressure equilibrium with its environment. 

Jets immediately start their expansion into the under-pressured ambient, transferring energy via radial propagating waves. 
These waves accelerate the ambient medium in the radial direction. 
As a consequence, some material is pushed out of the grid, slightly changing the conditions in the ambient. We have run tests with an extended grid to check the influence that this loss may have and found that the jet is unaffected, probably due to the large jet overpressure, which does not change significantly. 
For proofs in support of this and for a proper description of the pressure wave and its evolution, see Appendix \ref{app:pressure_wave}. 

We run the simulations for the necessary amount of time for the jets to reach equilibrium after the expansion is completed.
To determine when the steady solution is reached we use the radial velocity as the indicator.
Specifically, we consider equilibrium to have been reached in a certain model when $|v^r| < 0.01$ all over the grid, with the exception of jet expansion/recollimation associated with the achieved equilibrium configuration. 
An example of this is shown in Fig.\ \ref{fig:velx_equilibrium} for model FMH1\_m4. 
In the figure, the white contours highlight the levels of the tracer, i.e., the jet mass fraction.
In this case, outside of the jet, the grid shows radial velocities lower than 0.01 everywhere.

Figure \ref{fig:radial_profiles_FMH1} shows an example of radial equilibrium profiles at different locations along the jet and for different resolutions, for model FMH1. We show the radial distribution of different physical quantities at 
$t = 0$ and at the final step, at three different distances, 
injection (cell 1), half grid (cell $n_\mathrm{z}/2$), and $\mathrm{z}=\mathrm{z}_{\rm max}$ (cell $n_\mathrm{z}$).
The figure shows three sets of panels for the results at three different numerical resolutions: the upper set shows the results for a simulation with 10~cells/$r_j$, the middle set shows the case with 20~cells/$r_j$, while the bottom set shows the case with 40~cells/$r_j$.
For the latter case we show only the initial and final cell profiles, since we used a smaller grid of $[30\,r_j\times 30\,r_j]$ for computational time reasons.
The bottom right plot of the 10~cells/$r_j$ case shows a change of the axial field at injection, which implies a discontinuity with respect to the boundary condition. This problem, caused by numerical diffusion, disappears at 20 and also at 40~cells/$r_j$. Having found convergence between our results at 20~cells/$r_j$, we adopt this resolution for all our simulations.
In Appendix \ref{app:resolution_study} we show further proofs in support of this choice.

The plots show how the initial condition is 
preserved at injection, but the jet expands significantly, becoming much wider along the grid. 
These plots also show an increase in axial velocity, which appears to be displaced from the axis. 
We will discuss this result in the next Section.

\subsection{Jet acceleration and expansion profiles} \label{sec:acceleration}

\begin{figure*}[htpb]
    \centering
\begin{multicols}{4}
    \includegraphics[width=\linewidth]{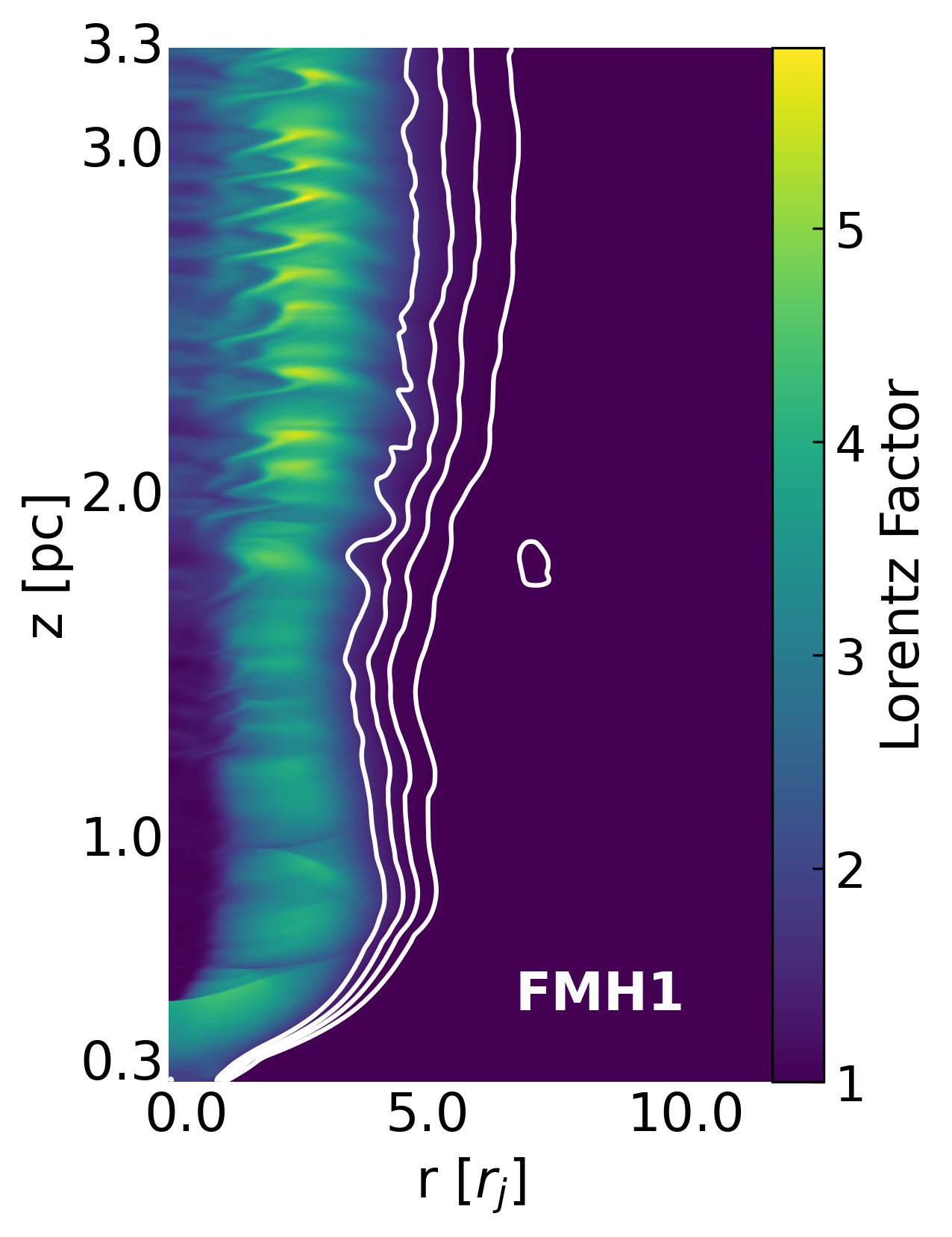}\par
    \includegraphics[width=\linewidth]{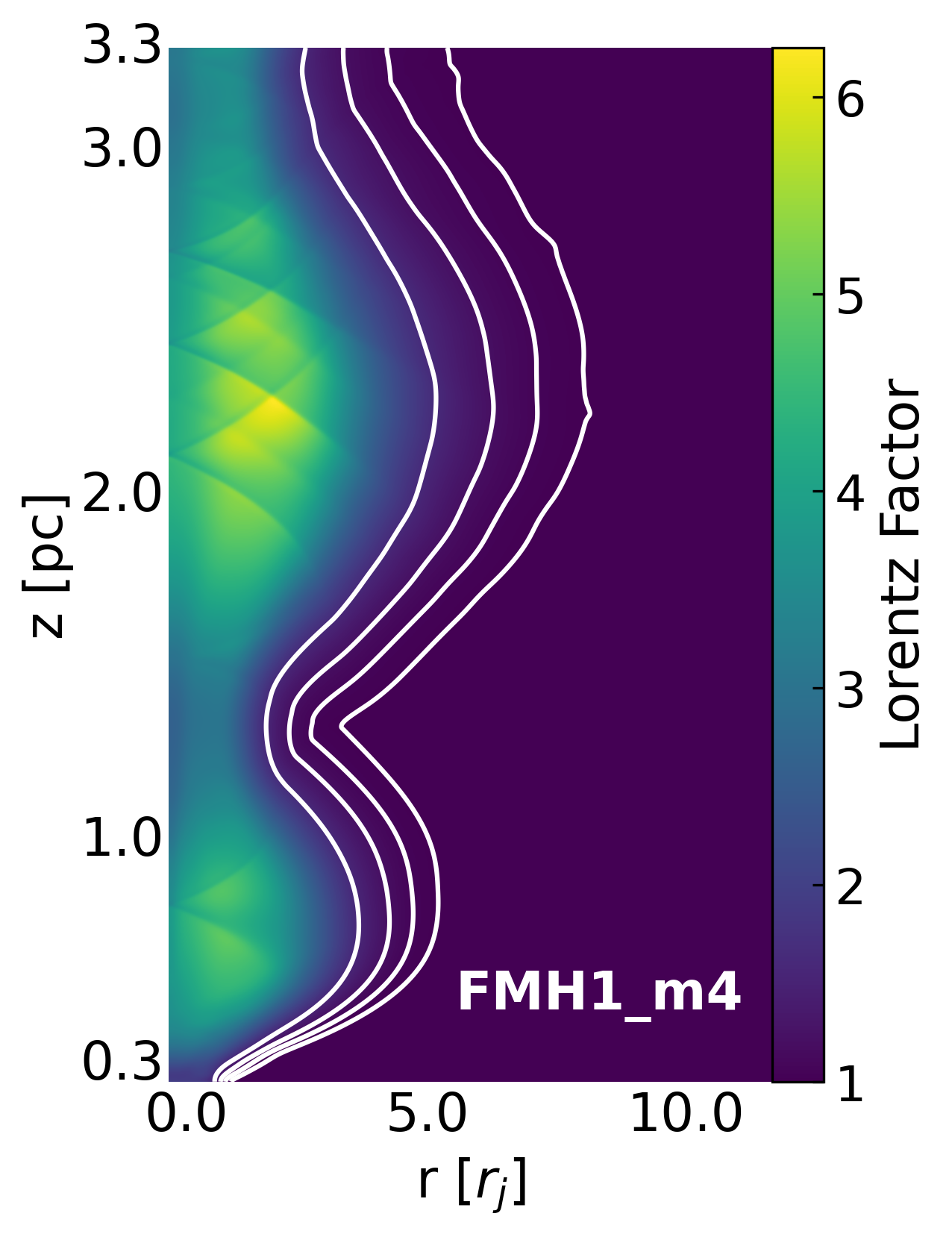}\par
    \includegraphics[width=\linewidth]{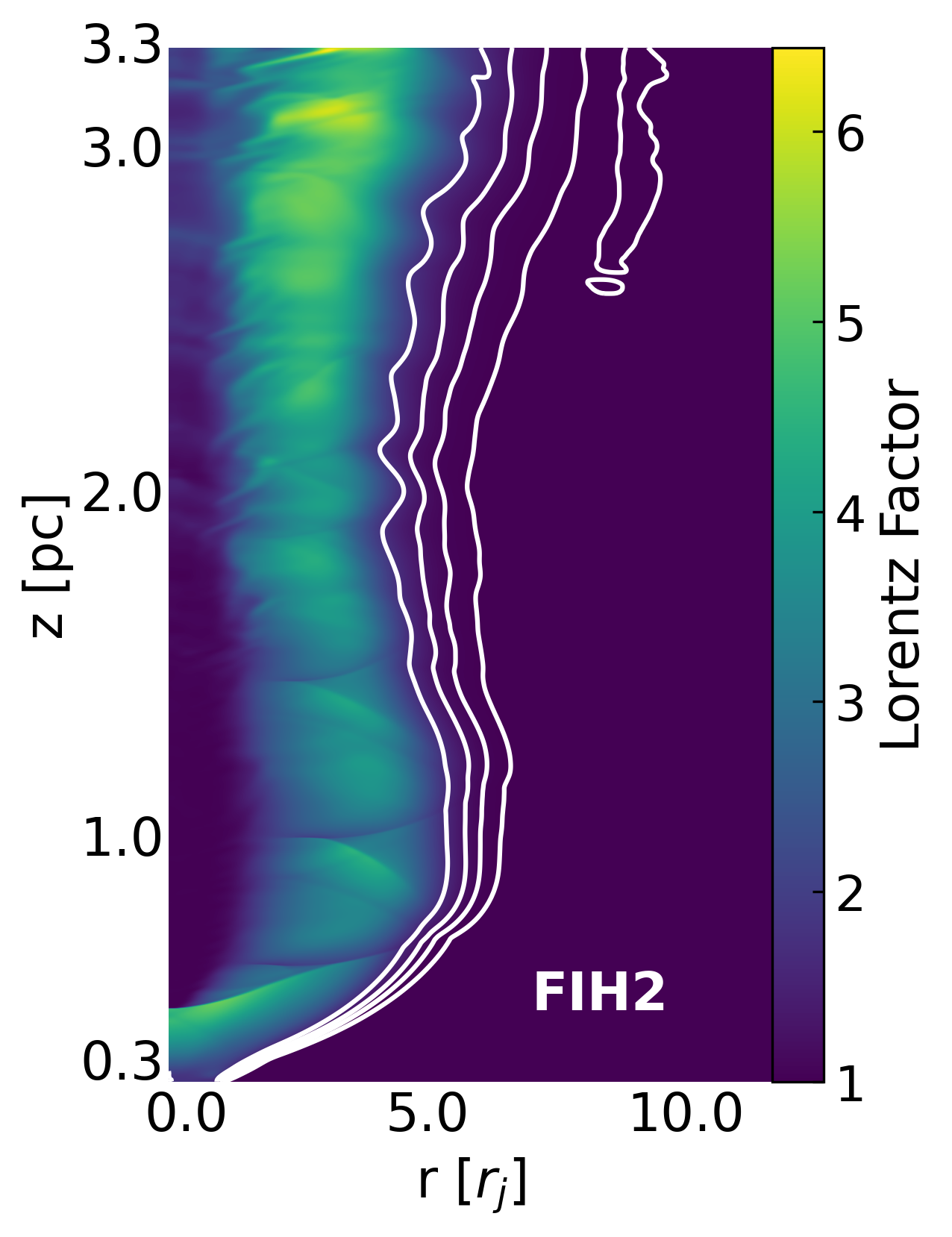}\par
    \includegraphics[width=\linewidth]{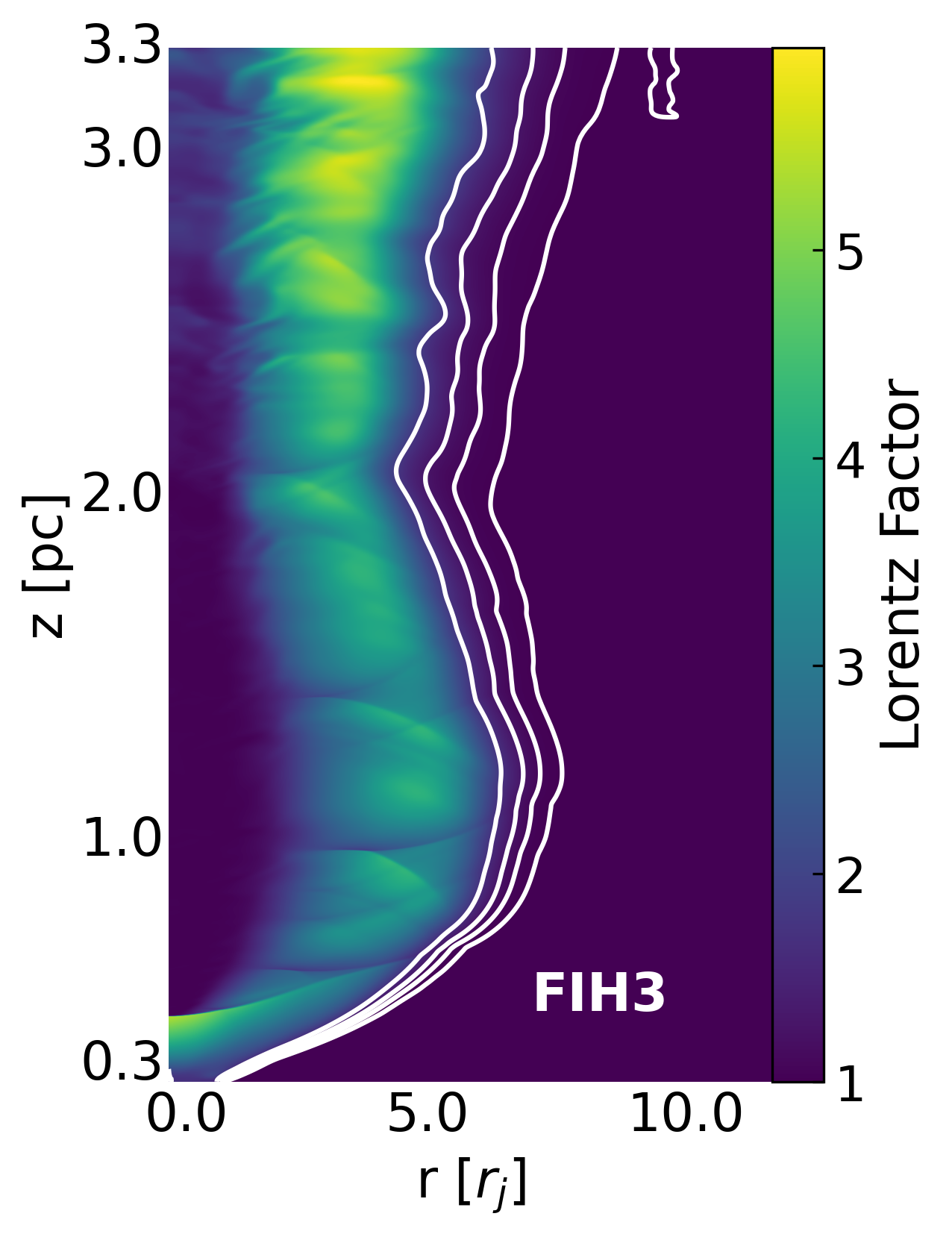}\par
\end{multicols}
\vspace{-0.8cm}
\begin{multicols}{4}
    \includegraphics[width=\linewidth]{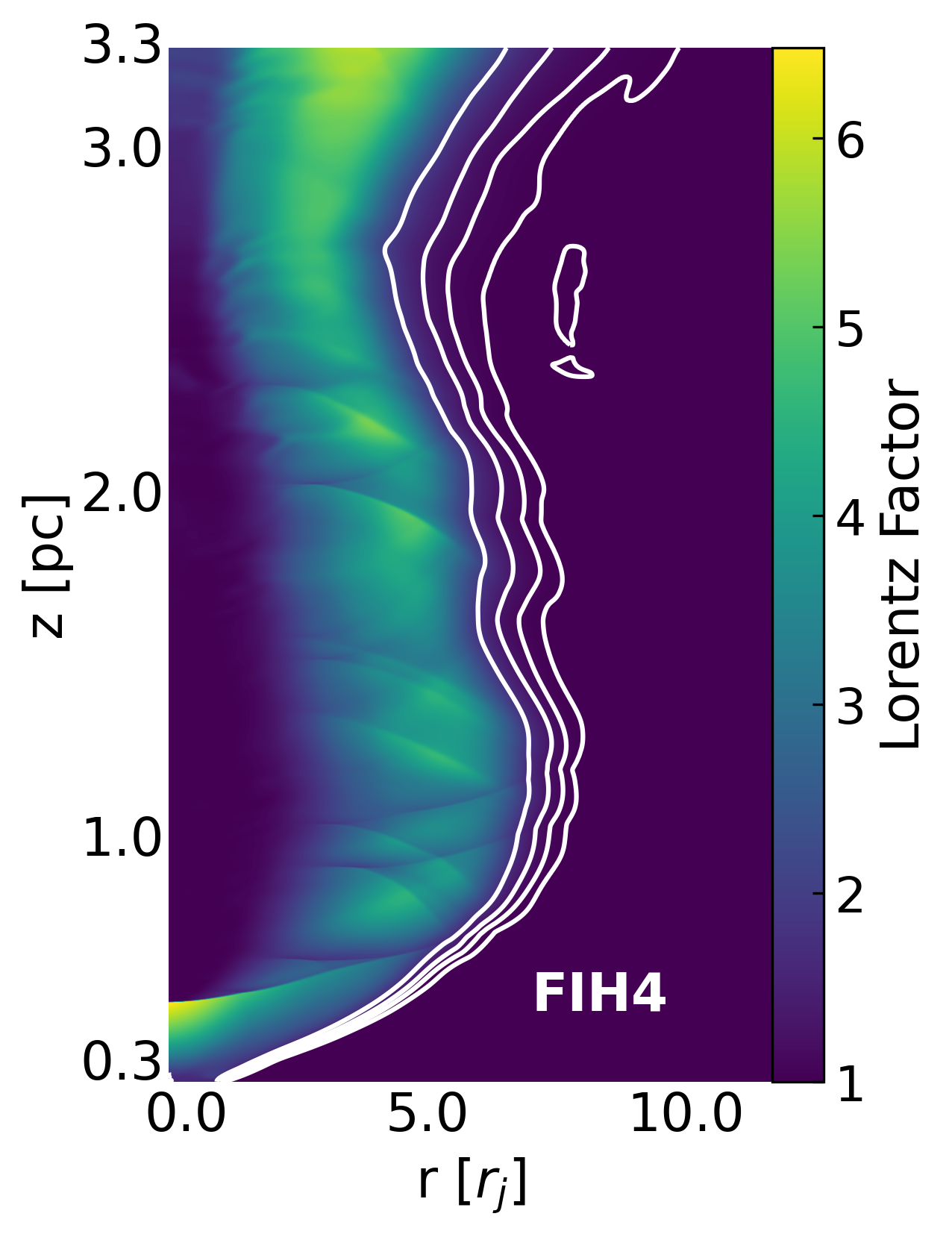}\par
    \includegraphics[width=\linewidth]{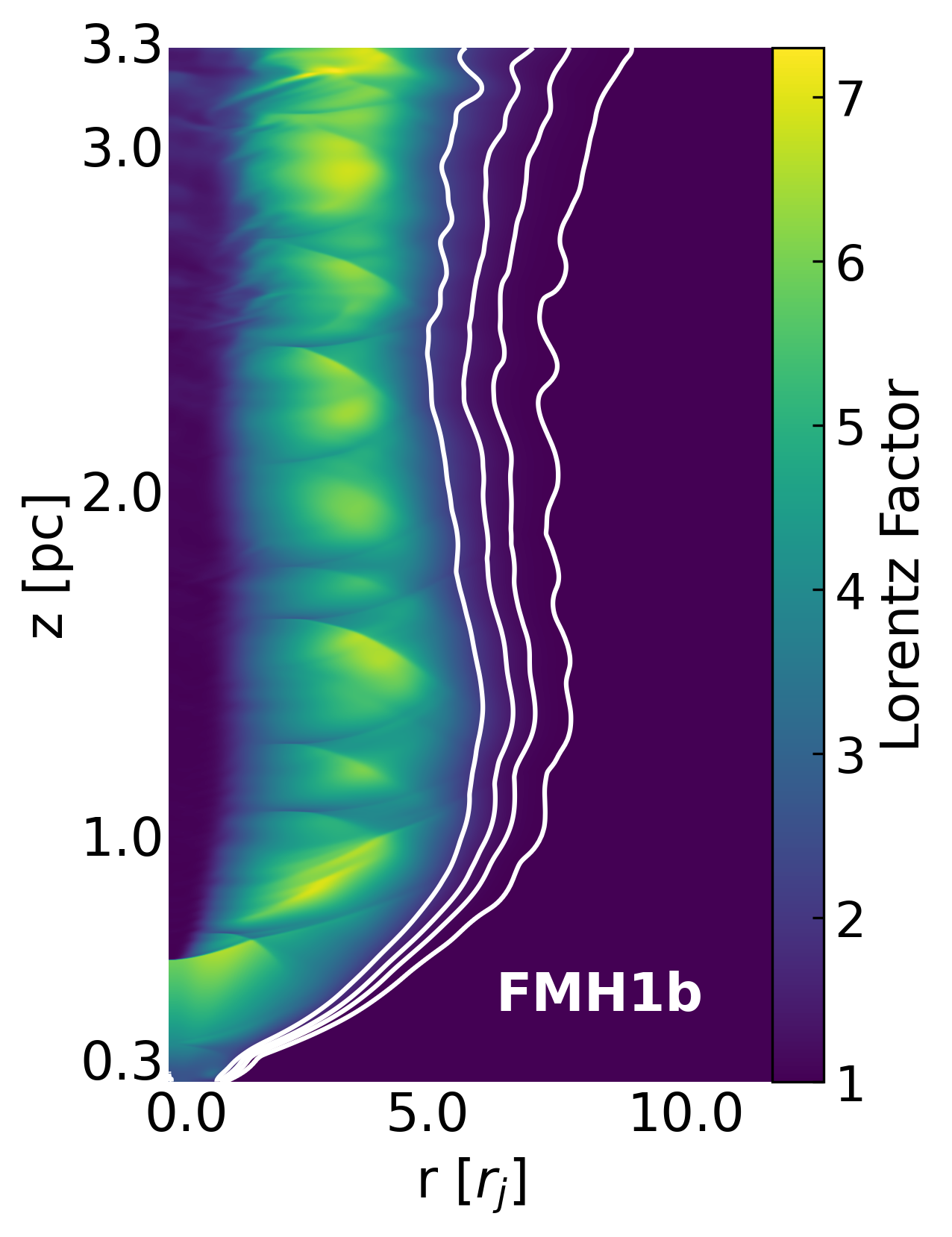}\par
    \includegraphics[width=\linewidth]{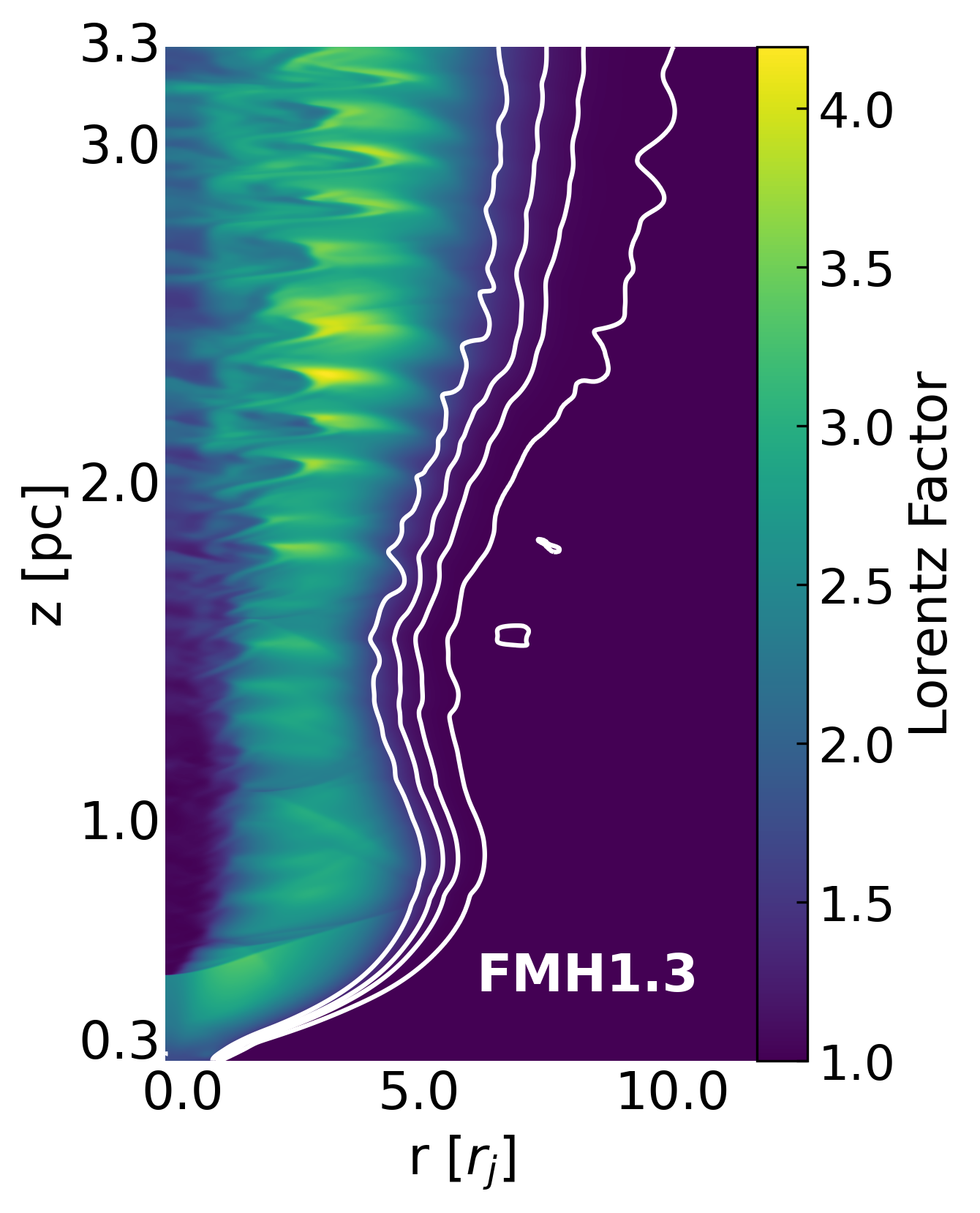}\par
    \includegraphics[width=\linewidth]{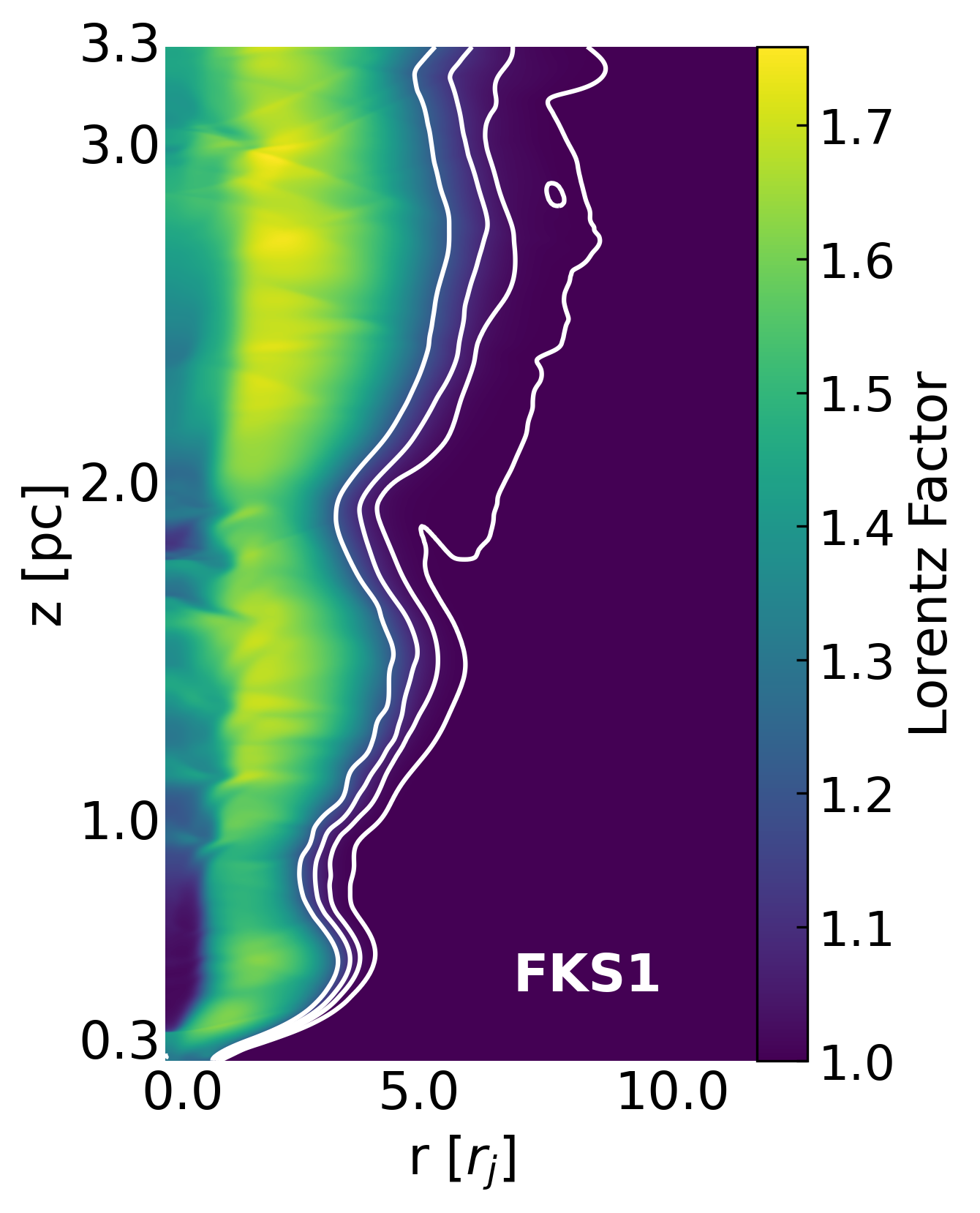}\par
\end{multicols}
\vspace{-0.8cm}
\begin{multicols}{4}
    \includegraphics[width=\linewidth]{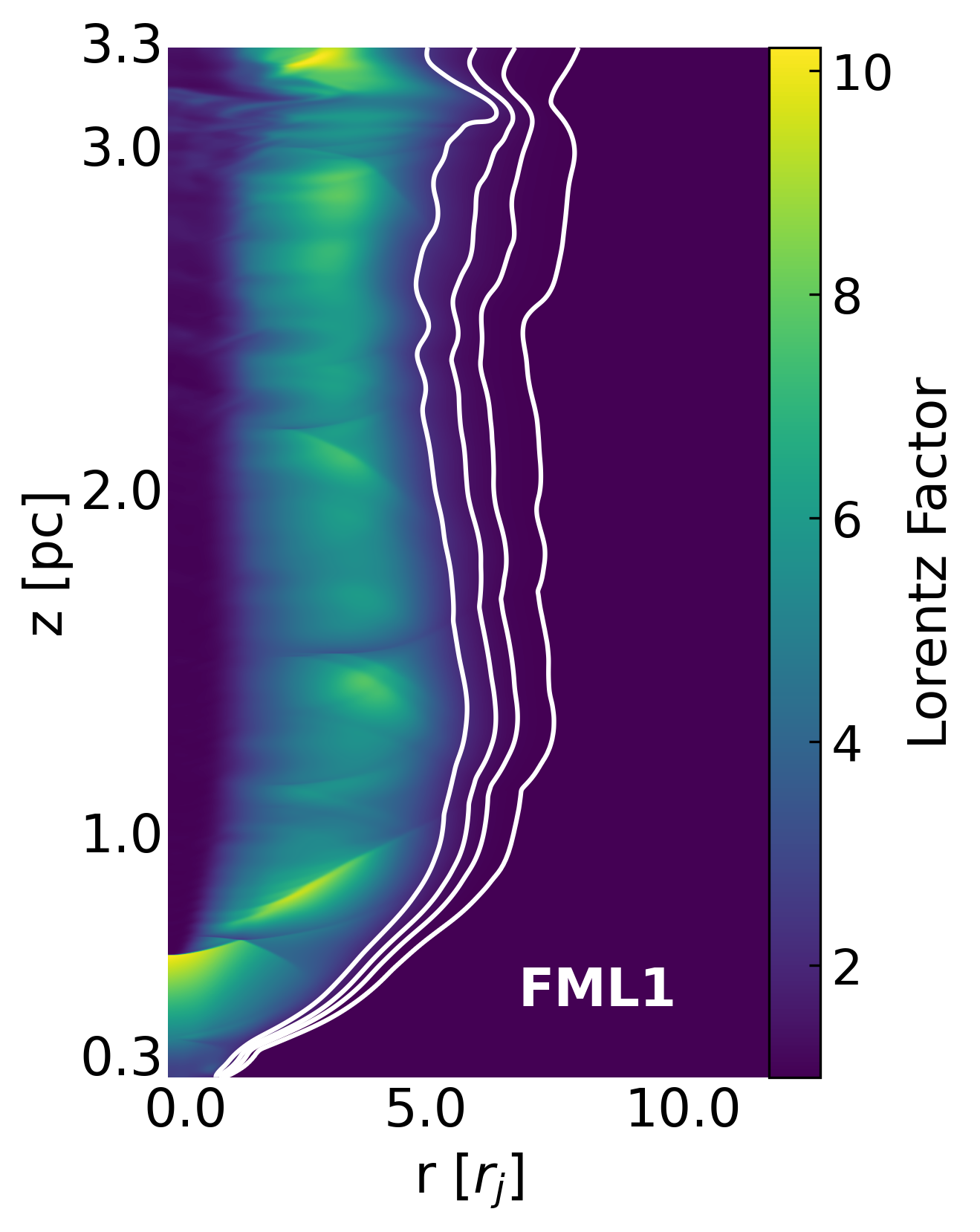}\par
    \includegraphics[width=\linewidth]{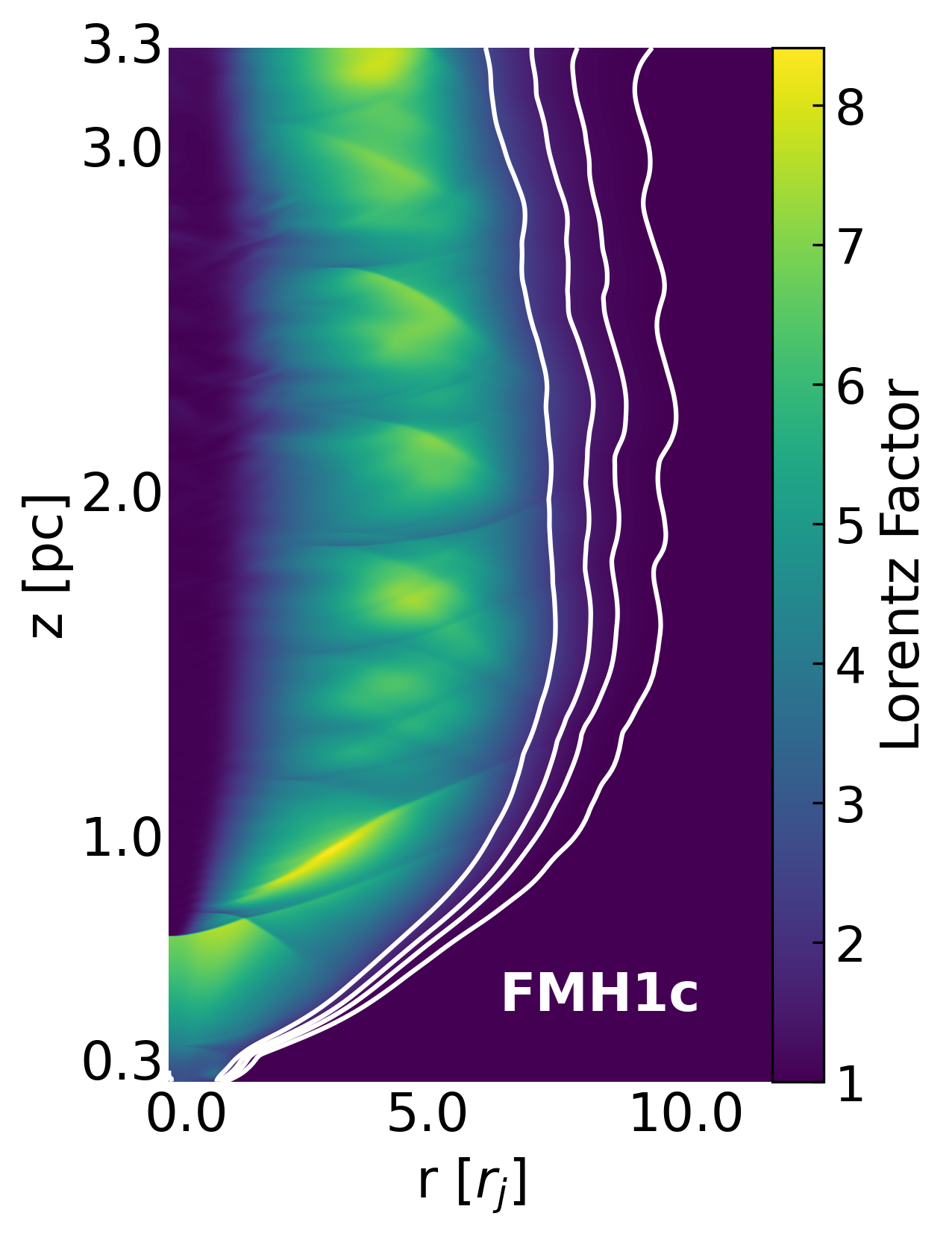}\par
    \includegraphics[width=\linewidth]{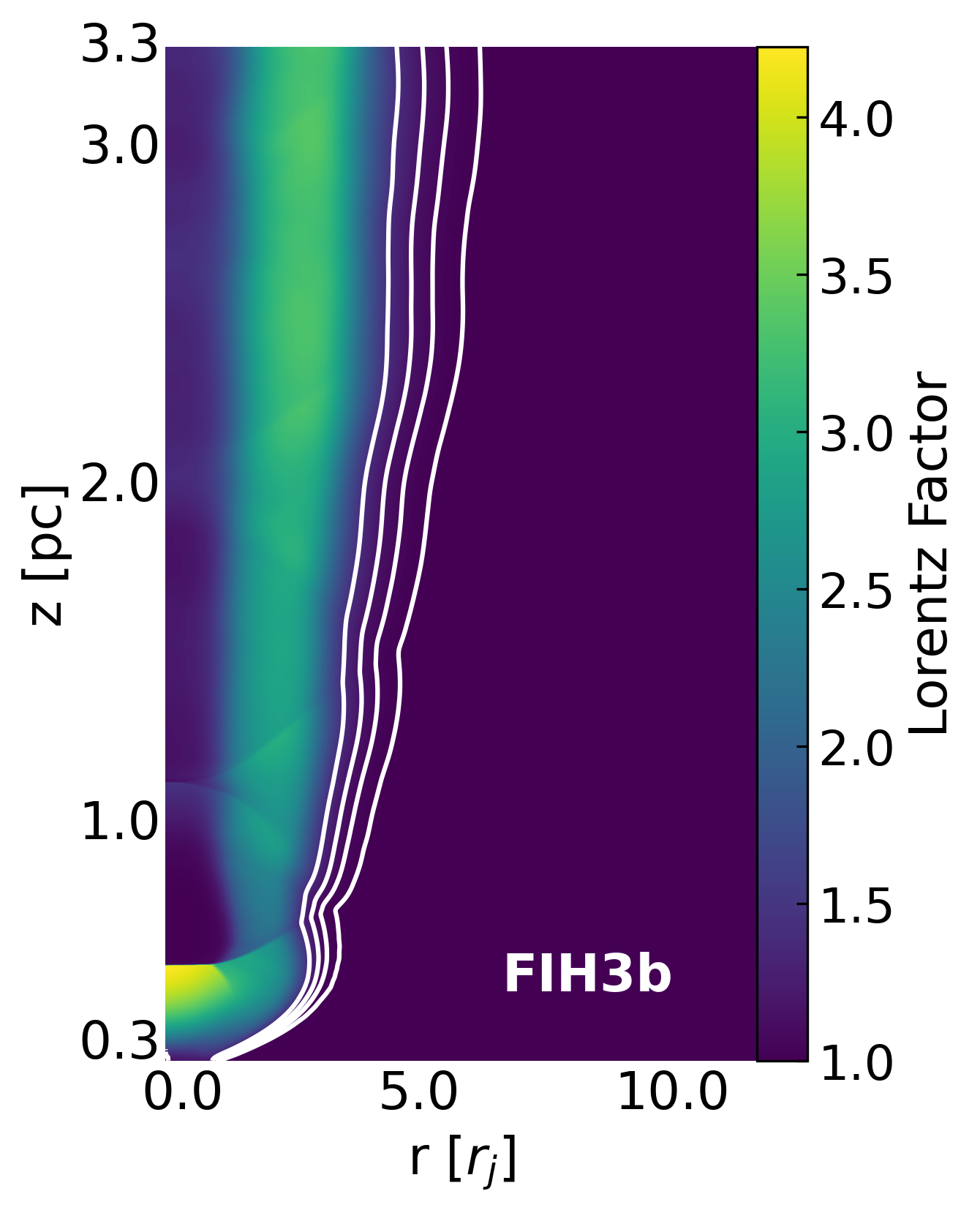}\par
    \includegraphics[width=\linewidth]{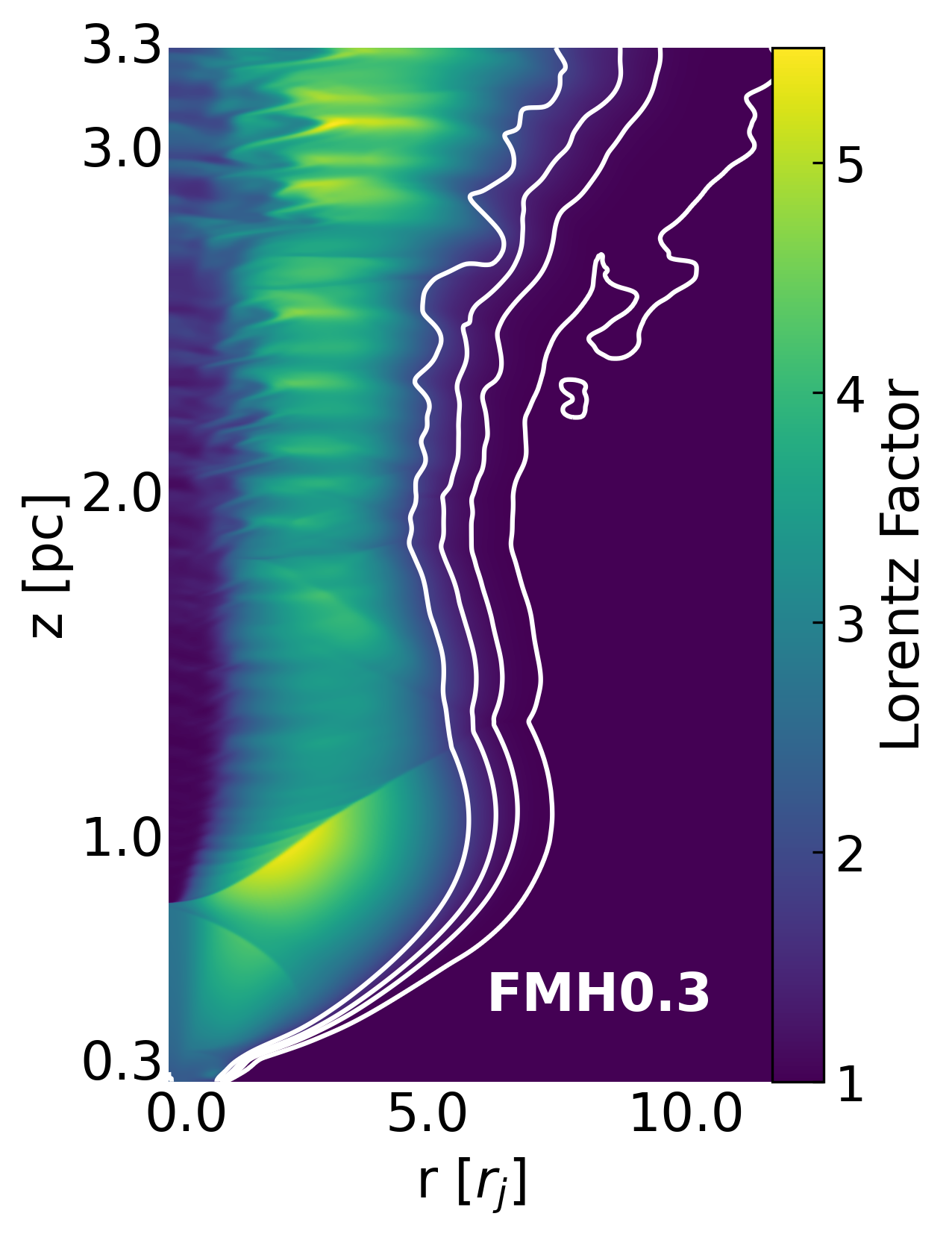}\par
\end{multicols}
\vspace{-0.8cm}
\begin{multicols}{4}
    \includegraphics[width=\linewidth]{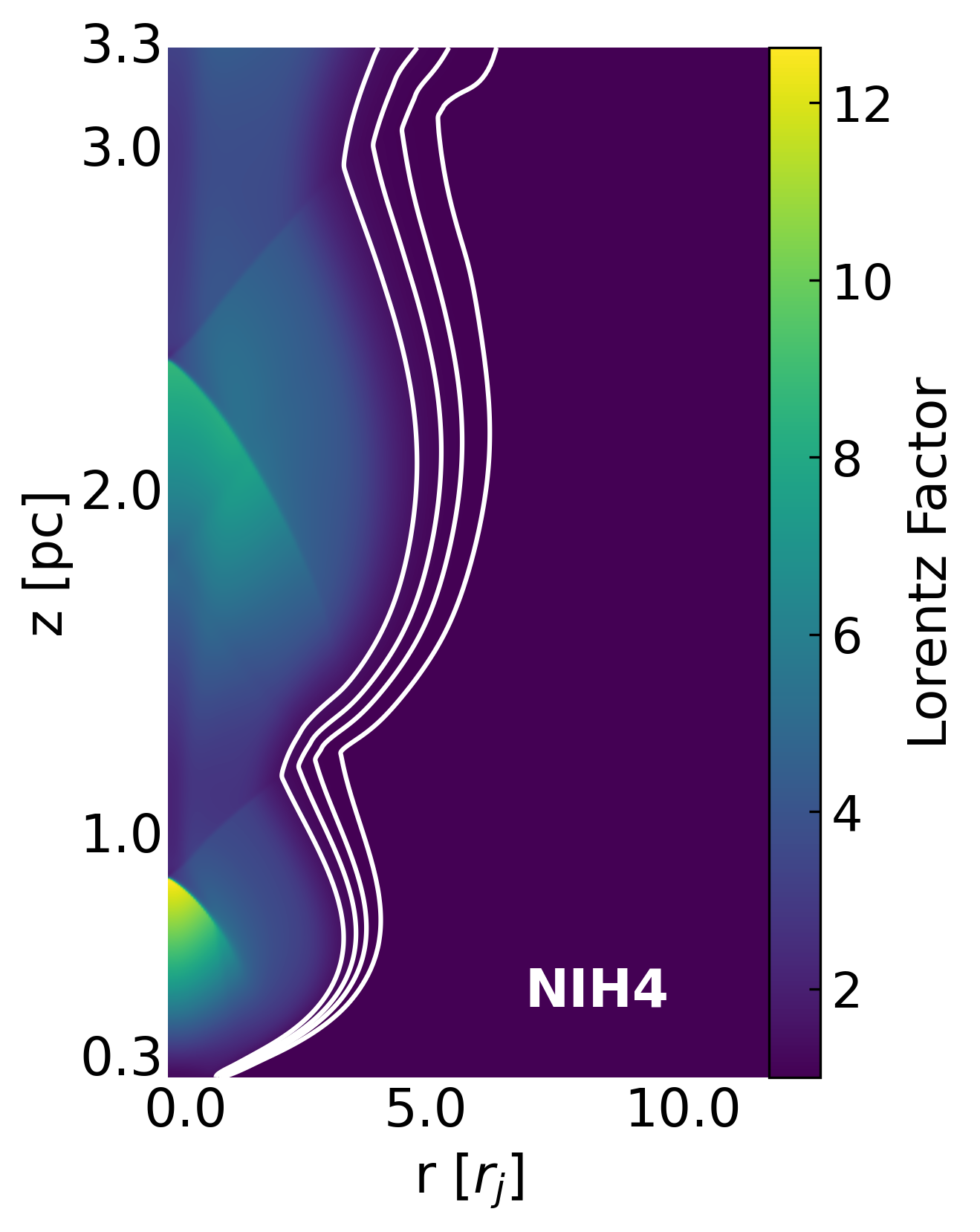}\par
    \includegraphics[width=\linewidth]{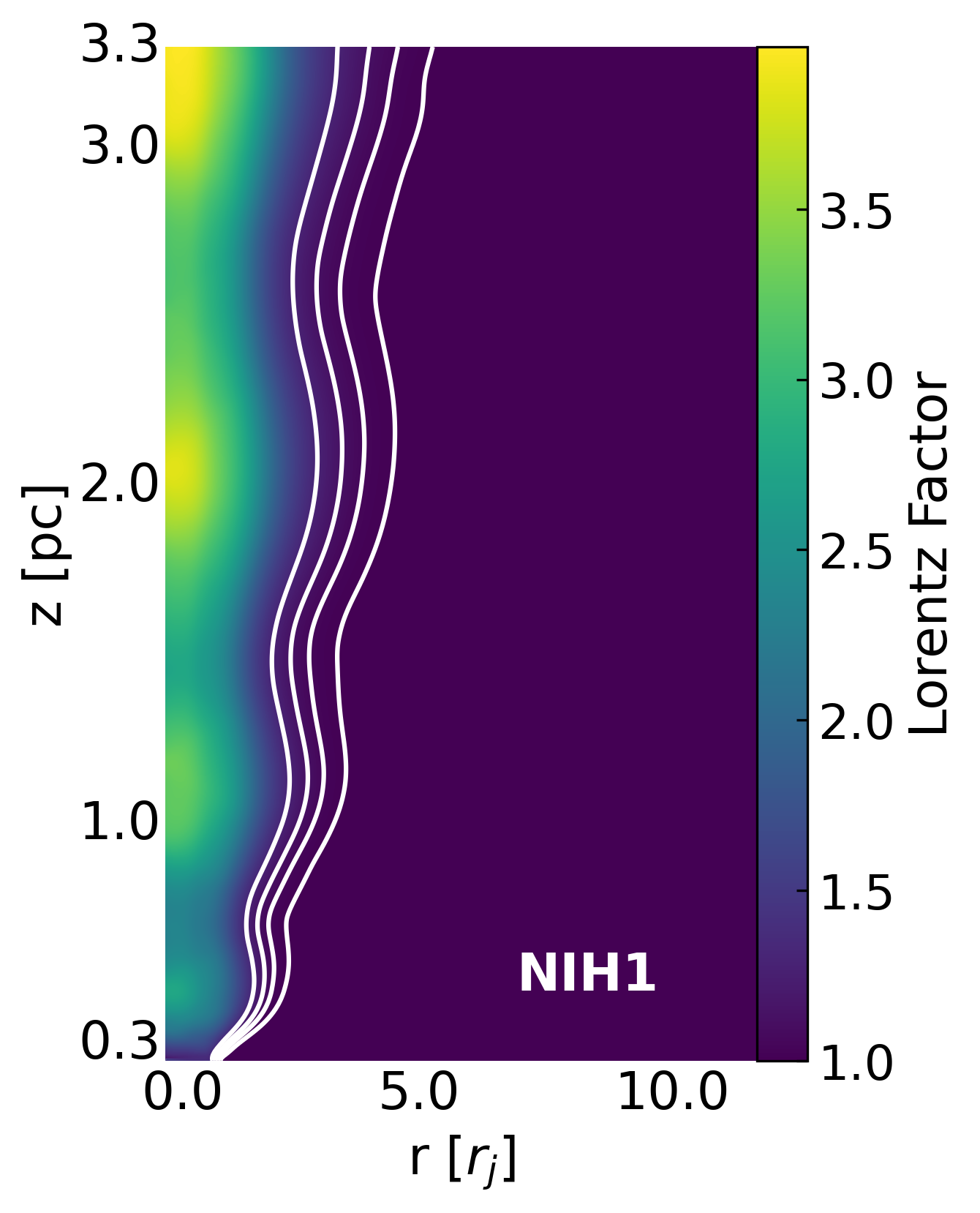}\par
\end{multicols}
    \caption{Lorentz factor maps for the simulated models. The white contours are representative of the tracer at levels of 0.2, 0.4, 0.6, 0.8. The results show the clear correlation between the the jet speed profiles and the intrinsic properties, with the main differences shown between force-free and non-force-free along with magnetically dominated or internally dominated models. 
    For details, see the text.}
    \label{fig:different_Lorentz_factor}
\end{figure*}

\begin{figure*}[h!]
    \centering
\begin{multicols}{4}
    \includegraphics[width=\linewidth]{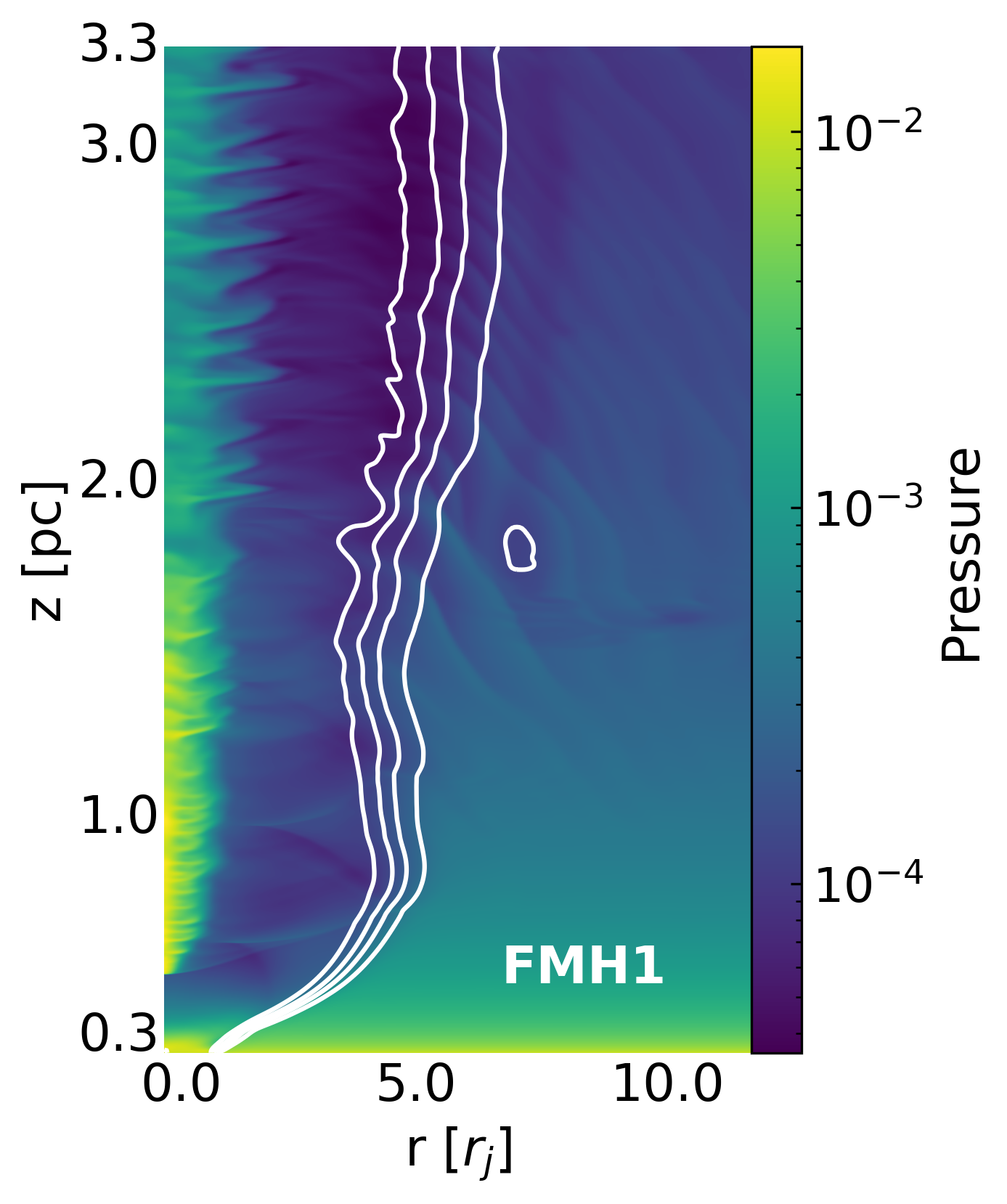}\par
    \includegraphics[width=\linewidth]{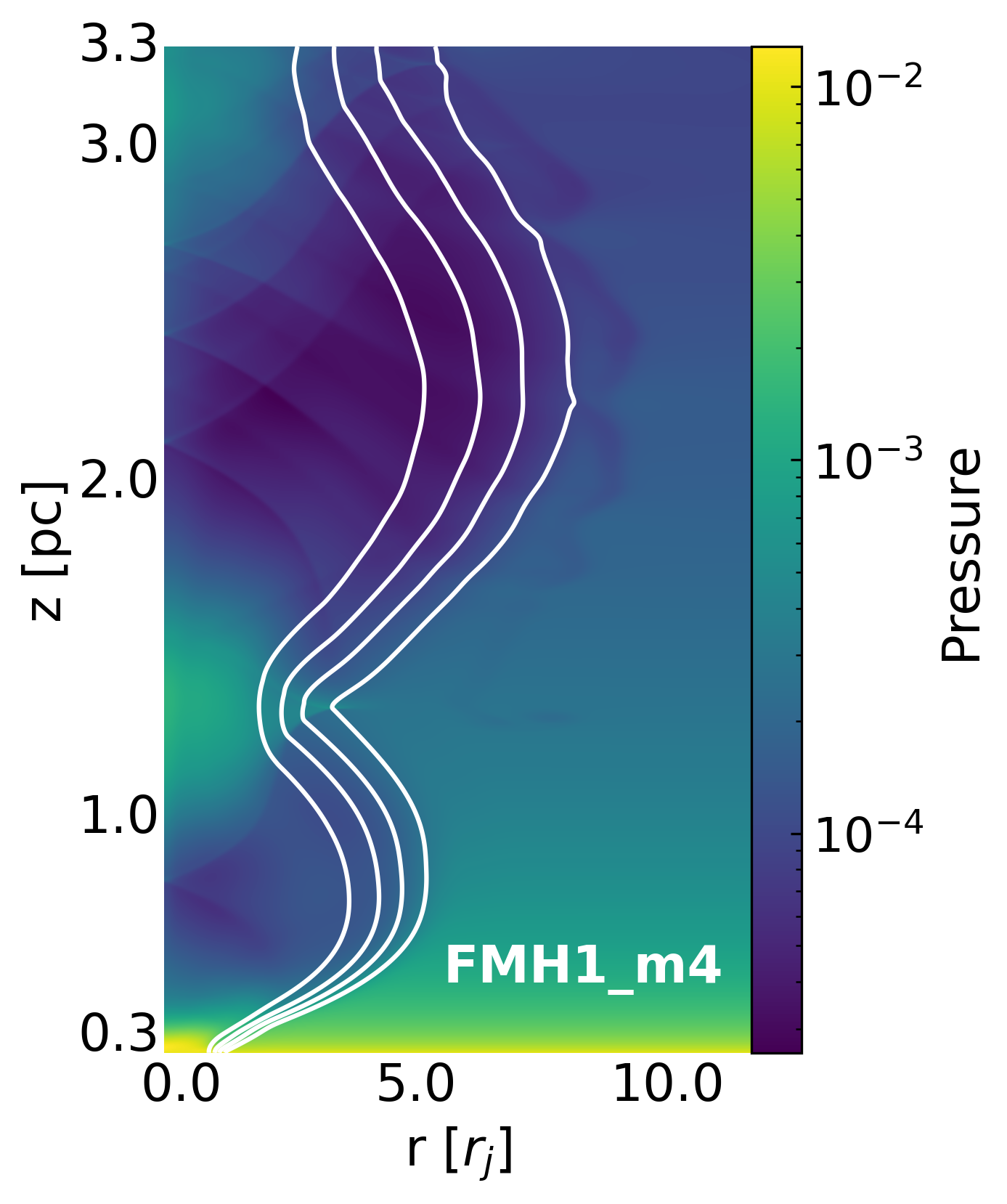}\par
    \includegraphics[width=\linewidth]{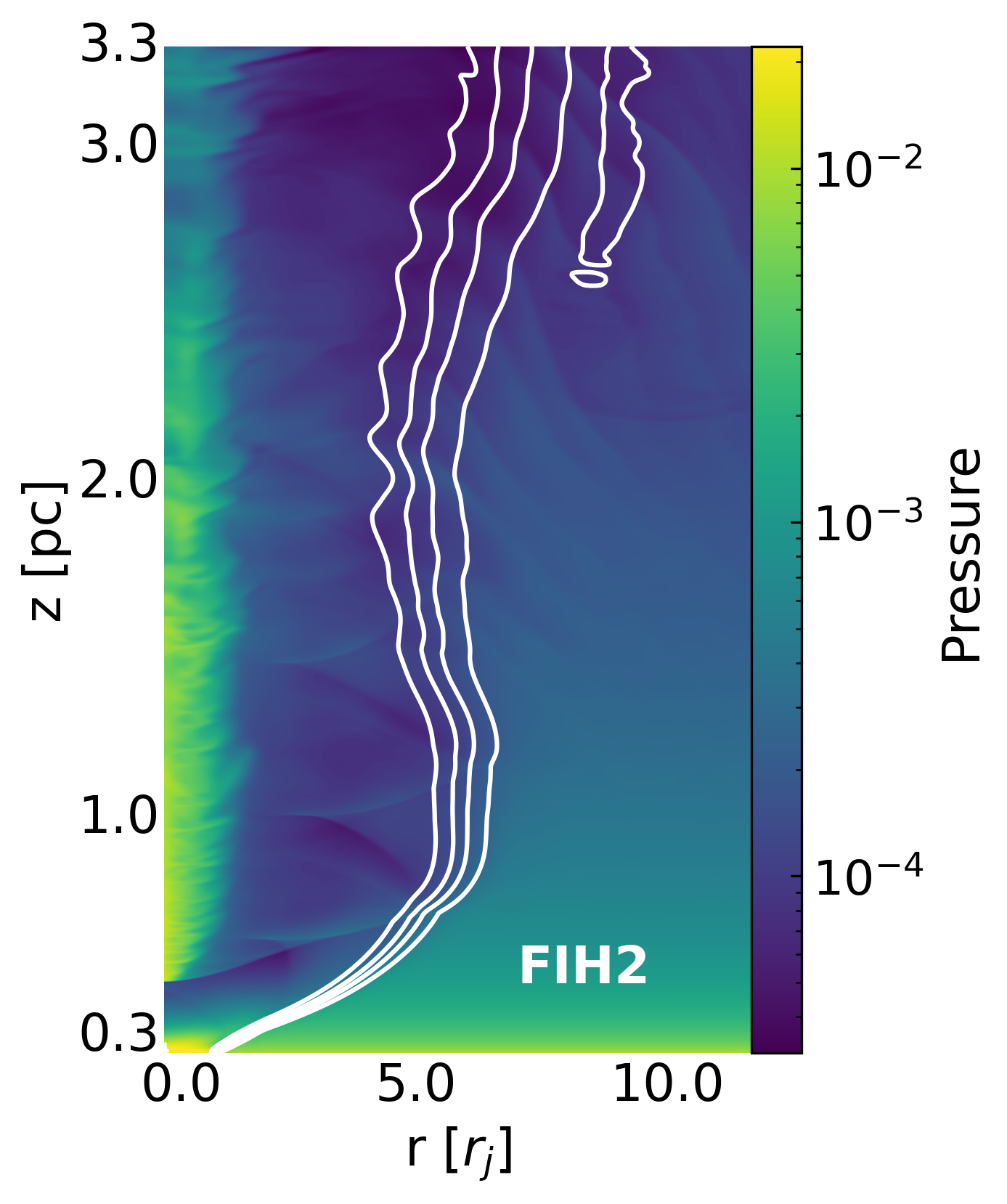}\par
    \includegraphics[width=\linewidth]{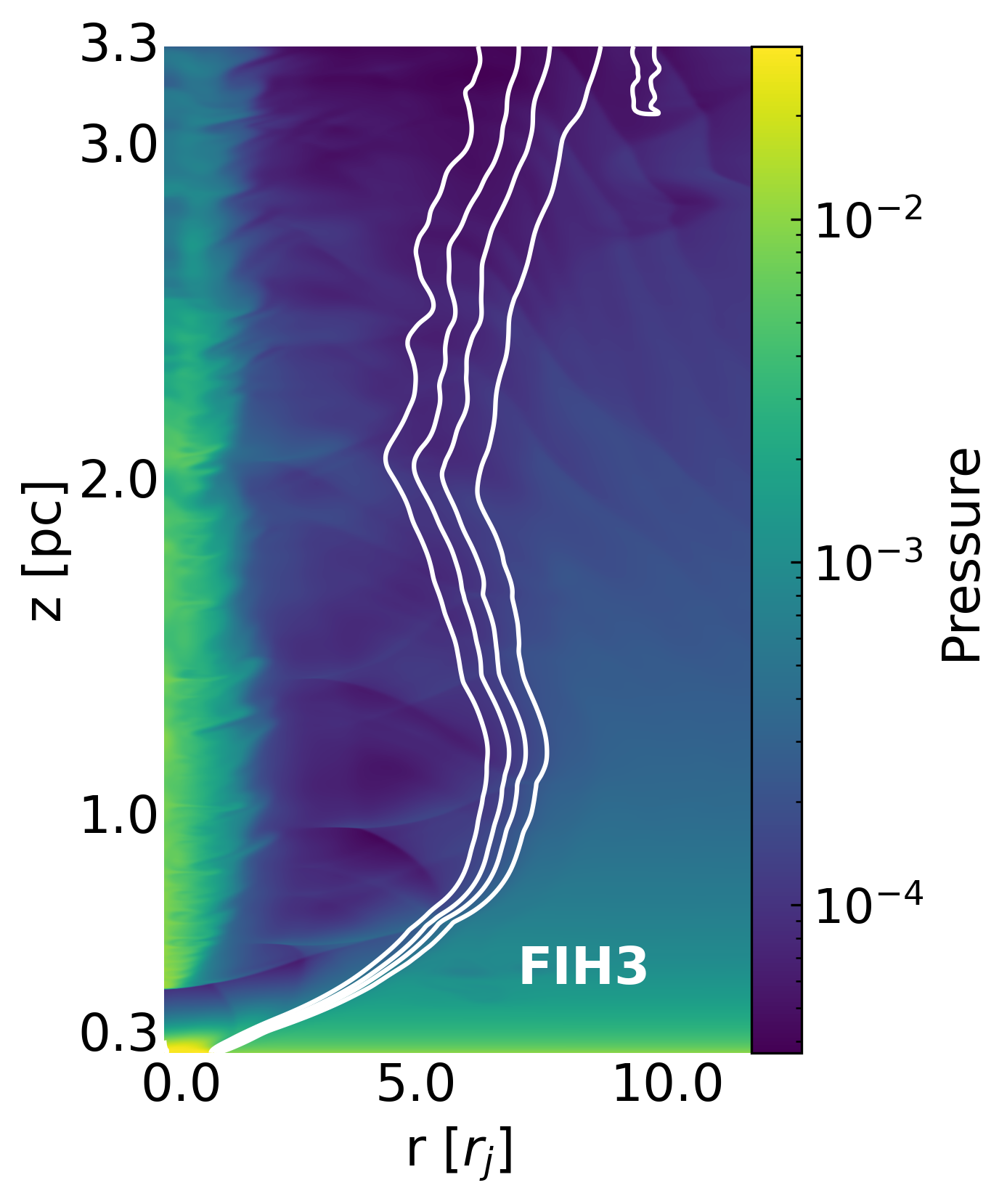}\par
\end{multicols}
\vspace{-0.8cm}
\begin{multicols}{4}
    \includegraphics[width=\linewidth]{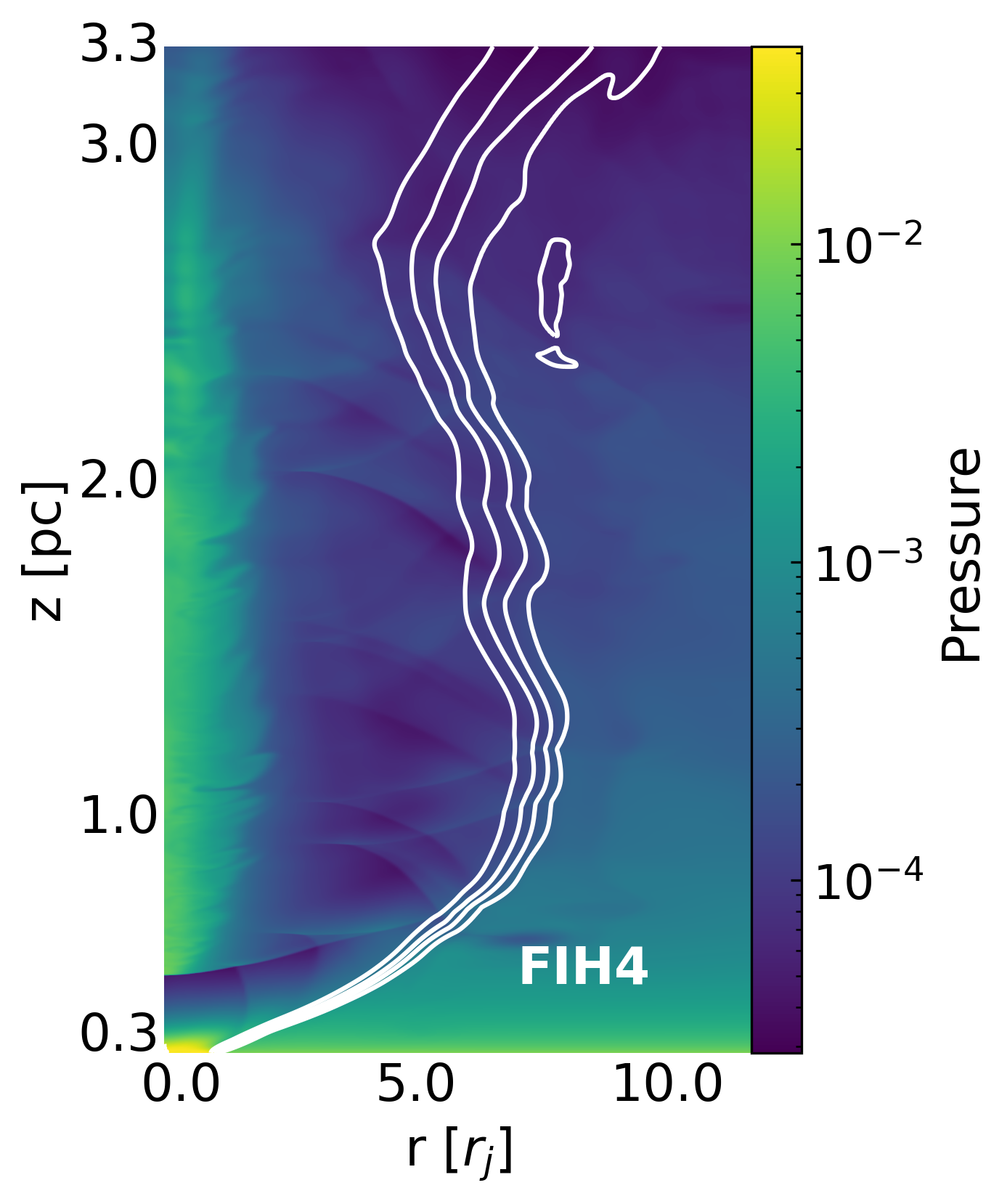}\par
    \includegraphics[width=\linewidth]{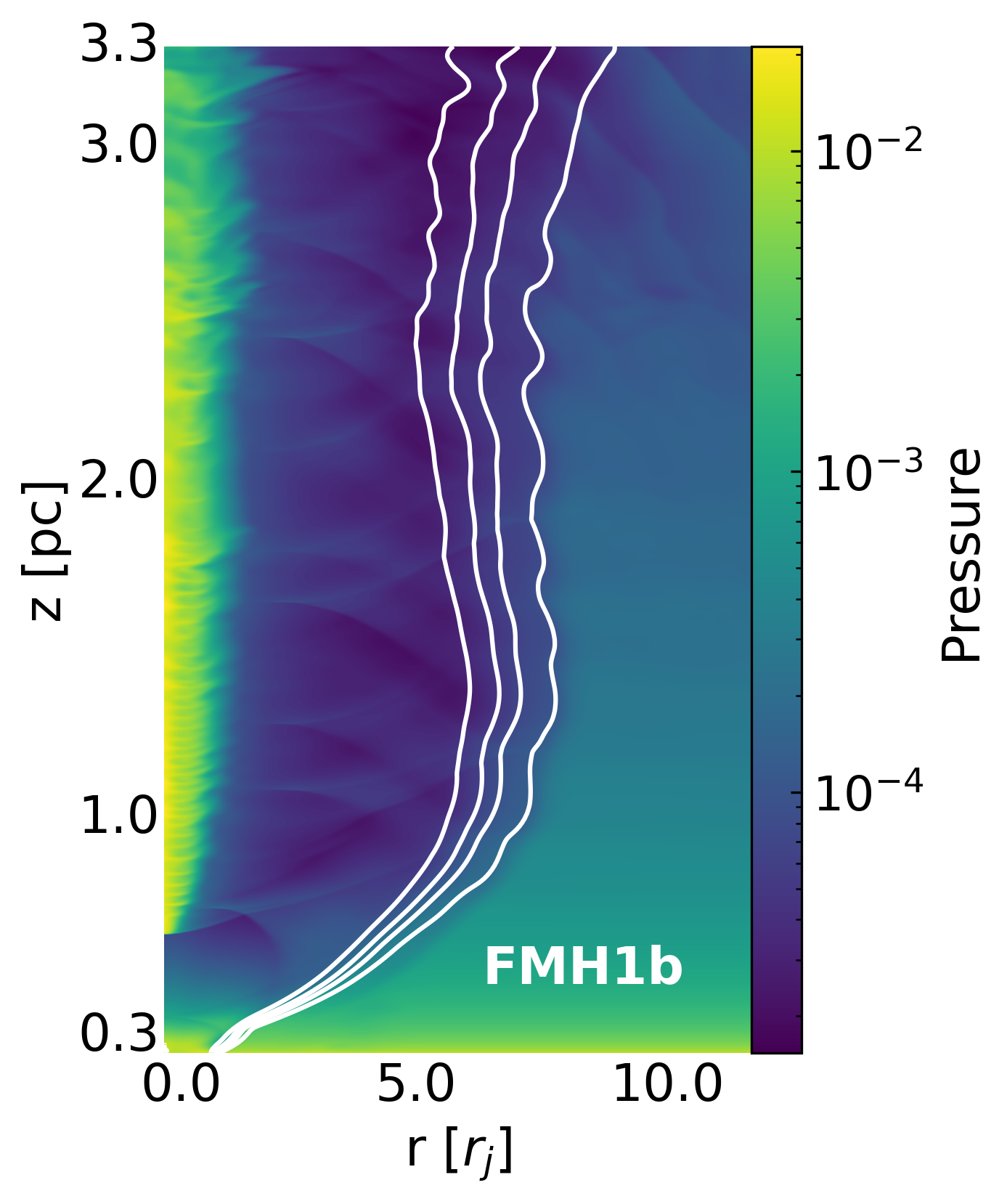}\par
    \includegraphics[width=\linewidth]{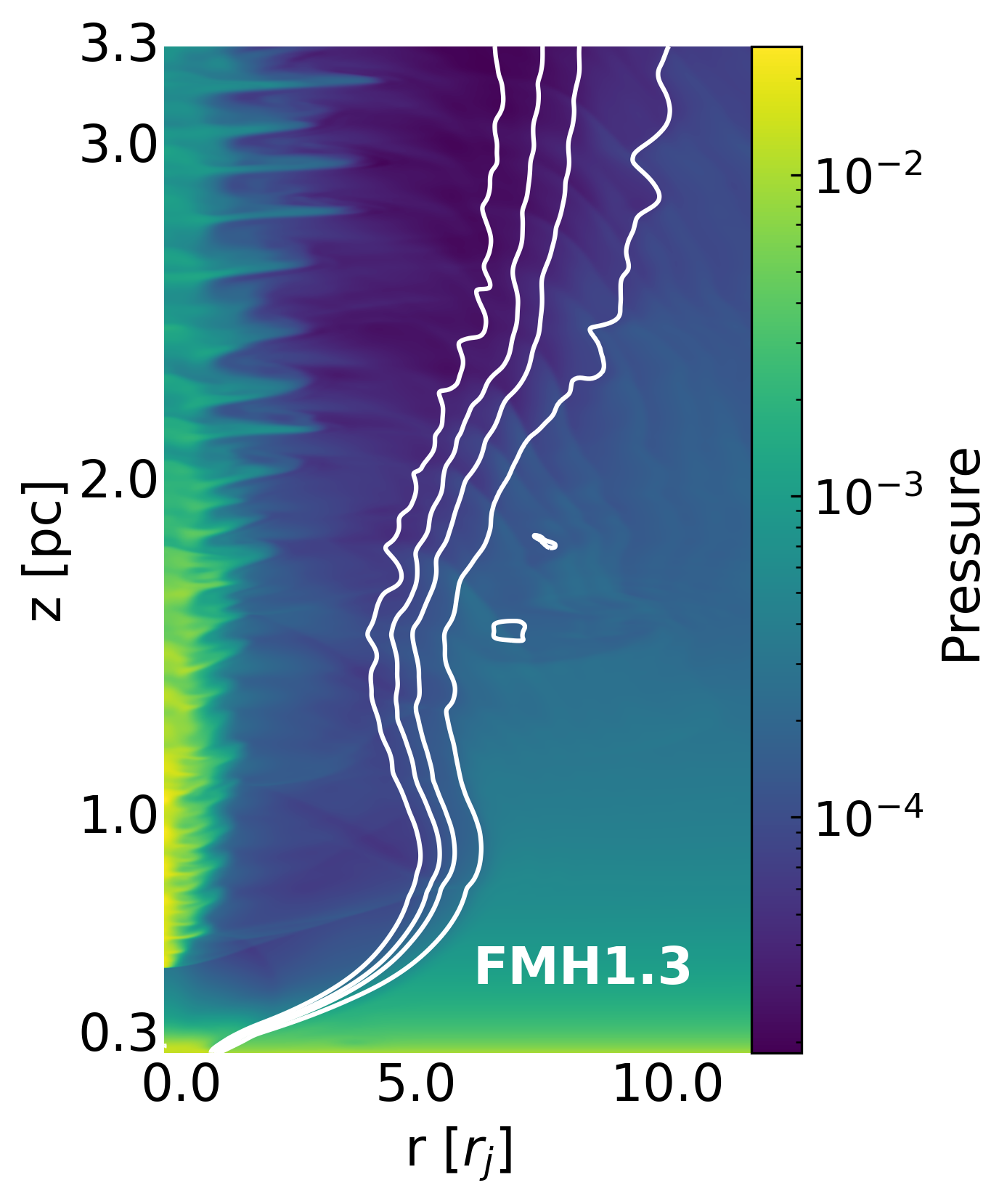}\par
    \includegraphics[width=\linewidth]{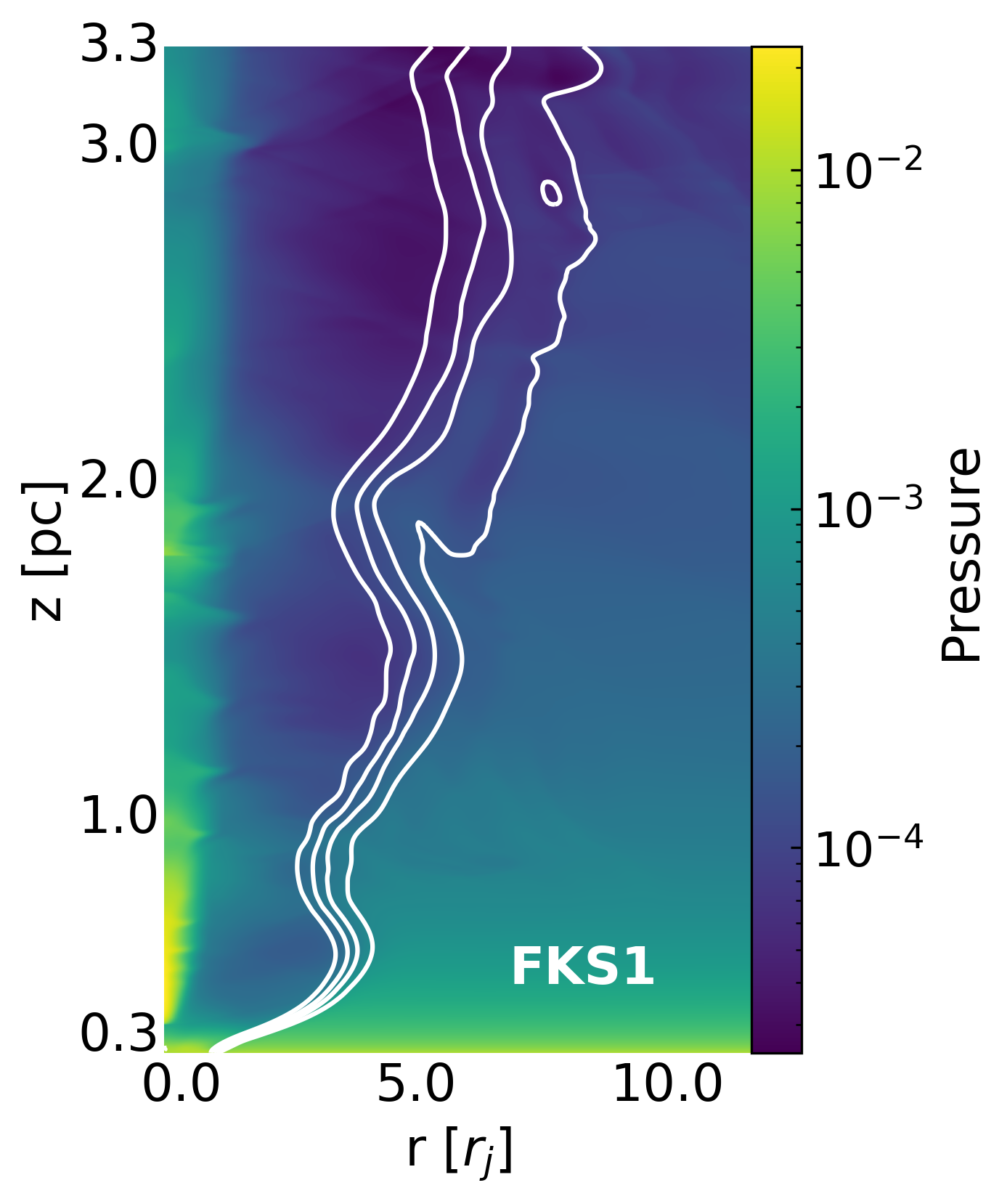}\par
\end{multicols}
\vspace{-0.8cm}
\begin{multicols}{4}
    \includegraphics[width=\linewidth]{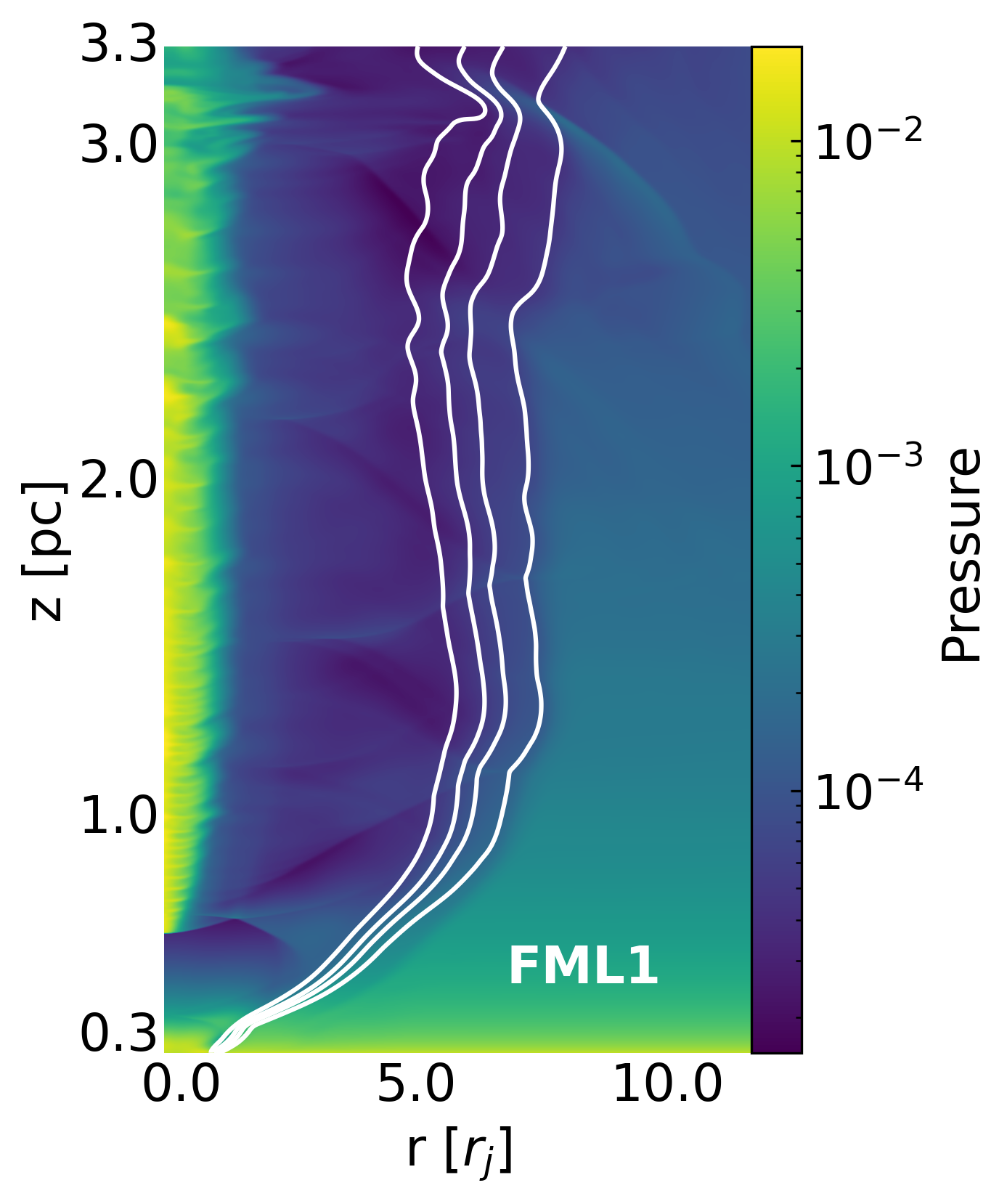}\par
    \includegraphics[width=\linewidth]{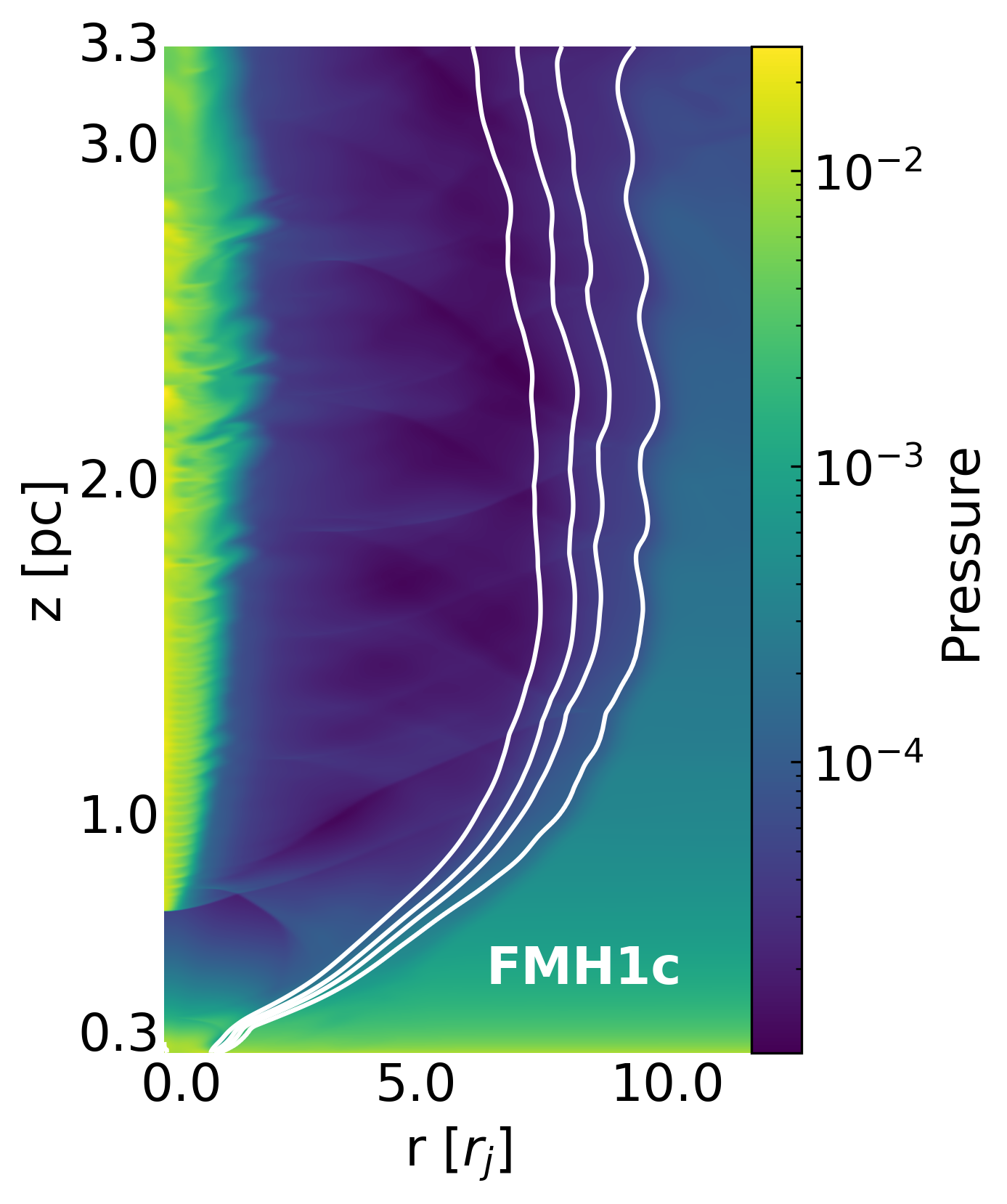}\par
    \includegraphics[width=\linewidth]{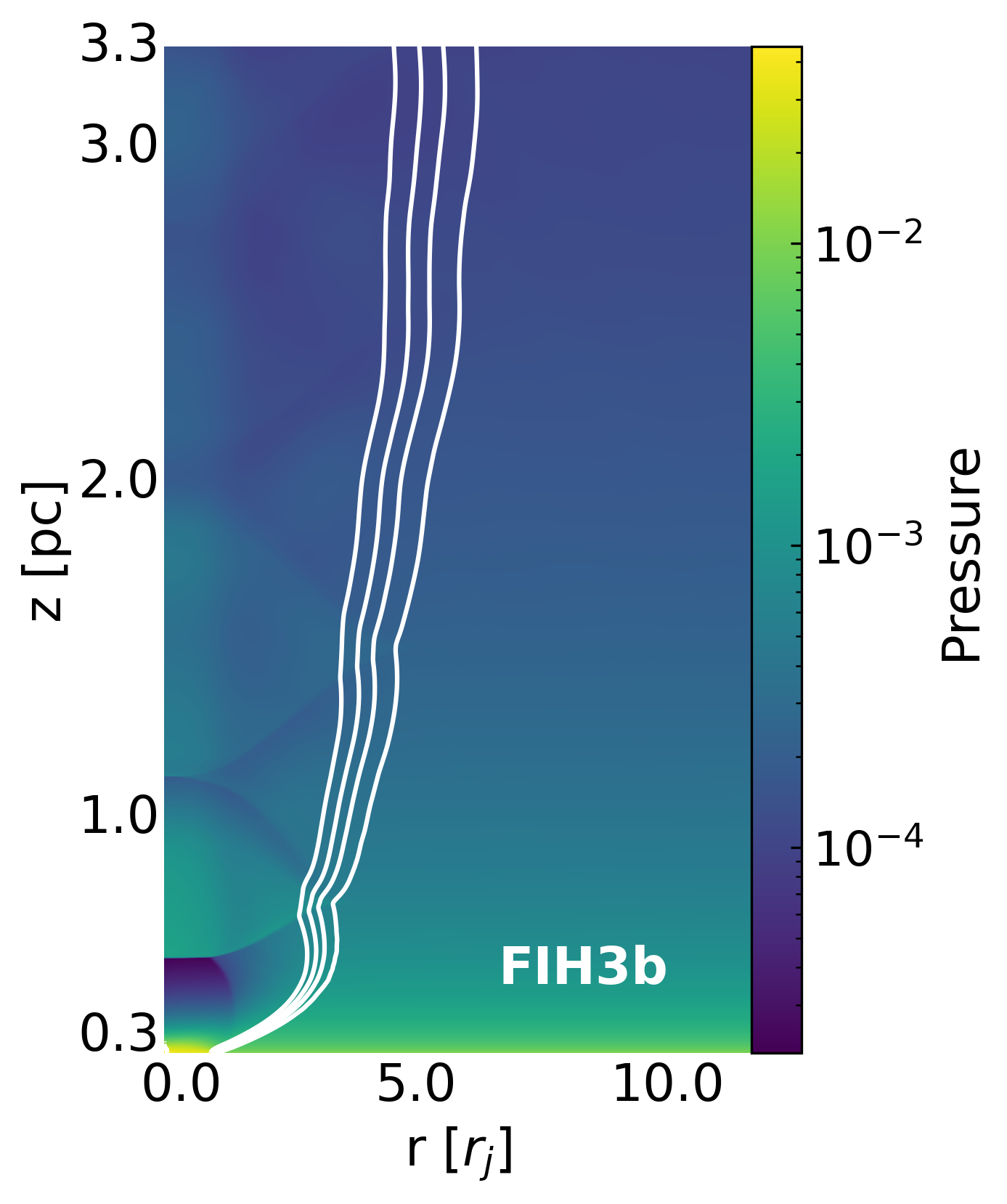}\par
    \includegraphics[width=\linewidth]{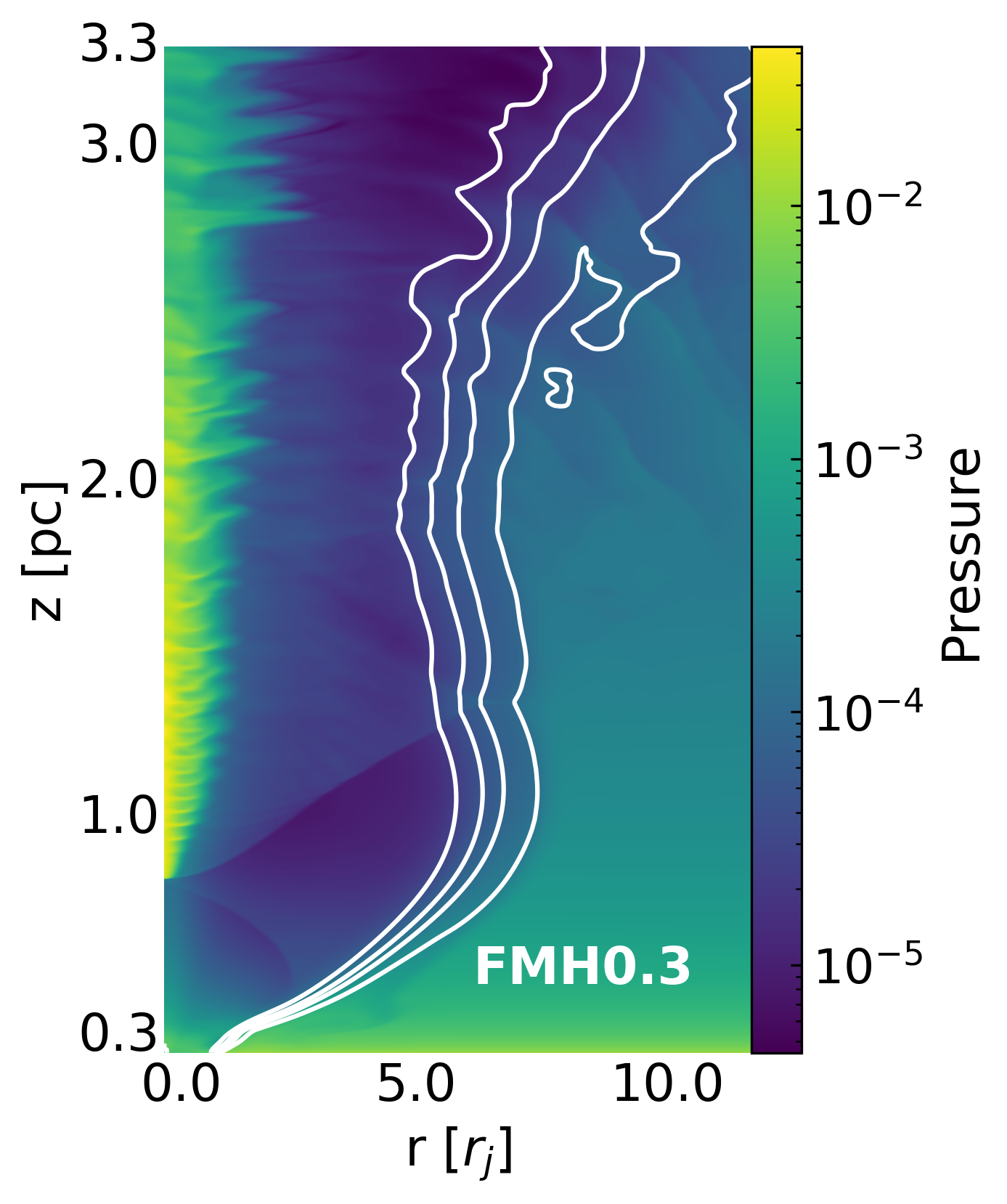}\par
\end{multicols}
\vspace{-0.8cm}
\begin{multicols}{4}
    \includegraphics[width=\linewidth]{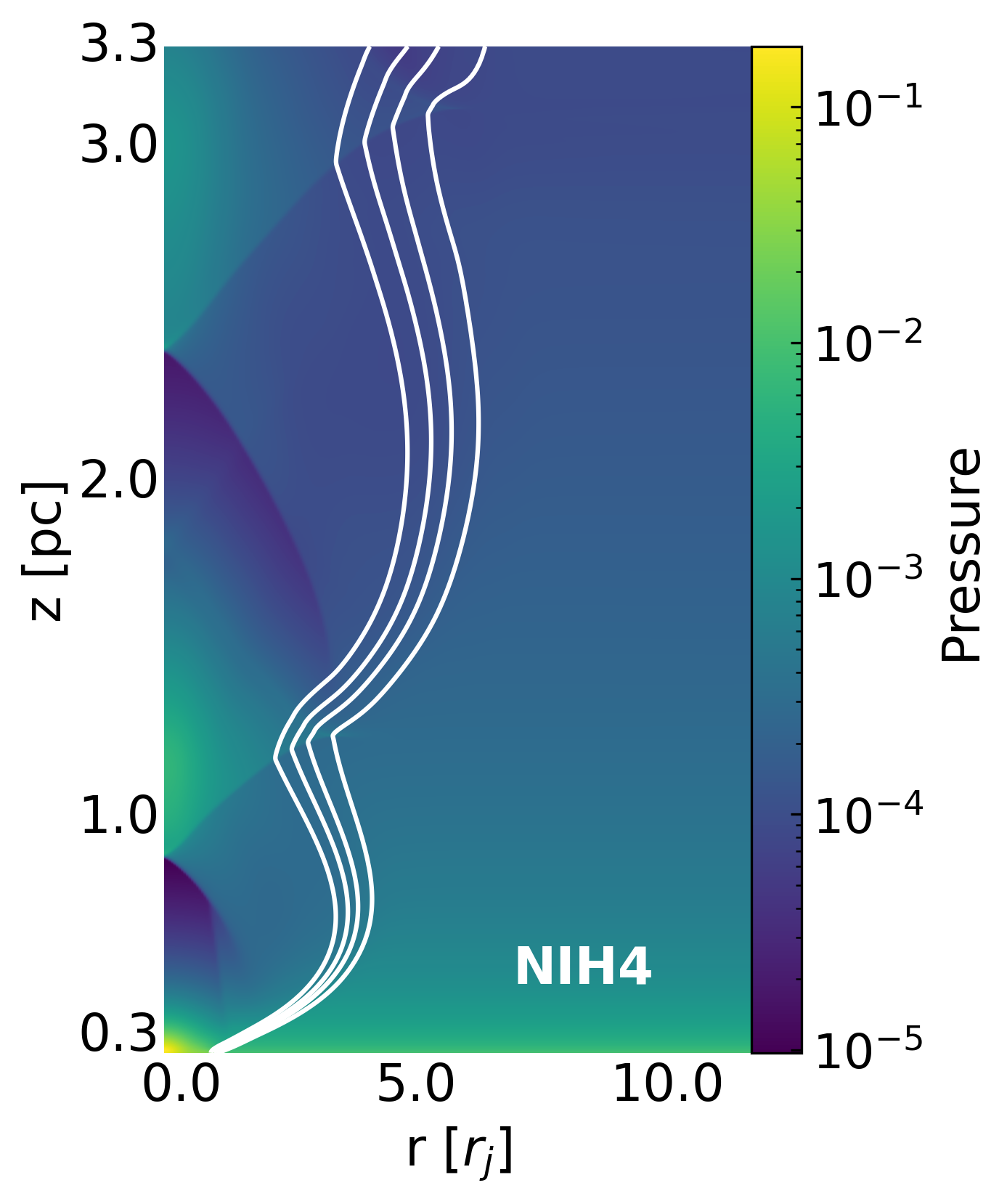}\par
    \includegraphics[width=\linewidth]{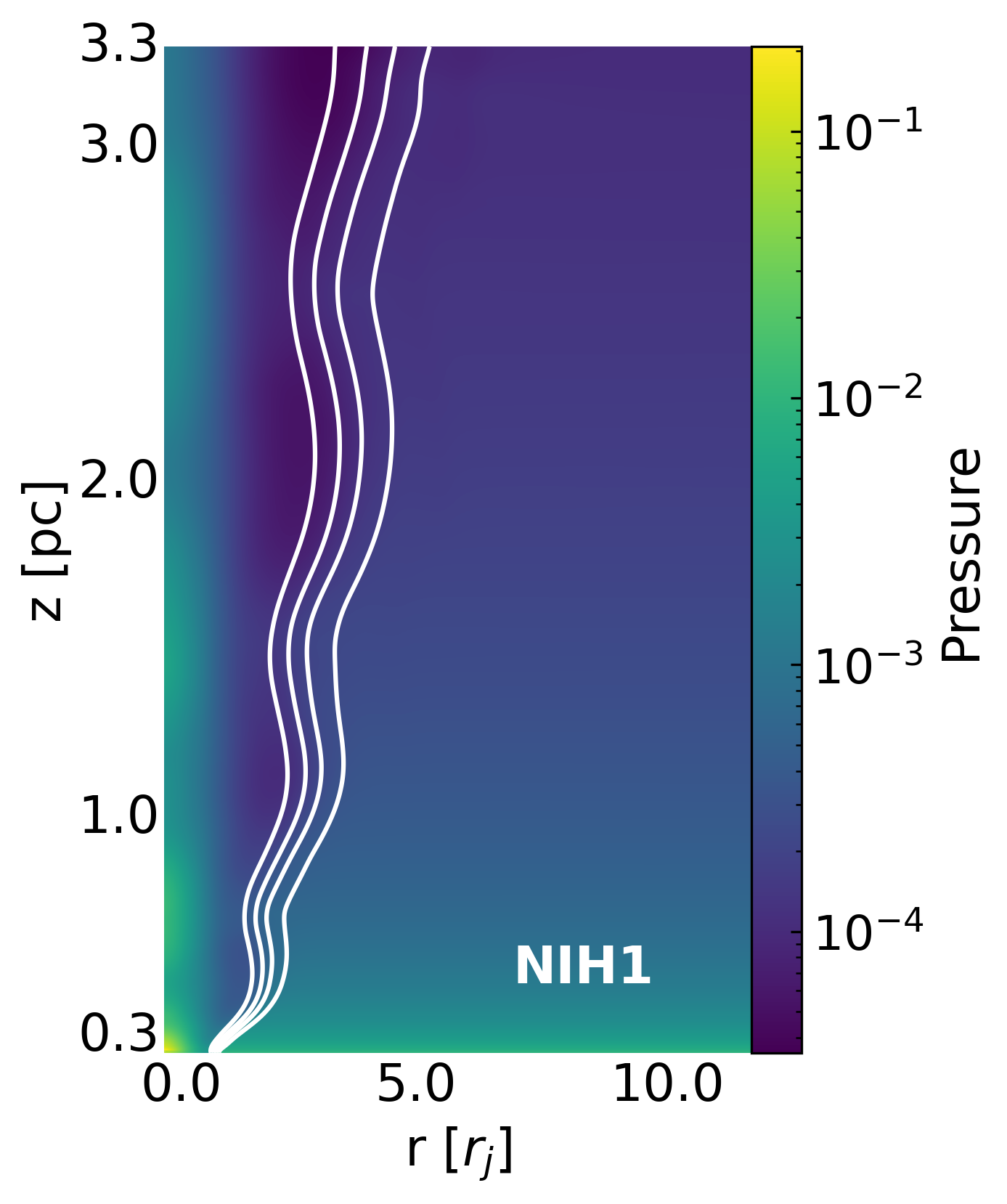}\par
\end{multicols}
    \caption{Pressure maps for the simulated models. The white contours are representative of the tracer at levels of 0.2, 0.4, 0.6, 0.8. As in Fig.\ \ref{fig:different_Lorentz_factor} the different jet nature leads to the different pressure profiles in the spine and in the shear layer regions.}
    \label{fig:different_pressure}
\end{figure*}

Figures~\ref{fig:different_Lorentz_factor} and \ref{fig:different_pressure} show the results obtained for all the different models.  
A clear relation between jet velocity profile and its intrinsic properties is inferred: jets developing strong internal shocks -Mach disks- present stronger acceleration in their outer, expanding layers, whereas those developing milder, conical shocks present a fast spine. 
Pressure maps (Fig.~\ref{fig:different_pressure}) show that the strong discontinuities observed in several models such as FMH1, FIH4, and FIH3b, are caused by the generation of a Mach disk in the jet.
The Mach disk formation is driven by the strong expansion,  due to the large angles formed by the shock waves with the jet axis \citep[see, e.g., ][and references therein]{marti2018}. 
In general, the main difference between the jets showing acceleration at the outer layers and those showing spine acceleration is their force-free versus non-force-free nature, or magnetically versus internal energy dominated. 

\begin{figure*}[htpb]
    \centering
\begin{multicols}{2}
    \includegraphics[width=0.7\linewidth]{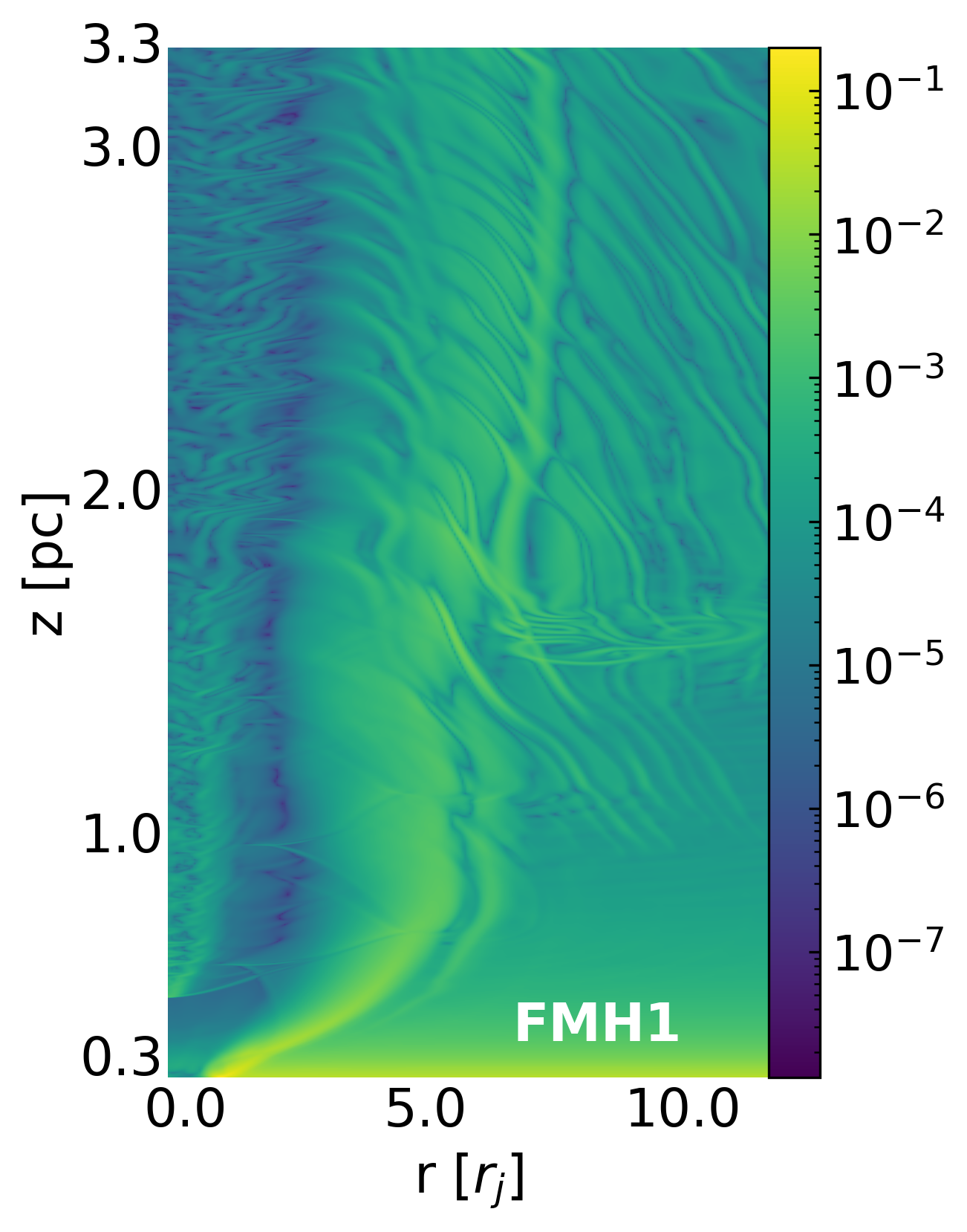}\par
    \includegraphics[width=0.7\linewidth]{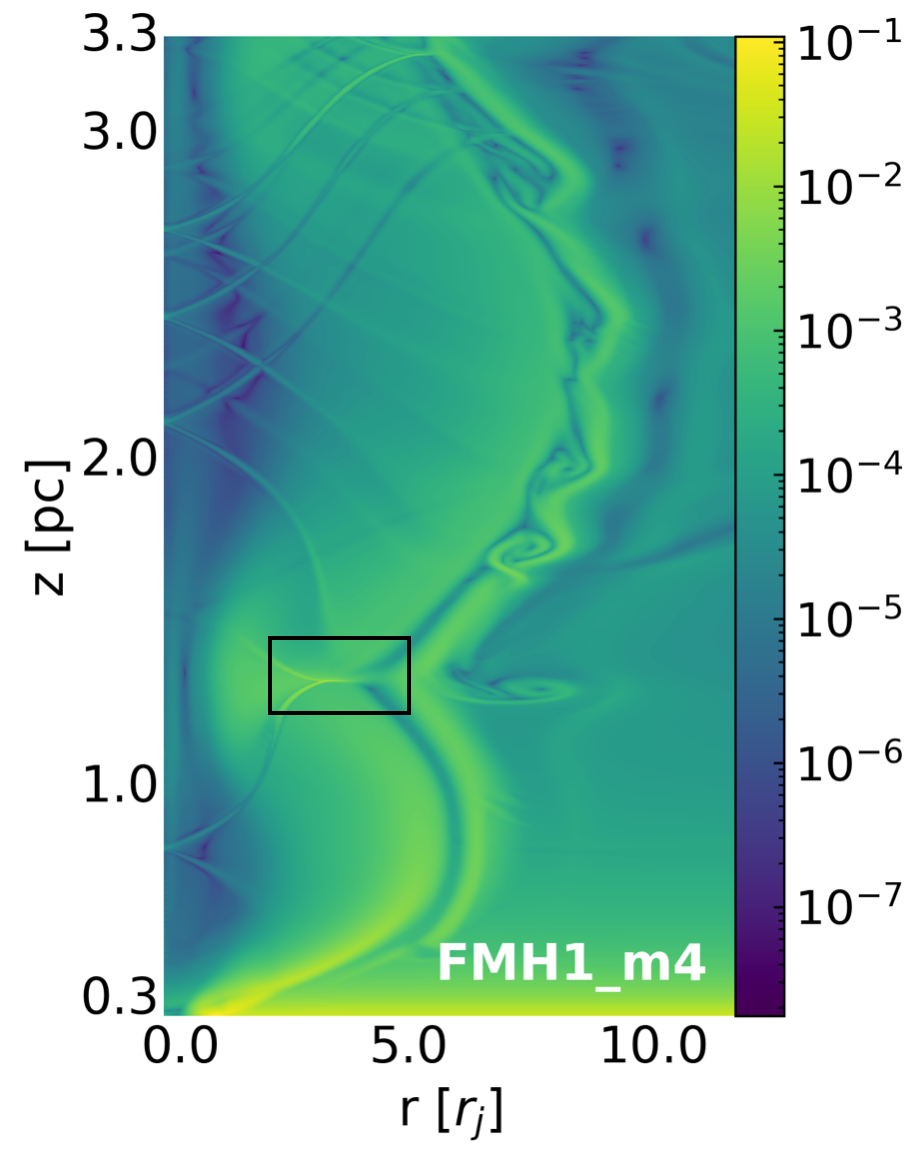}\par
\end{multicols}
\vspace{-1.0cm}
\begin{multicols}{2}
    \includegraphics[width=0.7\linewidth]{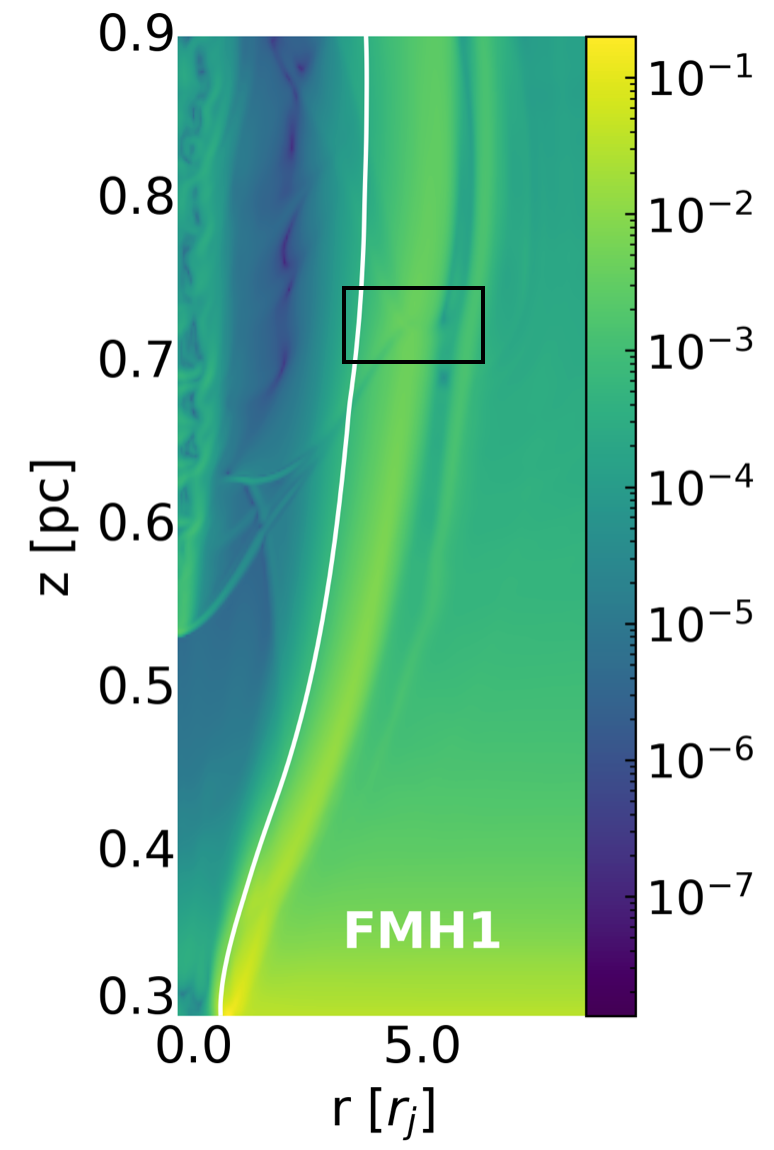}\par
    \includegraphics[width=0.7\linewidth]{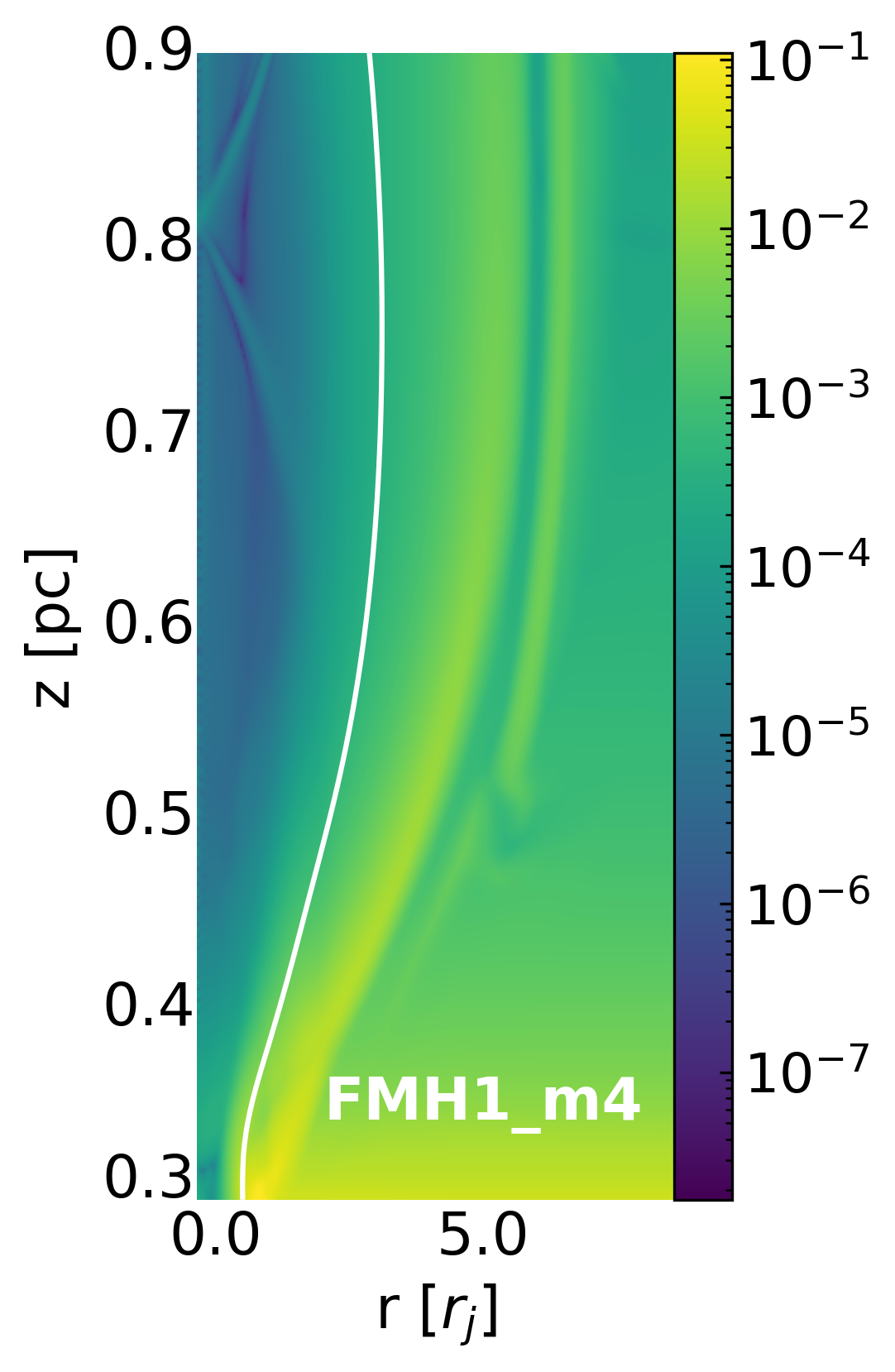}\par
\end{multicols}
    \caption{Modulus of the gradient of the rest-mass density (in logarithmic scale) for models FMH1 (left panels) and FMH1\_m4 (right panels).
    The upper panels show the entire grid, while in the bottom ones we show a zoom-in in the 0.3-0.9 pc region. The modulus of the gradient of the rest-mass density highlights the location of steep sound waves and shock waves.
    The black squares highlight the triple point associated with the incident/reflected/Mach shocks in the jet-ambient transition surface (for details, see Sect.~\ref{sec:shear_layer}).
    In the bottom panels, the contours represent the tracer at the level $f = 0.9$, showing the transition between the inner jet and the shear layer.}
    \label{fig:schlieren}
\end{figure*}

\begin{figure*}[t]
    \centering
    \includegraphics[width=\linewidth]{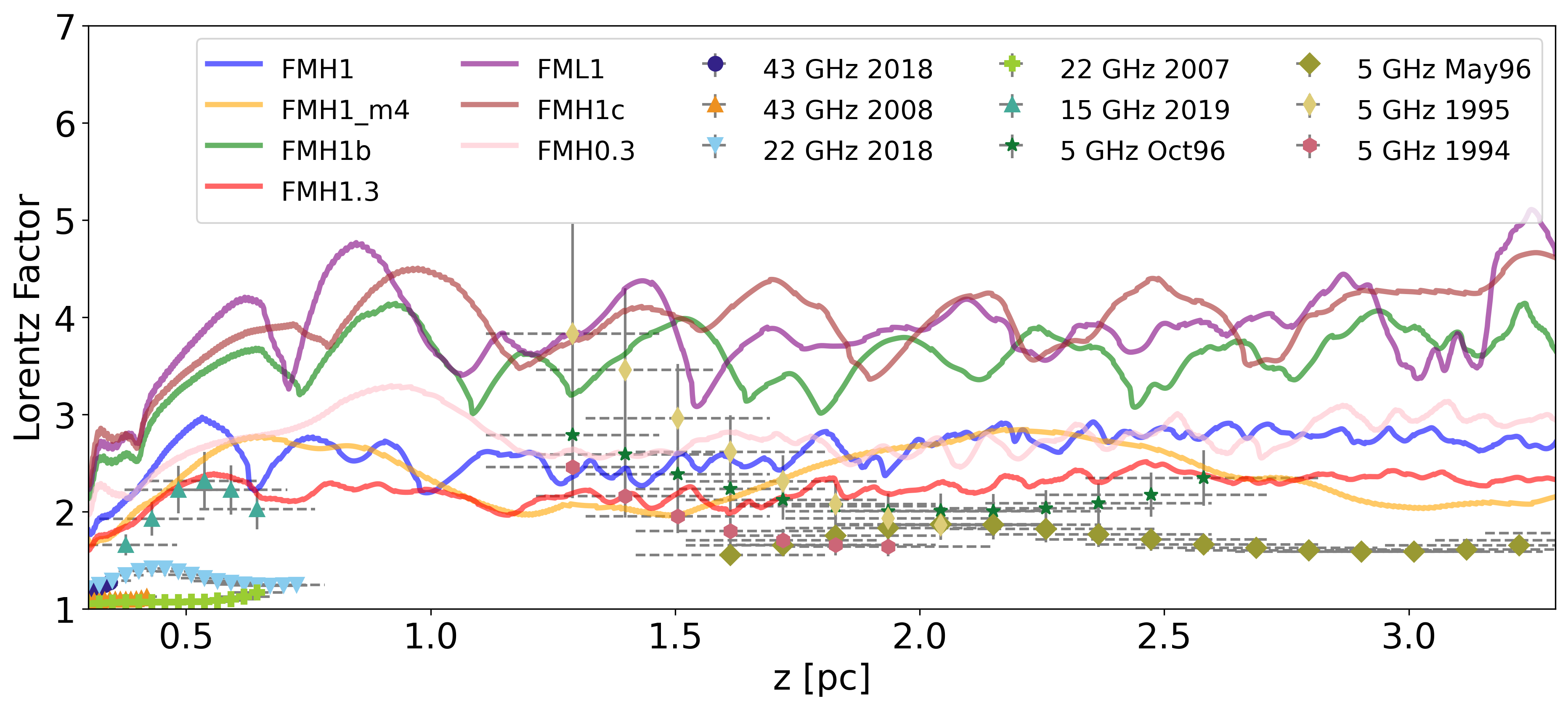}
    \includegraphics[width=\linewidth]{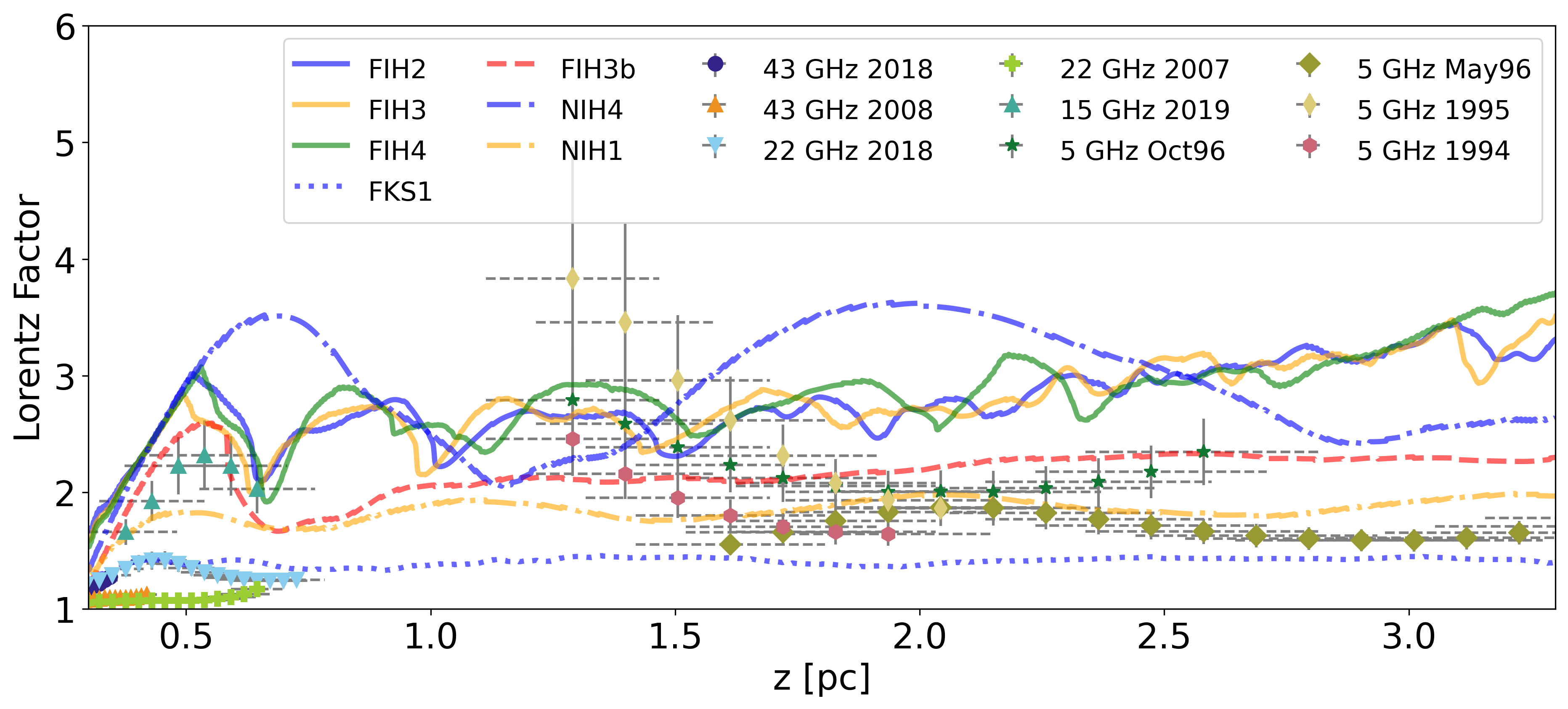}
    \caption{Radially-averaged Lorentz factor as a function of distance for all the simulated models (top panel: magnetically dominated jet models; bottom panel: thermally and kinetically dominated jet models). 
    The different data points are the Lorentz factor for different epochs and frequencies of NGC~315 inferred by \citet{Ricci_2022}. The continuous lines represent the force-free magnetically dominated models, the dashed lines the force-free thermally dominated ones, the dotted lines are for the kinetically dominated model, and the dashed-dotted lines for the non-force-free configurations.} 
    \label{fig:Lorentz_all_models}
\end{figure*}

\begin{figure*}[t]
    \centering
    \includegraphics[width=\linewidth]{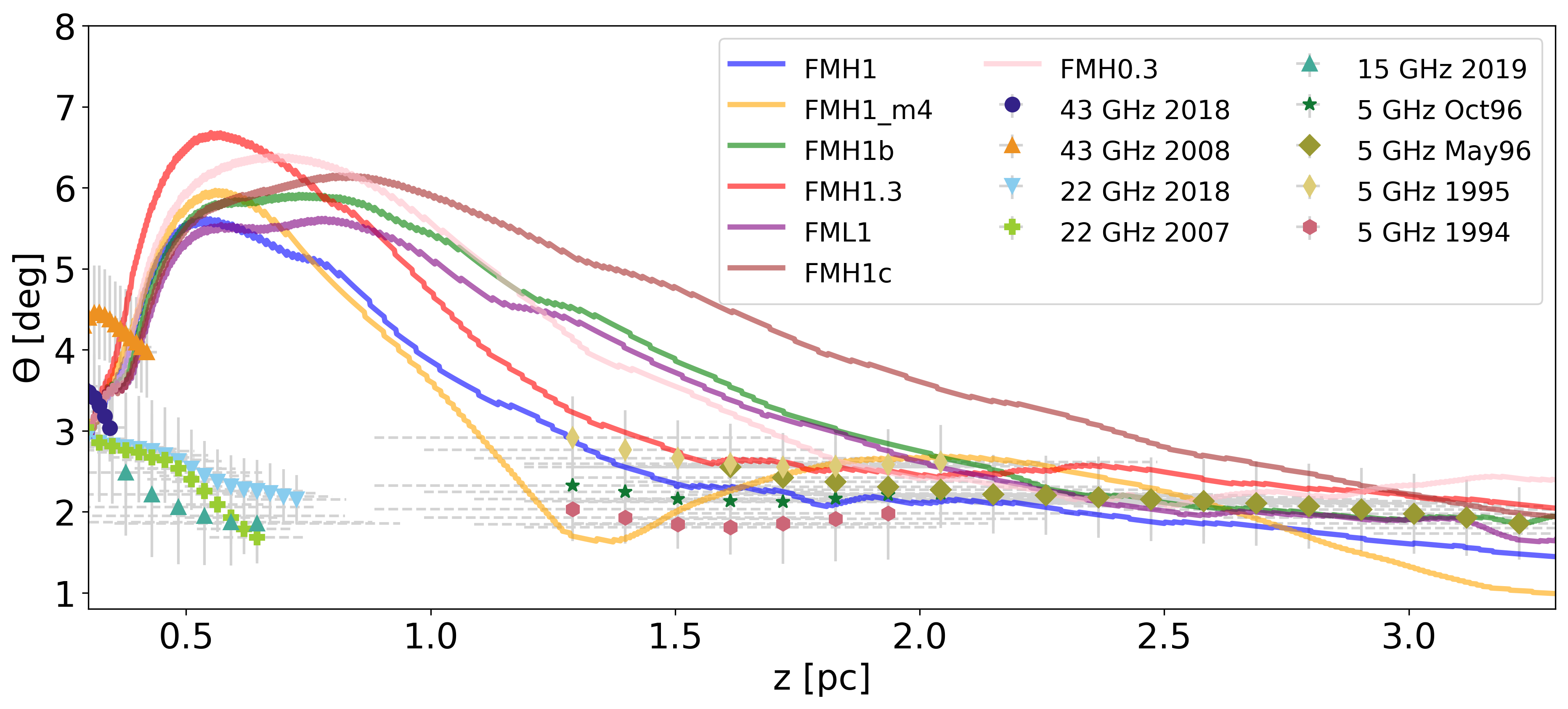}
    \includegraphics[width=\linewidth]{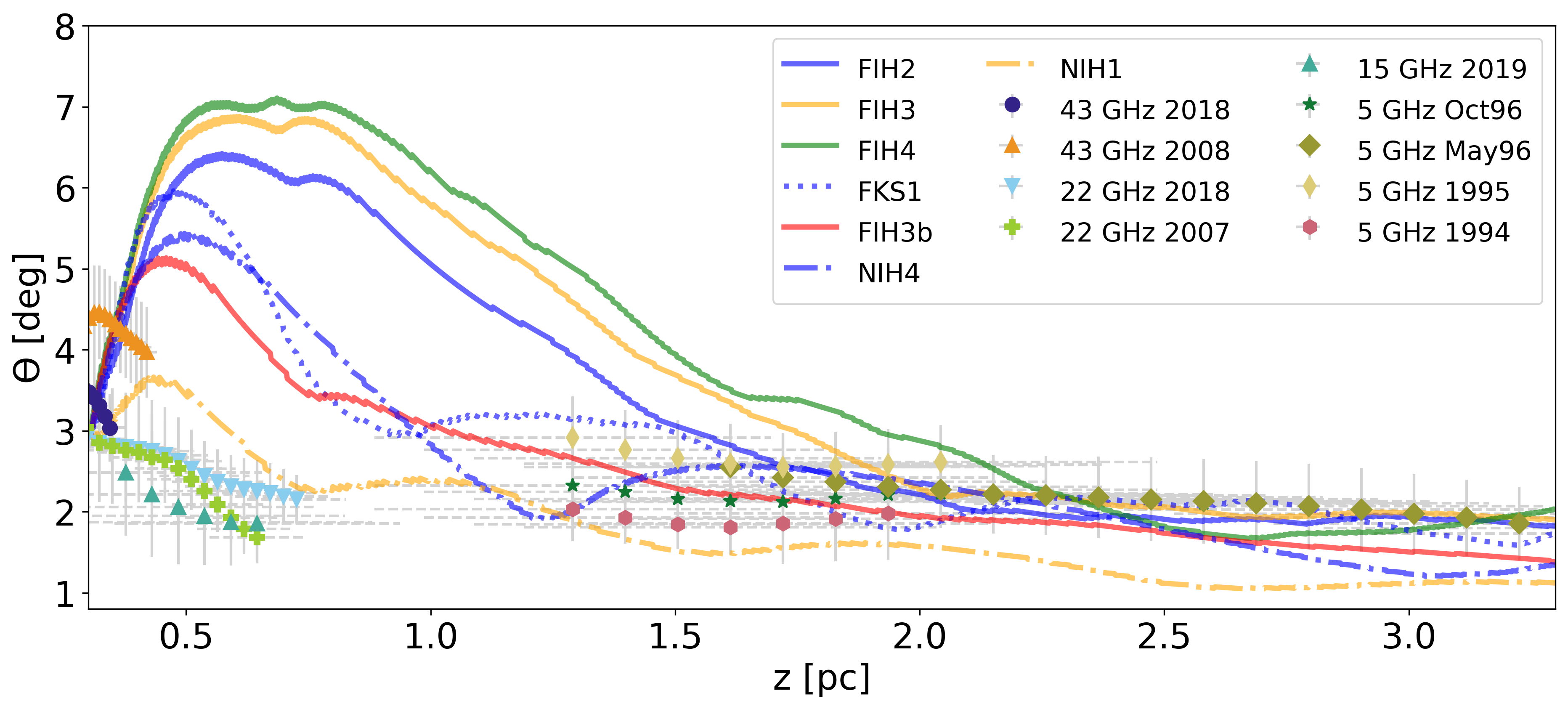}
    \caption{Half-opening angle as a function of distance for all the simulated models (top panel: magnetically dominated jet models; bottom panel: thermally and kinetically dominated jet models). The force-free magnetically dominated setups are shown in continuous lines, the force-free thermally dominated in dashed lines, the kinetically dominated in dotted lines, and the non-force-free in dashed-dotted lines. The data points represent the intrinsic half-opening angle profiles of NGC\,315 obtained by \cite{Ricci_2022} for multi-frequency and multi-epoch VLBI observations.}
    \label{fig:opening_all_models}
\end{figure*}

However, we observe that an increase in the width of the shear layer (FMH1\_m4 vs. FMH1 models) can completely change the jet acceleration pattern obtained. 
In the case of FMH1 (top, left panel), the jet expands until the point at which the extension of the planar shock wave at the center of the jet reaches its surface. 
Beyond this point, we observe a mild expansion/recollimation process. 
Downstream, in the region that is unaffected by the Mach disk, the jet develops instability patterns and locally dissipates some of the gained kinetic energy, which explains why the internal energy flux at the end of the grid is larger than at $\sim 0.5 \, \mathrm{pc}$ (see Sect.~\ref{sec:shear_layer} for the discussion on this effect). 
In FMH1\_m4 (top, second panel from the left), on the contrary, a Mach disk is not formed and the acceleration of the flow is concentrated towards its axis, unlike all the other force-free models. 
The corresponding pressure maps show that the main difference between both models is the delay in the wave propagation that the thicker shear layer causes in FMH1\_m4 with respect to FMH1: the outgoing waves have to cross a thicker portion of the jet before bouncing back, which delays the bounce and moves it downstream.
This delay relaxes the angle formed by the incident wave and the axis, changing the acceleration pattern and jet structure.
Figure~\ref{fig:schlieren} shows the modulus of the gradient of the rest-mass density in the $(r, \mathrm{z})$ plane\footnote{computed using centered finite differences.}, $| \nabla \rho |$,
of both FMH1 and FMH1\_m4 (top panels), and a zoom of the relevant region where the first recollimation shock forms (bottom panels).
The modulus of the gradient of the rest-mass density highlights the location of steep sound waves and shock waves. 
The images in Fig.~\ref{fig:schlieren} show how the different widths of the shear layer changes the wave structure of the models.
In model FMH1, the jet dynamics is dominated by instabilities both in the spine and at the jet/ambient medium transition on parsec scales, where we see the formation of vortexes typical of the growth of Kelvin-Helmholtz instabilities. 
On the contrary, in model FMH1\_m4 the jet pinches, but substantially fewer instabilities arise and the jet/ambient medium transition is smoother.
In the bottom panels, we can see that the opening angle of the transition between the inner jet and the shear layer is larger in FMH1. 
The bottom panels display contours for tracer $f=0.9$ to show the transition between the inner jet and the shear layer, which, at injection, occurs at 0.91$r_j$ in model FMH1, while it is at 0.82$r_j$ in model FMH1\_m4. 
We propose that this difference determines the two possible evolution patterns observed (see Sect.~\ref{sec:shear_layer} for details).
We have run other models with wider shear layers (with the same initial conditions as FIH2, FIH4, FMH1c, FIH3b) and we have observed the same behavior: thicker shear layers may result in a dramatic change in the acceleration pattern of force-free jets, avoiding the formation of Mach disks and favoring spine acceleration, as opposed to thin-layered models (see Appendix \ref{app:other_models}).

In all the force-free models (except FMH1\_m4), we observe the formation of the Mach disk with the resulting transversal velocity structure consisting of a slow spine surrounded by a fast sheath. 
In all cases, although this is better observed in model FMH1, the accelerated region slowly expands towards the jet axis, even if at the end of the grid none of them shows a completely accelerated jet cross-section. 
Instability modes develop at the interface between the spine and the outer accelerated layer.
This triggers dissipation and we thus expect it to contribute to increasing the internal energy of the region.
The fast development of instabilities is expected in the case of hot, slow flows \citep[see, e.g.,][]{Perucho_2004,Perucho_2005}. 

Non-force-free models show, in contrast, an acceleration pattern concentrated around the jet axis, i.e., they produce a fast spine. 
The main difference with respect to force-free models is the absence of a Mach disk. 
This acceleration pattern corresponds to that expected from the Bernoulli process. 
Highly overpressured jets, such as NIH4, show strong acceleration on the axis before the flow reaches a strong recollimation shock, where the flow is decelerated. 
The next expansion process produces a second acceleration region, but the Lorentz factor reached is $20\%$ smaller than the one prior to the first recollimation shock. 
In the case of NIH1, in thermal pressure equilibrium with the environment at injection, acceleration is milder, modulated by recollimation waves/shocks, with a progressively growing velocity from one acceleration region to the next.  


Figure \ref{fig:Lorentz_all_models} shows the radially-averaged Lorentz factor for all the simulated models.
For clarity, we divided the profiles into two separate plots: in the top panel we show the magnetically dominated models, while in the bottom panel the thermally dominated ones together with the kinetically dominated FKS1 are displayed.
The average Lorentz factor value is computed by weighting the value at each cell with its volume (given by a ring). 
As a discriminant to determine the jet width we use the tracer 
parameter with a value of 0.5 as a threshold, i.e., cells with tracer 
> 0.5 are considered to be part of the jet while cells with tracer < 0.5 are considered ambient medium. 
Along with the speed profiles, Fig.~\ref{fig:Lorentz_all_models} shows the data points for NGC\,315 in the frequency range 5-43 GHz and at different epochs inferred in \citet{Ricci_2022}.
These results were obtained based on the observed jet-to-counterjet intensity ratio, which provides an average velocity in a likely stratified jet structure \citep[hints of an edge-brightening and possible stratified velocity are given in][]{Park_2021}.
The comparison of the simulation profiles with the observational data is presented and discussed in Sect.~\ref{sec:comparison_NGC315}.
The plot shows that the simulations with different setups result in average jet acceleration from $\gamma \sim 1-2$ up to $\gamma \sim 2-5$ within a spatial propagation of only 3 pc. The weakest acceleration is observed for model FKS1, as expected for kinetically dominated models, whereas the strongest acceleration is obtained in models with the highest relative magnetic and/or internal fluxes at injection, such as FMH1c, FIH4 or FML1. 
It is also interesting to mention that force-free models accelerate larger volumes of jet plasma, as the outer regions of the jet represent larger rings in cylindrical coordinates.
All models show modulations in the average Lorentz factor produced, as stated above, by expansion and recollimation episodes, in which internal energy flux is invested into acceleration and the jet flow is then decelerated by shocks, where kinetic energy is converted into internal energy. 
As expected, this is more evident in significantly overpressured models, like FIH4, where the initial acceleration of the jet is followed by a drop in the average Lorentz factor. 
In the thin shear-layer, force-free jets, the difference in velocity between the plasma flowing along the spine, decelerated by the planar (Mach) shock, and that accelerated at outer radii triggers the development of instabilities. 
In models FMH1, FMH1.3, and FMH0.3 these instabilities develop to non-linear amplitude within the grid and, on the one hand, allow the gained axial momentum at the accelerated region to be shared with the inner jet spine and, on the other hand, dissipate part of the gained kinetic energy. 
Altogether, this results in a drop of the maximum average Lorentz factor achieved by the jet during its initial expansion, and a slow recovery with distance. 
Nevertheless, these models stabilize their Lorentz factors at values between 2 and 3. 

Figure \ref{fig:opening_all_models} shows the half-opening angle of the different models versus distance together with the observational points for NGC~315. 
We divided the plot as before, the upper panel for the magnetically dominated models, and the bottom panel for the thermally dominated (plus FKS1) ones.
In all models, we observe an initial increase of the opening angle due to the expansion of the jet towards the less-pressured ambient medium, followed by a decrease that starts at different distances, but always before $\sim 1 \, \mathrm{pc}$.
The fall in the opening angle profiles is associated with jet collimation, i.e., $r \propto \mathrm{z}^\alpha$, with $\alpha < 1$.
None of the models shows completed collimation on the simulated scales, i.e., $\alpha \approx 1$.
All models tend to final half-opening angles of $1-3^\circ$.
This plot is very clarifying regarding the causes for jet expansion in the different models: force-free models expand faster than non-force-free models, as can be seen by comparing cases FIH4 (green dashed line) and NIH4 (blue dashed line). 
This difference also allows us to group the models into two main behavior patterns: 
1) the models with the largest initial opening angles develop the Mach disk, while 2) those with smaller initial opening angles develop conical shocks. 

\begin{figure*}[htpb]
    \centering
\begin{multicols}{3}
    \includegraphics[width=\linewidth]{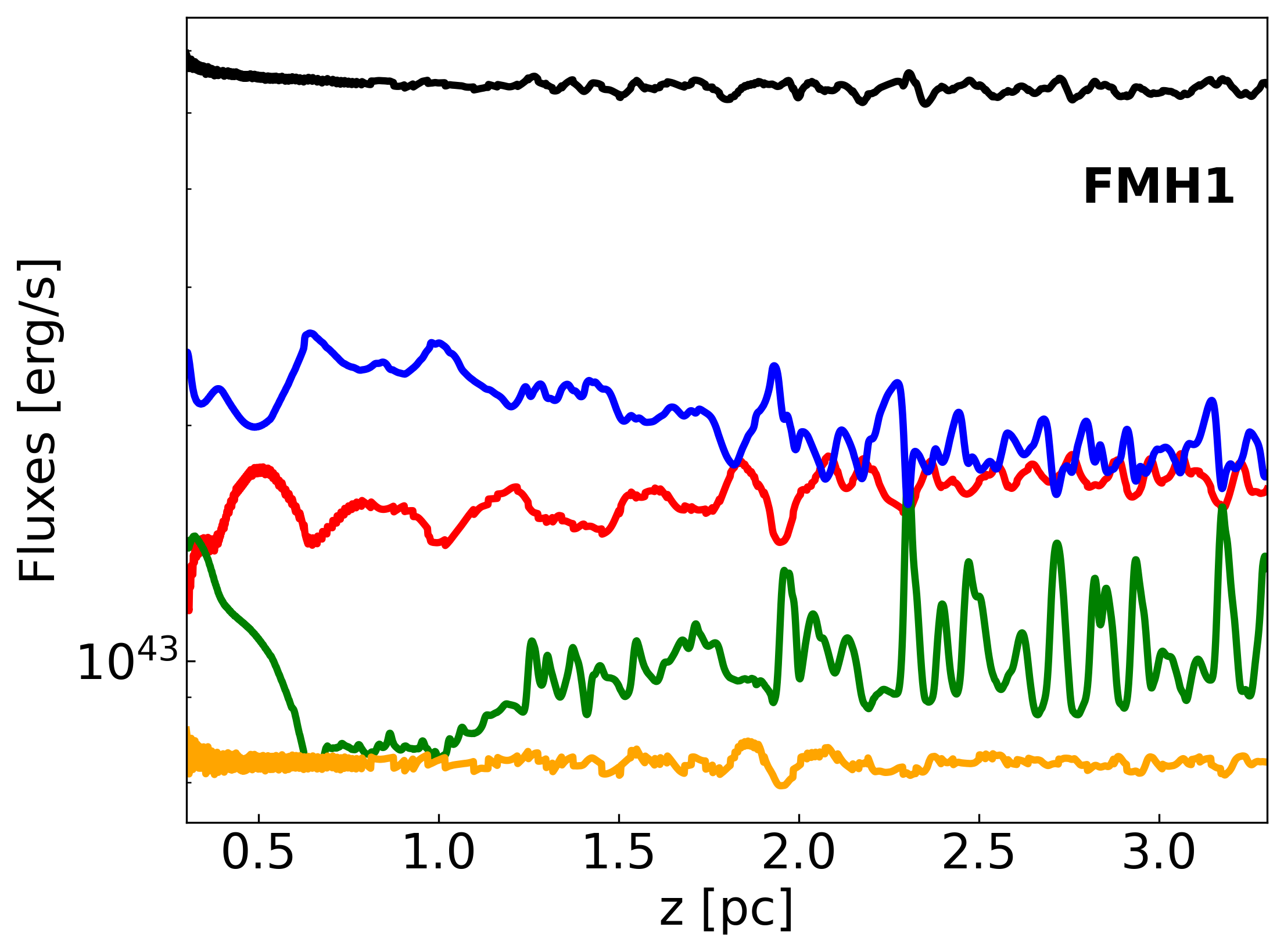}\par
    \includegraphics[width=\linewidth]{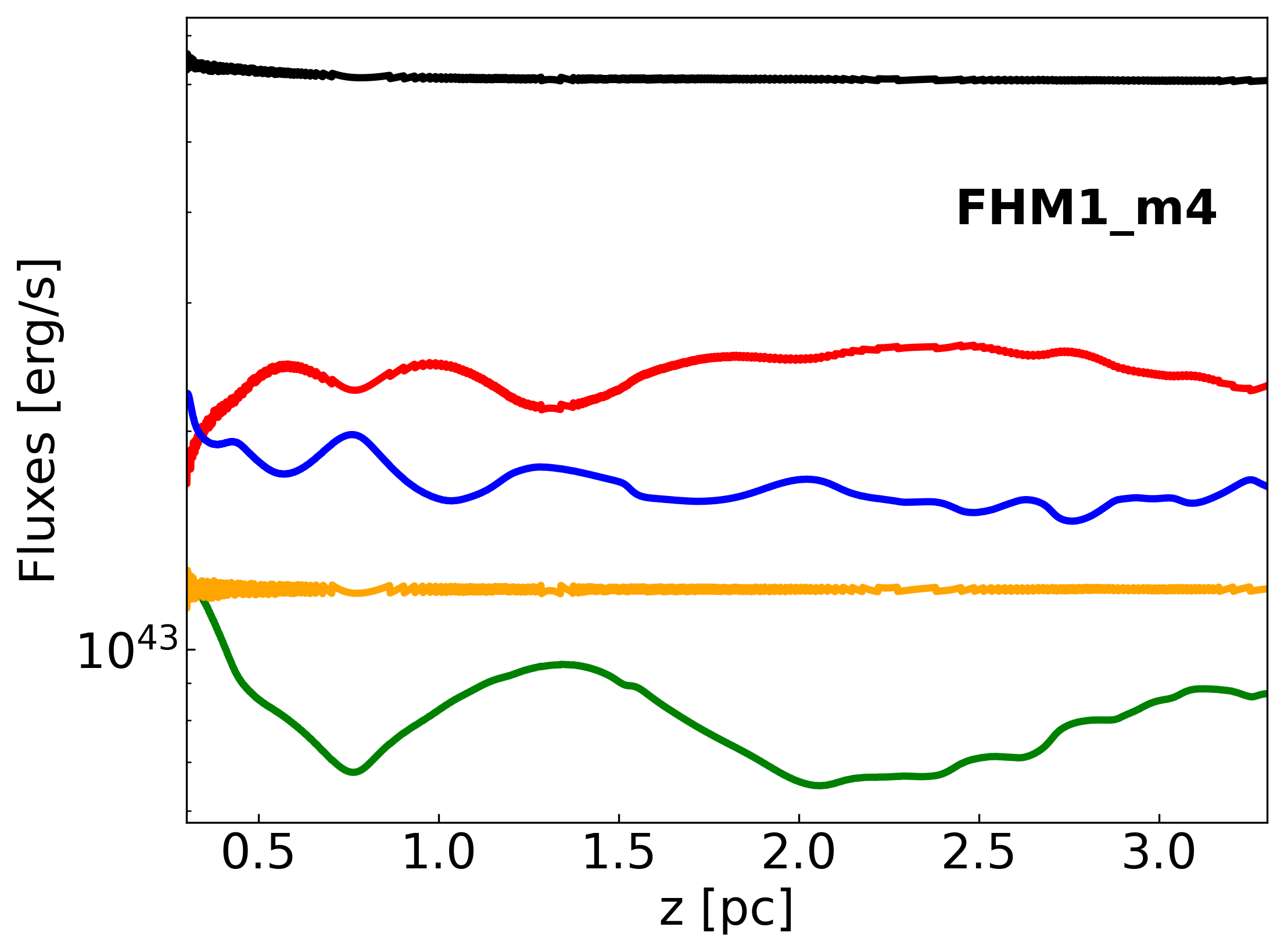}\par
    \includegraphics[width=\linewidth]{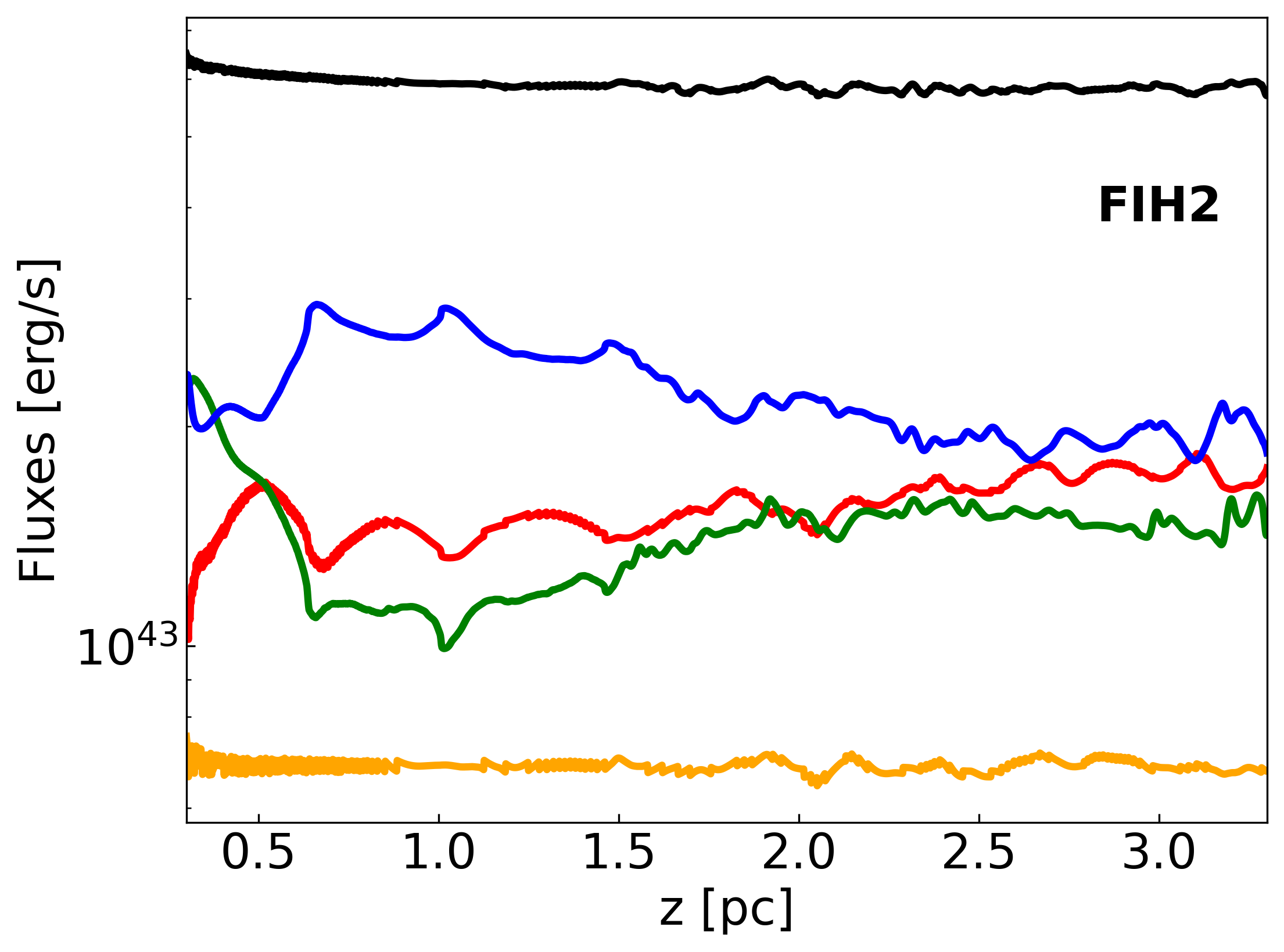}\par
\end{multicols}

\vspace{-0.8cm}

\begin{multicols}{3}
    \includegraphics[width=\linewidth]{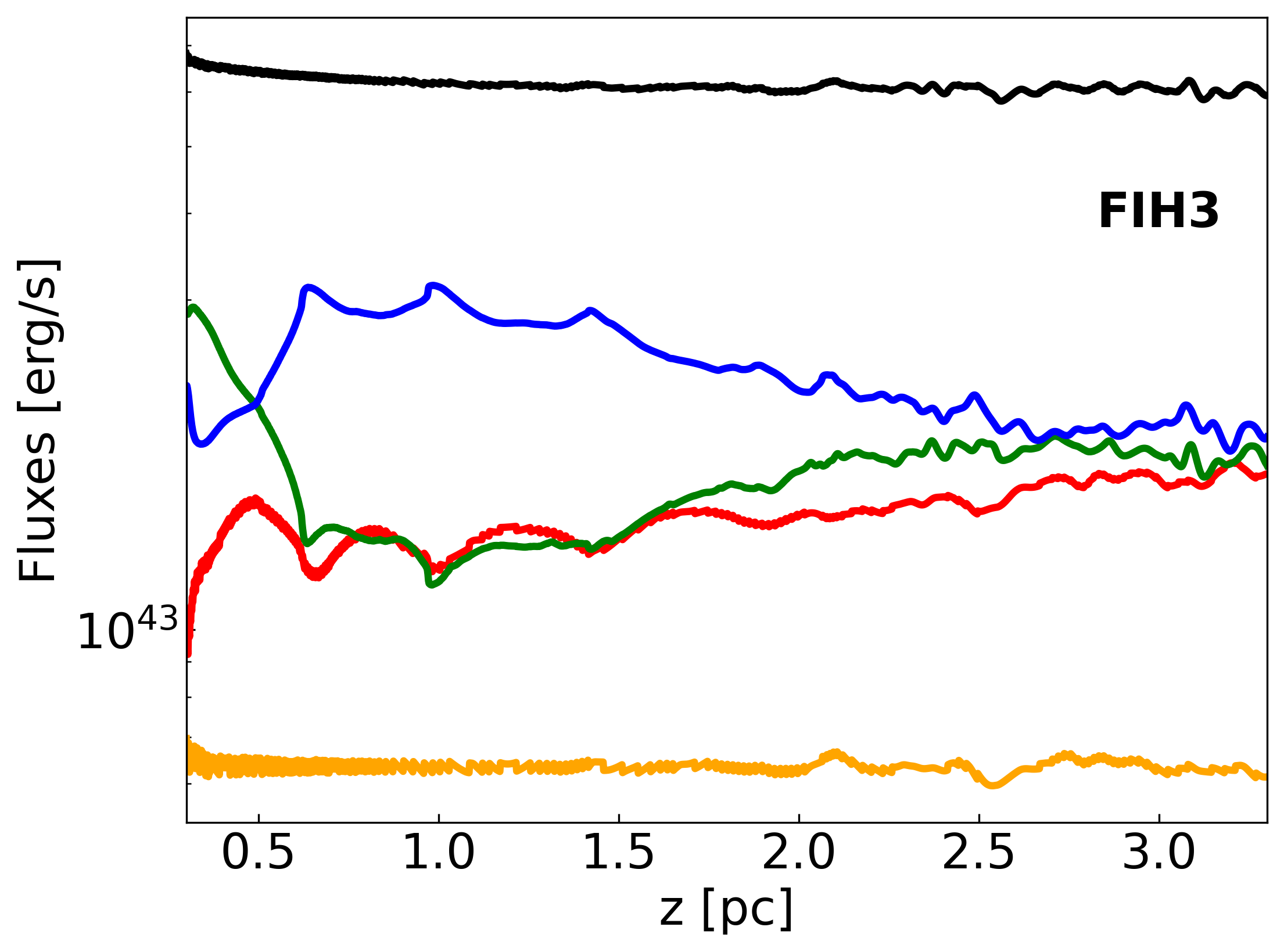}\par
    \includegraphics[width=\linewidth]{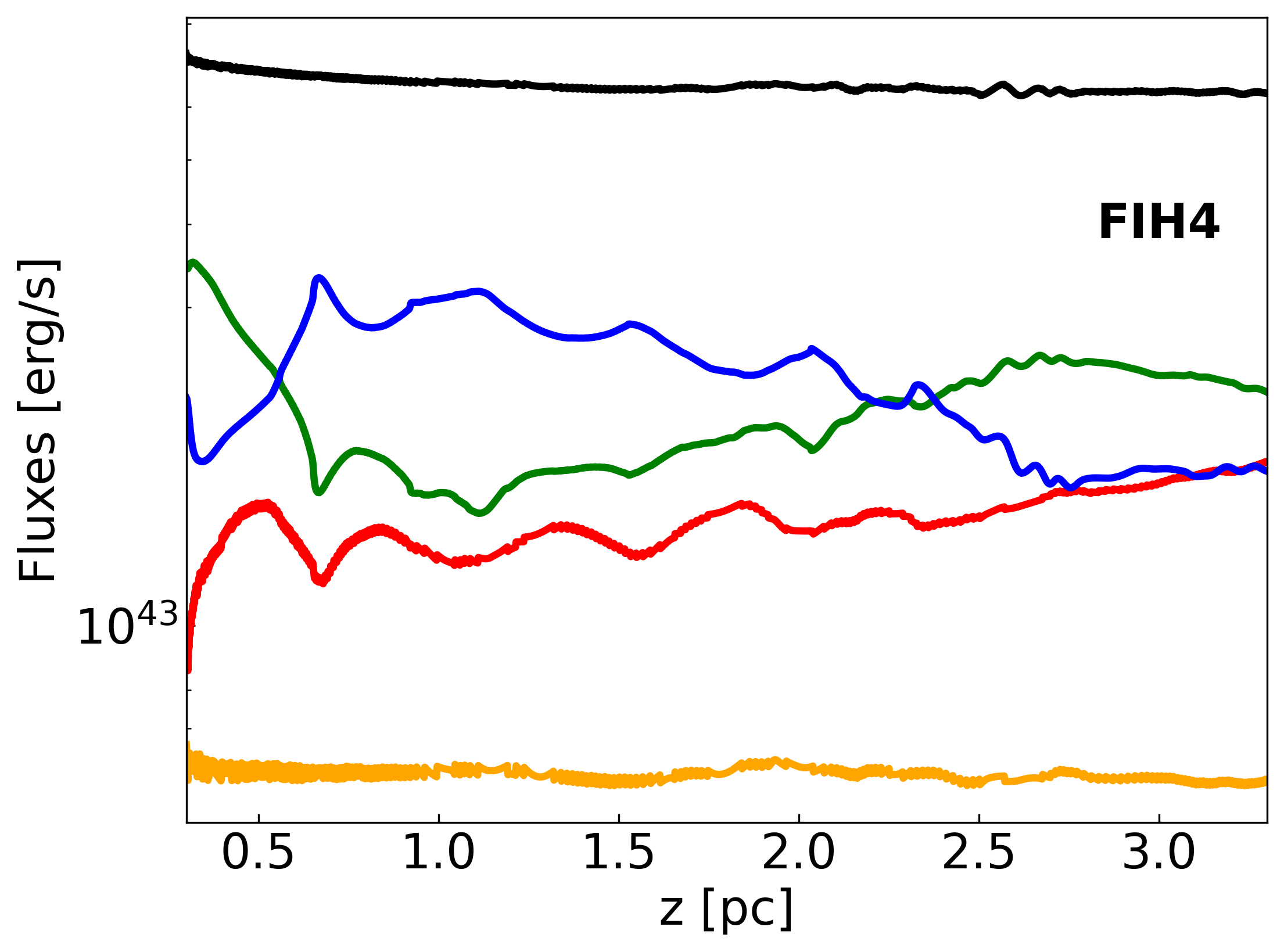}\par
    \includegraphics[width=\linewidth]{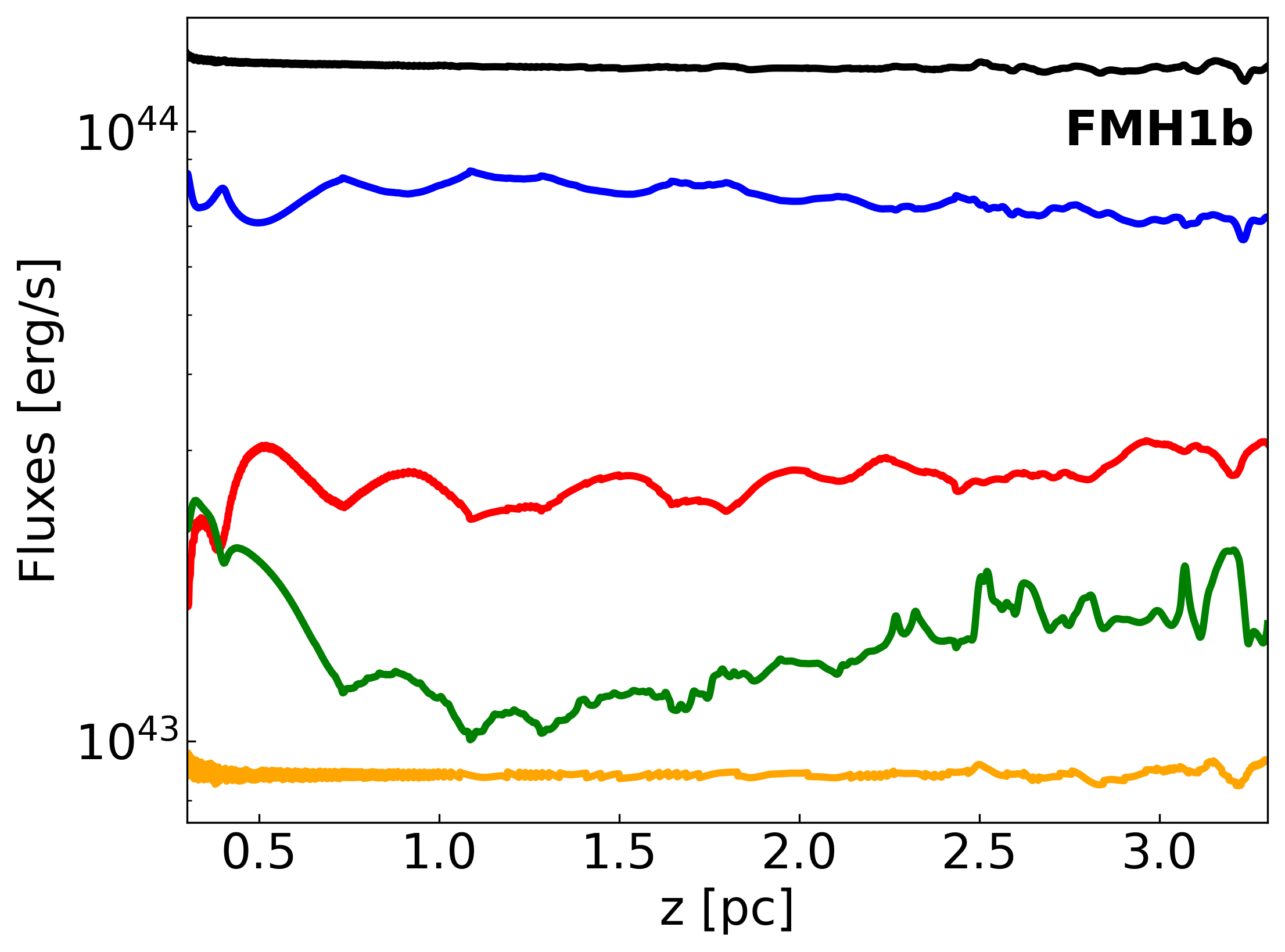}\par
\end{multicols}

\vspace{-0.8cm}

\begin{multicols}{3}
    \includegraphics[width=\linewidth]{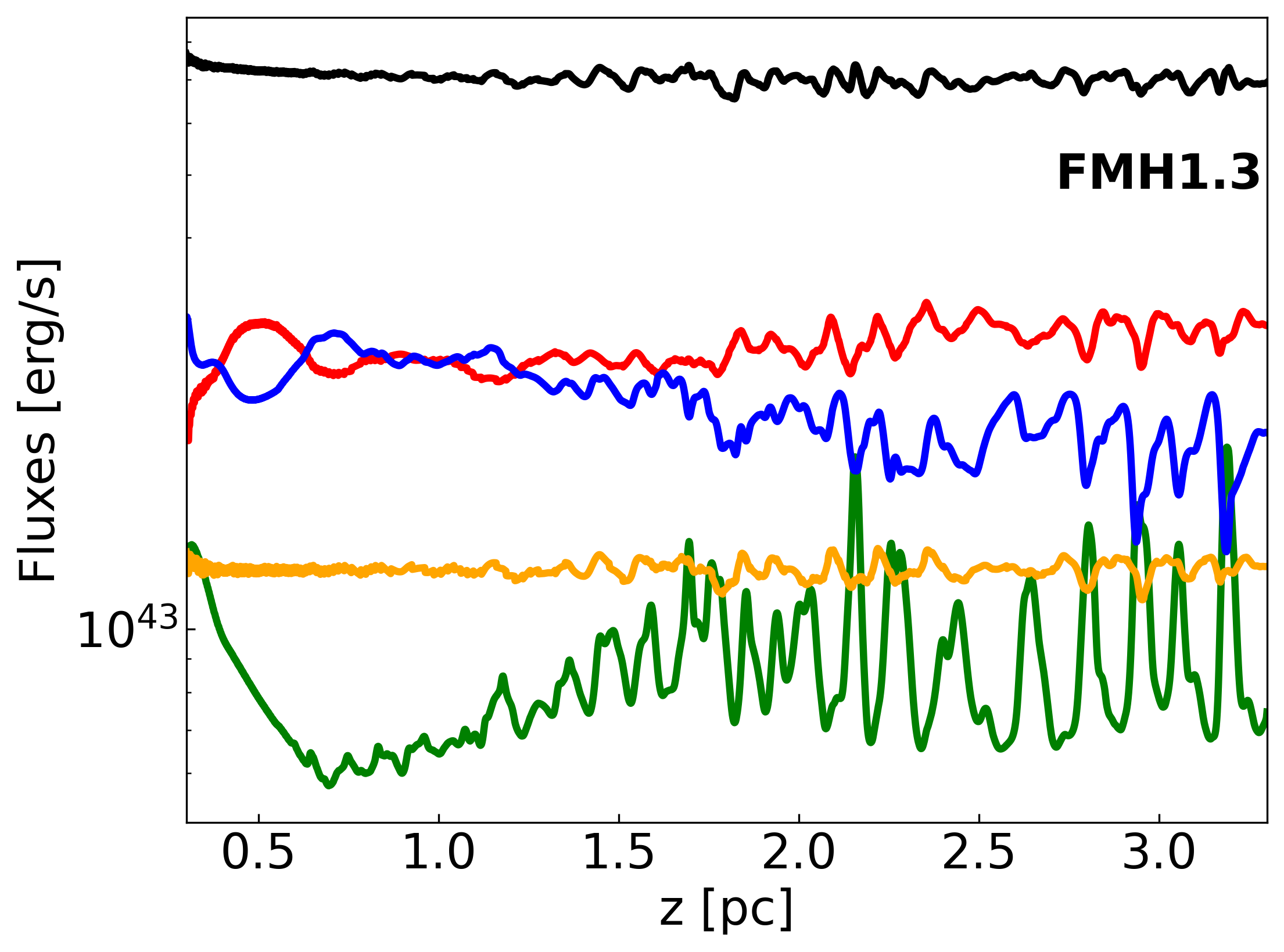}\par
    \includegraphics[width=\linewidth]{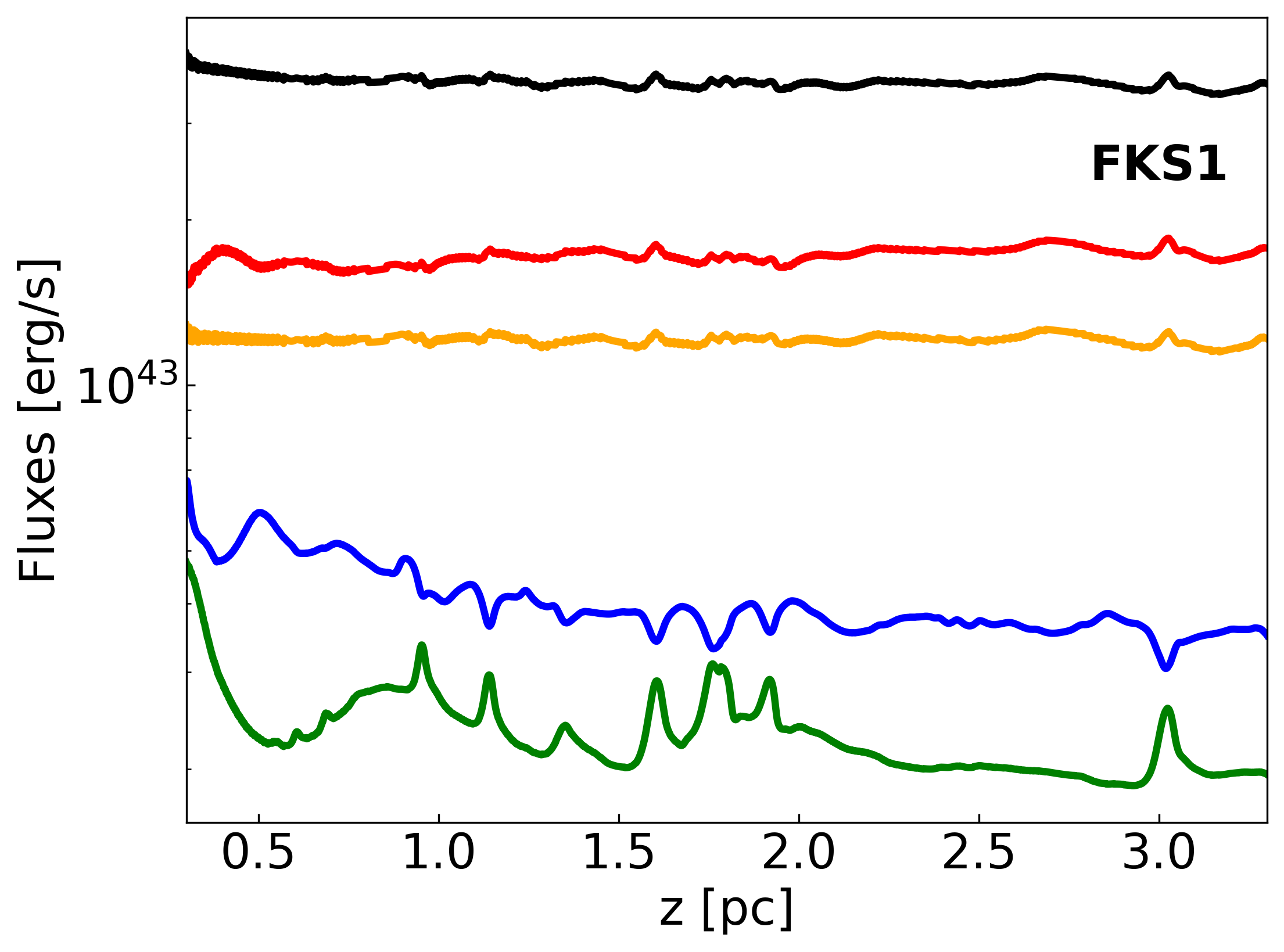}\par
    \includegraphics[width=\linewidth]{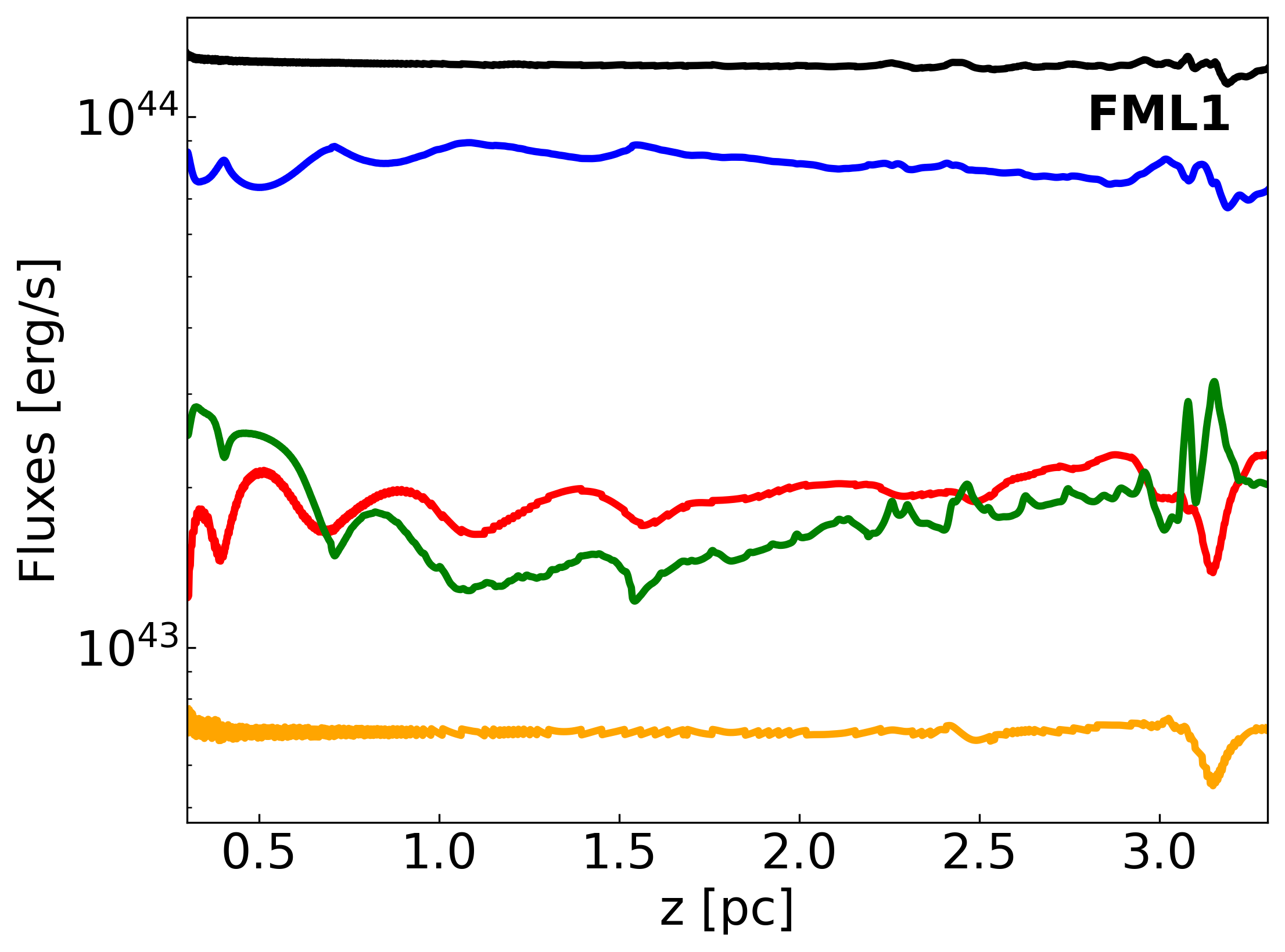}\par
\end{multicols}

\vspace{-0.8cm}

\begin{multicols}{3}
    \includegraphics[width=\linewidth]{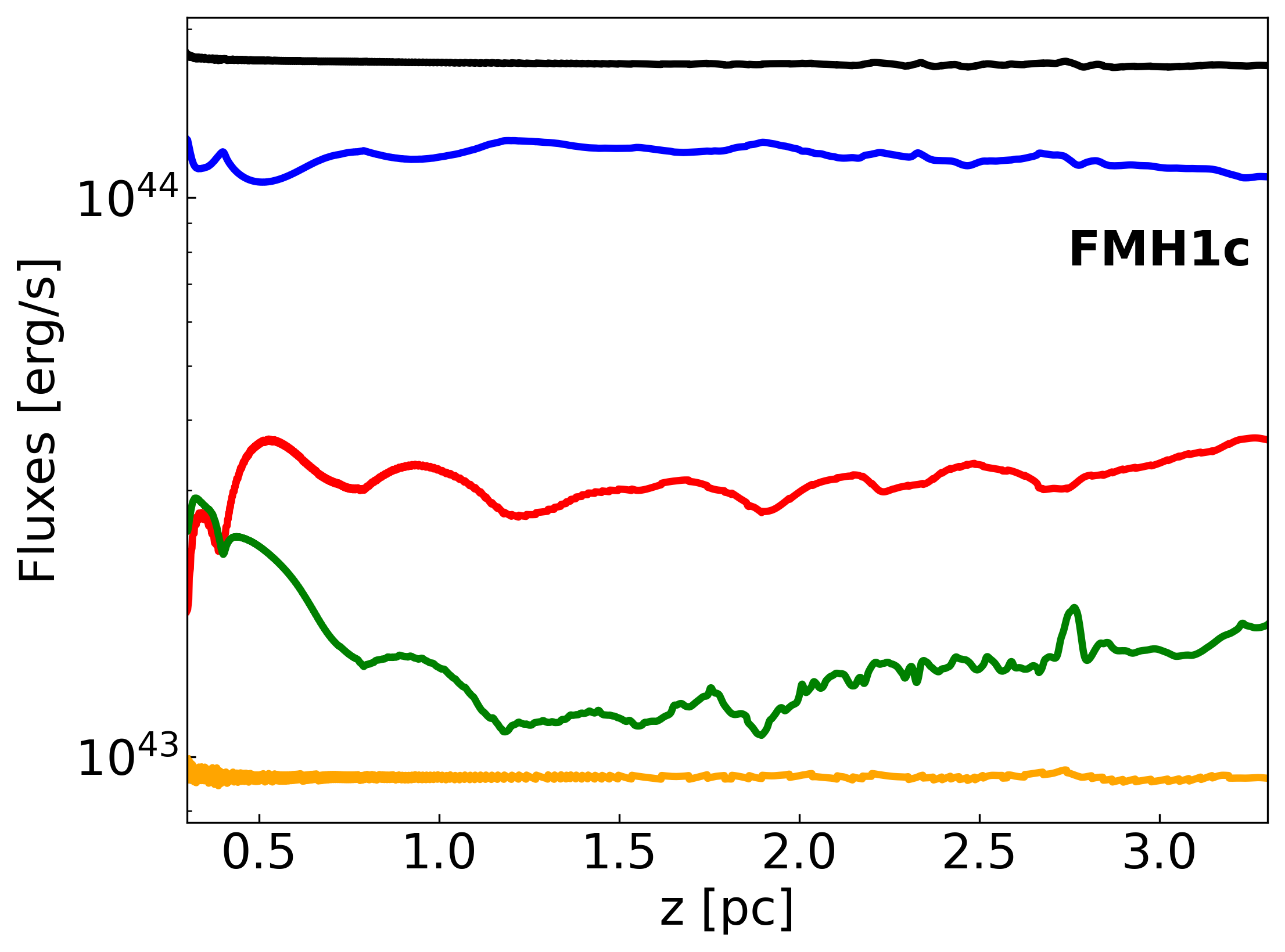}\par
    \includegraphics[width=\linewidth]{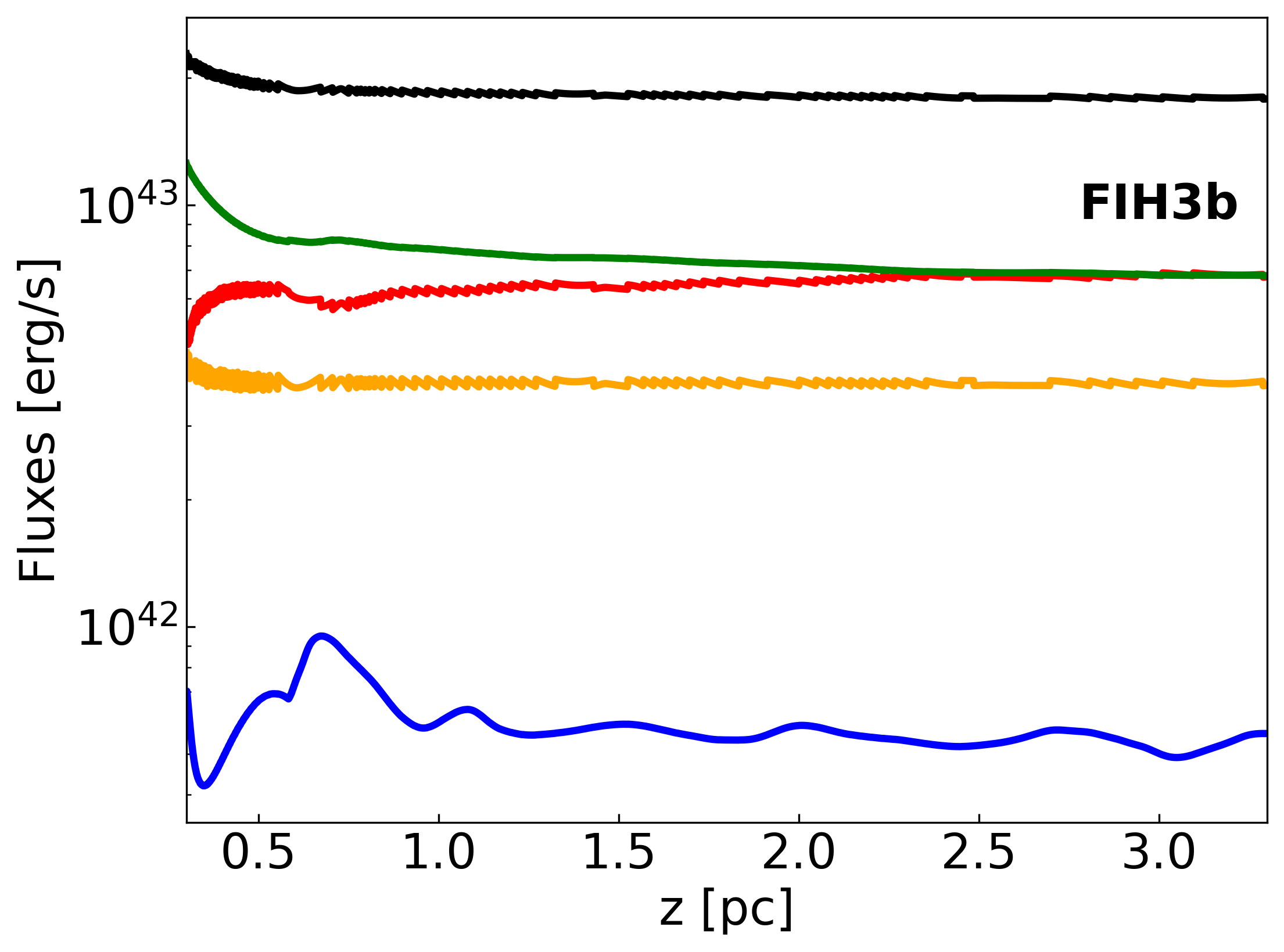}\par
    \includegraphics[width=\linewidth]{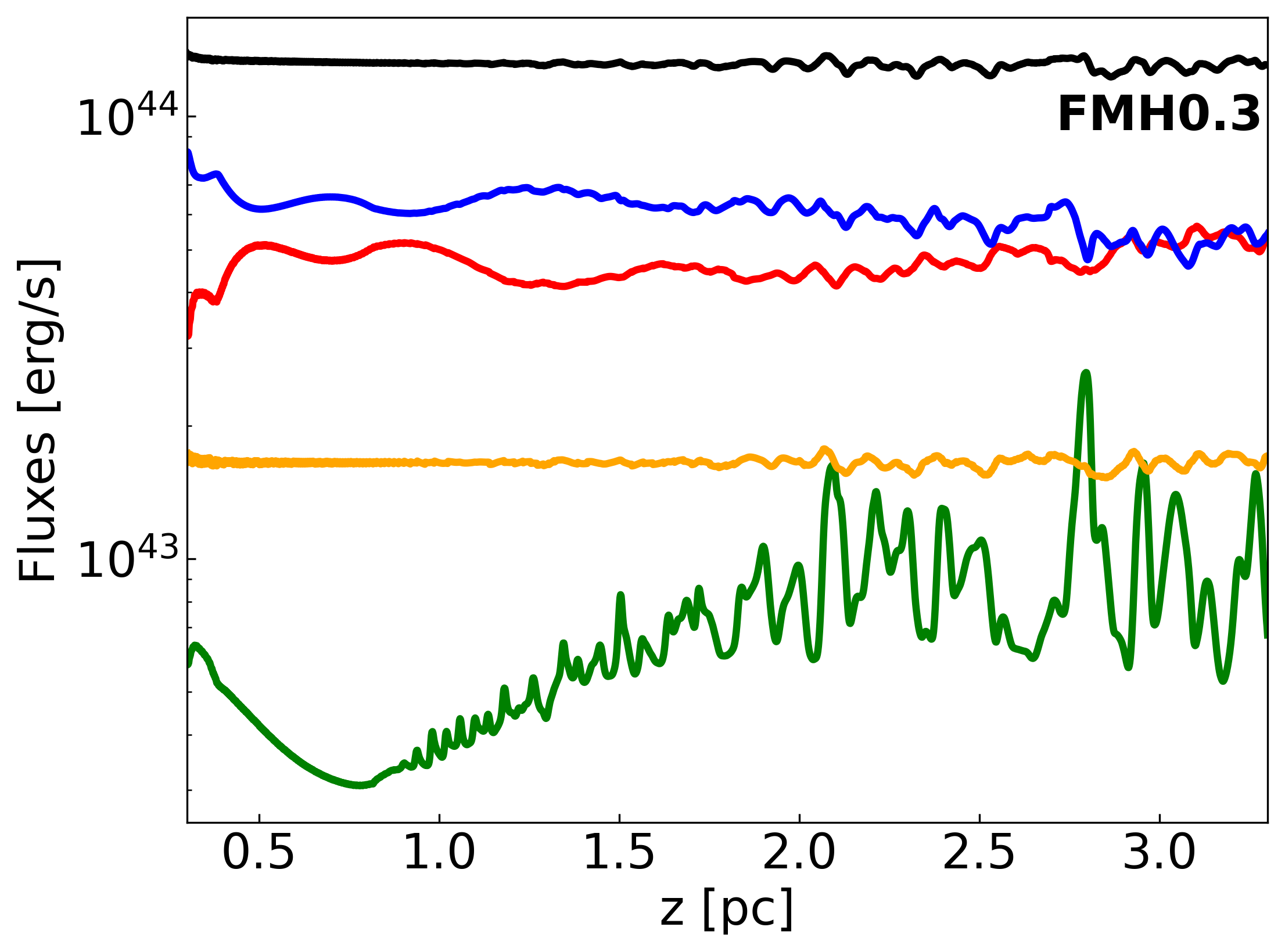}\par
\end{multicols}

\vspace{-0.8cm}

\begin{multicols}{3}
    \includegraphics[width=\linewidth]{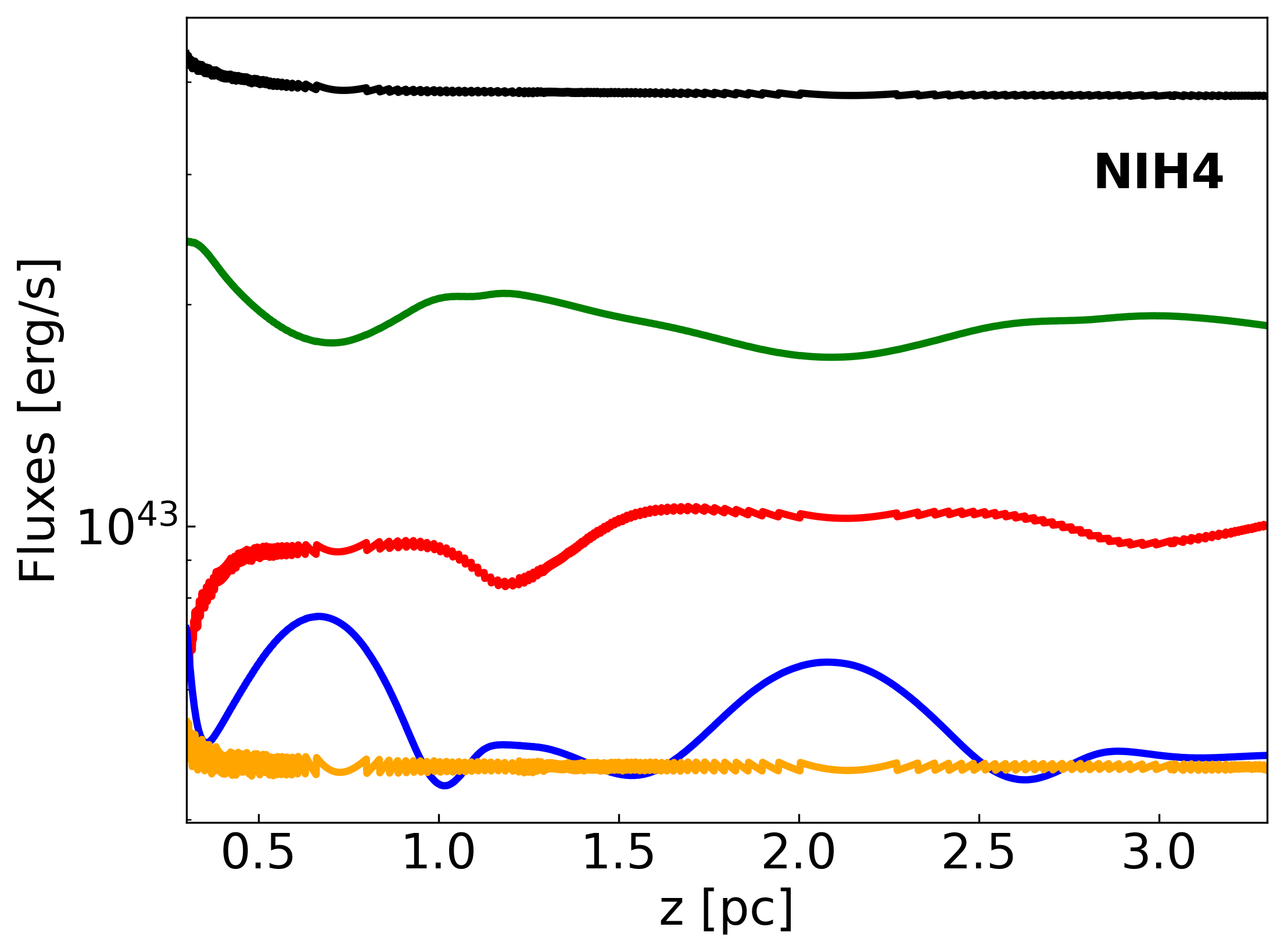}\par
    \includegraphics[width=\linewidth]{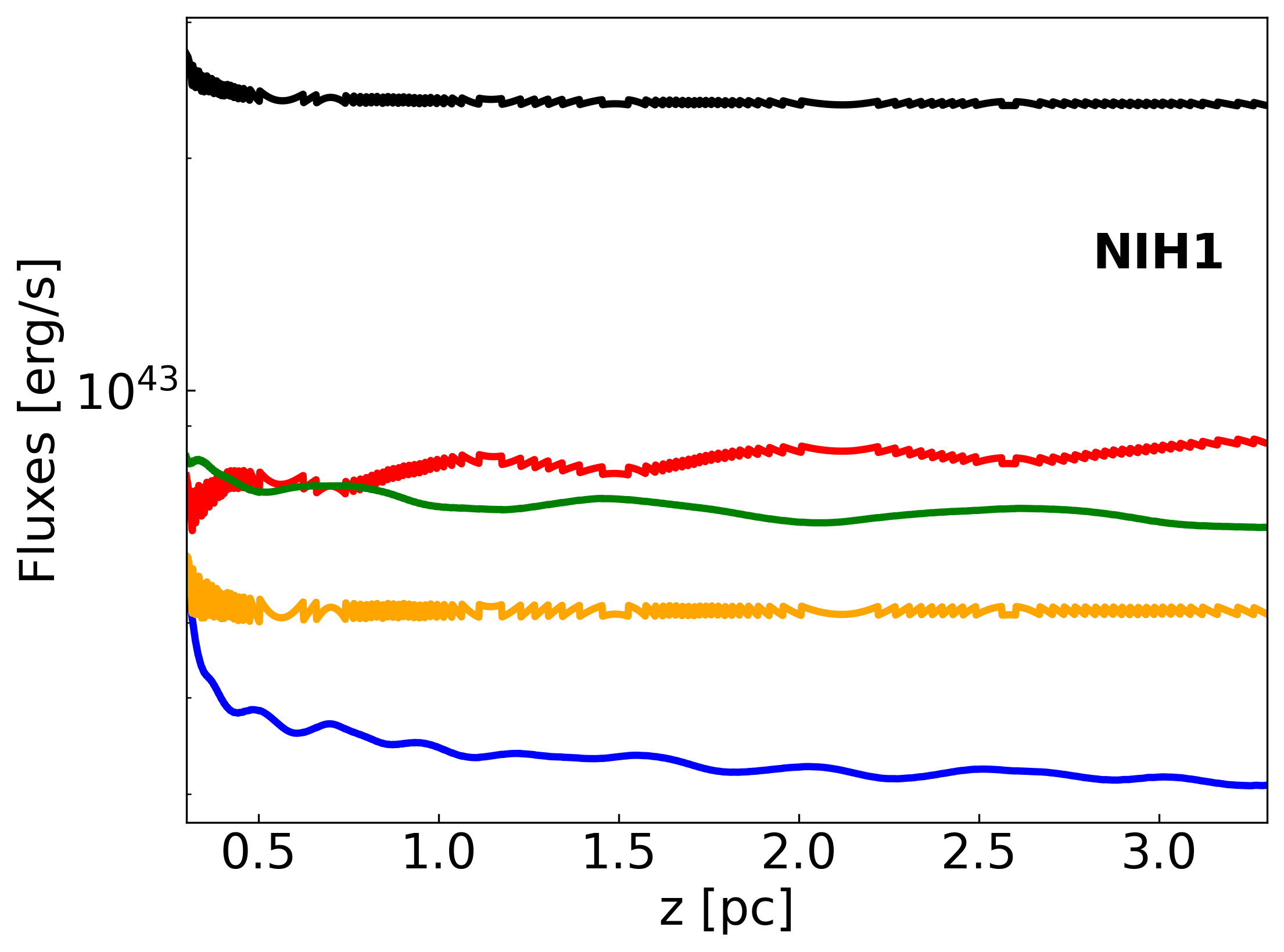}\par
    \includegraphics[width=0.9\linewidth]{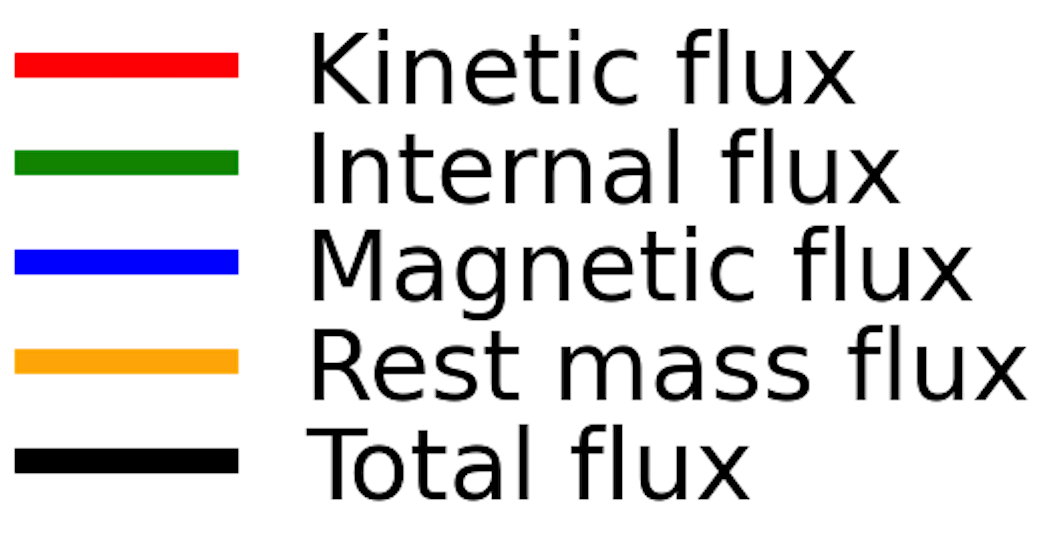}\par
\end{multicols}

    \caption{Evolution of the integrated energy fluxes along the axial direction for all the models reported in Table \ref{tab:setups}. 
    Black, red, green, blue and orange lines represent, respectively, the total, kinetic, internal, magnetic and rest-mass energy fluxes.
    Depending on the dominating energy channel, its dissipation in favor of the jet kinetic flux is visible in all the models. An exception is the kinetically dominated model FKS1, in which both the thermal and magnetical channels are dissipated to accelerate the jet, even if not dominant.}
    \label{fig:fluxes}
\end{figure*}

The opening angle at injection is also clearly correlated with initial jet overpressure, so force-free models with no thermal contribution to overpressure at injection ($p_j=p_a$) show smaller opening angles, albeit longer expansion regions (FMH1b, FMH1c and FML1). 
These models also show the largest average Lorentz factors (see Fig.\ \ref{fig:different_Lorentz_factor}), mainly driven by magnetic energy (as can be seen in Table~\ref{tab:fluxes} and Fig.~\ref{fig:fluxes}). 
Another relevant aspect to consider is that these models in (thermal) pressure equilibrium at injection also have larger initial velocities (see Table~\ref{tab:setups}), which causes smaller opening angles and therefore longer expansion lengths. 
Figure \ref{fig:opening_all_models} also shows the half-opening angles derived from the same observational results \citep{Ricci_2022}.
For the discussion of such comparison, we refer to Sect.~\ref{sec:comparison_NGC315}.

Figure \ref{fig:fluxes} shows the evolution of the different energy flux channels, together with the total one, along the $\mathrm{z}$ direction. 
These plots clearly show the correlation between the increase in kinetic energy flux and the drops in either magnetic or internal energy fluxes, or both. 
From the plots, it becomes evident that, at injection ($\mathrm{z}\,<\,0.6\,{\rm pc}$), a clear anti-correlation between internal energy and kinetic energy fluxes is only seen in internal energy dominated (FI and NI models). 
In the case of magnetically dominated models, it is the magnetic energy flux that shows anti-correlation with the kinetic energy flux, and we can observe regions in which the kinetic and internal energy fluxes are correlated (e.g., FMH1b). 
We can thus deduce that the dominant acceleration mechanism is controlled by the dominating energy flux, but that, in specific conditions, e.g., in expansions for the case of internal energy, the secondary energy flux can also contribute to acceleration. 
FKS1 represents an interesting case (Fig~\ref{fig:fluxes}, third row, central panel), in which 
the initial expansion results in a small increase of the kinetic energy flux favored by a drop in both internal and magnetic energy fluxes (although this jet reaches the smallest terminal values of the Lorentz factor; see Fig.~\ref{fig:different_Lorentz_factor}).

Finally, a slight global increase in magnetic flux is observed in the FMH1\_m4 model, beyond a distance of $\mathrm{z} = 2.5 \, \mathrm{pc}$.
The latter is in correspondence with the jet recollimating after an expansion phase (see Fig.~\ref{fig:different_Lorentz_factor}).
During the recollimation, the magnetic field is compressed and its value increases, but an increase in magnetic flux is typically not expected unless there is a boost in velocity due to internal energy, i.e., Bernoulli processes. However, this is not the case in the indicated region, so we propose such a small increase to be due to i) a physical process that transfers kinetic energy to magnetic energy as the reverse of magnetic acceleration, likely a consequence of local differential recollimation effects \citep[a similar behavior is observed in][Fig.\ 10, right panel]{Komissarov_2015}; ii) small inaccuracies arising during the summing of the different integrated quantities, or iii) numerical diffusion at the jet boundary, i.e., the magnetic field smearing out towards the outer cells. 
The exploration of such an effect is out of the scope of this paper and will be tackled by future works. 
Nevertheless, we remark that the fraction of such an increase is negligible with respect to the total jet and does not affect our results.

\section{Discussion} \label{sec:discussion}

Our results show that jet acceleration is driven by the dominant energy flux at injection, either magnetic or internal, but both are successful in increasing the jet Lorentz factor at expanding regions. 
However, there are significant differences in the acceleration patterns in both cases as shown in Sect.~\ref{sec:acceleration}.

In Sect.~\ref{sec:role_thermal} we discuss the role of thermal acceleration in relativistic jets.
The difference in the velocity structures due to the force-free versus non-force-free and thin versus thick shear layers are discussed in Sect.~\ref{sec:forcefreevsnonforce} and Sect.~\ref{sec:shear_layer}, respectively.
In Sects.~\ref{sec:comparison_NGC315} and \ref{sec:comparison_otherAGN} we focus on the comparison with the observational data and derive conclusions on how our results can provide insights for jet evolution. 
Finally, in Sect.~\ref{sec:discussion_jet_evolution}, we comment on the limitations of our study.

\subsection{The role of the thermal acceleration} \label{sec:role_thermal}

Jets produced by the Blandford-Znajek mechanism \citep{Blandford_1977} are probably dilute, hot, and significantly magnetized at sub-parsec scales, where entrainment has still not affected jet composition. 
Building upon the principle of Bernoulli, thermal acceleration is effective when the enthalpy $h$ is high enough (specifically, $h \gg c^2$). 
For a steady, relativistic flow, the principle is expressed by the law $h \gamma = \mathrm{const}$, establishing that the flow accelerates when $h$ decreases (as the flow expands). 
The process of thermal acceleration has been regularly seen in numerical simulations \citep[e.g.,][]{Perucho_2007b,Angles_2021}. 
By assuming that jets are hot at injection, we also found evidence for thermal acceleration at these scales: as highlighted in Sect.~\ref{sec:acceleration} and shown in Fig.~\ref{fig:fluxes}, we obtain the expected anti-correlation between thermal and kinetic energy fluxes during the evolution of our models. 
For instance, in the case of FIH3b model, in which the magnetic flux is an order of magnitude lower than the internal one and irrelevant to the acceleration of the jet, the jet is clearly accelerated by the Bernoulli process alone.
Interestingly, such model, and so thermal acceleration alone, is able to reach Lorentz factors higher than 2 on scales of $\sim 2-3 \, \mathrm{pc}$, matching thus the observed speed profile of NGC~315 (see Fig.~\ref{fig:Lorentz_all_models}).

Thermal acceleration can be relevant in relativistic jets at larger scales than those assigned to magnetic acceleration: if jets keep or gain internal energy on parsec scales, as it can be the case at recollimation shocks, it can act at the following conical expansions. In other words, even when jets are magnetically dominated at injection, once they are collimated and the magnetic acceleration stops, internal energy could still nourish the bulk acceleration.
The detection of jet flow acceleration on large scales, e.g., $\sim 100 \, \mathrm{pc}$ in blazar jets \citep{Homan_2015}, could be thus interpreted as an observational signature of thermal acceleration.

\subsection{Force-free versus non-force-free models}  \label{sec:forcefreevsnonforce}

One of the main triggers of the different acceleration patterns in our models is the force-free versus non-force-free nature of the flows. 
The reason lies in the cancellation of the field force in force-free models, eliminating the contribution of magnetic tension to jet collimation. 
Indeed, in the force-free configurations, the magnetic tension does not act as a collimating factor, and expansion is therefore enhanced, with half-opening angles reaching 7$^\circ$. 
On the contrary, in the non-force-free case, magnetic tension plays an active role in controlling expansion, which can only be temporarily overcome by considerable overpressure (as in, e.g., model NIH4). 
It is noticeable that the recollimation pattern in the non-force-free model NIH4 is, as a result, very different from that in FIH4, the force-free case: the angle formed by the shock wave and the jet axis is clearly smaller than in FIH4, and therefore NIH4 does not develop a Mach disk.

Because acceleration in the outer layers only takes place in force-free jets, which are, in general, magnetically dominated, we could derive the conclusion that this type of velocity pattern (slow spine and fast sheath) is a signature of magnetic acceleration in the case of thin shear layers, although FI models show that this does not strictly exclude internal energy as an accelerating mechanism. 
Fast spines would correspond, in contrast, either to internal energy-dominated flows or strongly sheared (possibly wind-shielded, two-flow) jets. 

\subsection{The role of shear layers} \label{sec:shear_layer}

Our simulations show that the different thicknesses of the shear layer can lead to a fundamental change in the acceleration pattern of force-free jets.
Fig.~\ref{fig:schlieren} shows that this may be caused by the difference in the opening angles of the region separating the inner jet and the shear layers (indicated by a white contour that stands for tracer $f=0.9$). 
Initially, thicker layers produce smaller opening angles and limit recollimation to conical shocks, allowing acceleration along the spine, even if modulated by the succession of expansions and recollimations. 
We propose that the direct consequence of the thicker shear layer is the delaying of the wave propagation.
As a result, the triple point, associated with the incident/reflected/Mach shocks in the jet-ambient interaction surface, which is visible in Fig.~\ref{fig:schlieren} (highlighted by the black squares) is found in model FMH1 at $\sim 0.7 \, \mathrm{pc}$ while in model FMH1\_m4 around $1.3 \, \mathrm{pc}$.
The role of the width of the shear layer has been confirmed by further simulations (see Appendix \ref{app:other_models}).

Comparing the maximum Lorentz factors reached by the different models (Fig.~\ref{fig:different_Lorentz_factor}), we observe that FMH1\_m4 reaches a slightly larger Lorentz factor than its thin-layered counterpart, FMH1. 
The most plausible explanation of this result is that the growth of the instabilities observed in model FMH1 dissipates some fraction of the kinetic energy of the flow, whereas the jet of model FMH1\_m4 is shielded from its environment by a thick shear layer, which, together with further expansion can contribute to preserving its collimation \citep[see, e.g.,][]{Perucho_2019}. 

Altogether, we can conclude that the role of shear layers at the collimation/acceleration region can be of great importance for jet evolution, with thick-layered jets producing collimated, stable, fast flows with velocity profiles consisting of a fast spine and a slower, shielding layer. 
On the contrary, thin-layered jets initially develop a slow, hot spine, surrounded by a sheath along which the flow is accelerated.
The boundary between these two regions, spine and sheath, is prone to the development of instabilities and can eventually dissipate part of the gained kinetic energy. 

We highlight, however, that this is a small-scale effect (see Sect.~\ref{sec:acceleration}, e.g., model FMH1)
and is expected to not be largely significant (see the small velocity differences between FMH1 and FMH1\_m4, as aforementioned) within the simulated grid because the instabilities have not reached the non-linear regime and have not developed mixing between the hot and slow spines and the faster and cold sheath.

\subsection{Comparison with NGC\,315} \label{sec:comparison_NGC315}

A one-to-one comparison between observations and simulations is, in general, not possible due to limitations of both. 
Radio maps can be affected by uncertainties in the calibration and imaging processes, as well as by time-variable and local effects that do not necessarily allow us to infer the physical conditions of the underlying plasma flow. 
In simulations, as discussed, some simplifications are unavoidable. Nonetheless, some general trends can be identified and discussed qualitatively, and we do this in the following by considering NGC~315.

Concerning the jet opening angles, Fig.~\ref{fig:opening_all_models} shows that most of our simulated models succeed in reproducing the terminal values at the end of the collimation region, as inferred from the 5~GHz observations at different epochs ($\theta \sim 1-3 \degree$). 
These opening angles also agree with the average values observed in radio galaxies \citep[e.g.][]{Pushkarev_2017}. 
In contrast, we observe that the opening angle in the innermost jet, described at the highest radio frequencies (15, 22, and 43 GHz), does not match our synthetic profiles.
The limitations suffered by numerical models forced us to inject super magnetosonic flows, as previously explained (Sect.~\ref{sec:jet_ambient_params_2D}). 
This also makes any comparison challenging at the compact scales: while the simulated jets open abruptly from the injection point at $\sim 0.3$~pc due to initial overpressure, the high-frequency data from the jets in NGC~315 indicate that such an opening happens on smaller scales and that the jet at 0.3~pc is going through the final part of its collimation before transiting to the conical shape \citep{Boccardi_2021}.
Nonetheless, as mentioned, the observational and simulated data reconcile on parsec scales.

Except for the innermost jet region, an agreement between observations and simulations can be found in the fact that they both show a co-spatiality of the acceleration and collimation processes. 
The simulations show acceleration episodes between $(0.5-1.0)$~pc and $(1.5-2.0)$~pc (Fig.~\ref{fig:Lorentz_all_models}), i.e., along the region where the opening angle is also decreasing (Fig.~\ref{fig:opening_all_models}). 
This is probably the last part of the accelerating region, which in NGC~315 extends within the inner $ \sim 1-2$~pc.
The co-spatiality agrees with the expectation, for FM models, of a magnetic acceleration being relevant when the jet is evolving with a parabolic or quasi-parabolic shape \citep{Komissarov_2007}.
Nonetheless, as mentioned in Sect.~\ref{sec:role_thermal}, thermal acceleration alone could in principle reproduce the observed velocities on parsec scales.

A comparison between the simulated and observed Lorentz factors may indicate a better agreement for models with jet power $\sim 10^{43} \, \mathrm{erg/s}$, consistent with the jet power estimated from observational constraints \citep[see, e.g.,][]{Morganti_2009, Ricci_2022}. 
Indeed, the average Lorentz factor of $\sim 2$ on scales of $\sim 2-3 \, \mathrm{pc}$ can be reproduced by models with such jet powers.
On the contrary, models with higher jet power ($\sim 10^{44} \, \mathrm{erg/s}$) seems to fail to align with the observational data: while they manage to reach the Lorentz factor peak of $\gamma \sim 4$ at the distance of $\sim 1.5 \, \mathrm{pc}$ (observed in only one epoch), they exhibit terminal Lorentz factors exceeding the maximum observed values by more than a factor of two. Remarkable examples are the highly-magnetized models, such as FML1, FMH1c, and FMH0.3.
As in the case of the opening angle, a meaningful comparison on sub-parsec scales is not possible: data at 22 GHz and 43 GHz show acceleration from very small velocities, out of the reach of our setups (see Sect.~\ref{sec:jet_ambient_params_2D}).

Another interesting aspect might be noticed from Fig.~\ref{fig:Lorentz_all_models}: the Lorentz factor evolution along the jet changes between epochs, even at the same frequency, e.g., the 1996 5~GHz epochs, which are separated by five months.
While taking into account that such variations can be a consequence of various local and observational effects, it is also possible that they are the result of intrinsic changes in the injection conditions.
Indeed, the jet velocities and variability in Lorentz factors recovered from the family of models with jet powers $\sim 10^{43} \mathrm{erg/s}$ could potentially match the observed time-dependent changes in the Lorentz factor.
The simulations we have presented here differ from each other by only a few parameters, within a range of plausible values derived from observational results (see, e.g., Fig.~\ref{fig:Magnetic_field_initial}). 
Thus, we could suggest that the different jet profiles inferred from the same frequencies but different epochs may arise from variations in the jet properties at the formation site, within timescales of a few months.
This might show how jets are far from being subject to regular injection and that conditions and, consequently, acceleration patterns, can behave in an extremely dynamic way. 
Further evidence for time-variability of the physical properties at the jet base is provided by the observed time-dependence of the core shift effect \citep{Plavin_2019}, as well as by the well-known flux density variability that characterizes the nuclear regions on timescales from hours to months.

\subsection{Comparison with other AGN jets} \label{sec:comparison_otherAGN}

As discussed, most of the force-free jets we have simulated are characterized by sheath acceleration and by a slow spine. 
This result appears at odds with observational constraints, which rather indicate the existence of a fast spine and a slow sheath. 
Such a velocity structure was suggested, for instance, to reconcile the observed spectral properties of FRI radio galaxies with their beamed parent population, the BL Lacs \citep{Chiaberge2000}, and appears required to explain the general anti-correlation between the observed Lorentz factor and the jet viewing angle \citep[see, e.g.,][Fig. 12]{Homan2021}.
Misaligned jets, which are seen from the side, are usually characterized by mildly relativistic speeds, as opposed to blazar jets, which are observed closer to the jet axis with Lorentz factors of the orders of tens \citep[e.g.,][]{Lister2019}. 
This is in agreement with the frequent observation of limb-brightening in misaligned jets \citep[see, among others, the recent cases of Centaurus A and NGC~315 by][respectively]{Janssen2021,Park_2021}, which can be interpreted as due to the Doppler de-boosting of the fast spine.
Note that this applies not only to the low-power FRI jets, but also to the powerful FRIIs \citep[see the case of Cygnus A, presented by][]{Boccardi2016}. 

Consequently, in this scenario, a match between our simulations and observations implies that jets would need to propagate with shear layers protecting the spine or with non-force-free magnetic field configurations, since jets with fast spine are produced in both families of models. 
On the opposite side, while faster shear layers with hot spines are likely not suggested from VLBI observations, such a radial velocity structure is not ruled out.
In this situation, the initial force-free, thin shear layer models are favored.

As recently shown by \cite{Boccardi_2021}, the properties of the outer jet sheath may vary depending on the accretion mode in the AGN. 
Specifically, an extended and possibly disk-launched jet sheath was observed in High Excitation Radio Galaxies (HERG), which are mostly characterized by FRII morphologies. 
On the other hand, Low Excitation Radio Galaxies (LERG), associated to both FRI and FRII morphologies, showed in comparison a narrow jet anchored in the very inner regions of the accretion flow. 
In future works, we aim to explore possible connections between HERG/LERG and the different acceleration patterns we observed in this paper.

Furthermore, in a future work, we plan to extrapolate synthetic spectral index maps from our models using the post-processing code published in \citet{Fromm_2018}.
In this way, we will be able to compare the simulated spectral properties with the observations and explore whether the different acceleration patterns we recover, and especially the outer layer acceleration, may result in characteristic signatures that can be related to the underlying magnetic field properties \citep[e.g.,][]{Kramer_2021}.

\subsection{A warning note on our setups: Jet evolution} 
\label{sec:discussion_jet_evolution}

When jets are initially launched in AGN, they trigger a bow shock that propagates through the ambient medium. 
Because the jet flow advances faster within the formed jet channel than this forward bow shock, the jets get surrounded by overpressured, shocked jet plasma (known as backflow or cocoon). 
Therefore, collimation must be, in this initial phase, controlled by both the pressure of this shocked environment and the jet toroidal field. 
As the jet expands and the bow and reverse shocks propagate to large distances, the pressure in the cavity formed by the jet falls continuously with time \citep[see, e.g.,][]{Perucho_2014b,Perucho_2019}.
Then, the gravitational potential of the galaxy, which acts in a dynamical time $\sim 1/\sqrt{G\,\rho_{DM}}$ (where $G$ is the gravitational constant and $\rho_{DM}$ is the dark matter density in the galaxy), allows the galactic pressure and density profiles to be recovered \citep[see, e.g.,][]{Perucho_2014b} and therefore, pressure should fall to the original ISM values, i.e., $\sim 10^{-10}\,{\rm dyn/cm^2}$. 

The expected jet energy fluxes and magnetic field intensity in this region ($F_j \sim 10^{43} - 10^{44} \, \mathrm{erg/s}$ and $B \sim 0.1-1 \, \mathrm{G}$) result in extremely large overpressure of the jet with respect to the ISM. 
Once the jet is accelerated to relativistic velocities, the opening angle is proportional to the inverse of the Lorentz factor \citep[see][]{Komissarov_2012}. 
The large initial overpressure poses a strong challenge to jet collimation and to numerical simulations aimed at studying this region.  
In our simulations, we decided to achieve overpressures as large as possible, which could allow free expansion, by introducing significant pressure gradients in distance. 
Once free expansion is allowed, the ambient pressure becomes relatively irrelevant, and this has allowed us to run our simulations. 
Establishing ambient media with realistic initial pressure causes the code to crash and makes such simulations impossible. 

An obvious question arises: how is it then possible that the jets manage to settle in such ambient media with a pressure that can be orders of magnitude smaller than the jet pressure? 
We would hypothesize that it is because the jet has adapted to its environment in a continuous, relatively slow way during the time it takes for its terminal shock to reach kiloparsec scales, as opposed to simulations. 
In addition, this probably makes a non-force-free magnetic field configuration necessary, for magnetic tension to play a significant role in jet collimation in the region where it is precisely still strong, i.e., the sub and trans-Alfvènic regions, until the jet is collimated by its own velocity. Furthermore, the collimating effect of the toroidal field also avoids the loss of internal energy via Bernoulli acceleration, allowing this energy budget to act at larger scales \citep[see also][]{Angles_2021}.

\section{Conclusions} \label{sec:Conclusions}

In this paper we have performed a numerical study of the jet acceleration in typical FR~I radio galaxies using the well-known source NGC~315 as a starting model.
We explored the evolution on sub-parsec and parsec scales of different axisymmetric jet models using a two-dimensional RMHD code, and compared our final results with the observational properties inferred on the same scales by means of VLBI observations.
Our results are summarized here.

\begin{itemize}

    \item All the simulated models show acceleration from initial Lorentz factor values of $\gamma \sim (1-2)$ up to maximum $\gamma \sim (2-5)$ within the simulated scales of $(0.3-3.3)$ pc (Fig.~\ref{fig:Lorentz_all_models}). 
    The flux budget evolution for the different simulated models reported in Fig.~\ref{fig:fluxes} clearly shows how the dominant energy flux at injection mainly drives the acceleration.
    Both the internal and magnetic energy channels are dissipated in turn to accelerate the jet, and so increase the kinetic energy of the beam. 
    Moreover, we notice how in very specific cases, such as during the expansion phase, the internal energy can be dissipated to increase the jet speed, even in magnetically dominated jets. 
    Our results on the acceleration mechanisms are mainly relevant in two different ways.
    i) They confirm the current magnetic acceleration paradigm \citep{Komissarov_2007, Komissarov_2012}. Indeed, in the case of cold, magnetically dominated jets the magnetic energy is converted into kinetic energy, thus accelerating the bulk flow.
    ii) They expand the current view on thermal acceleration in internally dominated jets.
    In contrast to what is proposed in \citep{Vlahakis_2003_a, Vlahakis_2003_b, Vlahakis_2004}, where the thermal acceleration is claimed to play a role only on compact scales before being overtaken by the magnetic acceleration, our findings indicated that when jets are thermodynamically relativistic \citep{Perucho_2017}, thermal acceleration remains relevant even at parsec scales.
    Furthermore, thermal acceleration enables to reach higher Lorentz factors than previously thought, aligning with those observed in AGN jets (in this specific case, in NGC~315).
    Remarkably, given that thermal acceleration can operate even when the jet undergoes conical expansion, our results suggest that jets can continue to accelerate on large scales, beyond the collimation region.

    \item We infer an explicit relation between the jet speed profiles and the intrinsic jet properties. 
    The different velocity structures show a dependence on the magnetic field configurations, i.e., force-free vs. non-force-free, and/or the dominating energy flux.
    When a Mach disk, a strong internal planar shock, is formed due to the fast expansion, the acceleration deviates towards the outer and expanding layers. 
    A mild acceleration on the spine is visible only towards the end of the grid of our simulation domain. 
    On the contrary, when jets develop milder, conical shocks, the acceleration is concentrated in the central spine and smoothly decreases toward the external medium.
    The Mach disk is prominently visible in the force-free and magnetically dominated scenario.
    However, a thicker shear layer 
    may delay the propagation and bouncing of the waves long enough to avoid the formation of the Mach disk and altering the downstream evolution of the jet, leading to a spine-accelerated structure.
    This result implies that the initial thickness of the shear layer, possibly associated with accretion disk-launched winds, 
    is a relevant element in understanding the evolution and propagation of relativistic jets.
    While the acceleration focused on the outer layers can be a signature of magnetically dominated jets with thin sheaths, this does not necessarily exclude hot jets.
    In the non-force-free configuration, the Mach disks are completely absent, and a classical expansion/recollimation pattern with spine acceleration is observed.
    
    \item The opening angle profiles shown in Fig.~\ref{fig:opening_all_models} highlight how the force-free models expand faster than the non-force-free counterpart. 
    This is attributed to the disengagement of the magnetic field force in the former scenario, which eliminates the contribution of the magnetic tension to the jet collimation. 
    From the comparison of the opening angle (Fig.~\ref{fig:opening_all_models}) and Lorentz factor (Fig.~\ref{fig:Lorentz_all_models}) profiles, we can distinguish two clear evolution cases: i) the models that develop strong shocks and have the acceleration focused in the outer layers are the ones showing the fastest growth and higher maxima in the opening angle and ii) those developing conical shocks with a downstream fast-spine structure are the ones with smaller slope and lower maxima.
    
    \item As seen in Figs.~\ref{fig:Lorentz_all_models} and \ref{fig:opening_all_models}, a number of models can reproduce both the Lorentz factor and opening angle profiles inferred for NGC\,315 employing cm- and mm-VLBI observations. 
    From the comparison with the observed jet speed in NGC~315, we suggest certain models are favored over others when it comes to modeling such a source.
    On the one hand, jets with power larger than $10^{44} \, \mathrm{erg/s}$ show Lorentz factors at 3 pc which are remarkably larger than the observed one, giving an indication of a possible upper limit on the jet power of NGC~315.
    On the other hand, we highlight how different models with jet powers $\sim 10^{43} \, \mathrm{erg/s}$ match the different acceleration profiles obtained across the multi-epoch observations at 5 GHz. 
    This may suggest how the small changes in the jet properties at the injection may lead to consistent variations in the observed terminal Lorentz factor. 
    Such variations can occur over time scales of a few months, indicating how the variations in the injection properties may drive the different observed speed profiles.
        
\end{itemize}

\begin{acknowledgements} 

We would like to thank the referee for the comments, which highly improved the quality of the paper.
We thank Eduardo Ros for his comments which improved the readability of the manuscript.
Moreover, we thank Giancarlo Mattia for the useful discussion on the possible effects of magnetic reconnection.
All the simulations were performed on the Cobra cluster of the Max Planck Society.
LR and BB acknowledge the financial support of a Otto Hahn research group from the Max Planck Society.
LR is funded by the Deutsche Forschungsgemeinschaft (DFG, German Research Foundation) – project number 443220636. MP and JMM acknowledge support by the Spanish Ministry of Science through Grants PID2019-105510GB-C31/AEI/10.13039/501100011033, PID2019-107427GB-C33, and from the Generalitat Valenciana through grants PROMETEU/2019/071 and ASFAE/2022/005. JLM acknowledges support from the Generalitat Valenciana through grant ASFAE/2022/005.

\end{acknowledgements}

\bibliographystyle{aa.bst}
\bibliography{bibliography}

\begin{appendix}

\section{Radial initial fluxes distribution} \label{app:radial_fluxes}

\begin{figure*}[h]
    \centering
\begin{multicols}{2}
    \includegraphics[width=\linewidth]{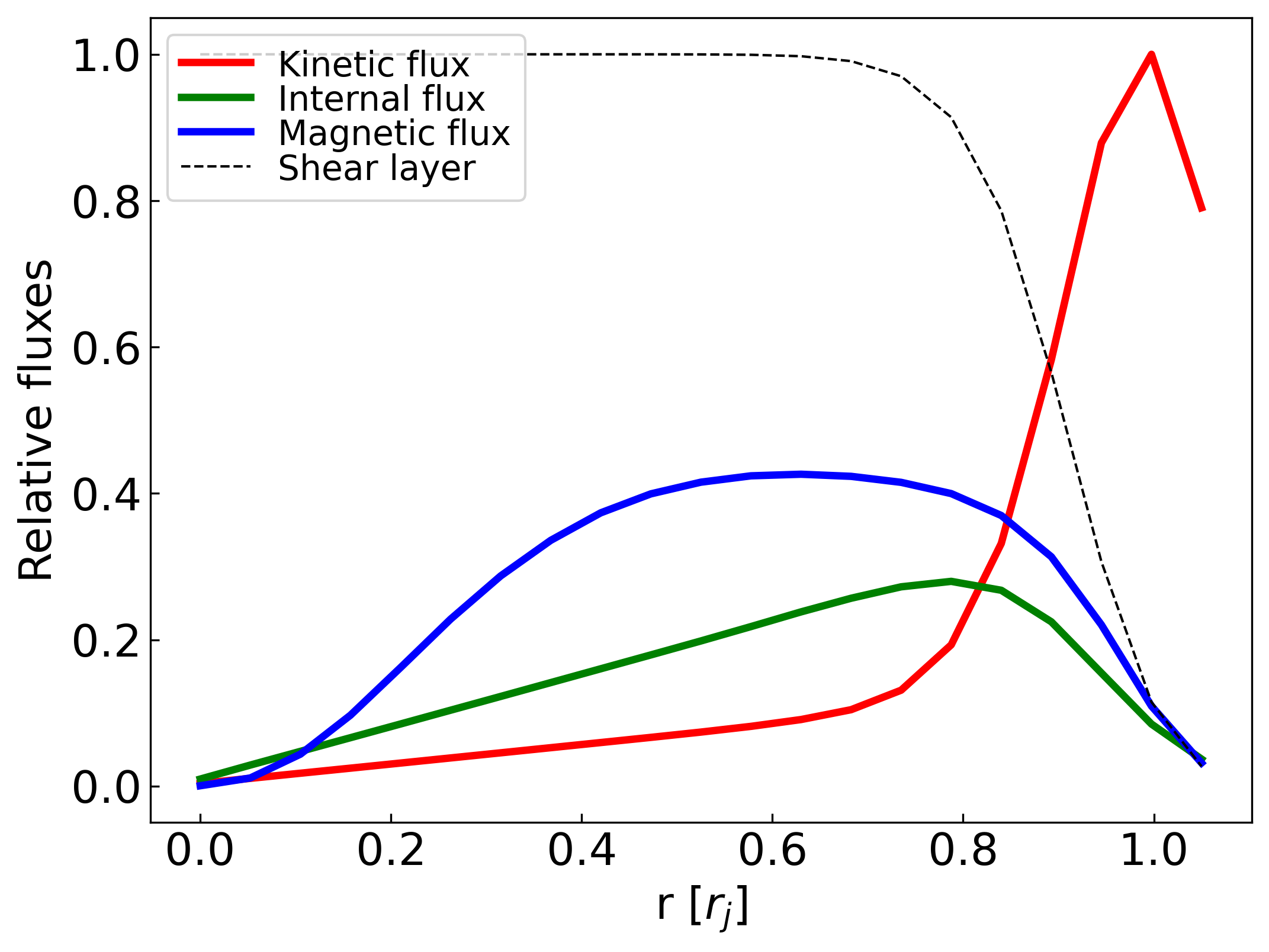}\par
    \includegraphics[width=\linewidth]{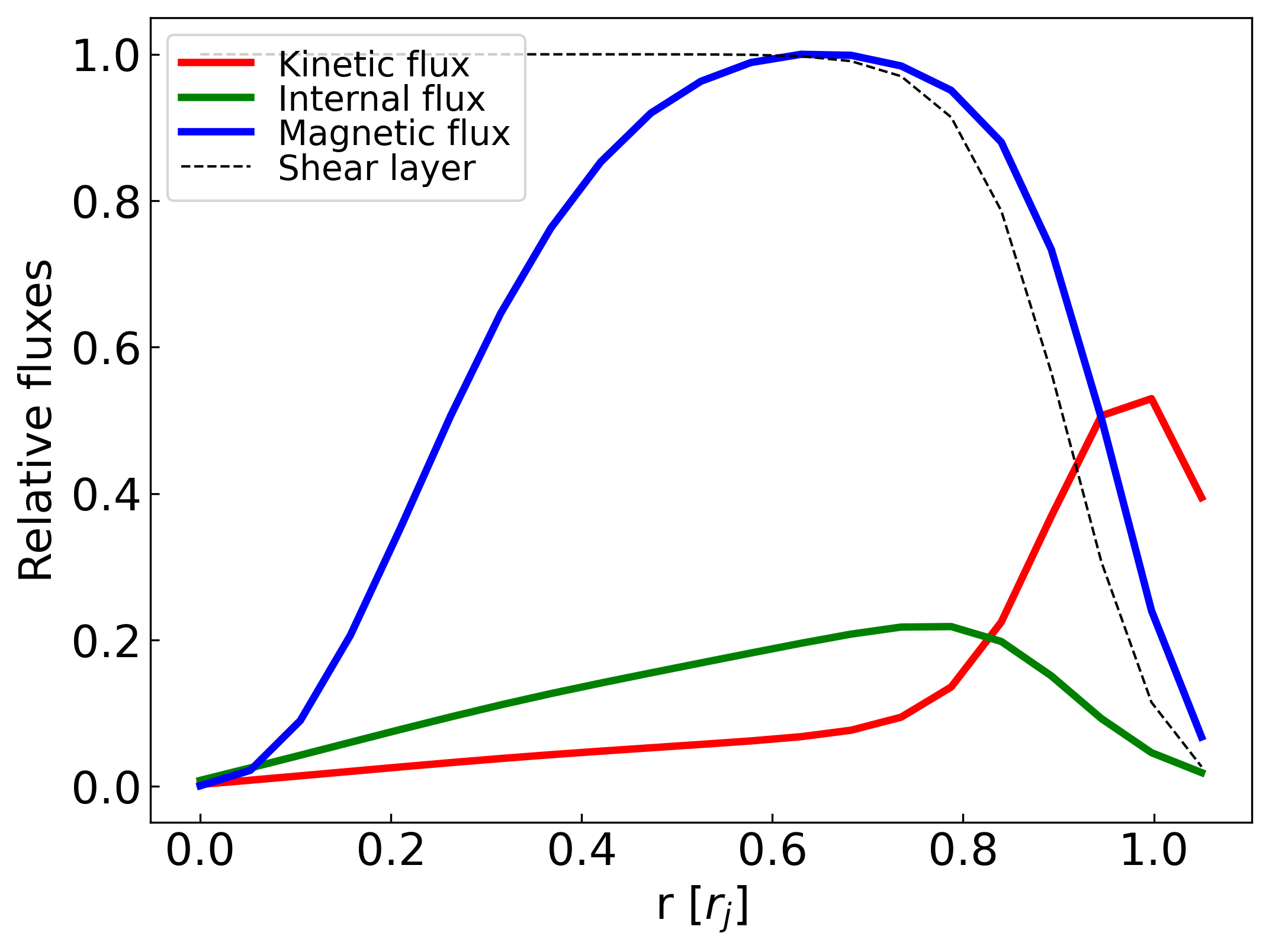}\par
\end{multicols}
    \caption{Relative initial radial distribution of the different energy channels for models FMH1 (left panel) and FMH1c (right panel). We highlight how the rest-mass contribution is absorbed into the kinetic energy. All the values are normalized to the maximum flux in each respective plot.}
    \label{fig:radial_fluxes}
\end{figure*}

The convolution of the jet with the shear layer alters the initial analytical distribution of the fluxes, leading to the differences between the modeled and simulated values reported in Table \ref{tab:fluxes}.

In Fig.~\ref{fig:radial_fluxes}, we show examples of the two main possibilities presented in this paper using models FMH1 and FMH1c.
On the one hand, in the lower power models (left panel), the shear layer carries a relevant fraction of the total jet power, in the form of kinetic energy (in Fig.~\ref{fig:radial_fluxes}, the kinetic energy absorbs the contribution of the rest mass).
This is the consequence of the density in the jet increasing in the shear layer to match the external one.
On the other hand, in the high-power, highly magnetized, models (right panel), the shear layer is magnetically dominated for the majority of its cross-section.
As shown in Sects.~\ref{sec:results} and \ref{sec:discussion}, these differences do not lead to differences in the steady solutions (models FMH1 and FMH1c both develop the Mach disk and evolve with a similar global structure).

\section{From initial conditions to equilibrium: pressure wave} \label{app:pressure_wave}

\begin{figure*}[h]
    \centering
\begin{multicols}{3}
    \includegraphics[width=\linewidth]{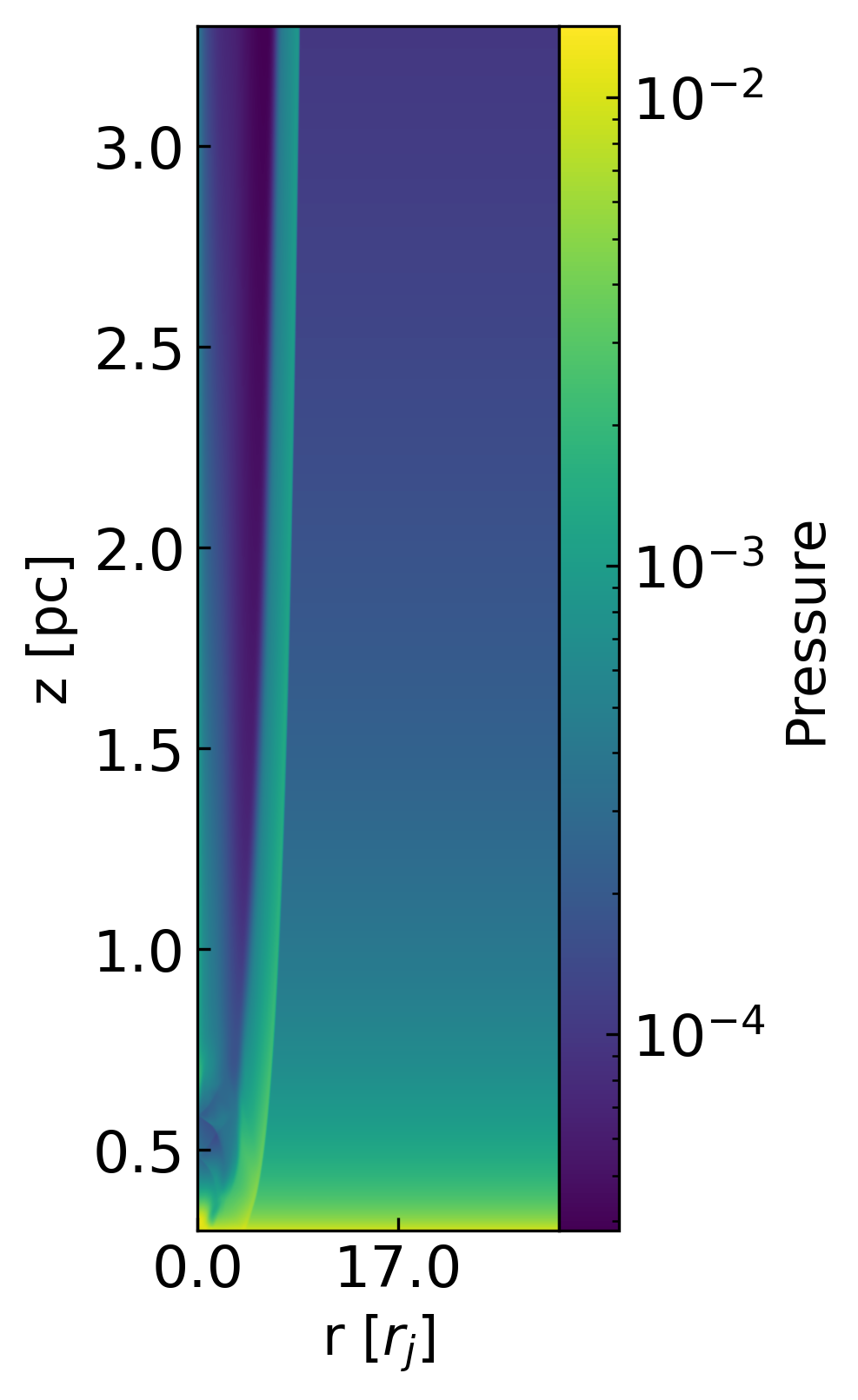}\par
    \includegraphics[width=\linewidth]{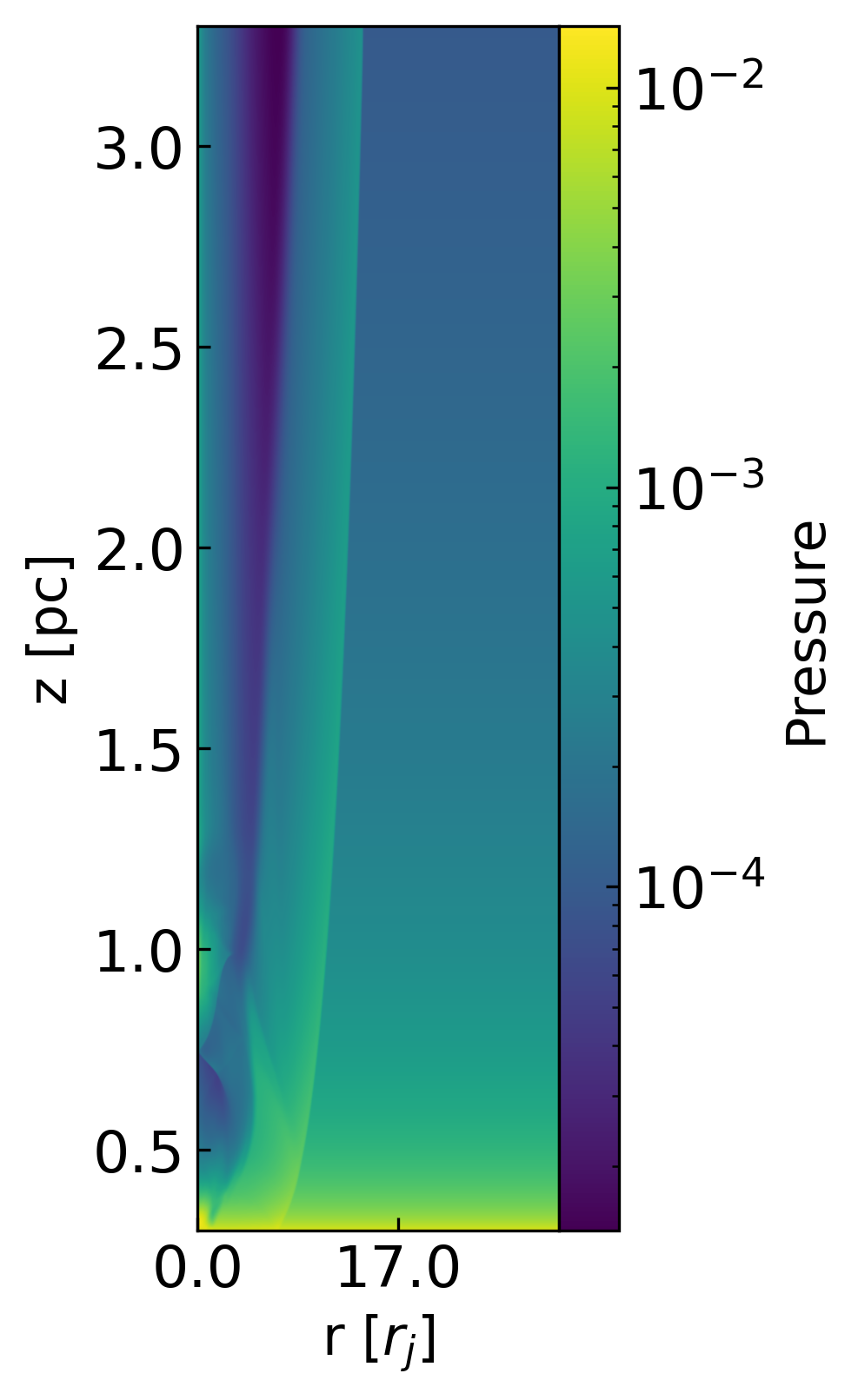}\par
    \includegraphics[width=\linewidth]{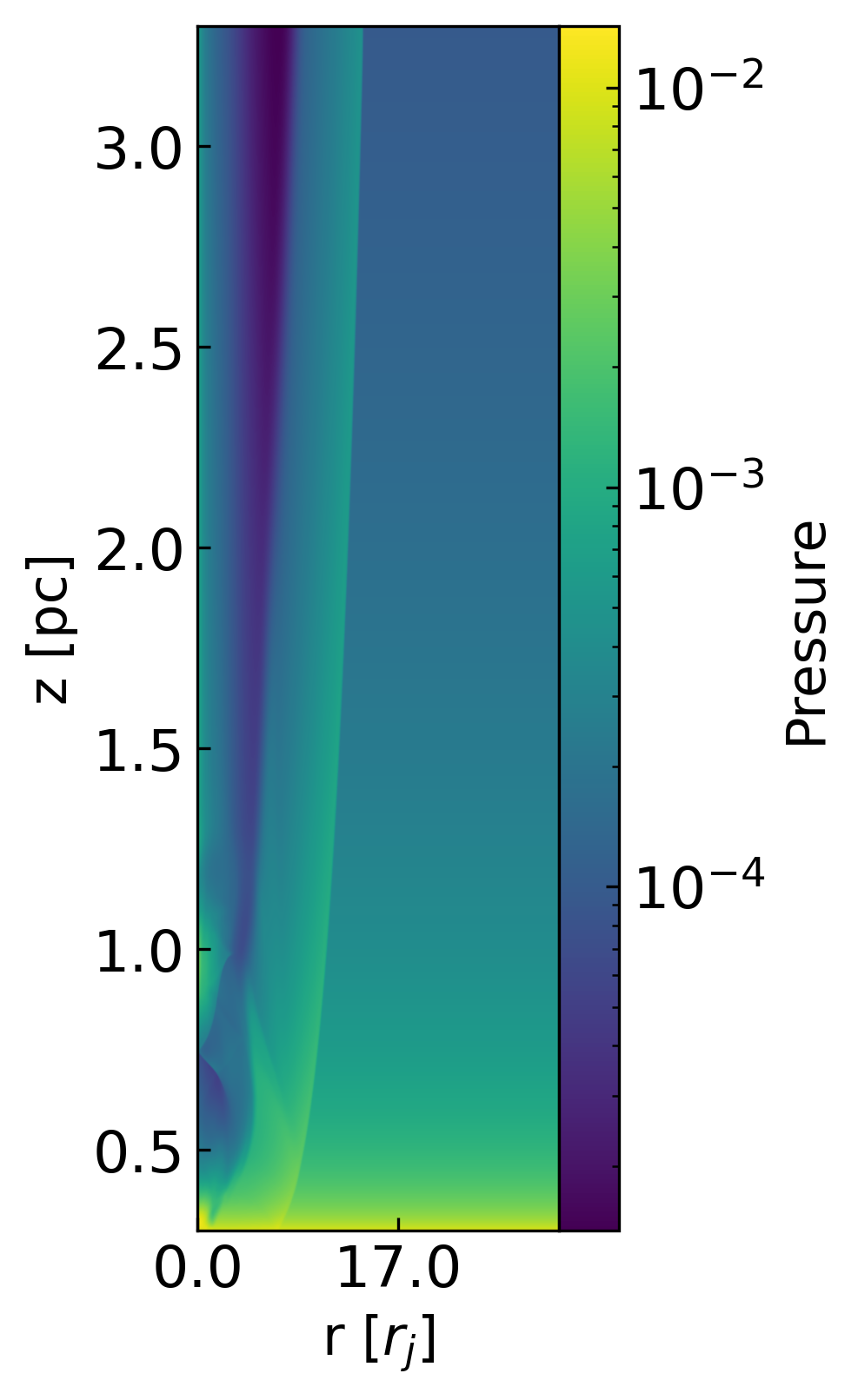}\par
\end{multicols}
    \caption{Pressure maps at different times for the model FMH1\_m4. Starting from left to right the time steps are t$=20 \, r_j/c$, t$=40 \, r_j/c$, and t$=70 \, r_j/c$. The pressure wave moving from the left to the right boundary is clearly visible.}
    \label{fig:pressure_wave}
\end{figure*}

\begin{figure}[h]
    \centering
    \includegraphics[width=0.85\linewidth]{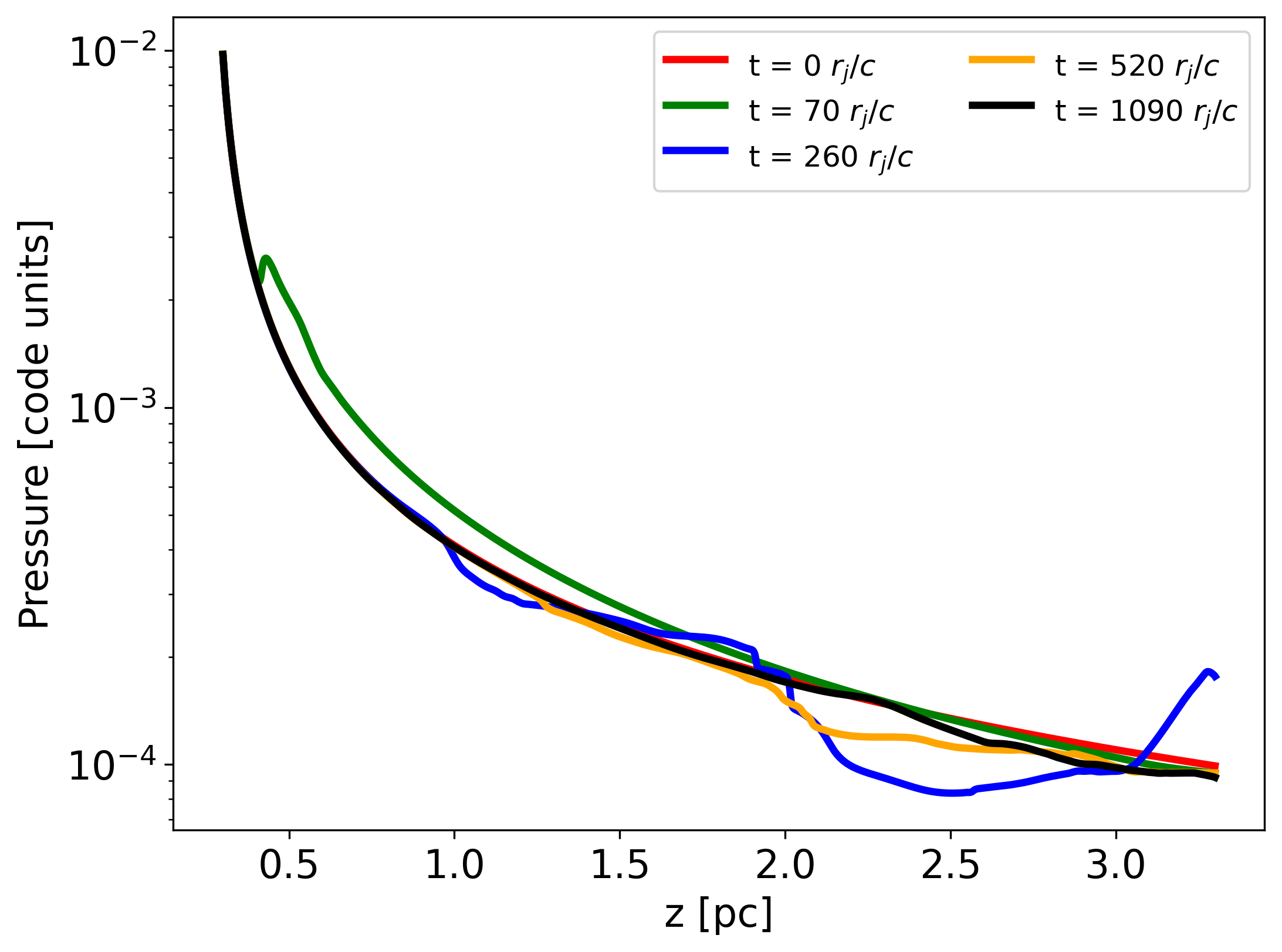}\par
    \caption{Left panel: ambient pressure evolution along the axial direction for the different time steps of t$ = 0 \, r_j/c$ (red line), t$ = 70 \, r_j/c$ (green line), t$ = 280 \, r_j/c$ (blue line), t$ = 520 \, r_j/c$ (orange line), and t$ = 1090 \, r_j/c$ (black line). In the intermediate time steps, the oscillations in the pressure values are due to the passage of the wave. After enough time, such oscillations are absorbed by the boundary conditions and the original profile is recovered.}
    \label{fig:pressure_wave_plots}
\end{figure}

\begin{figure*}[h]
    \centering
    \begin{multicols}{2}
    \includegraphics[width=0.95\linewidth]{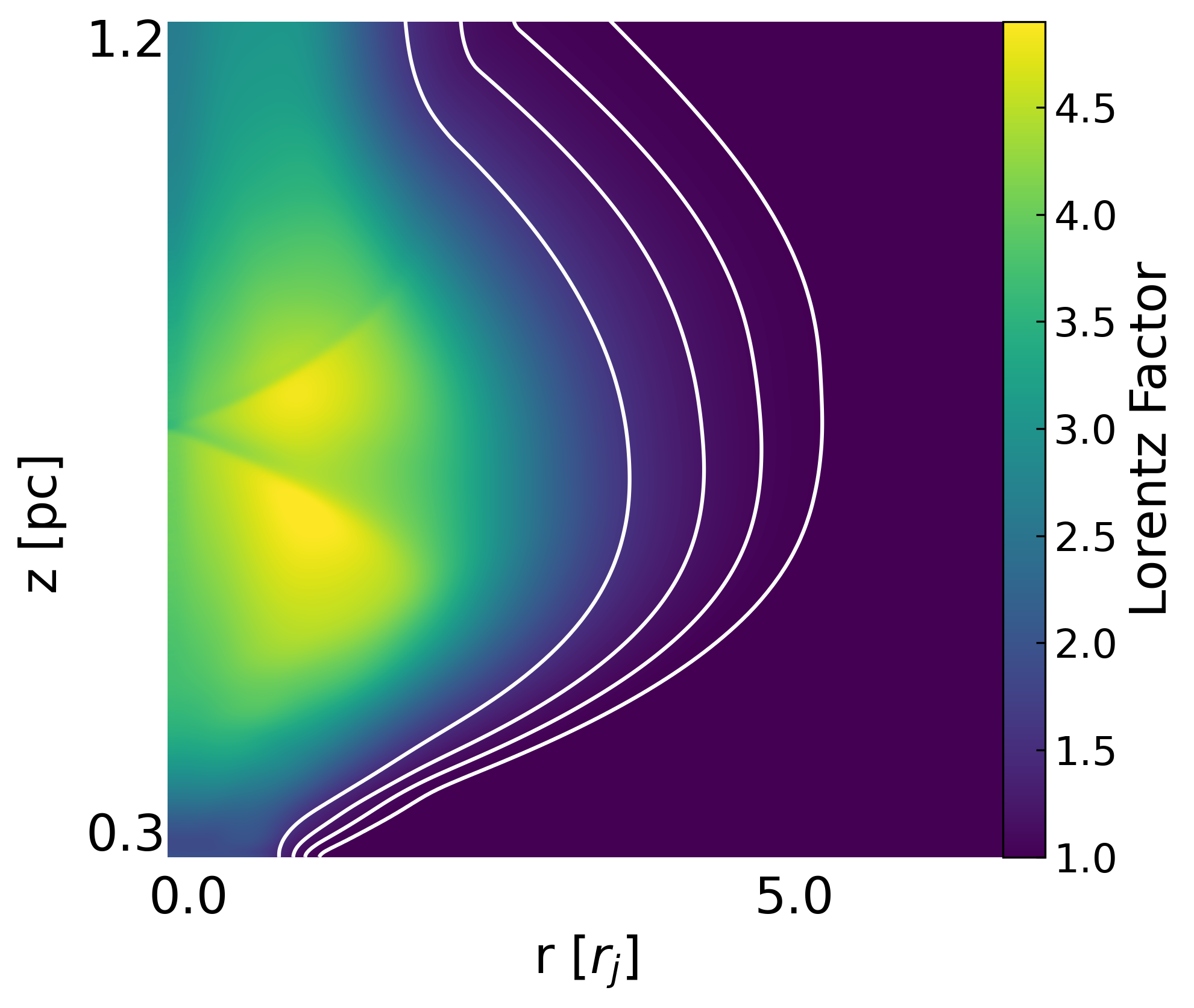}\par
    \includegraphics[width=\linewidth]{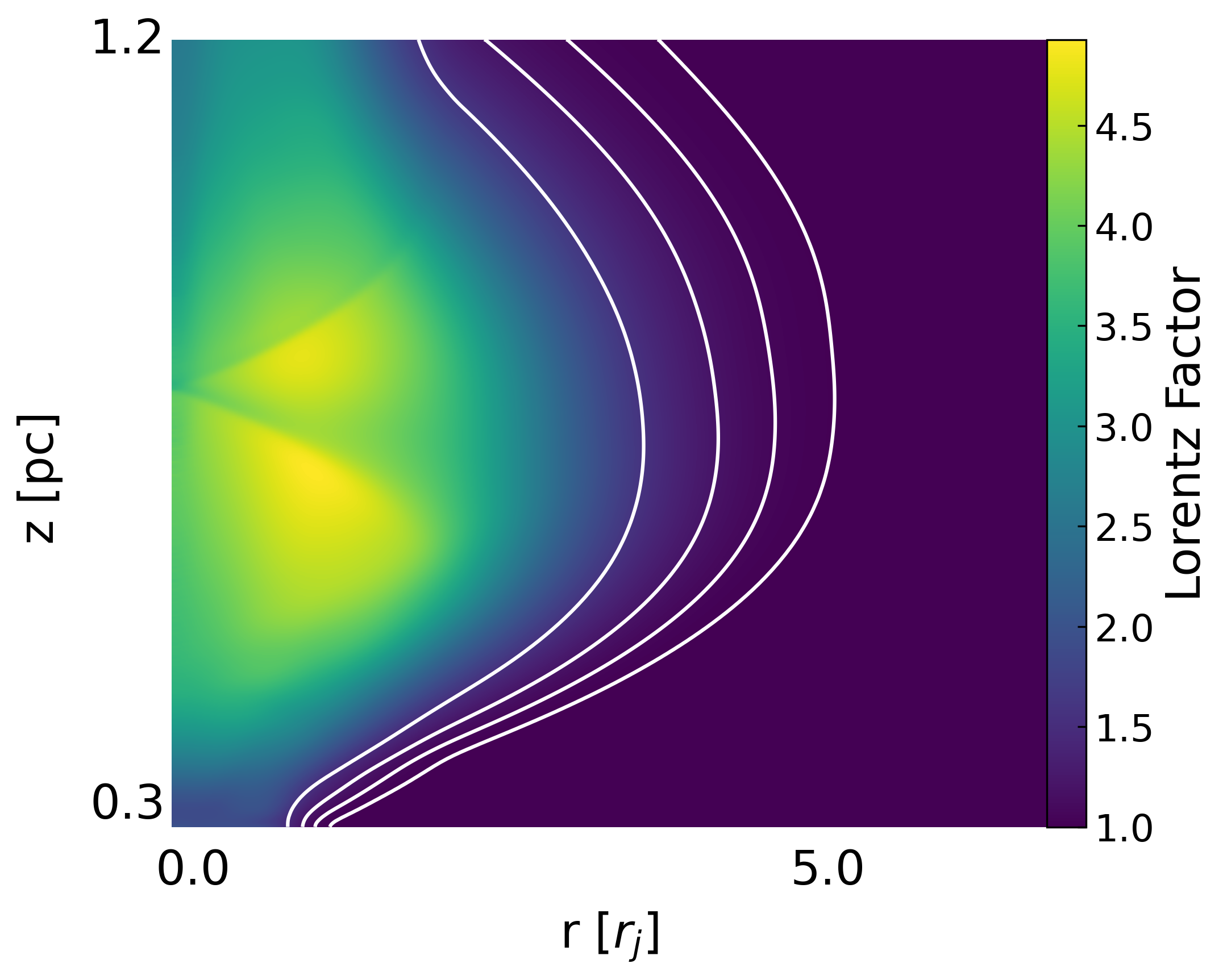}\par
    \end{multicols}
    \caption{Lorentz factor maps for FMH1\_m4 with the original grid (left plot) and the alternative grid of [$n_x$, $n_\mathrm{z}$] = [100, 30]. This comparison shows how the material lost in the right boundary does not affect the steady solutions.}
    \label{fig:pressure_wave_plots_2}
\end{figure*}

In Fig.\ \ref{fig:pressure_wave} we highlight the evolution of the pressure wave at the three different time steps of t$=20 \, r_j/c$, t$=40 \, r_j/c$, and t$=70 \, r_j/c$ for model FMH1\_m4.
As explained in Sect.\ \ref{sec:init_to_eq}, since the initial jet is overpressured with respect to the external medium, once the simulation starts the jet abruptly expands towards it, giving the material radial velocities and pushing a fraction of it outside of the right boundary.
As seen in Fig.~\ref{fig:pressure_wave_plots}, which shows the axial pressure evolution of the ambient for different time steps, the passage of the pressure wave alters the pressure profile which oscillates around the original one.
However, after a sufficient number of time steps, the lower and upper boundary conditions allow us to recover the initial, intended pressure ambient profile.
Moreover, to test whether this would lead to differences in our final, steady solutions we run model FMH1\_m4 with a larger grid of [$n_x$, $n_\mathrm{z}$] = [100, 30], in order to avoid the ambient material to leave the grid.
Fig.\ \ref{fig:pressure_wave_plots_2} shows the Lorentz factor for model FMH1\_m4 (same as Fig.\ \ref{fig:different_Lorentz_factor} but cut at 1.2 pc) and the counterpart with the larger grid on the x-axis. 
No major differences in the final profiles are seen, proving how the lost material dragged out by the pressure wave does not affect our results and thus our conclusions.

\section{Resolution study} \label{app:resolution_study}

\begin{figure*}[h]
    \centering
\begin{multicols}{2}
    \includegraphics[width=\linewidth]{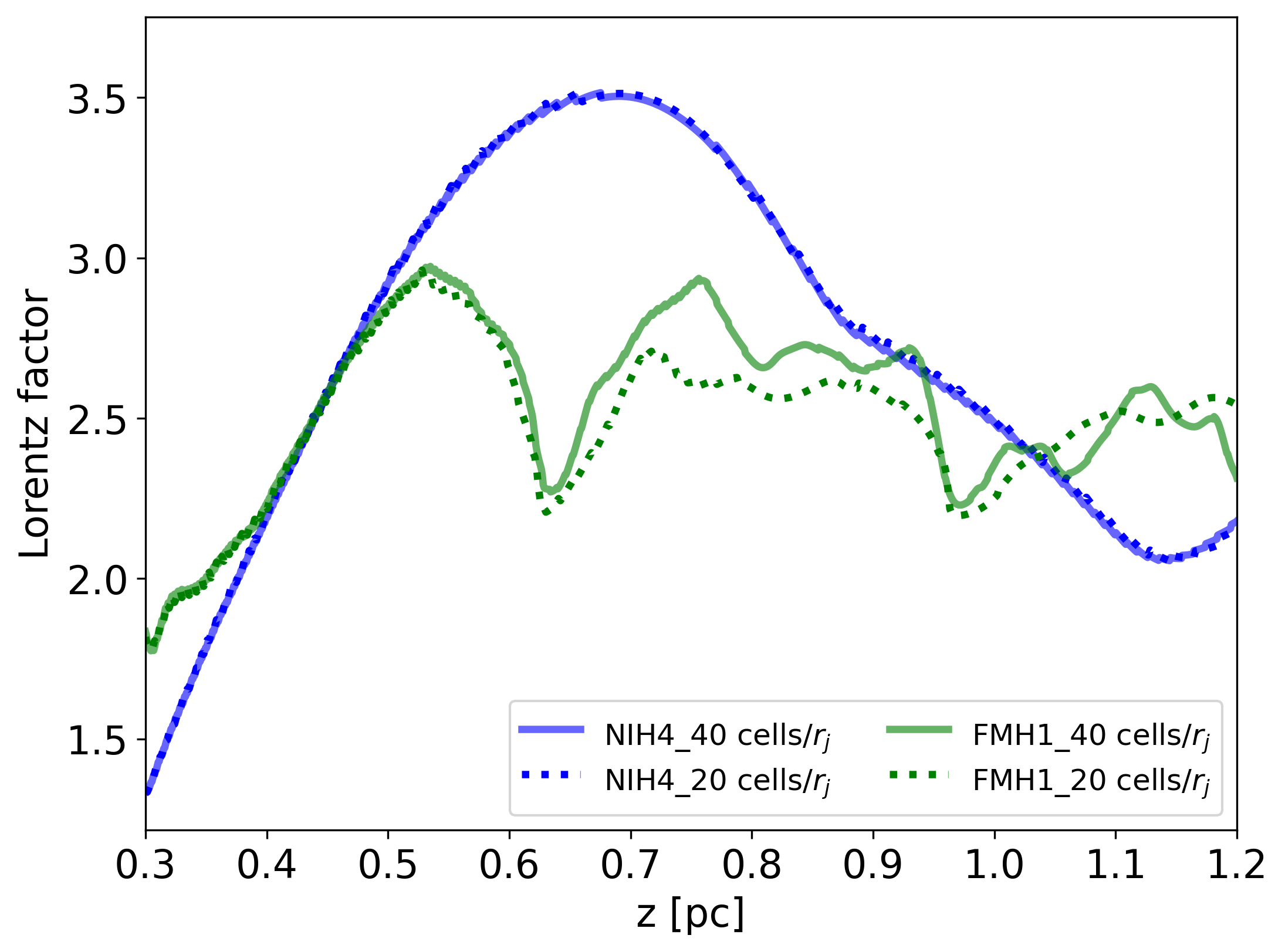}\par
    \includegraphics[width=\linewidth]{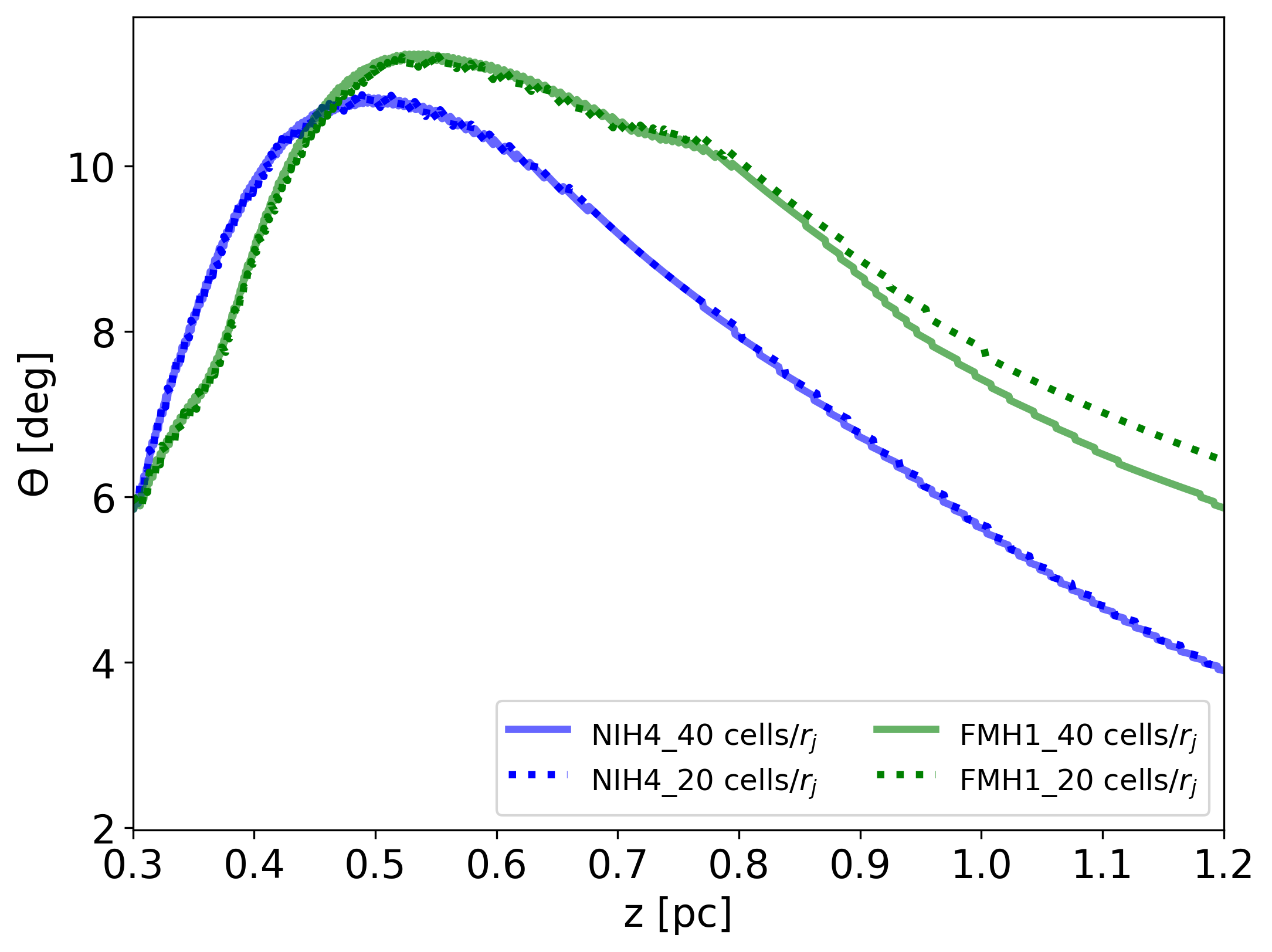}\par
\end{multicols}
    \caption{Left panel: Lorentz factor evolution up to 1.2 pc for models FMH1 and NIH4 at the resolutions of 20 and 40 cells/$r_j$. Right panel: half-opening angle obtained in the same models. Blue lines are for model NIH4 (continuous - 40 cells/$r_j$; dotted - 20 cells/$r_j$) while green lines are for model FMH1 (same line-style scheme). The comparison shows how the resolution of 20 cells/$r_j$ is enough to reach the saturation point, allowing us to use this resolution value for our simulations.}
    \label{fig:appendix_profiles}
\end{figure*}

For our simulations, we chose a resolution of 20 cells/$r_j$.
A first argument in support of this choice is given in Sect.~\ref{sec:init_to_eq} and in particular from the radial profiles shown in Fig.\ \ref{fig:radial_profiles_FMH1}, which show how the radial evolution of the different physical quantities do not differ between 20 and 40 points.
Further evidence in support of our chosen resolution is given in Fig.\ \ref{fig:appendix_profiles} which shows the evolution of both the Lorentz factor (left panel) and opening angles (right panel) for two different models (FMH1 and NIH4) at the two discussed resolutions of 20 and 40 points.
In order to save computational time, the simulations with 40 cells/$r_j$ have a smaller grid of [$n_x$, $n_\mathrm{z}$] = [30, 30], implying a physical distance of 0.3-1.2 pc. In the simulated distances, the two different resolutions lead to the same profiles in both tested models, confirming the validity of our choice of using the resolution of 20 cells/$r_j$.

\section{Further thicker shear layer models from Table \ref{tab:setups} } \label{app:other_models}

\begin{figure*}[htpb]
    \centering
\begin{multicols}{4}
    \includegraphics[width=\linewidth]{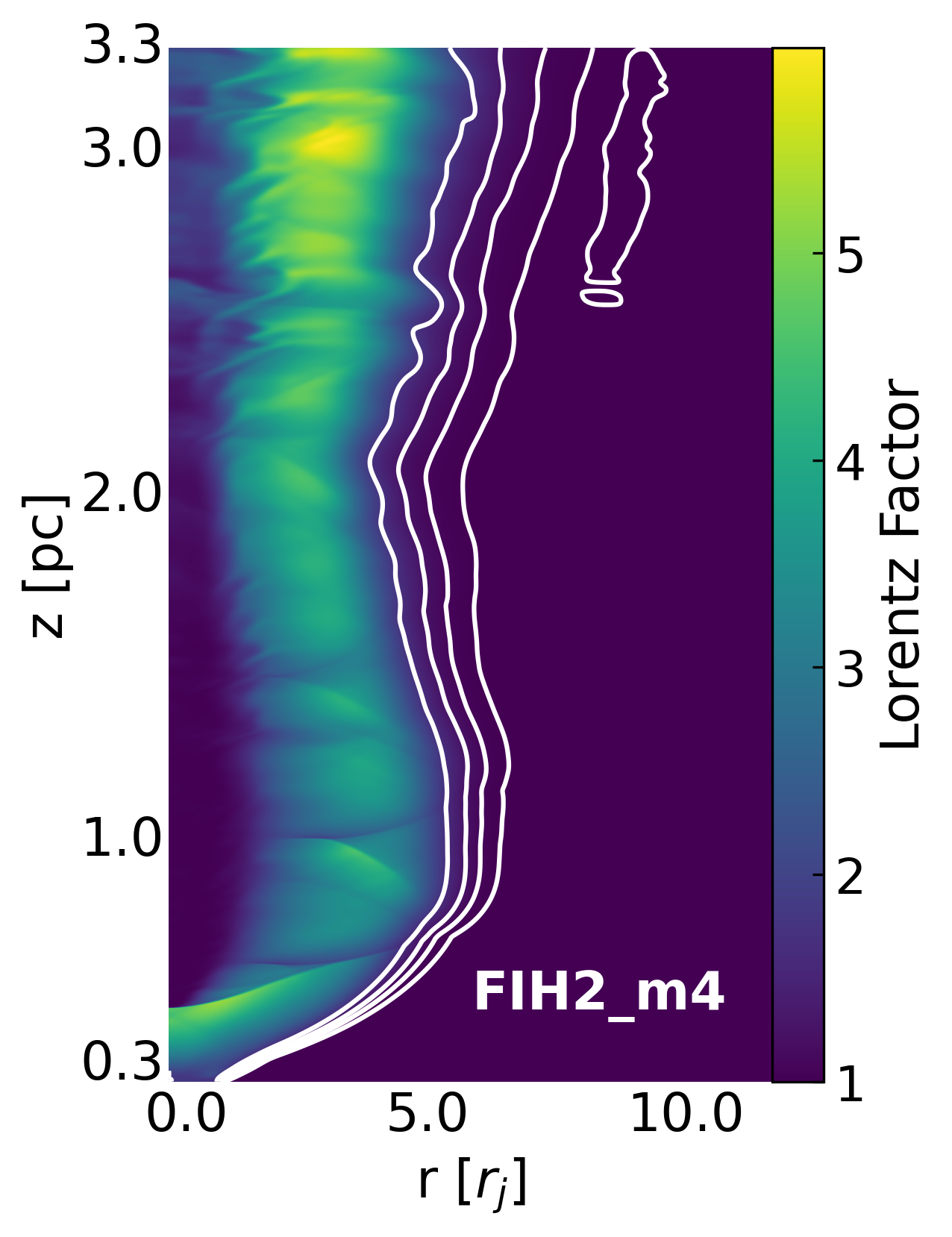}\par
    \includegraphics[width=\linewidth]{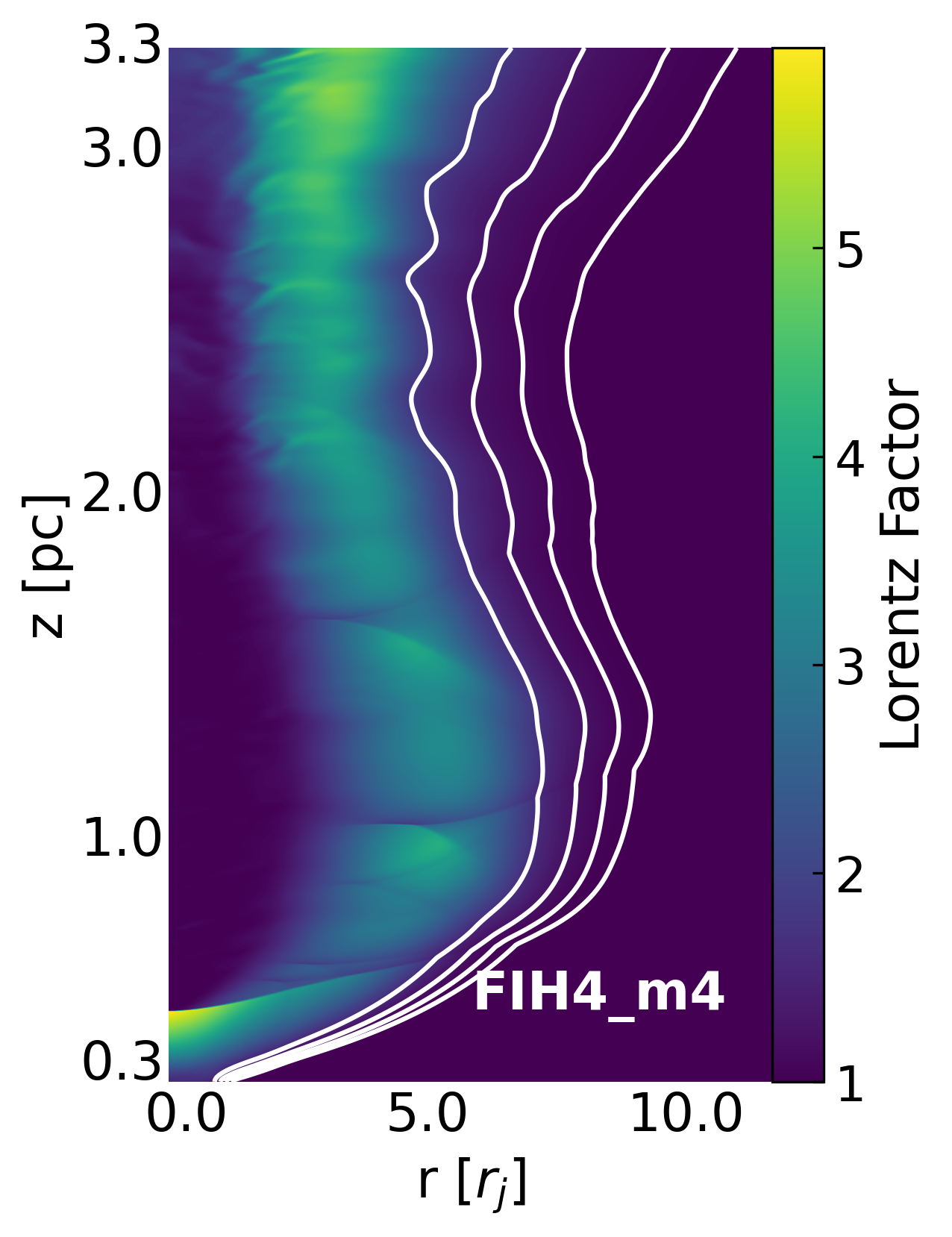}\par
    \includegraphics[width=\linewidth]{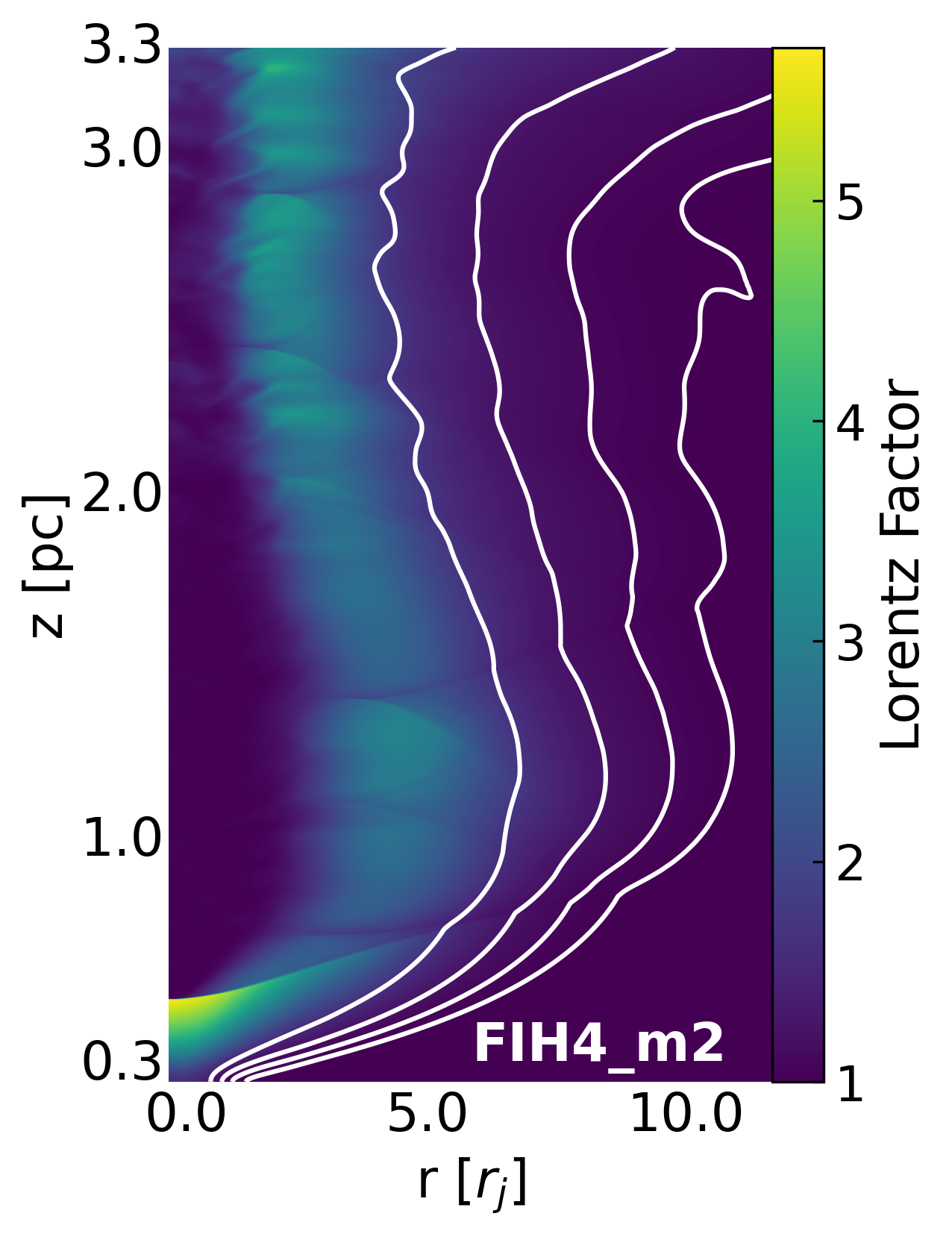}\par
    \includegraphics[width=\linewidth]{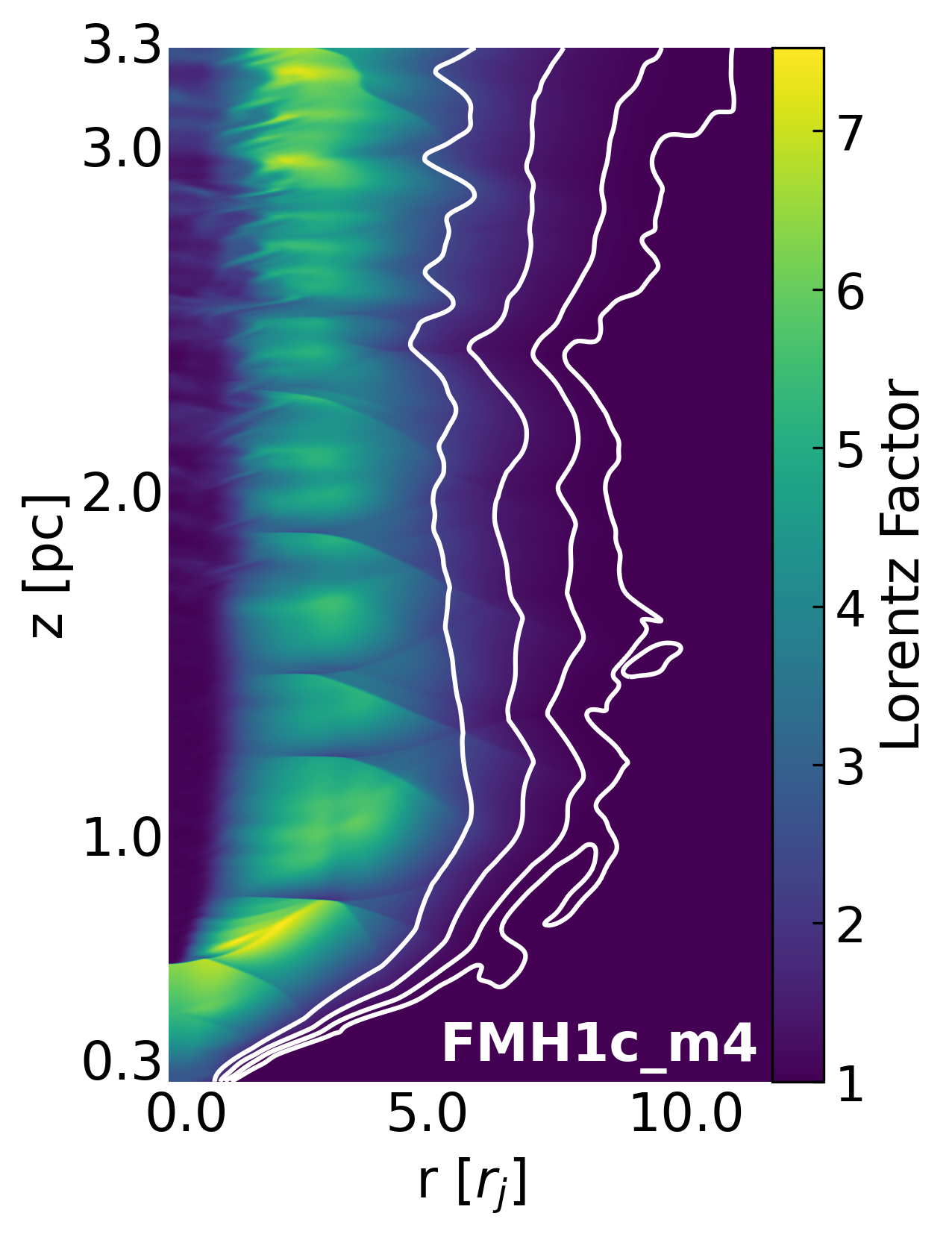}\par
\end{multicols}
\vspace{-0.8cm}
\begin{multicols}{4}
    \includegraphics[width=\linewidth]{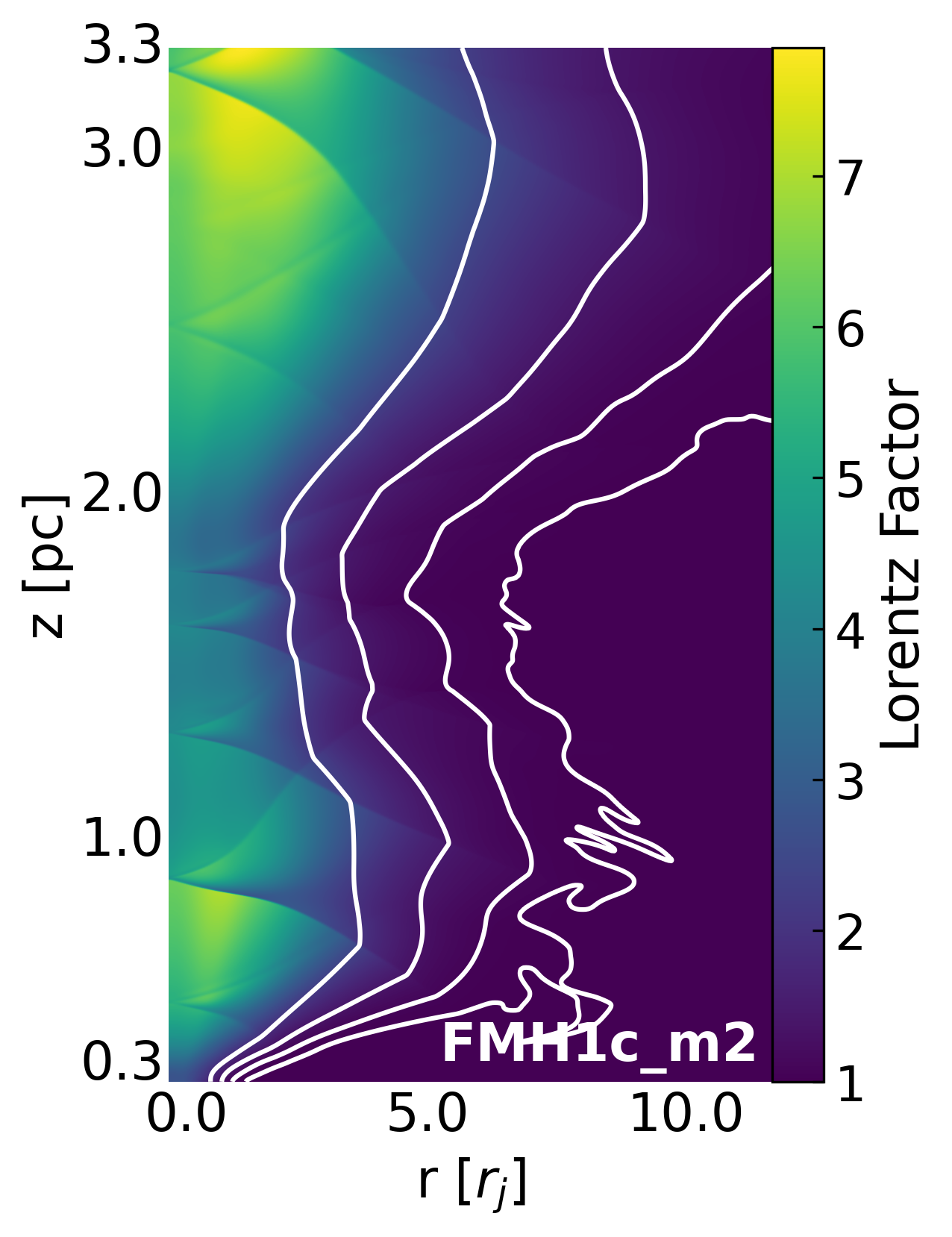}\par
    \includegraphics[width=\linewidth]{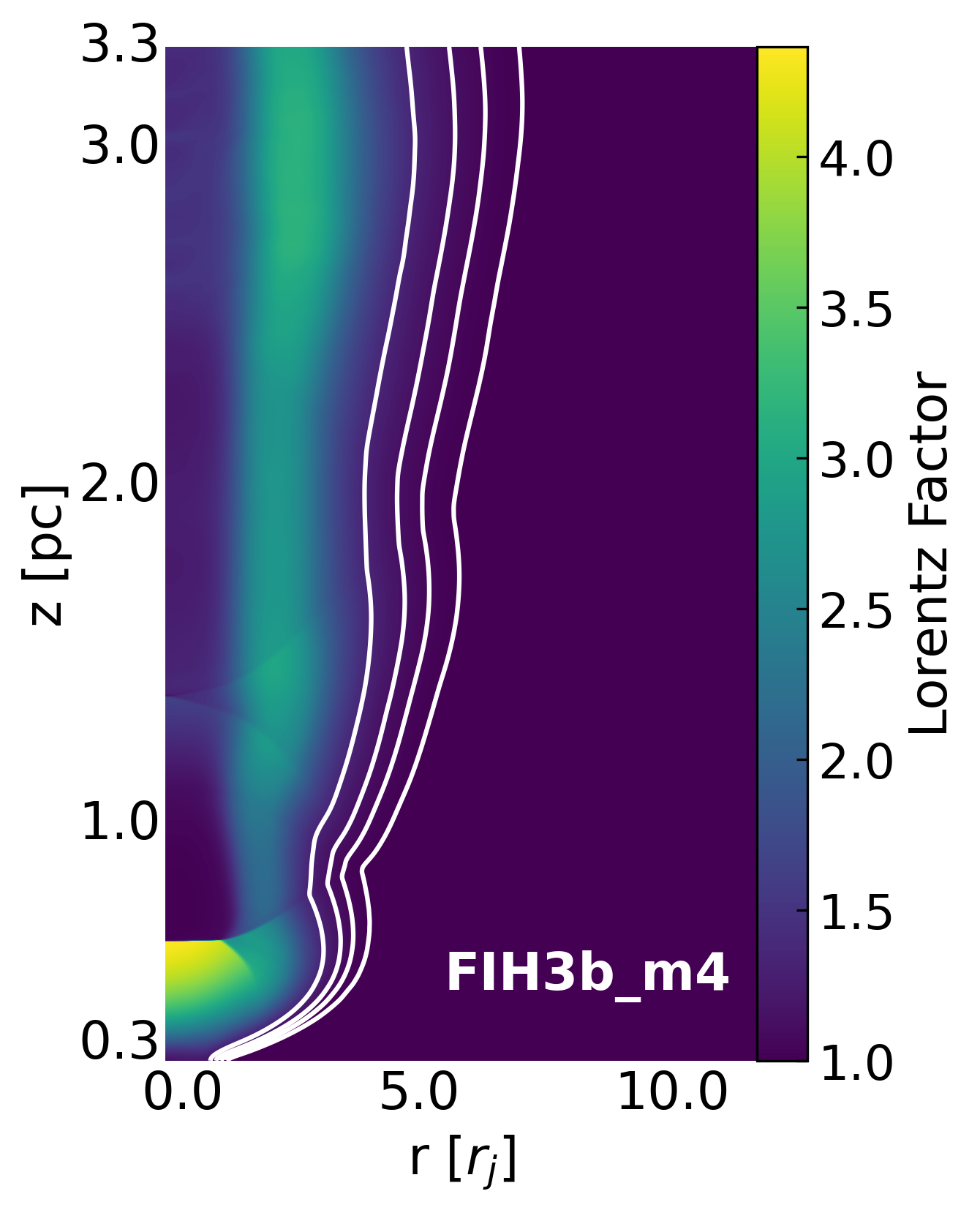}\par
    \includegraphics[width=\linewidth]{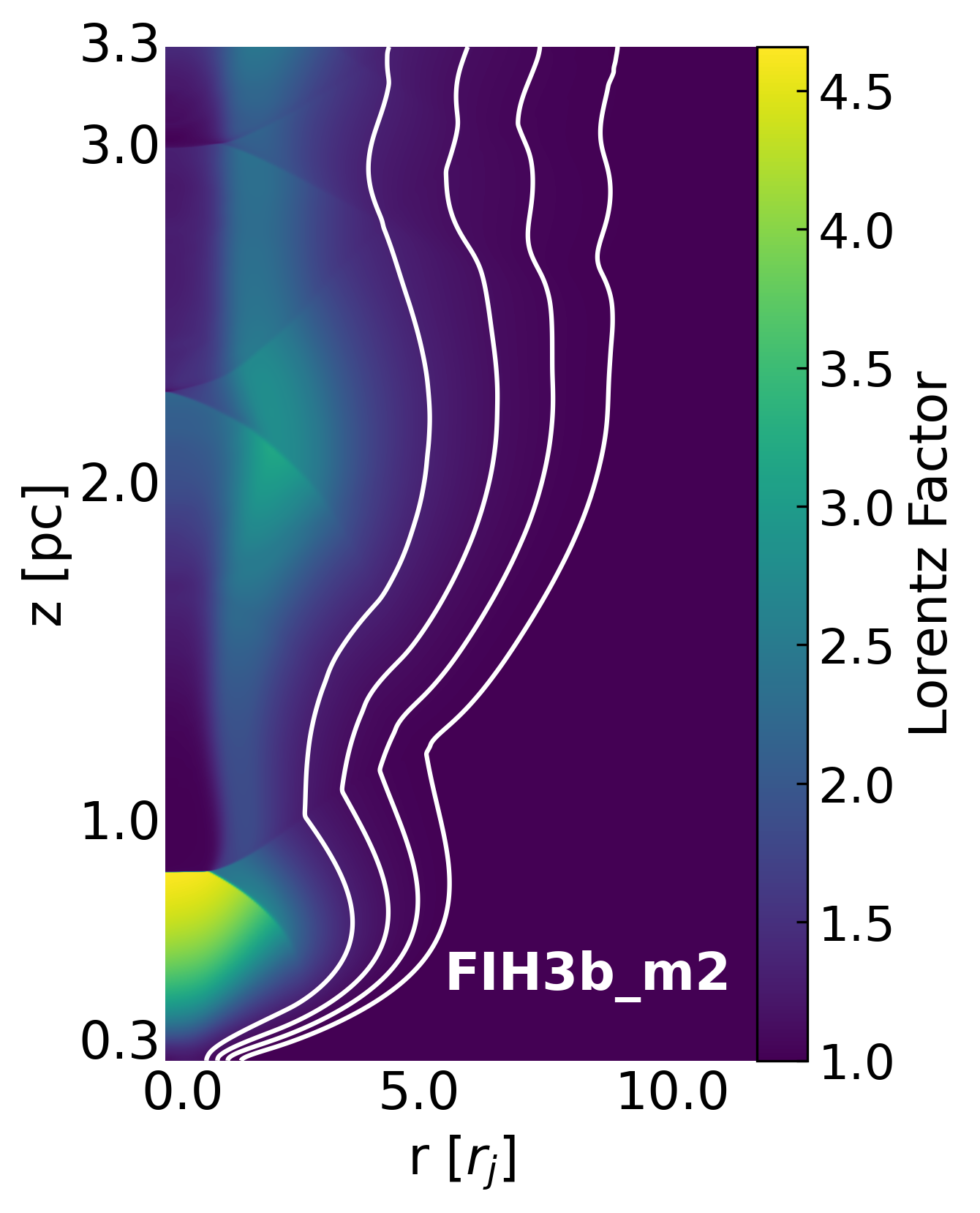}\par
\end{multicols}
    \caption{Lorentz factor maps for the other simulated models. The white contours represent the tracer at levels 0.2, 0.4, 0.6, 0.8.}
    \label{fig:others_different_Lorentz_factor}
\end{figure*}

In this Appendix, we report in Table \ref{tab:other_setups} the other simulated models that we did not include in the main text for the sake of the readability of the paper.
In detail, we report here the models for which we have tested the thicker shear layers with m = 2 and 4.
As shown in Fig.~\ref{fig:others_different_Lorentz_factor},
in all the cases here reported the increased width of the shear layer with m = 4 is not sufficient to avoid the formation of the Mach disks (see Sect.\ \ref{sec:results}). However, we highlight models FMH1c\_m4 and FMH1c\_m2.
In the former, the shear layer with m = 4 is not enough to lead to the morphology change we discussed in the main text.
Instead, the shear layer with m = 2 is thick enough to trigger the change in the acceleration pattern, from the outer layer to the internal spine.
This model supports what we proposed in Sect.~\ref{sec:shear_layer} concerning the role of the shear layer.

\begin{table*}[]
\caption{Initial conditions of the other simulated models.}
\centering
\begin{tabular}{c|c|c|c|c|c|c|c|c}

\hline
\begin{tabular}[c]{@{}c@{}}Model\\ code\end{tabular} & 
\begin{tabular}[c]{@{}c@{}}$p_a$\\ {[}code{]}\end{tabular} &
\begin{tabular}[c]{@{}c@{}}$p_j$\\ {[}$p_a${]}\end{tabular} & 
\begin{tabular}[c]{@{}c@{}}$\rho_j$\\ {[}$\rho_a${]}\end{tabular} & 
\begin{tabular}[c]{@{}c@{}}$B$\\ {[}G{]}\end{tabular} & 
\begin{tabular}[c]{@{}c@{}}$\phi_\mathrm{B}$\\ {[}deg{]}\end{tabular} &
\begin{tabular}[c]{@{}c@{}}$v_z$\\ {[}c{]}\end{tabular} &
\begin{tabular}[c]{@{}c@{}}$\mathcal{M}_\mathrm{ms}$\\ {}\end{tabular} &
\begin{tabular}[c]{@{}c@{}}$F_j$\\ {[}$\mathrm{10^{43} erg/s}${]}\end{tabular} \\
\hline

FIH2\_m4 & 0.01 & 2.0 & 0.01 & 0.46 & 68.9 & 0.84 & 1.05 & 5.6  \\ 
FIH4\_m4 & 0.01 & 4.0 & 0.01 & 0.52 & 66.4 & 0.79 & 1.00 & 6.7  \\ 
FIH4\_m2 & 0.01 & 4.0 & 0.01 & 0.52 & 66.4 & 0.79 & 1.00 & 6.7  \\ 
FMH1c\_m4 & 0.01 & 1.0 & 0.01 & 0.68 & 75.3 & 0.93 & 1.00 & 17.8 \\ 
FMH1c\_m2 & 0.01 & 1.0 & 0.01 & 0.68 & 75.3 & 0.93 & 1.00 & 17.8 \\ 
FIH3b\_m4 & 0.01 & 3.0 & 0.01 & 0.13 & 60.6 & 0.61 & 1.04 & 1.6  \\ 
FIH3b\_m2 & 0.01 & 3.0 & 0.01 & 0.13 & 60.6 & 0.61 & 1.04 & 1.6  \\ 
\hline 
\end{tabular}
\label{tab:other_setups}

\begin{flushleft} 

\textbf{Notes.} Column 1: model names; Column 2: initial ambient pressure in code units; Column 3: initial jet overpressure factor; Column 4: ratio between the initial jet density and the ambient one; Column 5: initial magnetic field strength in G; Column 6: initial magnetic pitch angle; Column 7: initial axial velocity; Column 8: average initial Magnetosonic number; Column 9: total jet flux in units of $\mathrm{10^{43} erg/s}$.

\end{flushleft}

\end{table*}

\end{appendix}

\end{document}